# Cox model inference for relative hazard and pure risk from stratified weight-calibrated case-cohort data

Short title (< 70 characters): Cox model inference from stratified weight-calibrated case-cohort data

Lola Etievant[1]*, Mitchell H. Gail[1]*

lola.etievant@nih.gov

gailm@mail.nih.gov

[1] National Cancer Institute, Division of Cancer Epidemiology and Genetics, Biostatistics Branch, 9609 Medical Center Drive, Rockville, MD 20850-9780.

* Corresponding authors

This work was supported by the Intramural Research Program of the Division of Cancer Epidemiology and Genetics, National Cancer Institute, National Institutes of Health.

**Table of contents**

**ABSTRACT**

The case-cohort design obtains complete covariate data only on cases and on a random sample (the subcohort) of the entire cohort. Subsequent publications described the use of stratification and weight calibration to increase efficiency of estimates of Cox model log relative hazards, and there has been some work estimating pure risk. Yet there are few examples of these options in the medical literature, and we could not find programs currently online to analyze these various options. We therefore present a unified approach and R software to facilitate such analyses. We used influence functions adapted to the various design and analysis options together with variance calculations that take the two-phase sampling into account. This work clarifies when the widely used "robust" variance estimate of Barlow is appropriate. The corresponding R software, CaseCohortCoxSurvival, facilitates analysis with and without stratification and/or weight calibration, for subcohort sampling with or without replacement. We also allow for phase-two data to be missing at random for stratified designs. We provide inference not only for log relative hazards in the Cox model, but also for cumulative baseline hazards and covariate-specific pure risks. We hope these calculations and software will promote wider use of more efficient and principled design and analysis options for case-cohort studies.

## 1. INTRODUCTION

Prentice[1] described the case-cohort design for time-to-response outcomes, in which one obtains covariate information on all cases (those with the event) and on a random subcohort (which may include some cases) from the entire study cohort. Two great advantages of this design are that hard-to-measure covariates need only be obtained for the cases and subcohort, which is much smaller than the entire study cohort, and the data from the subcohort can be used for several different types of time-to-response outcomes. There have been subsequent refinements and extensions of this design. Barlow[2] proposed a widely used "robust" variance estimator for log relative hazards (HR) based on the sum of squared influences. Borgan et al.[3] showed that a stratified case-cohort design had increased efficiency, and Samuelsen, Ånestad and Skrondaand[4] and Gray[5] noted that the "robust" variance estimate overestimated variances of log-relative hazard estimates with stratification when sampling without replacement. Breslow et al.[6,7] proposed survey weight calibration to improve efficiency of case-cohort estimates of relative hazard. Although much of this literature focused on estimation of log relative hazards, some authors considered estimation of cumulative baseline hazard and covariate-specific "pure" risk of an event[5,8,9, Chapters 16 and 17].

Sharp et al.[10] noted variability in the analysis and reporting of 32 case-cohort studies from 24 major medical and epidemiological journals. None of these analyses used weight calibration, some used an inappropriate "robust" variance estimate with stratified data, and various methods were used for missing covariate information. Our informal review of subsequent case-cohort publications also indicates that stratification, weight calibration, a principled approach to missing subcohort data, and analysis of pure risk are underutilized. This may be partly due to difficulty understanding the highly technical and varied methodologic literature and to lack of convenient software.

To facilitate wider use of improved design and analysis options for case-cohort data, we unify the various analytic options above by presenting empirical influence functions for log relative hazards and pure risk under a Cox proportional hazards model. These influence functions are adapted to the various design and analytic options above, and variance calculations acknowledge the phase-one sampling of the cohort from a superpopulation and the phase-two sampling of the subcohort. We develop software so that users can conveniently analyze case-cohort data with or without stratification and with or without weight calibration and can handle stratified case-cohort data with missing phase-two data.

We introduce notation in Section 2 and inference for the stratified case-cohort design in Section 3, which includes the unstratified design as a special case with one stratum. We describe weight calibration in Section 4, methods for missing phase-two data in Section 5, and software in Section 6. Sections 7 and 8 present simulations and a data illustration, where we investigate the comparative efficiencies associated with stratification and weight calibration for relative risks and covariate-specific pure risk, and how the "robust" variance estimate performs, with or without calibration. Concluding remarks are in Section 9. Most technical derivations and details are in Web Appendices.

## 2. NOTATION

We let $J$ be the number of strata in the whole cohort, $n^{(j)}$ be the number of subjects in stratum $j$, $j \in \{1, ..., J\}$. Then $n = \sum_{j=1}^{J} n^{(j)}$ is the number of subjects in the whole cohort. We further let $T_{i,j}$ be the event time (or age if the analysis is on the age scale) for subject $i$ in stratum $j$, and $C_{i,j}$ be the censoring time for subject $i$ in stratum $j$, $i \in \{1, ..., n^{(j)}\}, j \in \{1, ..., J\}$. Using the time-on-study scale, the at risk indicator for subject $i$ in stratum $j$ is $Y_{i,j}(t) = I(\tilde{T}_{i,j} \geq t)$, with $\tilde{T}_{i,j} = \min(T_{ij}, C_{ij})$. Using the age scale, $Y_{i,j}(t) = I(\tilde{T}_{i,j} \geq t > E_{i,j})$, with $E_{i,j}$ the entry age for subject $i$ in stratum $j$. Let $\tau$ the maximum follow-up time or maximum age for analyses on the age scale. With $N_{i,j}(t) = I(T_{i,j} \leq t, T_{ij} \leq C_{ij})$ indicating an observed event before or at $t$, $dN_{i,j}(t)$ indicates if individual $i$ in stratum $j$ fails (has the event) at time/age $t$. Finally, we let $\boldsymbol{X}_{i,j}$ be a vector of $p$ baseline covariates for subject $i$ in stratum $j$; $\boldsymbol{X}_{i,j}$ includes stratum indicators or stratum determinants.

We assume that failure follows the Cox proportional hazards model with hazard function $\lambda(t) = \lambda_0(t) \exp(\boldsymbol{\beta}'\boldsymbol{X})$, for covariates $\boldsymbol{X}$, and where $\lambda_0(t)$ is a baseline hazard function, i.e., the hazard for an individual with $\boldsymbol{X} = \boldsymbol{0}$. We further assume that $\lambda_0(t)$ is homogeneous across strata and we let $\Lambda_0(t) = \int_0^t \lambda_0(s) ds$ denote the cumulative baseline hazard.

Estimation from complete cohort data is reviewed in Web Appendix B.1 with corresponding influence functions in Web Appendix B.2.

## 3. STRATIFIED CASE-COHORT

### 3.1. Estimation of relative hazard, cumulative baseline hazard and pure risk

We assume that a fixed number of individuals, $m^{(j)}$, is sampled from stratum $j$ (of size $n^{(j)}$) in the cohort, without replacement and independently of case status, $j \in \{1, \ldots, J\}$. Sampling is performed independently across strata. The subcohort includes all the sampled subjects from the $J$ strata. In addition, we sample all the cases in the cohort, some of whom may have been included in the subcohort. All of these individuals constitute the stratified case-cohort, that we also call the *phase-two sample*, because it is a subset of the cohort, which is regarded as a *phase-one sample* from a super-population. We let $\xi_{i,j}$ be the sampling indicator of individual $i$ in stratum $j$ and $w_{i,j} = \begin{cases} \frac{n^{(j)}}{m^{(j)}} & \text{if } i \text{ is a non case in stratum } j \\ 1 & \text{if } i \text{ is a case in stratum } j \end{cases}$ be his/her known design weight, $i \in \{1, \ldots, n^{(j)}\}, j \in \{1, \ldots, J\}$. We assume that some of the covariates in $\boldsymbol{X}$ are only measured in the phase-two sample;

we call these "phase-two covariates". The stratum indicators are known for all members of the cohort and are not phase-two covariates. Because we sample all cases, $\xi_{i,j} w_{i,j} = 1$ for cases. Non-stratified case-cohort data correspond to the special case $J = 1$.

An estimate of the log-relative hazard $\boldsymbol{\beta}$ is obtained by solving the estimating equation

$$\boldsymbol{U}(\boldsymbol{\beta}) = \sum_{j=1}^{J} \sum_{i=1}^{n^{(j)}} \int_t \left\{ \boldsymbol{X}_{i,j} - \frac{\boldsymbol{S}_1(t;\boldsymbol{\beta})}{S_0(t;\boldsymbol{\beta})} \right\} dN_{i,j}(t) = 0, \tag{1}$$

with

$$S_0(t;\boldsymbol{\beta}) = \sum_{j=1}^{J} \sum_{k=1}^{n^{(j)}} w_{k,j} \, \xi_{k,j} \, Y_{k,j}(t) \exp(\boldsymbol{\beta}' \boldsymbol{X}_{k,j}), \tag{2}$$

$$\boldsymbol{S}_1(t;\boldsymbol{\beta}) = \sum_{j=1}^{J} \sum_{k=1}^{n^{(j)}} w_{k,j} \, \xi_{k,j} \, Y_{k,j}(t) \exp(\boldsymbol{\beta}' \boldsymbol{X}_{k,j}) \boldsymbol{X}_{k,j}, \tag{3}$$

and we also define

$$\boldsymbol{S}_2(t;\boldsymbol{\beta}) = \sum_{j=1}^{J} \sum_{k=1}^{n^{(j)}} w_{k,j} \xi_{k,j} Y_{k,j}(t) \exp(\boldsymbol{\beta}' \boldsymbol{X}_{k,j}) \boldsymbol{X}_{k,j} \boldsymbol{X}_{k,j}'. \tag{4}$$

Let $\widehat{\boldsymbol{\beta}}$ denote this solution. We then estimate the baseline hazard point mass at time $t$ non-parametrically[11] by

$$d\widehat{\Lambda}_0(t; \widehat{\boldsymbol{\beta}}) \equiv d\widehat{\Lambda}_0(t) = \frac{\sum_{j=1}^{J} \sum_{i=1}^{n^{(j)}} dN_{i,j}(t)}{S_0(t,\widehat{\boldsymbol{\beta}})}, \tag{5}$$

the cumulative baseline hazard up to time $t$ by

$$\widehat{\Lambda}_0(t; \widehat{\boldsymbol{\beta}}, \widehat{\lambda}_0) \equiv \widehat{\Lambda}_0(t) = \int_0^t d\widehat{\Lambda}_0(s), \tag{6}$$

and the pure covariate-specific risk for profile $x$ in the interval $(\tau_1, \tau_2]$ by

$$\hat{\pi}(\tau_1, \tau_2; \pmb{x}, \widehat{\pmb{\beta}}, d\widehat{\Lambda}_0) \equiv \hat{\pi}(\tau_1, \tau_2; \pmb{x}) = 1 - \exp\left\{-\int_{\tau_1}^{\tau_2} \exp(\widehat{\pmb{\beta}}'\pmb{x}) \, d\widehat{\Lambda}_0(s)\right\}. \tag{7}$$

For stratified case-cohort sampling, results in Lin[12] show that $n^{\frac{1}{2}}(\widehat{\pmb{\beta}} - \pmb{\beta})$ converges to a mean 0 Normal distribution with a variance corresponding to the sum of a normal variate from phase-one sampling and a normal variate from phase-two sampling (see also Borgan et al.[3]). Likewise, $n^{\frac{1}{2}}\{\widehat{\Lambda}_0(t) - \Lambda_0(t)\}$, for a fixed $t$, converges to a mean 0 Gaussian process representing the sum of two independent Gaussian processes. In Sections 3.3 and 4.3 we show how influence functions can be used to estimate the variances and covariances of $(\widehat{\pmb{\beta}} - \pmb{\beta})$ and $\{\widehat{\Lambda}_0(t) - \Lambda_0(t)\}$ for a fixed $t$.

### 3.2. Influence functions

We let $\pmb{\Delta}_{i,j}(\widehat{\pmb{\theta}})$ denote the influence of subject $i$ in stratum $j$ on one of the parameters $\widehat{\pmb{\theta}}$ from the set $\{\widehat{\pmb{\beta}}, d\widehat{\Lambda}_0(t), \widehat{\Lambda}_0(t), \hat{\pi}(\tau_1, \tau_2; \pmb{x})\}$, $i \in \{1, \ldots, n^{(j)}\}, j \in \{1, \ldots, J\}$. From these influences, we estimate the covariance matrix of the parameters by linearization[13] as

$$\text{var}(\widehat{\pmb{\theta}}) \approx \text{var}\left\{\sum_{j=1}^{J} \sum_{i=1}^{n^{(j)}} \pmb{\Delta}_{i,j}(\widehat{\pmb{\theta}})\right\}. \tag{8}$$

Following Graubard and Fears[14] and Pfeiffer and Gail[15] Section 4.6, we show in Web Appendix C.1 that an estimate of $\pmb{\Delta}_{i,j}(\widehat{\pmb{\theta}})$ is $\xi_{i,j} w_{i,j} \pmb{IF}_{i,j}^{(2)}(\widehat{\pmb{\theta}})$, where

$$\pmb{IF}_{i,j}^{(2)}(\widehat{\pmb{\beta}}) = \left[\sum_{l=1}^{J} \sum_{k=1}^{n^{(l)}} \int_t \left\{\frac{S_2(t;\widehat{\pmb{\beta}})}{S_0(t;\widehat{\pmb{\beta}})} - \frac{S_1(t;\widehat{\pmb{\beta}}) S_1(t;\widehat{\pmb{\beta}})'}{S_0(t;\widehat{\pmb{\beta}})^2}\right\} dN_{k,l}(t)\right]^{-1} \left[\int_t \left\{\pmb{X}_{i,j} - \frac{S_1(t;\widehat{\pmb{\beta}})}{S_0(t;\widehat{\pmb{\beta}})}\right\} \left\{dN_{i,j}(t) - \frac{Y_{i,j}(t) \exp(\widehat{\pmb{\beta}}' \pmb{X}_{i,j}) \sum_{l=1}^{J} \sum_{k=1}^{n^{(l)}} dN_{k,l}(t)}{S_0(t;\widehat{\pmb{\beta}})}\right\}\right], \tag{9}$$

$$IF_{i,j}^{(2)}\{d\widehat{\Lambda}_0(t)\} = \{S_0(t;\widehat{\boldsymbol{\beta}})\}^{-1}\{dN_{i,j}(t) - d\widehat{\Lambda}_0(t)\boldsymbol{S}_1(t;\widehat{\boldsymbol{\beta}})'\boldsymbol{IF}_{i,j}^{(2)}(\widehat{\boldsymbol{\beta}}) - \qquad(10)$$

$$d\widehat{\Lambda}_0(t)Y_{i,j}(t)\exp(\widehat{\boldsymbol{\beta}}'\boldsymbol{X}_{i,j})\},$$

$$IF_{i,j}^{(2)}\left\{\int_{\tau_1}^{\tau_2} d\widehat{\Lambda}_0(t)\right\} = \int_{\tau_1}^{\tau_2} IF_{i,j}^{(2)}\{d\widehat{\Lambda}_0(t)\}, \qquad(11)$$

and

$$IF_{i,j}^{(2)}\{\hat{\pi}(\tau_1,\tau_2;\boldsymbol{x})\} = \left\{\frac{\partial \hat{\pi}(\tau_1,\tau_2;\boldsymbol{x})}{\partial \boldsymbol{\beta}}\bigg|_{\boldsymbol{\beta}=\widehat{\boldsymbol{\beta}}}\right\}' \boldsymbol{IF}_{i,j}^{(2)}(\widehat{\boldsymbol{\beta}}) +$$

$$\left[\frac{\partial \hat{\pi}(\tau_1,\tau_2;\boldsymbol{x})}{\partial\{\int_{\tau_1}^{\tau_2} d\Lambda_0(t)\}}\bigg|_{d\Lambda_0(t)=d\widehat{\Lambda}_0(t)}\right]' IF_{i,j}^{(2)}\left\{\int_{\tau_1}^{\tau_2} d\widehat{\Lambda}_0(t)\right\}. \qquad(12)$$

Equations (9)-(12) depend on "phase-two covariates". Hence, we use the superscript 2 in $\boldsymbol{IF}_{i,j}^{(2)}(\widehat{\boldsymbol{\theta}})$. Equations (9)-(12) are estimates of the theoretical influences obtained by replacing $\widehat{\boldsymbol{\beta}}$ and $\widehat{\Lambda}_0$ by $\boldsymbol{\beta}$ and $\Lambda_0$. Equation (8) pertains to the theoretical influences, which are of order $O_P\left(n^{-\frac{1}{2}}\right)$, but because the difference between the estimated and theoretical influence is of order $O_P(n^{-1})$ the estimated influences can be used in equation (8)[13,16,17]. Hereafter, we let $\boldsymbol{\Delta}_{i,j}(\widehat{\boldsymbol{\theta}})$ denote the estimated influence.

### 3.3. Variance decomposition and estimation from influence functions

The variance $var(\widehat{\boldsymbol{\theta}}) \approx var\left\{\sum_{j=1}^{J}\sum_{i=1}^{n^{(j)}} \boldsymbol{\Delta}_{i,j}(\widehat{\boldsymbol{\theta}})\right\}$ can be decomposed as

$$var\left[E\left\{\sum_{j=1}^{J}\sum_{i=1}^{n^{(j)}} \boldsymbol{\Delta}_{i,j}(\widehat{\boldsymbol{\theta}})|\boldsymbol{C}_1\right\}\right] + E\left[var\left\{\sum_{j=1}^{J}\sum_{i=1}^{n^{(j)}} \boldsymbol{\Delta}_{i,j}(\widehat{\boldsymbol{\theta}})|\boldsymbol{C}_1\right\}\right], \qquad(13)$$

where $\boldsymbol{C}_1$ denotes the information from the whole cohort. The first component accounts for sampling the cohort from the "superpopulation" (phase-one component of variance), whereas the second component accounts for sampling the subcohort from the cohort (phase-two component of variance).

We let $w_{i,k,j}$ and $\sigma_{i,k,j}$ denote $E(\xi_{i,j}\xi_{k,j}|C_1)^{-1}$ and $cov(\xi_{i,j},\xi_{k,j}|C_1)$, respectively, $i,k \in \{1,\ldots,n^{(j)}\}, j \in \{1,\ldots,J\}$; they are be specified below. We know $w_{i,j}\mathbf{IF}^{(2)}_{i,j}(\widehat{\boldsymbol{\theta}})$ is fixed conditional on $C_1$ and $E(\xi_{i,j}w_{i,j}|C_1) = 1$. Thus $\text{var}\left\{\sum_{j=1}^{J}\sum_{i=1}^{n^{(j)}}\boldsymbol{\Delta}_{i,j}(\widehat{\boldsymbol{\theta}})\right\} = \text{var}\left\{\sum_{j=1}^{J}\sum_{i=1}^{n^{(j)}}\mathbf{IF}^{(2)}_{i,j}(\widehat{\boldsymbol{\theta}})\right\} + E\left\{\sum_{j=1}^{J}\sum_{i=1}^{n^{(j)}}\sum_{k=1}^{n^{(j)}}\sigma_{i,k,j}\,w_{i,j}w_{k,j}\mathbf{IF}^{(2)}_{i,j}(\widehat{\boldsymbol{\theta}})\mathbf{IF}^{(2)}_{k,j}(\widehat{\boldsymbol{\theta}})'\right\}$.

Because $\mathbf{IF}^{(2)}_{i,j}(\widehat{\boldsymbol{\theta}})\mathbf{IF}^{(2)}_{i,j}(\widehat{\boldsymbol{\theta}})'$ and $\mathbf{IF}^{(2)}_{i,j}(\widehat{\boldsymbol{\theta}})\mathbf{IF}^{(2)}_{k,j}(\widehat{\boldsymbol{\theta}})'$ can only be computed if individuals $i$ and $k$ in stratum $j$ are in the phase-two sample, we weight the contributions from the individuals in the phase-two sample by the "marginal" and "joint" design weights, $w_{i,j}$ and $w_{i,k,j}$, to estimate $\text{var}(\widehat{\boldsymbol{\theta}})$ by

$$\frac{n}{n-1}\sum_{j=1}^{J}\sum_{i=1}^{n^{(j)}}\xi_{i,j}\,w_{i,j}\,\mathbf{IF}^{(2)}_{i,j}(\widehat{\boldsymbol{\theta}})\mathbf{IF}^{(2)}_{i,j}(\widehat{\boldsymbol{\theta}})' + \tag{14}$$
$$\sum_{j=1}^{J}\sum_{i=1}^{n^{(j)}}\sum_{k=1}^{n^{(j)}}w_{i,k,j}\,\sigma_{i,k,j}\,w_{i,j}\,w_{k,j}\,\xi_{i,j}\,\xi_{k,j}\mathbf{IF}^{(2)}_{i,j}(\widehat{\boldsymbol{\theta}})\mathbf{IF}^{(2)}_{k,j}(\widehat{\boldsymbol{\theta}})'.$$

See Web Appendix C.2 for comparison with the very similar estimate of $\text{var}(\widehat{\boldsymbol{\beta}})$ by Samuelsen et al.[4].

Following Barlow[2], the "robust" variance estimate would be

$$\sum_{j=1}^{J}\sum_{i=1}^{n^{(j)}}\boldsymbol{\Delta}_{i,j}(\widehat{\boldsymbol{\theta}})\boldsymbol{\Delta}_{i,j}(\widehat{\boldsymbol{\theta}})' = \sum_{j=1}^{J}\sum_{i=1}^{n^{(j)}}\xi_{i,j}\,w_{i,j}\,w_{i,j}\,\mathbf{IF}^{(2)}_{i,j}(\widehat{\boldsymbol{\theta}})\mathbf{IF}^{(2)}_{i,j}(\widehat{\boldsymbol{\theta}})'. \tag{15}$$

With stratified data, Equation (15) is often too large (see also Section 7 and Web Tables 3-20 in Web Appendix E). Equation (15) minus Equation (14) is

$$\frac{1}{n-1}\sum_{j=1}^{J}\sum_{i=1}^{n^{(j)}} \xi_{i,j} w_{i,j} \boldsymbol{IF}_{i,j}^{(2)}(\widehat{\boldsymbol{\theta}}) \boldsymbol{IF}_{i,j}^{(2)}(\widehat{\boldsymbol{\theta}})' + \qquad (16)$$

$$\sum_{j=1}^{J}\sum_{i=1}^{n^{(j)}} \sum_{\substack{k=1,\\k\neq i}}^{n^{(j)}} w_{i,k,j} \sigma_{i,k,j}\, w_{i,j} w_{k,j} \xi_{i,j}\, \xi_{k,j} \boldsymbol{IF}_{i,j}^{(2)}(\widehat{\boldsymbol{\theta}}) \boldsymbol{IF}_{k,j}^{(2)}(\widehat{\boldsymbol{\theta}})'.$$

Because we sample *without replacement* in each stratum, we have $w_{i,k,j} = \frac{n^{(j)}(n^{(j)}-1)}{m^{(j)}(m^{(j)}-1)}$ if individuals $i$ and $k$ in stratum $j$ are both non-cases, and $w_{i,k,j} = w_{i,j} \times w_{k,j}$ otherwise, $i,k \in \{1,\ldots,n^{(j)}\}, k \neq i, j \in \{1,\ldots,J\}$. Recall that $w_{i,i,j} = w_{i,j} = \frac{n^{(j)}}{m^{(j)}}$ if individual $i$ in stratum $j$ is a non-case, and $w_{i,j} = 1$ if individual $i$ in stratum $j$ is a case. Then $\sigma_{i,k,j} = \frac{m^{(j)}}{n^{(j)}} \frac{m^{(j)}-1}{n^{(j)}-1} - \left(\frac{m^{(j)}}{n^{(j)}}\right)^2$ if individuals $i$ and $k$ in stratum $j$ are both non-cases, and $\sigma_{i,k,j} = 0$ otherwise, $i,k \in \{1,\ldots,n^{(j)}\}, k \neq i, j \in \{1,\ldots,J\}$. Similarly, if individual $i$ in stratum $j$ is a non-case, then $\sigma_{i,i,j} \equiv \sigma_{i,j} = \frac{m^{(j)}}{n^{(j)}}\left(1 - \frac{m^{(j)}}{n^{(j)}}\right)$, and $\sigma_{i,j} = 0$ otherwise. As a result, only the sampled non-cases contribute to the phase-two component of the variance in Equation (14). For sampling *with replacement* (i.e., Bernoulli sampling), individuals are sampled independently of each other. Then $w_{i,k,j} = w_{i,j} \times w_{k,j}$, and $\sigma_{i,k,j} = 0$ for any pair $(i,k)$ of distinct individuals in stratum $j$, $i,k \in \{1,\ldots,n^{(j)}\}, k \neq i,\ j \in \{1,\ldots,J\}$. In that case, the difference between the "robust" variance estimate in Equation (15)) and Equation (14)) reduces to $\frac{1}{n-1}\sum_{j=1}^{J}\sum_{i=1}^{n^{(j)}} \boldsymbol{\Delta}_{i,j}(\widehat{\boldsymbol{\theta}}) \boldsymbol{\Delta}_{i,j}(\widehat{\boldsymbol{\theta}})'$, which is negligible compared to Equation (14)) in large cohorts.

## 4. CALIBRATION OF THE DESIGN WEIGHTS

### 4.1. Calibration and choice of auxiliary variables

Breslow et al.[6,7] advocated "weight calibration" to improve the efficiency of case-cohort studies. First, one identifies auxiliary variables that are highly correlated with the influences on $\widehat{\boldsymbol{\theta}}$ and are known for the entire cohort. Then one perturbs the design weights to obtain calibrated weights that are close to the design weights but for which the observed sums of auxiliary variables in the phase-one sample equals the weighted sums in the phase-two sample with the calibrated weights. To obtain auxiliary variables, we follow Breslow et al.[6] and Shin et al.[18]. First, we use weighted regression in the phase-two sample to estimate the expected value of phase-two covariates given phase-one data. These expectations are used to impute the phase-two covariates for all members of the cohort, including those with measured phase-two covariates. The phase-one data used for imputation may consist of covariates in $\boldsymbol{X}$ and of phase-one proxies of the phase-two covariates that are measured on all cohort members. The auxiliary variables are (*i*) the influences for the log-relative hazard parameters estimated from the Cox model with imputed cohort data; and (*ii*) the products of total follow-up time (on the time interval for which pure risk is to be estimated) with the estimated relative hazard for the imputed cohort data, where the log-relative hazard parameters are estimated from the Cox model with case-cohort data and weights calibrated with (*i*). To standardize the weights, we also calibrate on a variable that is identically equal to 1. Additional details are in Web Appendix D.1; see also Breslow et al.[6] Shin et al.[18]. Other auxiliary variables have been proposed for $\widehat{\Lambda}_0(t)$[8], but in unreported simulations, the proposal by Shin et al. performed better; see also Web Appendix D.3.

We let $A_{i,j}$ be the vector of $q$ auxiliary variables for individual $i$ in stratum $j$, with calibrated weights $w^*_{i,j} = w_{i,j} \exp(\widehat{\boldsymbol{\eta}}' A_{i,j})$, $i \in \{1, \ldots, n^{(j)}\}$, $j \in \{1, \ldots, J\}$, that are obtained by solving $\sum_{j=1}^{J} \sum_{i=1}^{n^{(j)}} \{\xi_{i,j} w_{i,j} \exp(\boldsymbol{\eta}' A_{i,j}) A_{i,j} - A_{i,j}\} = 0$ for $\widehat{\boldsymbol{\eta}}$. See Web Appendix D.1.

### 4.2. Estimation of relative hazard, cumulative baseline hazard and pure risk using calibrated weights

An estimate of $\boldsymbol{\beta}$ solves $U^*(\boldsymbol{\beta}) = \sum_{j=1}^{J} \sum_{i=1}^{n^{(j)}} \int_t \xi_{i,j} w^*_{i,j} \left\{ X_{i,j} - \frac{S^*_1(t;\widehat{\boldsymbol{\eta}},\boldsymbol{\beta})}{S^*_0(t;\widehat{\boldsymbol{\eta}},\boldsymbol{\beta})} \right\} dN_{i,j}(t) = 0$, where $S^*_0(t;\widehat{\boldsymbol{\eta}}, \boldsymbol{\beta})$, $S^*_1(t;\widehat{\boldsymbol{\eta}}, \boldsymbol{\beta})$ and $S^*_2(t;\widehat{\boldsymbol{\eta}}, \boldsymbol{\beta})$ are obtained from Equations (2)-(4) with $w^*_{k,j}$ replacing $w_{k,j}$. Letting $\widehat{\boldsymbol{\beta}}^*(\widehat{\boldsymbol{\eta}}) \equiv \widehat{\boldsymbol{\beta}}^*$, we estimate the baseline hazard point mass at time $t$, $d\widehat{\Lambda}^*_0(t;\widehat{\boldsymbol{\eta}}, \widehat{\boldsymbol{\beta}}^*) \equiv d\widehat{\Lambda}^*_0(t)$, the cumulative baseline hazard up to time $t$, $\widehat{\Lambda}^*_0(t;\widehat{\boldsymbol{\eta}}, \widehat{\boldsymbol{\beta}}^*) \equiv \widehat{\Lambda}^*_0(t)$, and the pure risk for profile $x$ in the interval $(\tau_1, \tau_2]$, $\hat{\pi}^*(\tau_1, \tau_2; x, \widehat{\boldsymbol{\eta}}, \widehat{\boldsymbol{\beta}}^*, d\widehat{\Lambda}^*_0) \equiv \hat{\pi}^*(\tau_1, \tau_2; x)$, from Equations (5)-(7) with $S^*_0(t;\widehat{\boldsymbol{\eta}}, \boldsymbol{\beta})$ and $\widehat{\boldsymbol{\beta}}^*$ replacing $S_0(t, \widehat{\boldsymbol{\beta}})$ and $\widehat{\boldsymbol{\beta}}$. We do not calibrate the case weights in the numerator of the Breslow estimator because the event times are known for all cohort members[18–20].

### 4.3. Variance estimation from influence functions

We let $\Delta_{i,j}(\widehat{\boldsymbol{\theta}}^*)$ denote the influence of individual $i$ in stratum $j$ on one of the parameters $\widehat{\boldsymbol{\theta}}^*$ from the set $\{\widehat{\boldsymbol{\eta}}, \widehat{\boldsymbol{\beta}}^*, d\widehat{\Lambda}^*_0(t), \widehat{\Lambda}^*_0(t), \hat{\pi}^*(\tau_1, \tau_2; x)\}$, $i \in \{1, \ldots, n^{(j)}\}, j \in \{1, \ldots, J\}$, and use $\text{var}(\widehat{\boldsymbol{\theta}}^*) \approx \text{var}\left\{\sum_{j=1}^{J} \sum_{i=1}^{n^{(j)}} \Delta_{i,j}(\widehat{\boldsymbol{\theta}}^*)\right\}$. Following Shin et al., we can show that $\Delta_{i,j}(\widehat{\boldsymbol{\theta}}^*) = IF^{(1)}_{i,j}(\widehat{\boldsymbol{\theta}}^*) + \xi_{i,j} w_{i,j} IF^{(2)}_{i,j}(\widehat{\boldsymbol{\theta}}^*)$. The superscript 1 indicates that $IF^{(1)}_{i,j}(\widehat{\boldsymbol{\theta}}^*)$ depends only on variables measured on all cohort members. If individual $i$ in stratum $j$ is not in the phase-two sample, $\xi_{i,j} w_{i,j} IF^{(2)}_{i,j}(\widehat{\boldsymbol{\theta}}^*)$ is zero, but such an individual has an influence on $\widehat{\boldsymbol{\eta}}$ and hence $\widehat{\boldsymbol{\theta}}^*$ through $IF_{i,j}^{(1)}(\widehat{\theta}^*)$. Explicit forms of $IF_{i,j}^{(s)}(\widehat{\theta}^*)$, $s \in \{1,2\}$, are in APPENDIX A, Web Appendix A.2 and derived in Web Appendix D.2.

Because $IF_{i,j}^{(1)}(\widehat{\theta})$ is fixed conditional on $C_1$, a decomposition similar to Equation (13) yields

$$\text{var}\left\{\sum_{j=1}^{J}\sum_{i=1}^{n^{(j)}}\Delta_{i,j}(\widehat{\theta}^*)\right\} = \text{var}\left\{\sum_{j=1}^{J}\sum_{i=1}^{n^{(j)}}IF_{i,j}^{(1)}(\widehat{\theta}^*) + IF_{i,j}^{(2)}(\widehat{\theta}^*)\right\} + $$
$$\text{E}\left\{\sum_{j=1}^{J}\sum_{i=1}^{n^{(j)}}\sum_{k=1}^{n^{(j)}}\sigma_{i,k,j}\,w_{i,j}\,w_{k,j}IF_{i,j}^{(2)}(\widehat{\theta}^*)IF_{k,j}^{(2)}(\widehat{\theta}^*)'\right\}, \quad (17)$$

which can be estimated by

$$\frac{n}{n-1}\sum_{j=1}^{J}\sum_{i=1}^{n^{(j)}}\left\{IF_{i,j}^{(1)}(\widehat{\theta}^*)IF_{i,j}^{(1)}(\widehat{\theta}^*)' + 2\,\xi_{i,j}\,w_{i,j}\,IF_{i,j}^{(1)}(\widehat{\theta}^*)IF_{i,j}^{(2)}(\widehat{\theta}^*)' + \right.$$
$$\left.\xi_{i,j}\,w_{i,j}\,IF_{i,j}^{(2)}(\widehat{\theta}^*)IF_{i,j}^{(2)}(\widehat{\theta}^*)'\right\} + \quad (18)$$
$$\sum_{j=1}^{J}\sum_{i=1}^{n^{(j)}}\sum_{k=1}^{n^{(j)}}w_{i,k,j}\,\sigma_{i,k,j}\,\xi_{i,j}\,\xi_{k,j}\,w_{i,j}\,w_{k,j}\,IF_{i,j}^{(2)}(\widehat{\theta}^*)IF_{k,j}^{(2)}(\widehat{\theta}^*)'.$$

Finally, the robust variance estimate[2] is

$$\sum_{j=1}^{J}\sum_{i=1}^{n^{(j)}}\Delta_{i,j}(\widehat{\theta}^*)\Delta_{i,j}(\widehat{\theta}^*)' = \sum_{j=1}^{J}\sum_{i=1}^{n^{(j)}}\left\{IF_{i,j}^{(1)}(\widehat{\theta}^*)IF_{i,j}^{(1)}(\widehat{\theta}^*)' + \right.$$
$$\left.2\,\xi_{i,j}\,w_{i,j}\,IF_{i,j}^{(1)}(\widehat{\theta}^*)IF_{i,j}^{(2)}(\widehat{\theta}^*)' + \xi_{i,j}\,w_{i,j}\,w_{i,j}\,IF_{i,j}^{(2)}(\widehat{\theta}^*)IF_{i,j}^{(2)}(\widehat{\theta}^*)'\right\}. \quad (19)$$

and the difference between Equations (18) and (19) is

$$\frac{1}{n-1}\sum_{j=1}^{J}\sum_{i=1}^{n^{(j)}}\left\{IF_{i,j}^{(1)}(\widehat{\theta}^*)IF_{i,j}^{(1)}(\widehat{\theta}^*)' + 2\,\xi_{i,j}\,w_{i,j}\,IF_{i,j}^{(1)}(\widehat{\theta}^*)IF_{i,j}^{(2)}(\widehat{\theta}^*)' + \right.$$
$$\left.\xi_{i,j}\,w_{i,j}\,IF_{i,j}^{(2)}(\widehat{\theta}^*)IF_{i,j}^{(2)}(\widehat{\theta}^*)'\right\} + \quad (20)$$
$$\sum_{j=1}^{J}\sum_{i=1}^{n^{(j)}}\sum_{\substack{k=1,\\k\neq i}}^{n^{(j)}}w_{i,k,j}\,\sigma_{i,k,j}\,w_{i,j}\,w_{k,j}\,\xi_{i,j}\,\xi_{k,j}\,IF_{i,j}^{(2)}(\widehat{\theta}^*)IF_{k,j}^{(2)}(\widehat{\theta}^*)'.$$

For individuals $i$ and $k$ in stratum $j$ such that $\sigma_{i,k,j}$ and $\sigma_{i,j}$ are non-zero (i.e., non-cases), $\xi_{i,j}\, w_{i,j}\, \mathbf{IF}^{(2)}_{i,j}(\widehat{\boldsymbol{\theta}}^*)$ and $\xi_{k,j}\, w_{k,j}\, \mathbf{IF}^{(2)}_{k,j}(\widehat{\boldsymbol{\theta}}^*)$ are weighted residuals from a weighted linear regression on the auxiliary variables; see Web Appendix D.3. With good auxiliary variables for calibration, one can expect (*i*) the phase-two component of the variance and hence the total variance to be smaller; and (*ii*) the difference in Equation (20) to be smaller than the difference in Equation (16); see also Chapter 17 in Borgan et al..

## 5. MISSING DATA

### 5.1. Notation

Covariate information may be missing for individuals in phase-two. For example, stored blood samples from individuals in phase-two could have been previously used or lost. We assume such covariates are missing at random and we regard the set of individuals with complete covariate data as a *phase-three sample*. More precisely, let $V_{i,j}$ be the phase-three sampling indicator for subject $i$ in stratum $j$, $i \in \{1, \ldots, n^{(j)}\}, j \in \{1, \ldots, J\}$; we assume the Bernoulli indicators $V_{i,j}$ are mutually independent and independent of the phase-two indicators, $\xi_{i,j}$. Let $w^{(3)}_{i,j} \equiv \frac{1}{\pi^{(3)}_{i,j}}$ be the phase-three design weight, where $\pi^{(3)}_{i,j}$ is the phase-three design sampling probability. The overall sampling design weight of subject $i$ in stratum $j$ is $w_{i,j} = w^{(2)}_{i,j} \times w^{(3)}_{i,j}$.

The phase-three sampling probabilities may differ in $J^{(3)}$ exclusive and exhaustive subsets (phase-three strata) of the population that need not coincide with the $J$ phase-two strata. For example, cases may have a different probability of missingness from non-cases. Nonetheless, we index the members of the cohort as in Section 3. Web Appendix H describes analysis when the phase-three sampling probabilities are known. However, the $\pi_{i,j}^{(3)}$ are usually unknown and need to be estimated (Section 5.2).

## 5.2 Weight estimation

When the $\pi_{i,j}^{(3)}$ are unknown, $w_{i,j}^{(3)}$ can be estimated as follows, $i \in \{1, \ldots, n^{(j)}\}, j \in \{1, \ldots, J\}$. The $V_{i,j}$ are known for all members of the phase-two sample, and let $\boldsymbol{B}_{i,j}$ be a $J^{(3)} \times 1$ vector of indicator variables that take value 1 if subject $i$ in (phase-two) stratum $j$ is in the corresponding phase-three stratum, and 0 otherwise. Let $\exp(\widetilde{\boldsymbol{\gamma}})$ be the vector of $J^{(3)}$ estimated phase-three sampling weights that are obtained by solving the estimating equation $\sum_{j=1}^{J} \sum_{i=1}^{n^{(j)}} \xi_{i,j} \boldsymbol{B}_{i,j} - \exp(\boldsymbol{\gamma}' \boldsymbol{B}_{i,j}) \xi_{i,j} V_{i,j} \boldsymbol{B}_{i,j} = 0$. For example, if phase-three sampling is stratified on case status, we use weights $\widetilde{w}_{i,j}^{(3)} = \dfrac{\sum_{l=1}^{J} \sum_{\substack{k=1, \\ \text{non case}}}^{n^{(j)}} \xi_{k,l}}{\sum_{l=1}^{J} \sum_{\substack{k=1, \\ \text{non case}}}^{n^{(j)}} \xi_{k,l} V_{k,l}}$ if subject $i$ in stratum $j$ is a non-case, and $\widetilde{w}_{i,j}^{(3)} = \dfrac{\sum_{l=1}^{J} \sum_{\substack{k=1, \\ \text{case}}}^{n^{(j)}} \xi_{k,l}}{\sum_{l=1}^{J} \sum_{\substack{k=1, \\ \text{case}}}^{n^{(j)}} \xi_{k,l} V_{k,l}}$ if subject $i$ in stratum $j$ is a case. Finally, we estimate $\text{var}(V_{i,j}) \equiv \sigma_{i,j}^{(3)}$ by $\widetilde{\sigma}_{i,j}^{(3)} = \dfrac{1}{\widetilde{w}_{i,j}^{(3)}} \left( 1 - \dfrac{1}{\widetilde{w}_{i,j}^{(3)}} \right)$.

## 5.3 Estimation of relative hazard, cumulative baseline hazard and pure risk

We obtain the log-relative hazard estimate $\widetilde{\boldsymbol{\beta}}(\widetilde{\boldsymbol{\gamma}}) \equiv \widetilde{\boldsymbol{\beta}}$ from solving for $\boldsymbol{\beta}$ in the estimating equation $\sum_{j=1}^{J} \sum_{i=1}^{n^{(j)}} \int_t V_{i,j} \widetilde{w}_{i,j}^{(3)} \left\{ \boldsymbol{X}_{i,j} - \dfrac{\widetilde{\boldsymbol{S}}_1(t; \widetilde{\boldsymbol{\gamma}}, \boldsymbol{\beta})}{\widetilde{S}_0(t; \widetilde{\boldsymbol{\gamma}}, \boldsymbol{\beta})} \right\} dN_{i,j}(t) = 0$. We let $\widetilde{w}_{k,j} = w_{i,j}^{(2)} \times \widetilde{w}_{i,j}^{(3)}$ and compute $\widetilde{S}_0(t; \widetilde{\boldsymbol{\gamma}}, \boldsymbol{\beta}), \widetilde{\boldsymbol{S}}_1(t; \widetilde{\boldsymbol{\gamma}}, \boldsymbol{\beta})$ and $\widetilde{\boldsymbol{S}}_2(t; \widetilde{\boldsymbol{\gamma}}, \boldsymbol{\beta})$ from Equations (2)-(4) by substituting $\xi_{k,j} V_{k,j}$ for $\xi_{k,j}$ and $\widetilde{w}_{k,j}$ for $w_{k,j}$. We estimate the baseline hazard point mass at time $t$, $d\widetilde{\Lambda}_0(t; \widetilde{\boldsymbol{\gamma}}, \widetilde{\boldsymbol{\beta}}) \equiv d\widetilde{\Lambda}_0(t)$, the cumulative baseline hazard up to time $t$, $\widetilde{\Lambda}_0(t; \widetilde{\boldsymbol{\gamma}}, \widetilde{\boldsymbol{\beta}}) \equiv \widetilde{\Lambda}_0(t)$, and the pure risk for profile $\boldsymbol{x}$ in the interval $(\tau_1, \tau_2]$, $\tilde{\pi}(\tau_1, \tau_2; x, \tilde{\gamma}, \tilde{\beta}, d\tilde{\Lambda}_0) \equiv \tilde{\pi}(\tau_1, \tau_2; x)$, from Equations (5)-(7) with $\tilde{S}_0(t; \tilde{\gamma}, \beta)$ and $\tilde{\beta}$ replacing $S_0(t, \hat{\beta})$ and $\hat{\beta}$.

If a case with missing covariate data occurs at a time $t$ when no other member of the phase-three sample is at risk, the contribution to the Breslow estimate of cumulative baseline hazard is undefined. One option is to restrict the risk projection interval by increasing $\tau_1$ or decreasing $\tau_2$ to avoid such times. If there are only a small number of such times, we recommend ignoring them in all calculations.

### 5.4 Influence functions

Let $\Delta_{i,j}(\tilde{\theta})$ denote the influence of subject $i$ in stratum $j$ on one of the parameters $\tilde{\theta}$ from the set $\{\tilde{\gamma}, \tilde{\beta}, d\tilde{\Lambda}_0(t), \tilde{\Lambda}_0(t), \tilde{\pi}(\tau_1, \tau_2; x)\}$, $i \in \{1, \ldots, n^{(j)}\}, j \in \{1, \ldots, J\}$. We can show that $\Delta_{i,j}(\tilde{\theta}) = \xi_{i,j} \, IF^{(2)}_{i,j}(\tilde{\theta}) + \xi_{i,j} \, V_{i,j} \exp(\tilde{\gamma}' B_{i,j}) IF^{(3)}_{i,j}(\tilde{\theta})$. Explicit forms of $IF^{(s)}_{i,j}(\tilde{\theta})$, $s \in \{2,3\}$, are given in the APPENDIX B, Web Appendix A.3 and derived in Web Appendix G.1. The superscript 3 emphasizes that $IF^{(3)}_{i,j}(\tilde{\theta})$ involves variables that are measured only on individuals in the phase-three sample. Thus $\xi_{i,j} V_{i,j} \, IF^{(3)}_{i,j}(\tilde{\theta})$ is zero if individual $i$ in stratum $j$ is not in the phase-three sample. However, such an individual affects $\tilde{\theta}$ through her/his influence on $\tilde{\gamma}$ via $\xi_{i,j} IF^{(2)}_{i,j}(\tilde{\theta})$, as he/she is used to estimate the phase-three sampling weights.

### 5.5 Variance decomposition and estimation from influence functions

From $\mathrm{var}(\tilde{\theta}) \approx \mathrm{var}\left\{\sum_{j=1}^{J} \sum_{i=1}^{n^{(j)}} \Delta_{i,j}(\tilde{\theta})\right\}$, $\tilde{\theta} \in \{\tilde{\gamma}, \tilde{\beta}, d\tilde{\Lambda}_0(t), \tilde{\Lambda}_0(t), \tilde{\pi}(\tau_1, \tau_2; x)\}$, the variance $\mathrm{var}(\tilde{\theta})$ can be decomposed as

$$\text{var}\left(\text{E}\left[\text{E}\left\{\sum_{j=1}^{J}\sum_{i=1}^{n^{(j)}}\Delta_{i,j}(\widetilde{\boldsymbol{\theta}})\,|C_1,C_2\right\}|C_1\right]\right)+$$

$$\text{E}\left(\text{var}\left[\text{E}\left\{\sum_{j=1}^{J}\sum_{i=1}^{n^{(j)}}\Delta_{i,j}(\widetilde{\boldsymbol{\theta}})\,|C_1,C_2\right\}|C_1\right]\right)+ \quad (21)$$

$$\text{E}\left(\text{E}\left[\text{var}\left\{\sum_{j=1}^{J}\sum_{i=1}^{n^{(j)}}\Delta_{i,j}(\widetilde{\boldsymbol{\theta}})\,|C_1,C_2\right\}|C_1\right]\right),$$

where $C_1$ denotes the information from the whole cohort, and $C_2$ denotes the information from the phase-two sample. The three terms correspond respectively to sampling from the "superpopulation", sampling the subcohort from the cohort, and sampling the phase-three sample from the phase-two sample.

We estimate $\text{var}(\widetilde{\boldsymbol{\theta}})$ by

$$\frac{n}{n-1}\sum_{j=1}^{J}\sum_{i=1}^{n^{(j)}}\frac{1}{w_{i,j}^{(2)}}\Big\{\xi_{i,j}\,\boldsymbol{IF}_{i,j}^{(2)}(\widetilde{\boldsymbol{\theta}})\boldsymbol{IF}_{i,j}^{(2)}(\widetilde{\boldsymbol{\theta}})' + 2\,\xi_{i,j}\,V_{i,j}\,\widetilde{w}_{i,j}^{(3)}\,\boldsymbol{IF}_{i,j}^{(2)}(\widetilde{\boldsymbol{\theta}})\boldsymbol{IF}_{i,j}^{(3)}(\widetilde{\boldsymbol{\theta}})' +$$

$$\xi_{i,j}\,V_{i,j}\,\widetilde{w}_{i,j}^{(3)}\boldsymbol{IF}_{i,j}^{(3)}(\widetilde{\boldsymbol{\theta}})\boldsymbol{IF}_{i,j}^{(3)}(\widetilde{\boldsymbol{\theta}})'\Big\} + \sum_{j=1}^{J}\sum_{i=1}^{n^{(j)}}\sigma_{i,j}^{(2)}w_{i,j}^{(2)}\Big\{\xi_{i,j}\,\boldsymbol{IF}_{i,j}^{(2)}(\widetilde{\boldsymbol{\theta}})\boldsymbol{IF}_{i,j}^{(2)}(\widetilde{\boldsymbol{\theta}})' +$$

$$2\,\xi_{i,j}\,V_{i,j}\,\widetilde{w}_{i,j}^{(3)}\,\boldsymbol{IF}_{i,j}^{(2)}(\widetilde{\boldsymbol{\theta}})\boldsymbol{IF}_{i,j}^{(3)}(\widetilde{\boldsymbol{\theta}})' + \xi_{i,j}\,V_{i,j}\,\widetilde{w}_{i,j}^{(3)}\boldsymbol{IF}_{i,j}^{(3)}(\widetilde{\boldsymbol{\theta}})\boldsymbol{IF}_{i,j}^{(3)}(\widetilde{\boldsymbol{\theta}})'\Big\} + \quad (22)$$

$$\sum_{j=1}^{J}\sum_{i=1}^{n^{(j)}}\sum_{\substack{k=1,\\k\neq i}}^{n^{(j)}}\sigma_{i,k,j}^{(2)}\,w_{i,k,j}^{(2)}\Big\{\xi_{i,j}\,\boldsymbol{IF}_{i,j}^{(2)}(\widetilde{\boldsymbol{\theta}}) + \xi_{i,j}\,V_{i,j}\,\widetilde{w}_{i,j}^{(3)}\boldsymbol{IF}_{i,j}^{(3)}(\widetilde{\boldsymbol{\theta}})\Big\}\Big\{\xi_{k,j}\,\boldsymbol{IF}_{k,j}^{(2)}(\widetilde{\boldsymbol{\theta}}) +$$

$$\xi_{k,j}\,V_{k,j}\,\widetilde{w}_{k,j}^{(3)}\boldsymbol{IF}_{k,j}^{(3)}(\widetilde{\boldsymbol{\theta}})\Big\}' + \sum_{j=1}^{J}\sum_{i=1}^{n^{(j)}}\widetilde{\sigma}_{i,j}^{(3)}\widetilde{w}_{i,j}^{(3)}\,\xi_{i,j}\,V_{i,j}\,\widetilde{w}_{i,j}^{(3)}\,\widetilde{w}_{i,j}^{(3)}\boldsymbol{IF}_{i,j}^{(3)}(\widetilde{\boldsymbol{\theta}})\,\boldsymbol{IF}_{i,j}^{(3)}(\widetilde{\boldsymbol{\theta}})'.$$

See Web Appendix G.2 for details. Estimation of the phase-three weights is accounted for in the variance via a part of $\xi_{i,j}\,\boldsymbol{IF}_{i,j}^{(2)}(\widetilde{\boldsymbol{\theta}})$.

## 6. SOFTWARE: CaseCohortCoxSurvival ON CRAN

"Dfbetas", which approximate influences and are available from survival software, can be used to estimate the variance of $\widehat{\boldsymbol{\beta}}$ from unstratified case-cohort data[21], and a similar code was given to estimate the variance of $\widehat{\boldsymbol{\beta}}$ (which corresponds to Estimate II in Borgan et al.[3]) for stratified case-cohort designs[4]. The cch function from the CRAN package survival[22] deals with Estimate I, in addition to Estimate II of Borgan et al.[3]. The CRAN package cchs[23] was created for Estimate III of Borgan et al.[3], but we do not consider Estimate III, which is less efficient than Estimate II. The twophase function from the CRAN package survey[24] estimates $\boldsymbol{\beta}$ and its variance from a phase-two sample, and thus from a unstratified or stratified case-cohort data. The previous papers did not discuss pure risk. SAS code was presented for Estimate III for stratified case-cohort studies and pure risk[25], but except for cchs[23], survival[22] and survey[24], we have been unable to find active online procedures, even at sites mentioned in the original articles. More general survey software can accommodate weight calibration in addition to stratification[24], but R code showing how to use these more general programs to estimate $\boldsymbol{\beta}$ are no longer online[6,7], nor is the CRAN package NestedCohort[26]. Thus, there is a need for convenient software to allow for stratification, weight calibration and missing phase-two data.

We have created a CRAN package called CaseCohortCoxSurvival to facilitate such analyses. Details will be provided elsewhere, but we present illustrative script in Table 4 for analysis of mortality data from Golestan, Iran in Section 8.

## 7. SIMULATIONS

### 7.1 Simulation designs

We compared how well the methods in Sections 3.3 and 4.3 estimate the variance of the log-relative hazard and of pure risk estimates in simulated cohorts. We also evaluated the gain in precision from using calibrated weights rather than the design weights. We considered a range of scenarios, defined by the models described below and by parameter values in Web Tables 1 and 2 of Web Appendix E.1.

We simulated cohorts with $n \in \{5 \times 10^3, 10^4\}$ and used time on study as the time scale. We simulated three covariates $\boldsymbol{X} = (X_1, X_2, X_3)'$: $X_1 \sim \mathcal{N}(0,1)$, $X_2$ takes values in $\Omega_{X_2} = \{0,1,2\}$ with respective probabilities $\{p_{0|X_1}, p_{1|X_1}, p_{2|X_1}\}$, given in Web Table 1 in Web Appendix E.1, and $X_3 \sim \mathcal{N}(\alpha_1 \times X_1 + \alpha_2 \times X_2, 1)$, where $\mathcal{N}(a, b)$ denotes the Normal distribution with mean $a$ and variance $b$. We simulated failure time $T$ from a Cox proportional model with hazard $\lambda(t; \boldsymbol{X}) = \lambda_0 \times \exp(\beta_1 X_1 + \beta_2 X_2 + \beta_3 X_3)$, where the baseline hazard $\lambda_0 = \frac{p_Y}{\mathrm{E}\{\exp(\beta_1 X_1 + \beta_2 X_2 + \beta_3 X_3)\} \times 10}$, $p_Y \in \{0.02, 0.05, 0.1\}$, is a constant calculated to have approximately 98%, 95% or 90% 10-year pure survival probability. Parameters $\alpha_1, \alpha_2, \beta_1, \beta_2$ and $\beta_3$ are in Web Table 2 in Web Appendix E.1. Cohort entry time, $E$, was uniform on the first 5 years, and we assumed the time to censoring by loss to follow-up, $C$, had an exponential distribution with hazard $\frac{10}{-\log(0.98)}$, corresponding to a chance of loss to follow-up of 2% in 10 years. We assumed $T$, $E$ and $C$ were mutually independent. We let $\tilde{T} = \min(T, 10 - E, C)$ be the observed time. The total follow-up time on time interval $(\tau_1, \tau_2]$ was thus $\max\{0, \min(\tilde{T}, \tau_2) - \tau_1\}$.

We sampled from four strata defined by $W = 0 \times I(X_1 \geq 0, X_2 = 0) + 1 \times I(X_1 < 0, X_2 < 2) + 2 \times I(X_1 \geq 0, X_2 > 0) + 3 \times I(X_1 < 0, X_2 = 2)$, where stratum "0" is low risk, strata "1" and "2" are both medium risk, stratum "3" is high risk, and $I(\ )$ is the indicator function. We sampled without replacement fixed numbers of individuals, $m^{(j)}$, independently across strata, $j \in \{0,1,2,3\}$. The $m^{(j)}$ depended on the expected numbers of failures and of individuals in the strata via $m^{(j)} = \left\lfloor \frac{\lambda_0 \times 10 \times \mathrm{E}\{\exp(\beta_1 X_1 + \beta_2 X_2 + \beta_3 X_3)|W=j\}}{1 - \lambda_0 \times 10 \times \mathrm{E}\{\exp(\beta_1 X_1 + \beta_2 X_2 + \beta_3 X_3)|W=j\}} \times \mathrm{E}(n^{(j)}) \times K + \frac{1}{2} \right\rfloor$, where $K \in \{2, 4\}$ is the number of non-

cases we wish to sample for each case, and $\lfloor \rfloor$ is the floor function. The phase-two sample consisted of the subcohort and unsampled cases. Design weights were computed as in Section 3.3. Calibration of the weights was performed against the auxiliary variables proposed in Section 4.1, with the covariates in the full cohort imputed from three prediction models using $W$ and proxies $\widetilde{X} = (\widetilde{X}_1, \widetilde{X}_2, \widetilde{X}_3)'$. More precisely, we assumed that the $\widetilde{X}$ were measured on all cohort members, and we used $\widetilde{X}_1 = X_1 + \varepsilon_1$, $\widetilde{X}_3 = X_3 + \varepsilon_3$, with $\varepsilon_1$ and $\varepsilon_3$ independently distributed as normal $\mathcal{N}(0, 0.75^2)$, so that $\text{corr}(\widetilde{X}_1, X_1) = \text{corr}(\widetilde{X}_3, X_3) = 0.8$. We used $\widetilde{X}_2 = \{2 \times I(X_2 = 0) + 1 \times I(X_2 = 2) + 0 \times I(X_2 = 1)\} \times I(\varepsilon_2 \in [0, 0.2[) + \{2 \times I(X_2 = 1) + 1 \times I(X_2 = 0) + 0 \times I(X_2 = 2)\} \times I(\varepsilon_2 \in ]0.9, 1]) + X_2 \times I(\varepsilon_2 \in [0.1, 0.9])$, with $\varepsilon_2$ uniformly distributed on $[0,1]$, so that $\widetilde{X}_2$ was equal to $X_2$ for approximately 80% of the observations. To develop imputation equations for the phase-two covariates, we performed weighted linear regressions of $X_1$ on $\widetilde{X}_1$ and $W$, of $X_3$ on $\widetilde{X}_1$ and $\widetilde{X}_3$, and a weighted multinomial logistic regression of $X_2$ on $\widetilde{X}_1$, $\widetilde{X}_2$ and $W$.

For each scenario, we simulated 5,000 cohorts. We estimated the log-relative hazard $\boldsymbol{\beta} = (\beta_1, \beta_2, \beta_3)'$ and pure risks $\pi(\tau_1, \tau_2; \boldsymbol{x})$ in time interval $(\tau_1, \tau_2] = (0, 8]$ and for covariate profiles $\boldsymbol{x} \in \{(-1, 1, -0.6)', (1, -1, 0.6)', (1, 1, 0.6)'\}$, using the following sampling designs and methods of analysis: the stratified case-cohort with design weights (SCC); the stratified case-cohort with calibrated weights (SCC.Calib); the unstratified case-cohort with design weights (USCC); and the unstratified case-cohort with calibrated weights (USCC.Calib). We then estimated their variance. For each simulated realization, we obtained the variance estimate $\widehat{V}$ for SCC from Equation (14)) and the robust variance estimate ($\widehat{V}_{\text{Robust}}$) from Equation (15). For SCC.Calib, we used $\widehat{V}$ in Equation (18) and $\widehat{V}_{\text{Robust}}$ in Equation (19). For USCC and USCC.Calib, we used the variance estimates in Equations (14)), (15), (18) and (19) with $J = 1$. As a point of reference, we also estimated these parameters using the data from the whole cohort (Cohort).

## 7.2 Simulation results

The simulation results for the scenario with $n = 10{,}000$, $p_Y = 0.02$ and $K = 2$ are displayed in Figure 1, Table 1 and Table 2; see Web Tables 3 to 20 in Web Appendix E.2 for other scenarios. The robust variance formula overestimated the variance (Table 1) and yielded supra-nominal confidence interval coverage (Table 2) for most log-relative hazards and pure risks with stratified designs, and for pure risk with unstratified designs. Weight calibration led to smaller variances, and robust variance estimates where approximately valid with calibrated weights (Table 1 and Table 2). Because they properly accounted for the sampling features, the variance estimates in Equation (14)) and Equation (18) (for design weights or calibrated weights, respectively) yielded proper coverage in all designs (Table 2), except for $\log\{\pi(\tau_1, \tau_2; \boldsymbol{x})\}$ when $\boldsymbol{x} = (-1, 1, -0.6)'$, for which the full cohort analysis also had supra-nominal coverage. As shown in Figure 1, stratification and/or weight calibration improved efficiency. Moreover, the unstratified case-cohort with weight calibration was nearly as efficient as the stratified case-cohort with weight calibration, and both were considerably more efficient than analyses with design weights. With design weights, stratification improved efficiency compared to the unstratified case-cohort design.

A few remarks follow. First, all three covariates in $\boldsymbol{X}$ were only measured in the phase-two sample. A stronger increase of efficiency would be obtained for covariates available on the whole cohort (see Section 8). With weaker proxies, the efficiency gain would be more modest, and robust variance estimates may be too large (see Web Appendix E.5). Second, we log transformed the pure risks to improve coverage based on asymptotic normal theory. Third, some authors used *post-stratified* weights instead of design weights, by having a separate stratum for cases and excluding cases from the strata with non-cases[3,4]. This approach improved the precision of estimates with SCC and USCC negligibly (variance ratios of 1.01 or less), compared to using design weights (Web Appendix E.4).

Simulations concerning missing phase-two data showed that Equation (22) in Section 5.5 and a simpler formula that ignores variability in the estimated weights (Web Appendix H.3) yielded nominal confidence interval coverage of log relative hazards and pure risk (Web Appendix I), but in non-reported simulations with larger proportions missing, the latter overestimated the variance. We therefore recommend using Equation (22), as is computed in CaseCohortCoxSurvival.

## 8 DATA ANALYSIS

The Golestan Cohort included 49,819 individuals aged 36-81 and recruited in 2003-2009[27]. To reduce computation, we randomly sampled $n = 30,000$ individuals and analyzed this subset. We used the age-scale and assumed a Cox proportional hazards model predicting mortality from baseline variables: $x_1 =$ indicator of male gender, $x_2 =$ wealth score, $x_3 =$ indicator of former smoker (cigarettes, nass, or opium), $x_4 =$ indicator of current smoker, $x_5 =$ indicator of morbidity, $x_6 = x_1 x_3$ and $x_7 = x_1 x_4$. We used "never smoker" as the reference category, and morbidity was a binary indicator with value 1 if the individuals had at least one of the following morbidities at baseline: cardiovascular disease, cerebrovascular accident, hypertension, diabetes, chronic obstructive pulmonary disease, tuberculosis, cancer. The wealth score had been computed from information such as house ownership and number and type of household appliances; see Islami et al. (2009). We also estimated the pure risk in interval $(\tau_1, \tau_2] = (52,66]$ and for covariate profiles $x \in \{(0, -0.4, 0, 1, \mathbf{0}_3)', (0, 0.4, 0, 1, \mathbf{0}_3)', (\mathbf{0}_4, 1, \mathbf{0}_2)', \mathbf{0}_7'\}$, where $\mathbf{0}_a$ is the $a \times 1$ vector of zeros, and where for example $(0, -0.4, 0, 1, \mathbf{0}_3)'$ corresponds to the profile of a currently smoking woman with a low wealth score, while $(\mathbf{0}_4, 1, \mathbf{0}_2)'$ corresponds to a never-smoking woman with morbidity at baseline. We assumed that age, gender, smoking status, morbidity, residence (urban, rural), ethnicity (Turkmen, others), marital status (unmarried, married, widowed, divorced/separated, other), education (nil, less than 5$^{th}$, 6$^{th}$-8$^{th}$, 9$^{th}$-12$^{th}$, College), socioeconomic status (low, low to medium, medium to high, high), death status and follow-up time were known for everybody in the cohort, but the wealth score was available only for individuals in phase-two. We sampled 33, 42, 192, 246, 57, 62, 313, 382, 82, 86, 391, 477, 565, 770, 1934 and 2949 individuals respectively in the 16 strata defined by gender (male, female), residence and four baseline age categories ([36,45), [45,50), [50,55) and [55,81)), so that we expected approximately one non-case per case in each stratum. We estimated the log-relative hazards and pure risks using SCC, SCC.Calib, USCC and USCC.Calib (see notation and methods of analysis in Section 5). We used gender, socioeconomic status, age at baseline, marital status, ethnicity, education and residence as proxies to impute the wealth score for the entire cohort and then calibrated the design weights. We also analyzed the whole cohort ($n = 30{,}000$).

Table 3 displays the estimation results for log-relative hazard and pure risk parameters, respectively. When using design weights, robust variance estimates were larger for the log-relative hazards of covariates $x_1$ and $x_2$, for all the pure risks in the stratified design, and for the pure risks with profiles $x \in \{(0, -0.4, 0, 1, \mathbf{0}_3)', (\mathbf{0}_4, 1, \mathbf{0}_2)', \mathbf{0}_7'\}$ in the unstratified design. In the stratified design, $\hat{V}_{\text{Robust}}$ agreed well with $\hat{V}$ for 5 of the 7 log relative hazard parameters, possibly because stratification was only based on $x_1$.

Weight calibration improved efficiency, and robust variance estimates were very close to $\hat{V}$ for all parameters. Notably, calibration led to estimates with almost as much precision as with the full cohort, not only for covariates that were available on the whole cohort, but also for wealth score, for which there were good proxies.

For a few parameters, variances estimated using SCC.Calib or USCC.Calib were slightly smaller than with the whole cohort. To see if this reflected an unusual feature of our data, we replicated our analysis over 2,500 cohorts of size $n = 20,000$ sampled with replacement from the Golestan dataset. The empirical variances and the mean estimated variances from the full cohort were smaller than from the calibration procedures; see Web Appendix F.

To illustrate how easily such analyses can be performed with the CaseCohortCoxSurvival CRAN package, we present a script and a pseudo-code for SCC and SCC.Calib in Table 4- and Web Appendix J, respectively.

## 9 DISCUSSION

We presented a unified approach to analysis of case-cohort data that allows the practitioner to take advantage of various options and improvements in design and analysis since the landmark paper of Prentice[1]. We used influence functions adapted to the various design and analysis options together with variance calculations that take two-phase sampling into account. We developed corresponding software, CaseCohortCoxSurvival, that facilitates analysis with and without stratification and/or weight calibration, for subcohort sampling with or without replacement. We allow for phase-two data to be missing at random for stratified designs. We provide inference not only for log relative hazards in the Cox model, but also for covariate-specific cumulative hazards and pure risks. We hope these calculations and software will promote wider and more principled design and analysis of case-cohort data, for which there is a need[10]. Convenient software of the type we describe does not appear to be available online (Section 6).

We found that weight calibration improves efficiency with stratified or unstratified sampling of the subcohort, in line with previous findings for unstratified designs. We found theoretically and empirically that the robust variance estimate[2] is nearly unbiased if the covariances of the phase-two sampling indicators, $\sigma_{i,k,j}$, $i \neq k$, are zero, as when the subcohort members are sampled with replacement (Table 5). For sampling without replacement, these covariances are negative, which tend to bias the robust variance estimate upward. This has been noted for log-relative hazards in stratified designs[4,5], but we also found this bias for pure risk in unstratified designs. With weight calibration based on strong predictors of phase-two covariates, the robust variance had little bias (Table 5). Nonetheless, we recommend our influence-based approach with complete variance decomposition for theoretical and empirical reasons. In addition, and as previously recommended[10], we stress the practical importance of describing the design fully in publications, including stratification details and whether or not the subcohort was sampled with replacement.

The methods we presented used design weights. Borgan et al.[3] and Samuelsen et al.[4] recommended weights that are post-stratified into a case stratum and multiple non-case strata. In our simulations, there was less than 2% increase in efficiency from post-stratification. Using the influences we derived for design weights with post-stratified weights (with cases in one stratum and non-cases in the original strata) yielded confidence intervals with nominal coverage (results not shown). Thus, the influence functions we provide can be used for such post-stratification. Further efficiency gains might be obtained by post-stratifying on time intervals in which follow-up ends[4,29,30] or on other features[9, Chapter 6, Section 16.4.5].

An alternative approach to sampling is to select the subcohort sample size in each stratum such that the expected number of non-cases is a multiple of the observed number of cases. Using the influences we gave for design weights and substituting post-stratified weights yielded valid variance estimates and coverage of confidence intervals in unreported simulations, unless the number of cases and non-cases in a stratum is small (e.g. fewer than 10 cases and 20 non-cases).

As discussed by Keogh et al.[31], likelihood-based methods for missing data and imputation can increase efficiency of case-cohort analyses, but, unlike stratification and weight calibration, they yield biased risk model estimates if imputation models are mis-specified. Indeed, a key advantage of weight calibration is that poor imputation models reduce the efficiency gains, but do not bias estimates of risk model parameters[32]. Weight calibrated estimators are in the class of augmented inverse-probability weighted estimators that are similarly robust[32,33].

In Section 5.3, we suggested modifications for times when a case had missing covariate data and no other member of the phase-three sample was at risk when the case failed. An alternative would be to weight the numerator of the Breslow estimator and only use event times $t$ from cases with complete covariate data, namely $d\widetilde{\Lambda}_0(t) = \frac{\sum_{j=1}^{J}\sum_{i=1}^{n^{(j)}} V_{i,j}\widetilde{w}_{i,j}^{(3)} dN_{i,j}(t)}{\widetilde{S}_0(t;\widetilde{\gamma},\widetilde{\beta})}$. Unreported simulations showed this led to biased estimates of pure risks, however.

This paper dealt with covariates measured at baseline. Although the influences for log-relative hazards apply equally to time-varying covariates, modifications are needed for pure-risks, and computational challenges arise for large cohorts. Moreover, pure risk estimates are uninterpretable unless the time-varying covariates are "external"[34]. We have assumed a common baseline hazard across strata. If strata are defined by covariates in the risk model and/or the time scale, there is no need for stratum effects in the Cox model, but otherwise stratum indicators could be included and analyzed like other covariates known for all cohort members. A stratified Cox model with different baseline hazards in each stratum would require modifications of the influences given in this paper.

**ACKNOWLEDGMENTS**

We thank Dr. Barry Graubard for insightful comments and Dr. Arash Etemadi and GEMShare for documenting and sharing data from the Golestan Cohort. This work was supported by the

**DATA AVAILABILITY STATEMENT**

R code and functions used for the simulations in Section 7 are available in the Supporting Information of this article and on GitHub at https://github.com/Etievant/CaseCohort. The R package CaseCohortCoxSurvival integrating the various functions on GitHub is available on CRAN[link]. We are not authorized to release the clinical data used in Section 8.

# APPENDIX A   INFLUENCES FOR STRATIFIED CASE-COHORT WITH CALIBRATED WEIGHTS

$$IF_{i,j}^{(1)}(\widehat{\boldsymbol{\eta}}) = \left\{\sum_{l=1}^{J}\sum_{k=1}^{n^{(j)}} \xi_{k,l} w_{k,l} \exp(\widehat{\boldsymbol{\eta}}' \boldsymbol{A}_{k,l}) \boldsymbol{A}_{k,l} \boldsymbol{A}_{k,l}'\right\}^{-1} \boldsymbol{A}_{i,j},$$

$$IF_{i,j}^{(2)}(\widehat{\boldsymbol{\eta}}) = -\exp(\widehat{\boldsymbol{\eta}}' \boldsymbol{A}_{i,j}) \left\{\sum_{l=1}^{J}\sum_{k=1}^{n^{(j)}} \xi_{k,l} w_{k,l} \exp(\widehat{\boldsymbol{\eta}}' \boldsymbol{A}_{k,l}) \boldsymbol{A}_{k,l} \boldsymbol{A}_{k,l}'\right\}^{-1} \boldsymbol{A}_{i,j},$$

$$IF_{i,j}^{(1)}(\widehat{\boldsymbol{\beta}}^*) = \left\{\sum_{l=1}^{J}\sum_{k=1}^{n^{(j)}} \xi_{k,l} w_{k,l} \exp(\widehat{\boldsymbol{\eta}}' \boldsymbol{A}_{k,l}) \boldsymbol{Z}_{k,l} \boldsymbol{A}_{k,l}'\right\} IF_{i,j}^{(1)}(\widehat{\boldsymbol{\eta}}),$$

and $IF_{i,j}^{(2)}(\widehat{\boldsymbol{\beta}}^*) = \exp(\widehat{\boldsymbol{\eta}}' \boldsymbol{A}_{i,j}) \boldsymbol{Z}_{i,j} + \left\{\sum_{l=1}^{J}\sum_{k=1}^{n^{(j)}} \xi_{k,l} w_{k,l} \exp(\widehat{\boldsymbol{\eta}}' \boldsymbol{A}_{k,l}) \boldsymbol{Z}_{k,l} \boldsymbol{A}_{k,l}'\right\} IF_{i,j}^{(2)}(\widehat{\boldsymbol{\eta}})$, with

$$\boldsymbol{Z}_{i,j} = \left[\sum_{l=1}^{J}\sum_{k=1}^{n^{(l)}} \int_t \xi_{k,l} w_{k,l} \exp(\widehat{\boldsymbol{\eta}}' \boldsymbol{A}_{k,l}) \left\{\frac{\boldsymbol{S}_2^*(t;\widehat{\boldsymbol{\eta}},\widehat{\boldsymbol{\beta}}^*)}{S_0^*(t;\widehat{\boldsymbol{\eta}},\widehat{\boldsymbol{\beta}}^*)} - \frac{\boldsymbol{S}_1^*(t;\widehat{\boldsymbol{\eta}},\widehat{\boldsymbol{\beta}}^*)\boldsymbol{S}_1^*(t;\widehat{\boldsymbol{\eta}},\widehat{\boldsymbol{\beta}}^*)'}{S_0^*(t;\widehat{\boldsymbol{\eta}},\widehat{\boldsymbol{\beta}}^*)^2}\right\} dN_{k,l}(t)\right]^{-1} \left[\int_t \left\{\boldsymbol{X}_{i,j} - \frac{\boldsymbol{S}_1^*(t;\widehat{\boldsymbol{\eta}},\widehat{\boldsymbol{\beta}}^*)}{S_0^*(t;\widehat{\boldsymbol{\eta}},\widehat{\boldsymbol{\beta}}^*)}\right\} \left\{dN_{i,j}(t) - \frac{Y_{i,j}(t)\exp(\widehat{\boldsymbol{\beta}}^{*'} \boldsymbol{X}_{i,j})\sum_{l=1}^{J}\sum_{k=1}^{n^{(l)}} \xi_{k,l} w_{k,l} \exp(\widehat{\boldsymbol{\eta}}' \boldsymbol{A}_{k,l}) dN_{k,l}(t)}{S_0^*(t;\widehat{\boldsymbol{\eta}},\widehat{\boldsymbol{\beta}}^*)}\right\}\right].$$

$$IF_{i,j}^{(1)}\{d\widehat{\Lambda}_0^*(t)\} = \left\{\sum_{l=1}^{J}\sum_{k=1}^{n^{(l)}} \xi_{k,l} w_{k,l} \exp(\widehat{\boldsymbol{\eta}}' \boldsymbol{A}_{k,l}) H_{k,l}(t) \boldsymbol{A}_{k,l}'\right\} IF_{i,j}^{(1)}(\widehat{\boldsymbol{\eta}}),$$

$$IF_{i,j}^{(2)}\{d\widehat{\Lambda}_0^*(t)\} = \{S_0^*(t;\widehat{\boldsymbol{\eta}},\widehat{\boldsymbol{\beta}}^*)\}^{-1} \Big[dN_{i,j}(t) +$$

$$\exp(\widehat{\boldsymbol{\eta}}' \boldsymbol{A}_{i,j}) H_{i,j}(t) + \left\{\sum_{l=1}^{J}\sum_{k=1}^{n^{(l)}} \xi_{k,l} w_{k,l} \exp(\widehat{\boldsymbol{\eta}}' \boldsymbol{A}_{k,l}) H_{k,l}(t) \boldsymbol{A}_{k,l}'\right\} IF_{i,j}^{(2)}(\widehat{\boldsymbol{\eta}})\Big],$$

with $H_{i,j}(t) = -\{S_0^*(t;\widehat{\boldsymbol{\eta}},\widehat{\boldsymbol{\beta}}^*)\}^{-1} d\widehat{\Lambda}_0^*(t) \{\boldsymbol{S}_1^*(t;\widehat{\boldsymbol{\eta}},\widehat{\boldsymbol{\beta}}^*)' \boldsymbol{Z}_{i,j} + K_{i,j}(t)\},$

and $K_{i,j}(t) = Y_{i,j}(t)\exp(\widehat{\boldsymbol{\beta}}^{*'} \boldsymbol{X}_{i,j}).$

Finally, for any $s \in \{1,2\}$, $IF_{i,j}^{(s)}\{\int_{\tau_1}^{\tau_2} d\widehat{\Lambda}_0^*(t)\} = \int_{\tau_1}^{\tau_2} IF_{i,j}^{(s)}\{d\widehat{\Lambda}_0^*(t)\}$, and $IF_{i,j}^{(s)}\{\widehat{\pi}^*(\tau_1,\tau_2;x)\} =$

$$\left\{\frac{\partial \widehat{\pi}(\tau_1,\tau_2;x)}{\partial \boldsymbol{\beta}}\bigg|_{\boldsymbol{\beta}=\widehat{\boldsymbol{\beta}}^*}\right\} IF_{i,j}^{(s)}(\widehat{\boldsymbol{\beta}}^*) + \left[\frac{\partial \widehat{\pi}(\tau_1,\tau_2;x)}{\partial \{\int_{\tau_1}^{\tau_2} d\Lambda_0(t)\}}\bigg|_{d\Lambda_0(t)=d\widehat{\Lambda}_0^*(t)}\right] IF_{i,j}^{(s)}\left\{\int_{\tau_1}^{\tau_2} d\widehat{\Lambda}_0^*(t)\right\}.$$

# APPENDIX B    INFLUENCES FOR STRATIFIED CASE-COHORT WITH MISSING COVARIATE INFORMATION AND ESTIMATED DESIGN SAMPLING PHASE-THREE WEIGHTS

$$IF_{i,j}^{(2)}(\widetilde{\gamma}) = \left\{\sum_{l=1}^{J}\sum_{k=1}^{n^{(l)}} \xi_{k,l}\, V_{k,l}\, \exp(\widetilde{\gamma}' B_{k,l})\, B_{k,l} B_{k,l}'\right\}^{-1} B_{i,j},$$

and $IF_{i,j}^{(3)}(\widetilde{\gamma}) = -\left\{\sum_{l=1}^{J}\sum_{k=1}^{n^{(l)}} \xi_{k,l} V_{k,l} \exp(\widetilde{\gamma}' B_{k,l}) B_{k,l} B_{k,l}'\right\} B_{i,j}.$

Then $IF_{i,j}^{(2)}(\widetilde{\beta}) = \left\{\sum_{l=1}^{J}\sum_{k=1}^{n^{(l)}} \xi_{k,l} V_{k,l} \exp(\widetilde{\gamma}' B_{k,l}) \widetilde{Z}_{k,l} B_{k,l}'\right\} IF_{i,j}^{(2)}(\widetilde{\gamma}),$

and $IF_{i,j}^{(3)}(\widetilde{\beta}) = \widetilde{Z}_{i,j} + \left\{\sum_{l=1}^{J}\sum_{k=1}^{n^{(l)}} \xi_{k,l} V_{k,l} \exp(\widetilde{\gamma}' B_{k,l}) \widetilde{Z}_{k,l} B_{k,l}'\right\} IF_{i,j}^{(3)}(\widetilde{\gamma}),$ with

$$\widetilde{Z}_{i,j} = w_{i,j}^{(2)} \times \left[\sum_{l=1}^{J}\sum_{k=1}^{n^{(l)}} \int_t \xi_{k,l}\, V_{k,l}\, w_{k,l}^{(2)} \exp(\widetilde{\gamma}' B_{k,l}) \left\{\frac{\tilde{S}_2(t;\widetilde{\gamma},\widetilde{\beta})}{\tilde{S}_0(t;\widetilde{\gamma},\widetilde{\beta})} - \frac{\tilde{S}_1(t;\widetilde{\gamma},\widetilde{\beta})\tilde{S}_1(t;\widetilde{\gamma},\widetilde{\beta})'}{\tilde{S}_0(t;\widetilde{\gamma},\widetilde{\beta})^2}\right\} dN_{k,l}(t)\right]^{-1} \times$$

$$\left[\int_t \left\{X_{i,j} - \frac{\tilde{S}_1(t;\widetilde{\gamma},\widetilde{\beta})}{\tilde{S}_0(t;\widetilde{\gamma},\widetilde{\beta})}\right\}\left\{dN_{i,j}(t) - \frac{Y_{i,j}(t)\exp(\widetilde{\beta}' X_{i,j})\sum_{l=1}^{J}\sum_{k=1}^{n^{(l)}} \xi_{k,l} V_{k,l} w_{k,l}^{(2)} \exp(\widetilde{\gamma}' B_{k,l}) dN_{k,l}(t)}{\tilde{S}_0(t;\widetilde{\gamma},\widetilde{\beta})}\right\}\right].$$

$IF_{i,j}^{(2)}\{d\widetilde{\Lambda}_0(t)\} = \{\tilde{S}_0(t;\widetilde{\gamma},\widetilde{\beta})\}^{-1} dN_{i,j}(t) +$

$\left\{\sum_{l=1}^{J}\sum_{k=1}^{n^{(j)}} \xi_{k,l} V_{k,l} \exp(\widetilde{\gamma}' B_{k,l}) \widetilde{H}_{k,l}(t) B_{k,l}'\right\} IF_{i,j}^{(2)}(\widetilde{\gamma}),$

and    $IF_{i,j}^{(3)}\{d\widetilde{\Lambda}_0(t)\} = \widetilde{H}_{i,j}(t) + \left\{\sum_{l=1}^{J}\sum_{k=1}^{n^{(j)}} \xi_{k,l} V_{k,l} \exp(\widetilde{\gamma}' B_{k,l}) \widetilde{H}_{k,l}(t) B_{k,l}'\right\} IF_{i,j}^{(3)}(\widetilde{\gamma}),$    with

$\widetilde{H}_{i,j}(t) = -\tilde{S}_0(t;\widetilde{\gamma},\widetilde{\beta})^{-1} d\widetilde{\Lambda}_0(t)\left\{\tilde{S}_1(t;\widetilde{\gamma},\widetilde{\beta})' \widetilde{Z}_{i,j} + \widetilde{K}_{i,j}(t)\right\},$

and $\widetilde{K}_{i,j}(t) = w_{i,j}^{(2)} Y_{i,j}(t) \exp(\widetilde{\beta}' X_{i,j}).$

Finally, for any $s \in \{2,3\}$, $IF_{i,j}^{(s)}\left\{\int_{\tau_1}^{\tau_2} \mathrm{d}\widetilde{\Lambda}_0(t)\right\} = \int_{\tau_1}^{\tau_2} IF_{i,j}^{(2)}\{\mathrm{d}\widetilde{\Lambda}_0(t)\}$ and $IF_{i,j}^{(s)}\{\hat{\pi}^*(\tau_1,\tau_2;x)\} =$

$$\left\{\frac{\partial \widetilde{\pi}(\tau_1,\tau_2;x)}{\partial \boldsymbol{\beta}}_{|\boldsymbol{\beta}=\widetilde{\boldsymbol{\beta}}}\right\} IF_{i,j}^{(s)}(\widehat{\boldsymbol{\beta}}^*) + \left[\frac{\partial \widetilde{\pi}(\tau_1,\tau_2;x)}{\partial\left\{\int_{\tau_1}^{\tau_2} \mathrm{d}\Lambda_0(t)\right\}}_{|\mathrm{d}\Lambda_0(t)=\mathrm{d}\widetilde{\Lambda}_0(t)}\right] IF_{i,j}^{(s)}\left\{\int_{\tau_1}^{\tau_2} \mathrm{d}\widetilde{\Lambda}_0(t)\right\}.$$

**SUPPORTING INFORMATION**

Web Appendices referenced in Sections 2, 3, 4, 5, 7 and 8 are available with this paper. R code and functions used for the simulations in Web Appendices E and I are available in the Supporting Information of this article and on GitHub at https://github.com/Etievant/CaseCohort.

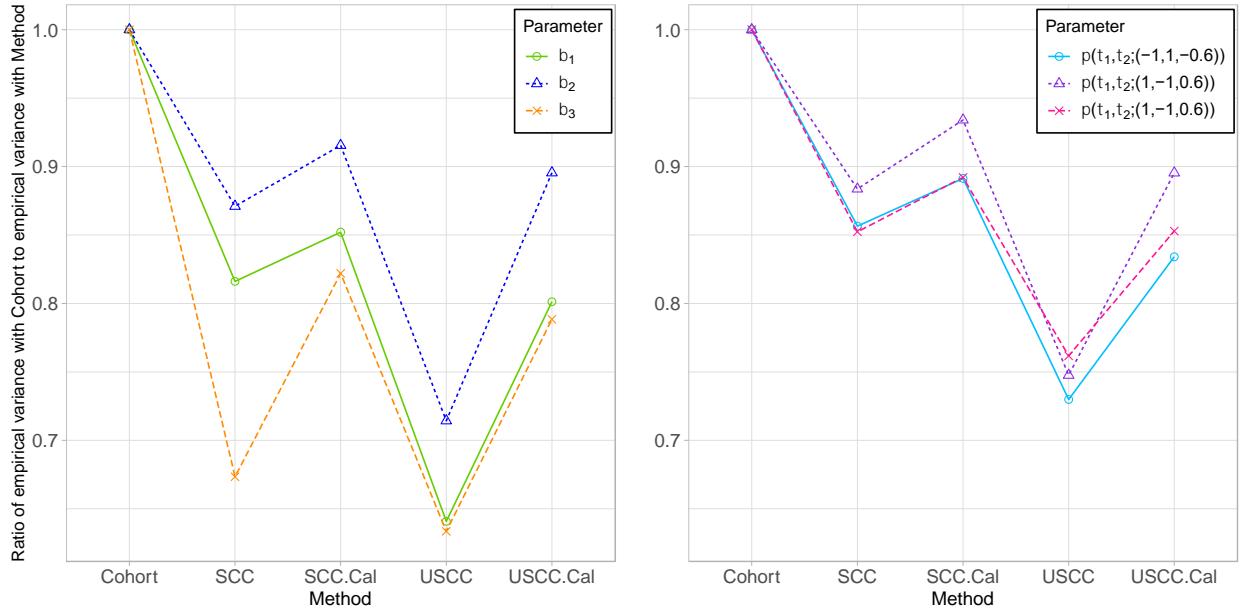

Figure 1- Ratio of empirical variance of log-relative hazard and pure risk estimates with the whole cohort to that when using different sampling designs and methods of analysis. The results are obtained from 5,000 simulated cohorts with $n = 10{,}000$, $p_Y = 0.02$, $K = 2$. The variance ratio is a measure of relative efficiency.

| Parameter | Cohort | SCC | | SCC.Calib | | USCC | | USCC.Calib | |
|---|---|---|---|---|---|---|---|---|---|
| | | $\hat{V}_{\text{Robust}}$ | $\hat{V}$ | $\hat{V}_{\text{Robust}}$ | $\hat{V}$ | $\hat{V}_{\text{Robust}}$ | $\hat{V}$ | $\hat{V}_{\text{Robust}}$ | $\hat{V}$ |
| $\beta_1$ | 0.0069 (0.007) | 0.0102 (0.0085) | 0.0087 | 0.0085 (0.0082) | 0.0083 | 0.0106 (0.0108) | 0.0106 | 0.0087 (0.0087) | 0.0087 |
| $\beta_2$ | 0.0097 (0.01) | 0.0139 (0.0115) | 0.0114 | 0.0109 (0.0109) | 0.0109 | 0.014 (0.014) | 0.014 | 0.0109 (0.0112) | 0.0109 |
| $\beta_3$ | 0.0068 (0.0069) | 0.0102 (0.0103) | 0.0102 | 0.0085 (0.0084) | 0.0085 | 0.0107 (0.0109) | 0.0107 | 0.0087 (0.0088) | 0.0087 |
| $\log\{\pi(\tau_1,\tau_2;x)\}$, $x=(-1,1,-0.6)'$ | 0.0122 (0.0119) | 0.0172 (0.0136) | 0.014 | 0.0142 (0.0133) | 0.0137 | 0.0181 (0.0158) | 0.0159 | 0.0145 (0.014) | 0.0145 |
| $x=(1,-1,0.6)'$ | 0.062 (0.0618) | 0.0861 (0.0688) | 0.0697 | 0.0684 (0.0662) | 0.0681 | 0.086 (0.0823) | 0.0837 | 0.0696 (0.0696) | 0.0696 |
| $x=(1,1,0.6)'$ | 0.0277 (0.0274) | 0.0379 (0.0326) | 0.0333 | 0.0319 (0.0309) | 0.0316 | 0.0386 (0.0361) | 0.0363 | 0.0328 (0.0325) | 0.0328 |

Table 1- Mean of estimated variances of log-relative hazard and pure risk estimates, from using different sampling designs, methods of analysis and variance estimation, over 5,000 simulated cohorts with $n = 10,000$, $p_Y = 0.02$, $K = 2$. The corresponding empirical variances are displayed between parentheses.

| Parameter | Cohort | SCC | | SCC.Calib | | USCC | | USCC.Calib | |
|---|---|---|---|---|---|---|---|---|---|
| | | $\hat{V}_{\text{Robust}}$ | $\hat{V}$ | $\hat{V}_{\text{Robust}}$ | $\hat{V}$ | $\hat{V}_{\text{Robust}}$ | $\hat{V}$ | $\hat{V}_{\text{Robust}}$ | $\hat{V}$ |
| $\beta_1$ | 0.9476 | 0.9668* | 0.9524 | 0.9538 | 0.95 | 0.947 | 0.947 | 0.9452 | 0.945 |
| $\beta_2$ | 0.9486 | 0.9716* | 0.9522 | 0.9514 | 0.9512 | 0.9554 | 0.9554 | 0.9506 | 0.9506 |
| $\beta_3$ | 0.9422* | 0.9492 | 0.9494 | 0.9474 | 0.9474 | 0.9454 | 0.9454 | 0.9476 | 0.9478 |
| $\log\{\pi(\tau_1,\tau_2;x)\}$, $x=(-1,1,-0.6)'$ | 0.956* | 0.973* | 0.9568* | 0.96* | 0.9546 | 0.9656* | 0.9526 | 0.955 | 0.955 |
| $\log\{\pi(\tau_1,\tau_2;x)\}$, $x=(1,-1,0.6)'$ | 0.952 | 0.9722* | 0.9518 | 0.9522 | 0.9516 | 0.954 | 0.9522 | 0.9472 | 0.947 |
| $\log\{\pi(\tau_1,\tau_2;x)\}$, $x=(1,1,0.6)'$ | 0.952 | 0.9666* | 0.9542 | 0.9532 | 0.9516 | 0.9574* | 0.95 | 0.9502 | 0.9502 |

Table 2- Coverage of 95% CIs of log-relative hazard and pure risk estimates, from using different sampling designs, methods of analysis and variance estimation, over 5,000 simulated cohorts with $n = 10,000$, $p_Y = 0.02$, $K = 2$. * indicates coverage outside the expected interval [0.9440; 0.9560].

| Parameter | Cohort | SCC | | SCC.Calib | | USCC | | USCC.Calib | |
|---|---|---|---|---|---|---|---|---|---|
| | | $\hat{V}_{\text{Robust}}$ | $\hat{V}$ | $\hat{V}_{\text{Robust}}$ | $\hat{V}$ | $\hat{V}_{\text{Robust}}$ | $\hat{V}$ | $\hat{V}_{\text{Robust}}$ | $\hat{V}$ |
| $\beta_1$ | 0.0013 | 0.0018 | 0.0016 | 0.0013 | 0.0013 | 0.0025 | 0.0025 | 0.0013 | 0.0013 |
| $\beta_2$ | 0.005 | 0.0075 | 0.0072 | 0.0055 | 0.0055 | 0.0104 | 0.0104 | 0.0061 | 0.0061 |
| $\beta_3$ | 0.0173 | 0.0291 | 0.0291 | 0.0178 | 0.0178 | 0.0394 | 0.0394 | 0.017 | 0.017 |
| $\beta_4$ | 0.0029 | 0.0057 | 0.0057 | 0.0029 | 0.0029 | 0.007 | 0.007 | 0.003 | 0.003 |
| $\beta_5$ | 0.0009 | 0.0013 | 0.0013 | 0.0008 | 0.0008 | 0.002 | 0.002 | 0.0009 | 0.0009 |
| $\beta_6$ | 0.0217 | 0.0344 | 0.0344 | 0.0219 | 0.0219 | 0.0497 | 0.0497 | 0.0213 | 0.0213 |
| $\beta_7$ | 0.0044 | 0.0077 | 0.0076 | 0.0043 | 0.0043 | 0.0103 | 0.0103 | 0.0045 | 0.0045 |
| $\pi(\tau_1, \tau_2; x)$, $x = (0, -0.4, 0, 1, \mathbf{0}_3)'$ | 0.00012 | 0.00023 | 0.00022 | 0.00012 | 0.00012 | 0.00031 | 0.00030 | 0.00013 | 0.00012 |
| $x = (0, 0.4, 0, 1, \mathbf{0}_3)'$ | 5.10E-05 | 9.10E-05 | 8.80E-05 | 5.10E-05 | 5.10E-05 | 0.00010 | 0.00010 | 5.70E-05 | 5.70E-05 |
| $x = (\mathbf{0}_4, 1, \mathbf{0}_2)'$ | 2.50E-05 | 4.00E-05 | 3.50E-05 | 2.50E-05 | 2.50E-05 | 4.50E-05 | 4.30E-05 | 2.70E-05 | 2.70E-05 |
| $x = \mathbf{0}_7'$ | 6.70E-06 | 1.10E-05 | 7.90E-06 | 6.90E-06 | 6.80E-06 | 1.10E-05 | 9.80E-06 | 6.90E-06 | 6.90E-06 |

Table 3- Estimated variances of log-relative hazard and pure risk parameters from using different sampling designs, methods of analysis and variance estimation, in the Golestan Cohort ($n = 30{,}000$). $\beta_p$ denotes the log-relative hazard parameter of covariate $x_p$, $p \in \{1, \ldots, 7\}$.

```
library(survival)
library(CaseCohortCoxSurvival)
library(dplyr)

# Load the data set ----------------------------------------------------
load("Golestancohort.RData")

# Estimation using the stratified case cohort with design weights -----------
caseCohortCoxSurvival(data = Golestancohort, status = "status", times = 
c("agebegin", "ageend"), cox.phase1 = c("x1", "x3", "x4", "x5", "x6", "x7"), 
cox.phase2 = c("x2"), strata = "gender.residence.age", subcohort = "phase2", 
Tau1 = 52, Tau2 = 66, risk.data = c(0,0,1,0,0,0,-0.4))

# Estimation using the stratified case cohort with calibrated weights --------
caseCohortCoxSurvival(data = Golestancohort, status = "status", times = 
c("agebegin", "ageend"), cox.phase1 = c("x1", "x3", "x4", "x5", "x6", "x7"), 
cox.phase2 = c("x2"), strata = "gender.residence.age", subcohort = "phase2", 
Tau1 = 52, Tau2 = 66, risk.data = c(0,0,1,0,0,0,-0.4), calibrated = TRUE, 
predictors.cox.phase2 = c("x1", "ses", "age", "maritalstatus", "ethnicity", 
"education", "residence"))
```

Table 4- R script to obtain variance estimates $\hat{V}$ for SCC and SCC.Calib in Table 3 with the CaseCohortCoxSurvival R package.

| Parameter | Approximate validity of robust variance estimate | | | | |
|---|---|---|---|---|---|
| | Sampling with replacement | Sampling without replacement | | | |
| | | Unstratified sampling with design weights | Unstratified sampling with calibrated weights | Stratified sampling with design weights | Stratified sampling with calibrated weights* |
| Relative hazard | Yes | Yes | Yes | No | Yes* |
| Pure risk | Yes | No | Yes* | No | Yes* |

Table 5- Sampling designs and methods of analysis for which the robust variance estimator is approximately valid for relative hazard and pure risk estimates. We categorize the robust variance estimate as approximately valid if theoretical calculations indicate that the bias is negligible and/or if in simulations the means of the robust variance estimates were close to the means of the two-phase variance procedures we describe. * Holds if the phase-one covariates are good predictors of covariates measured only in phase-two. If the proxies of phase-two covariates are too weak, the robust variance estimate may be inappropriate.

# Supporting Information for Cox model inference for relative hazard and pure risk from stratified weight-calibrated case-cohort data

Lola Etievant[1]*, Mitchell H. Gail[1]*

lola.etievant@nih.gov

gailm@mail.nih.gov

[1] National Cancer Institute, Division of Cancer Epidemiology and Genetics, Biostatistics Branch, 9609 Medical Center Drive, Rockville, MD 20850-9780.

* Corresponding authors

**Table of content**

# Web Appendix A. EXPLICIT FORMS OF THE INFLUENCE FUNCTIONS

In Web Appendix A we recall the influence functions for the stratified case-cohort design with design weights (A.1), with calibrated weights (A.2), and with allowance for missing phase-two data (A.3). These results cover the unstratified case-cohort design by setting the number of strata to 1. Derivations and details are in subsequent Web Appendices.

## A.1 Influence functions for stratified case-cohort with design weights

Let $\Delta_{i,j}(\widehat{\boldsymbol{\theta}})$ denote the influence of subject $i$ in stratum $j$ on $\widehat{\boldsymbol{\theta}}$, $i \in \{1, \dots, n^{(j)}\}$, $j \in \{1, \dots, J\}$, $\widehat{\boldsymbol{\theta}} \in \{\widehat{\boldsymbol{\beta}}, d\widehat{\Lambda}_0(t), \widehat{\Lambda}_0(t), \widehat{\pi}(\tau_1, \tau_2; x)\}$. We can show $\Delta_{i,j}(\widehat{\boldsymbol{\theta}}) = \xi_{i,j} w_{i,j} IF_{i,j}^{(2)}(\widehat{\boldsymbol{\theta}})$, with $IF_{i,j}^{(2)}(\widehat{\boldsymbol{\theta}})$ involving variables that are measured only for individuals in the phase-two sample. Here, we give explicit forms of $IF_{i,j}^{(2)}(\widehat{\boldsymbol{\theta}})$; see Web Appendix C.1 for derivation of the influence functions. We have

$$IF_{i,j}^{(2)}(\widehat{\boldsymbol{\beta}}) = \left[\sum_{l=1}^{J}\sum_{k=1}^{n^{(l)}} \int_t \left\{\frac{S_2(t;\widehat{\boldsymbol{\beta}})}{S_0(t;\widehat{\boldsymbol{\beta}})} - \frac{S_1(t;\widehat{\boldsymbol{\beta}}) S_1(t;\widehat{\boldsymbol{\beta}})'}{S_0(t;\widehat{\boldsymbol{\beta}})^2}\right\} dN_{k,l}(t)\right]^{-1} \left[\int_t \left\{X_{i,j} - \frac{S_1(t;\widehat{\boldsymbol{\beta}})}{S_0(t;\widehat{\boldsymbol{\beta}})}\right\}\left\{dN_{i,j}(t) - \frac{Y_{i,j}(t)\exp(\widehat{\boldsymbol{\beta}}'X_{i,j}) \sum_{l=1}^{J}\sum_{k=1}^{n^{(l)}} dN_{k,l}(t)}{S_0(t;\widehat{\boldsymbol{\beta}})}\right\}\right],$$

with $S_0(t;\boldsymbol{\beta}) = \sum_{l=1}^{J}\sum_{k=1}^{n^{(l)}} w_{k,l}\, \xi_{k,l}\, Y_{k,l}(t) \exp(\boldsymbol{\beta}'X_{k,l})$,

$S_1(t;\boldsymbol{\beta}) = \sum_{l=1}^{J}\sum_{k=1}^{n^{(l)}} w_{k,l}\, \xi_{k,l}\, Y_{k,l}(t) \exp(\boldsymbol{\beta}'X_{k,l})X_{k,l}$,

and $S_2(t;\boldsymbol{\beta}) = \sum_{l=1}^{J}\sum_{k=1}^{n^{(l)}} w_{k,l}\xi_{k,l}Y_{k,l}(t)\exp(\boldsymbol{\beta}'X_{k,l})X_{k,l}X_{k,l}'$,

and we have

$$IF_{i,j}^{(2)}\{d\widehat{\Lambda}_0(t)\} = \{S_0(t;\widehat{\boldsymbol{\beta}})\}^{-1}\{dN_{i,j}(t) - d\widehat{\Lambda}_0(t)S_1(t;\widehat{\boldsymbol{\beta}})' IF_{i,j}^{(2)}(\widehat{\boldsymbol{\beta}}) - d\widehat{\Lambda}_0(t)Y_{i,j}(t)\exp(\widehat{\boldsymbol{\beta}}'X_{i,j})\}.$$

Note, $IF_{i,j}^{(2)}\{d\widehat{\Lambda}_0(t)\}$ is a linear combination of the increments $dN_{i,j}(t)$ and $d\widehat{\Lambda}_0(t)$. Thus we can write $IF_{i,j}^{(2)}\left\{\int_{\tau_1}^{\tau_2} d\widehat{\Lambda}_0(t)\right\} = \int_{\tau_1}^{\tau_2} IF_{i,j}^{(2)}\{d\widehat{\Lambda}_0(t)\}.$ Finally $IF_{i,j}^{(2)}\{\hat{\pi}(\tau_1, \tau_2; \boldsymbol{x})\} =$

$$\left\{\frac{\partial \hat{\pi}(\tau_1,\tau_2;\boldsymbol{x})}{\partial \boldsymbol{\beta}}\bigg|_{\boldsymbol{\beta}=\widehat{\boldsymbol{\beta}}}\right\} \boldsymbol{IF}_{i,j}^{(2)}(\widehat{\boldsymbol{\beta}}) + \left[\frac{\partial \hat{\pi}(\tau_1,\tau_2;\boldsymbol{x})}{\partial \left\{\int_{\tau_1}^{\tau_2} d\Lambda_0(t)\right\}}\bigg|_{d\Lambda_0(t)=d\widehat{\Lambda}_0(t)}\right] IF_{i,j}^{(2)}\left\{\int_{\tau_1}^{\tau_2} d\widehat{\Lambda}_0(t)\right\}.$$

## A.2 Influence functions for stratified case-cohort with calibrated weights

Let $\boldsymbol{\Delta}_{i,j}(\widehat{\boldsymbol{\theta}}^*)$ denote the influence of subject $i$ in stratum $j$ on $\widehat{\boldsymbol{\theta}}^*$, $i \in \{1, \dots, n^{(j)}\}, j \in \{1, \dots, J\}$, $\widehat{\boldsymbol{\theta}}^* \in \{\widehat{\boldsymbol{\eta}}, \widehat{\boldsymbol{\beta}}^*, d\widehat{\Lambda}_0^*(t), \widehat{\Lambda}_0^*(t), \hat{\pi}^*(\tau_1, \tau_2; \boldsymbol{x})\}$. We can show that $\boldsymbol{\Delta}_{i,j}(\widehat{\boldsymbol{\theta}}^*) = \boldsymbol{IF}_{i,j}^{(1)}(\widehat{\boldsymbol{\theta}}^*) + \xi_{i,j} w_{i,j} \boldsymbol{IF}_{i,j}^{(2)}(\widehat{\boldsymbol{\theta}}^*)$, with $\boldsymbol{IF}_{i,j}^{(1)}(\widehat{\boldsymbol{\theta}}^*)$ only involving variables that are measured on all cohort members, and $\boldsymbol{IF}_{i,j}^{(2)}(\widehat{\boldsymbol{\theta}}^*)$ involving variables that are measured only for individuals in the phase-two sample. Here, we give explicit forms of $\boldsymbol{IF}_{i,j}^{(s)}(\widehat{\boldsymbol{\theta}}^*)$, $s \in \{1,2\}$; see Web Appendix D.2 for derivation of the influence functions. We have

$$\boldsymbol{IF}_{i,j}^{(1)}(\widehat{\boldsymbol{\eta}}) = \left\{\sum_{l=1}^{J}\sum_{k=1}^{n^{(j)}} \xi_{k,l} w_{k,l} \exp(\widehat{\boldsymbol{\eta}}' \boldsymbol{A}_{k,l}) \boldsymbol{A}_{k,l} \boldsymbol{A}_{k,l}'\right\}^{-1} \boldsymbol{A}_{i,j},$$

$$\boldsymbol{IF}_{i,j}^{(2)}(\widehat{\boldsymbol{\eta}}) = -\exp(\widehat{\boldsymbol{\eta}}' \boldsymbol{A}_{i,j})\left\{\sum_{l=1}^{J}\sum_{k=1}^{n^{(j)}} \xi_{k,l} w_{k,l} \exp(\widehat{\boldsymbol{\eta}}' \boldsymbol{A}_{k,l}) \boldsymbol{A}_{k,l} \boldsymbol{A}_{k,l}'\right\}^{-1} \boldsymbol{A}_{i,j},$$

$$\boldsymbol{IF}_{i,j}^{(1)}(\widehat{\boldsymbol{\beta}}^*) = \left\{\sum_{l=1}^{J}\sum_{k=1}^{n^{(j)}} \xi_{k,l} w_{k,l} \exp(\widehat{\boldsymbol{\eta}}' \boldsymbol{A}_{k,l}) \boldsymbol{Z}_{k,l} \boldsymbol{A}_{k,l}'\right\} \boldsymbol{IF}_{i,j}^{(1)}(\widehat{\boldsymbol{\eta}}),$$

and $\boldsymbol{IF}_{i,j}^{(2)}(\widehat{\boldsymbol{\beta}}^*) = \exp(\widehat{\boldsymbol{\eta}}' \boldsymbol{A}_{i,j}) \boldsymbol{Z}_{i,j} + \left\{\sum_{l=1}^{J}\sum_{k=1}^{n^{(j)}} \xi_{k,l} w_{k,l} \exp(\widehat{\boldsymbol{\eta}}' \boldsymbol{A}_{k,l}) \boldsymbol{Z}_{k,l} \boldsymbol{A}_{k,l}'\right\} \boldsymbol{IF}_{i,j}^{(2)}(\widehat{\boldsymbol{\eta}})$, with

$$\boldsymbol{Z}_{i,j} = \left[\sum_{l=1}^{J}\sum_{k=1}^{n^{(l)}}\int_t \xi_{k,l} w_{k,l} \exp(\widehat{\boldsymbol{\eta}}' \boldsymbol{A}_{k,l})\left\{\frac{S_2^*(t;\widehat{\boldsymbol{\eta}},\widehat{\boldsymbol{\beta}}^*)}{S_0^*(t;\widehat{\boldsymbol{\eta}},\widehat{\boldsymbol{\beta}}^*)} - \frac{S_1^*(t;\widehat{\boldsymbol{\eta}},\widehat{\boldsymbol{\beta}}^*)S_1^*(t;\widehat{\boldsymbol{\eta}},\widehat{\boldsymbol{\beta}}^*)'}{S_0^*(t;\widehat{\boldsymbol{\eta}},\widehat{\boldsymbol{\beta}}^*)^2}\right\} dN_{k,l}(t)\right]^{-1} \left[\int_t \left\{\boldsymbol{X}_{i,j} - \frac{S_1^*(t;\widehat{\boldsymbol{\eta}},\widehat{\boldsymbol{\beta}}^*)}{S_0^*(t;\widehat{\boldsymbol{\eta}},\widehat{\boldsymbol{\beta}}^*)}\right\}\left\{dN_{i,j}(t) - \frac{Y_{i,j}(t)\exp(\widehat{\boldsymbol{\beta}}^{*\prime}\boldsymbol{X}_{i,j})\sum_{l=1}^{J}\sum_{k=1}^{n^{(l)}}\xi_{k,l} w_{k,l}\exp(\widehat{\boldsymbol{\eta}}'\boldsymbol{A}_{k,l})dN_{k,l}(t)}{S_0^*(t;\widehat{\boldsymbol{\eta}},\widehat{\boldsymbol{\beta}}^*)}\right\}\right],$$

and $S_0^*(t;\widehat{\boldsymbol{\eta}},\boldsymbol{\beta}) = \sum_{l=1}^{J}\sum_{k=1}^{n^{(l)}} \xi_{k,l} w_{k,l} \exp(\widehat{\boldsymbol{\eta}}' \boldsymbol{A}_{k,l}) Y_{k,l}(t) \exp(\boldsymbol{\beta}' \boldsymbol{X}_{k,l}),$

$$S_1^*(t;\widehat{\boldsymbol{\eta}},\boldsymbol{\beta}) = \sum_{l=1}^J \sum_{k=1}^{n^{(l)}} \xi_{k,l}\, w_{k,l} \exp(\widehat{\boldsymbol{\eta}}'\boldsymbol{A}_{k,l})\, Y_{k,l}(t) \exp(\boldsymbol{\beta}'\boldsymbol{X}_{k,l}) \boldsymbol{X}_{k,l},$$

and $S_2^*(t;\widehat{\boldsymbol{\eta}},\boldsymbol{\beta}) = \sum_{l=1}^J \sum_{k=1}^{n^{(l)}} \xi_{k,l}\, w_{k,l} \exp(\widehat{\boldsymbol{\eta}}'\boldsymbol{A}_{k,l})\, Y_{k,l}(t) \exp(\boldsymbol{\beta}'\boldsymbol{X}_{k,l}) \boldsymbol{X}_{k,l}\boldsymbol{X}_{k,l}'.$

Then $IF_{i,j}^{(1)}\{d\widehat{\Lambda}_0^*(t)\} = \left\{\sum_{l=1}^J \sum_{k=1}^{n^{(l)}} \xi_{k,l} w_{k,l} \exp(\widehat{\boldsymbol{\eta}}'\boldsymbol{A}_{k,l}) H_{k,l}(t) \boldsymbol{A}_{k,l}'\right\} IF_{i,j}^{(1)}(\widehat{\boldsymbol{\eta}}),$

and

$$IF_{i,j}^{(2)}\{d\widehat{\Lambda}_0^*(t)\} = \{S_0^*(t;\widehat{\boldsymbol{\eta}},\widehat{\boldsymbol{\beta}}^*)\}^{-1}\Big[dN_{i,j}(t) +$$

$$\exp(\widehat{\boldsymbol{\eta}}'\boldsymbol{A}_{i,j}) H_{i,j}(t) + \left\{\sum_{l=1}^J \sum_{k=1}^{n^{(l)}} \xi_{k,l} w_{k,l} \exp(\widehat{\boldsymbol{\eta}}'\boldsymbol{A}_{k,l}) H_{k,l}(t) \boldsymbol{A}_{k,l}'\right\} IF_{i,j}^{(2)}(\widehat{\boldsymbol{\eta}})\Big],$$

with $H_{i,j}(t) = -\{S_0^*(t;\widehat{\boldsymbol{\eta}},\widehat{\boldsymbol{\beta}}^*)\}^{-1} d\widehat{\Lambda}_0^*(t)\{S_1^*(t;\widehat{\boldsymbol{\eta}},\widehat{\boldsymbol{\beta}}^*)'\boldsymbol{Z}_{i,j} + K_{i,j}(t)\},$

and $K_{i,j}(t) = Y_{i,j}(t) \exp(\widehat{\boldsymbol{\beta}}^{*\prime}\boldsymbol{X}_{i,j}).$

Note, $H_{i,j}(t)$ is a multiple of $d\widehat{\Lambda}_0^*(t)$ and thus $IF_{i,j}^{(s)}\{d\widehat{\Lambda}_0^*(t)\}$ is a linear combination of $d\widehat{\Lambda}_0^*(t)$ and $dN_{i,j}(t)$. Hence $IF_{i,j}^{(s)}\left\{\int_{\tau_1}^{\tau_2} d\widehat{\Lambda}_0^*(t)\right\} = \int_{\tau_1}^{\tau_2} IF_{i,j}^{(s)}\{d\widehat{\Lambda}_0^*(t)\}$, $s \in \{1,2\}$.

Finally, for any $s \in \{1,2\}$, $IF_{i,j}^{(s)}\{\widehat{\pi}^*(\tau_1,\tau_2;\boldsymbol{x})\} = \left\{\frac{\partial \widehat{\pi}(\tau_1,\tau_2;\boldsymbol{x})}{\partial \boldsymbol{\beta}}\Big|_{\boldsymbol{\beta}=\widehat{\boldsymbol{\beta}}^*}\right\} IF_{i,j}^{(s)}(\widehat{\boldsymbol{\beta}}^*) +$

$$\left[\frac{\partial \widehat{\pi}(\tau_1,\tau_2;\boldsymbol{x})}{\partial\{\int_{\tau_1}^{\tau_2} d\Lambda_0(t)\}}\Big|_{d\Lambda_0(t)=d\widehat{\Lambda}_0^*(t)}\right] IF_{i,j}^{(s)}\left\{\int_{\tau_1}^{\tau_2} d\widehat{\Lambda}_0^*(t)\right\}.$$

## A.3     Influences functions for stratified case-cohort with missing covariate information and estimated design sampling phase-three weights

We let $\boldsymbol{\Delta}_{i,j}(\widetilde{\boldsymbol{\theta}})$ denote the influence of subject $i$ in stratum $j$ on $\widetilde{\boldsymbol{\theta}}$, $i \in \{1,\ldots,n^{(j)}\}, j \in \{1,\ldots,J\}$, $\widetilde{\boldsymbol{\theta}} \in \{\widetilde{\boldsymbol{\gamma}}, \widetilde{\boldsymbol{\beta}}, d\widetilde{\Lambda}_0(t), \widetilde{\Lambda}_0(t), \widetilde{\pi}(\tau_1,\tau_2;\boldsymbol{x})\}$. We can show that $\boldsymbol{\Delta}_{i,j}(\widetilde{\boldsymbol{\theta}}) = \xi_{i,j}\, IF_{i,j}^{(2)}(\widetilde{\boldsymbol{\theta}}) + \xi_{i,j} V_{i,j} \exp(\widetilde{\boldsymbol{\gamma}}'\boldsymbol{B}_{i,j}) IF_{i,j}^{(3)}(\widetilde{\boldsymbol{\theta}})$, with $IF_{i,j}^{(3)}(\widetilde{\boldsymbol{\theta}})$ involving variables that are measured only on individuals in the phase-three sample. Here, we give explicit forms of $IF_{i,j}^{(s)}(\widetilde{\boldsymbol{\theta}})$, $s \in \{2,3\}$; see Web Appendix G.1 for derivation of the influence functions. We have

$$IF_{i,j}^{(2)}(\widetilde{\boldsymbol{\gamma}}) = \left\{\sum_{l=1}^{J}\sum_{k=1}^{n^{(l)}} \xi_{k,l} V_{k,l} \exp(\widetilde{\boldsymbol{\gamma}}'\boldsymbol{B}_{k,l}) \boldsymbol{B}_{k,l}\boldsymbol{B}_{k,l}'\right\}^{-1} \boldsymbol{B}_{i,j},$$

and $IF_{i,j}^{(3)}(\widetilde{\boldsymbol{\gamma}}) = -\left\{\sum_{l=1}^{J}\sum_{k=1}^{n^{(l)}} \xi_{k,l} V_{k,l} \exp(\widetilde{\boldsymbol{\gamma}}'\boldsymbol{B}_{k,l}) \boldsymbol{B}_{k,l}\boldsymbol{B}_{k,l}'\right\} \boldsymbol{B}_{i,j}$.

Then $IF_{i,j}^{(2)}(\widetilde{\boldsymbol{\beta}}) = \left\{\sum_{l=1}^{J}\sum_{k=1}^{n^{(l)}} \xi_{k,l} V_{k,l} \exp(\widetilde{\boldsymbol{\gamma}}'\boldsymbol{B}_{k,l})\widetilde{\boldsymbol{Z}}_{k,l}\boldsymbol{B}_{k,l}'\right\} IF_{i,j}^{(2)}(\widetilde{\boldsymbol{\gamma}})$,

and $IF_{i,j}^{(3)}(\widetilde{\boldsymbol{\beta}}) = \widetilde{\boldsymbol{Z}}_{i,j} + \left\{\sum_{l=1}^{J}\sum_{k=1}^{n^{(l)}} \xi_{k,l} V_{k,l} \exp(\widetilde{\boldsymbol{\gamma}}'\boldsymbol{B}_{k,l})\widetilde{\boldsymbol{Z}}_{k,l}\boldsymbol{B}_{k,l}'\right\} IF_{i,j}^{(3)}(\widetilde{\boldsymbol{\gamma}})$, with

$$\widetilde{\boldsymbol{Z}}_{i,j} = w_{i,j}^{(2)} \times \left[\sum_{l=1}^{J}\sum_{k=1}^{n^{(l)}}\int_t \xi_{k,l} V_{k,l} w_{k,l}^{(2)} \exp(\widetilde{\boldsymbol{\gamma}}'\boldsymbol{B}_{k,l})\left\{\frac{\widetilde{S}_2(t;\widetilde{\boldsymbol{\gamma}},\widetilde{\boldsymbol{\beta}})}{\widetilde{S}_0(t;\widetilde{\boldsymbol{\gamma}},\widetilde{\boldsymbol{\beta}})} - \frac{\widetilde{S}_1(t;\widetilde{\boldsymbol{\gamma}},\widetilde{\boldsymbol{\beta}})\widetilde{S}_1(t;\widetilde{\boldsymbol{\gamma}},\widetilde{\boldsymbol{\beta}})'}{\widetilde{S}_0(t;\widetilde{\boldsymbol{\gamma}},\widetilde{\boldsymbol{\beta}})^2}\right\}dN_{k,l}(t)\right]^{-1} \times$$

$$\left[\int_t\left\{\boldsymbol{X}_{i,j} - \frac{\widetilde{S}_1(t;\widetilde{\boldsymbol{\gamma}},\widetilde{\boldsymbol{\beta}})}{\widetilde{S}_0(t;\widetilde{\boldsymbol{\gamma}},\widetilde{\boldsymbol{\beta}})}\right\}\left\{dN_{i,j}(t) - \frac{Y_{i,j}(t)\exp(\widetilde{\boldsymbol{\beta}}'\boldsymbol{X}_{i,j})\sum_{l=1}^{J}\sum_{k=1}^{n^{(l)}} \xi_{k,l} V_{k,l} w_{k,l}^{(2)} \exp(\widetilde{\boldsymbol{\gamma}}'\boldsymbol{B}_{k,l})dN_{k,l}(t)}{\widetilde{S}_0(t;\widetilde{\boldsymbol{\gamma}},\widetilde{\boldsymbol{\beta}})}\right\}\right],$$

$$\widetilde{S}_0(t;\widetilde{\boldsymbol{\gamma}},\boldsymbol{\beta}) = \sum_{l=1}^{J}\sum_{k=1}^{n^{(l)}} \xi_{k,l} V_{k,l} w_{k,l}^{(2)} \exp(\widetilde{\boldsymbol{\gamma}}'\boldsymbol{B}_{k,l}) Y_{k,l}(t) \exp(\boldsymbol{\beta}'\boldsymbol{X}_{k,l}),$$

$$\widetilde{S}_1(t;\widetilde{\boldsymbol{\gamma}},\boldsymbol{\beta}) = \sum_{l=1}^{J}\sum_{k=1}^{n^{(l)}} \xi_{k,l} V_{k,l} w_{k,l}^{(2)} \exp(\widetilde{\boldsymbol{\gamma}}'\boldsymbol{B}_{k,l}) Y_{k,l}(t) \exp(\boldsymbol{\beta}'\boldsymbol{X}_{k,l}) \boldsymbol{X}_{k,l},$$

and $\widetilde{S}_2(t;\widetilde{\boldsymbol{\gamma}},\widetilde{\boldsymbol{\beta}}) = \sum_{l=1}^{J}\sum_{k=1}^{n^{(l)}} \xi_{k,l} V_{k,l} w_{k,l}^{(2)} \exp(\widetilde{\boldsymbol{\gamma}}'\boldsymbol{B}_{k,l}) Y_{k,l}(t) \exp(\widetilde{\boldsymbol{\beta}}'\boldsymbol{X}_{k,l}) \boldsymbol{X}_{k,l}\boldsymbol{X}_{k,l}'$.

In addition,

$$IF_{i,j}^{(2)}\{d\widetilde{\Lambda}_0(t)\} = \{\widetilde{S}_0(t;\widetilde{\boldsymbol{\gamma}},\widetilde{\boldsymbol{\beta}})\}^{-1}dN_{i,j}(t) +$$

$$\left\{\sum_{l=1}^{J}\sum_{k=1}^{n^{(j)}} \xi_{k,l} V_{k,l} \exp(\widetilde{\boldsymbol{\gamma}}'\boldsymbol{B}_{k,l})\widetilde{H}_{k,l}(t)\boldsymbol{B}_{k,l}'\right\} IF_{i,j}^{(2)}(\widetilde{\boldsymbol{\gamma}}),$$

and $IF_{i,j}^{(3)}\{d\widetilde{\Lambda}_0(t)\} = \widetilde{H}_{i,j}(t) + \left\{\sum_{l=1}^{J}\sum_{k=1}^{n^{(j)}} \xi_{k,l} V_{k,l} \exp(\widetilde{\boldsymbol{\gamma}}'\boldsymbol{B}_{k,l})\widetilde{H}_{k,l}(t)\boldsymbol{B}_{k,l}'\right\} IF_{i,j}^{(3)}(\widetilde{\boldsymbol{\gamma}})$, with

$$\widetilde{H}_{i,j}(t) = -\widetilde{S}_0(t;\widetilde{\boldsymbol{\gamma}},\widetilde{\boldsymbol{\beta}})^{-1}d\widetilde{\Lambda}_0(t)\left\{\widetilde{S}_1(t;\widetilde{\boldsymbol{\gamma}},\widetilde{\boldsymbol{\beta}})'\widetilde{\boldsymbol{Z}}_{i,j} + \widetilde{K}_{i,j}(t)\right\},$$

and $\widetilde{K}_{i,j}(t) = w_{i,j}^{(2)} Y_{i,j}(t) \exp(\widetilde{\boldsymbol{\beta}}'\boldsymbol{X}_{i,j})$.

Note, $\widetilde{H}_{i,j}(t)$ is a multiple of $d\widetilde{\Lambda}_0(t)$ and thus $IF_{i,j}^{(s)}\{d\widetilde{\Lambda}_0(t)\}$ is a linear combination of $d\widetilde{\Lambda}_0(t)$ and $dN_{i,j}(t)$. Hence $IF_{i,j}^{(s)}\left\{\int_{\tau_1}^{\tau_2} d\widetilde{\Lambda}_0(t)\right\} = \int_{\tau_1}^{\tau_2} IF_{i,j}^{(2)}\{d\widetilde{\Lambda}_0(t)\}, s \in \{2,3\}$.

Finally, for any $s \in \{2,3\}$, $IF_{i,j}^{(s)}\{\hat{\pi}^*(\tau_1,\tau_2;\boldsymbol{x})\} = \left\{\frac{\partial \widetilde{\pi}(\tau_1,\tau_2;\boldsymbol{x})}{\partial \boldsymbol{\beta}}_{|\boldsymbol{\beta}=\widetilde{\boldsymbol{\beta}}}\right\} \boldsymbol{IF}_{i,j}^{(s)}(\widehat{\boldsymbol{\beta}}^*) +$

$\left[\frac{\partial \widetilde{\pi}(\tau_1,\tau_2;\boldsymbol{x})}{\partial\left\{\int_{\tau_1}^{\tau_2} d\Lambda_0(t)\right\}}_{|d\Lambda_0(t)=d\widetilde{\Lambda}_0(t)}\right] IF_{i,j}^{(s)}\left\{\int_{\tau_1}^{\tau_2} d\widetilde{\Lambda}_0(t)\right\}.$

# Web Appendix B.  ESTIMATION WITH COMLETE COVARIATE DATA FOR THE WHOLE COHORT

## B.1 Parameters estimation

Here we review methods for analyzing cohort data with complete covariates $X$ measured on all cohort members. The estimate of the log-relative hazard, $\widehat{\boldsymbol{\beta}}_c$, is obtained by maximizing the partial log-likelihood

$$\sum_{j=1}^{J}\sum_{i=1}^{n^{(j)}} \int_t \left[\log\{Y_{i,j}(t)\exp(\boldsymbol{\beta}'X_{i,j})\} - \log\left\{\sum_{l=1}^{J}\sum_{k=1}^{n^{(l)}} Y_{k,l}(t)\exp(\boldsymbol{\beta}'X_{k,l})\right\}\right]dN_{i,j}(t),$$

or equivalently by solving in $\boldsymbol{\beta}$ the estimating equation

$$\boldsymbol{U}_c(\boldsymbol{\beta}) = \sum_{j=1}^{J}\sum_{i=1}^{n^{(j)}} \int_t \left\{X_{i,j} - \frac{\sum_{l=1}^{J}\sum_{k=1}^{n^{(l)}} Y_{k,l}(t)\exp(\boldsymbol{\beta}'X_{k,l})X_{k,l}}{\sum_{l=1}^{J}\sum_{k=1}^{n^{(l)}} Y_{k,l}(t)\exp(\boldsymbol{\beta}'X_{k,l})}\right\} dN_{i,j}(t) = 0.$$

The baseline hazard point mass at time $t$ is estimated non-parametrically (Breslow, 1974) as

$$d\widehat{\Lambda}_{0,c}(t;\widehat{\boldsymbol{\beta}}_c) \equiv d\widehat{\Lambda}_{0,c}(t) = \frac{\sum_{j=1}^{J}\sum_{i=1}^{n^{(j)}} dN_{i,j}(t)}{\sum_{j=1}^{J}\sum_{i=1}^{n^{(j)}} Y_{i,j}(t)\exp(\widehat{\boldsymbol{\beta}}'_c X_{i,j})},$$ and the cumulative baseline hazard estimate

up to time $t$ is estimated as $\widehat{\Lambda}_{0,c}(t;\widehat{\boldsymbol{\beta}}_c,\widehat{\lambda}_{0,c}) \equiv \widehat{\Lambda}_{0,c}(t) = \int_0^t d\widehat{\Lambda}_{0,c}(s)$.

Finally, the pure risk for covariate profile $\boldsymbol{x}$ in $(\tau_1,\tau_2]$ (i.e., the probability of experiencing the event within the interval $(\tau_1,\tau_2]$ when having profile $X = \boldsymbol{x}$), $0 \leq \tau_1 < \tau_2 \leq \tau$, is be estimated as

$$\widehat{\pi}_c(\tau_1,\tau_2;\boldsymbol{x},\widehat{\boldsymbol{\beta}}_c,\widehat{\lambda}_{0,c}) \equiv \widehat{\pi}_c(\tau_1,\tau_2;\boldsymbol{x}) = 1 - \exp\left\{-\int_{\tau_1}^{\tau_2} \exp(\widehat{\boldsymbol{\beta}}'_c\boldsymbol{x})\,d\widehat{\Lambda}_{0,c}(t)\right\}.$$

## B.2 Influence functions

Let $\boldsymbol{\Delta}_{i,j}(\widehat{\boldsymbol{\theta}}_c)$ denote the influence of subject $i$ in stratum $j$ on $\widehat{\boldsymbol{\theta}}_c$, $i \in \{1,\ldots,n^{(j)}\}, j \in \{1,\ldots,J\}$, $\widehat{\boldsymbol{\theta}}_c \in \{\widehat{\boldsymbol{\beta}}_c, d\widehat{\Lambda}_{0,c}(t), \widehat{\Lambda}_{0,c}(t), \widehat{\pi}_c(\tau_1,\tau_2;\boldsymbol{x})\}$. From Reid and Crépeau, (1985),

$$\Delta_{i,j}(\widehat{\boldsymbol{\beta}}_c) = \left[\sum_{l=1}^{J}\sum_{k=1}^{n^{(l)}}\int_t \left\{\frac{S_2^c(t;\widehat{\boldsymbol{\beta}}_c)}{S_0^c(t;\widehat{\boldsymbol{\beta}}_c)} - \frac{S_1^c(t;\widehat{\boldsymbol{\beta}}_c)\,S_1^c(t;\widehat{\boldsymbol{\beta}}_c)'}{S_0^c(t;\widehat{\boldsymbol{\beta}}_c)^2}\right\}dN_{k,l}(t)\right]^{-1} \times \left[\int_t \left\{\boldsymbol{X}_{i,j} - \frac{S_1^c(t;\widehat{\boldsymbol{\beta}}_c)}{S_0^c(t;\widehat{\boldsymbol{\beta}}_c)}\right\} \times \right.$$

$$\left. \left\{dN_{i,j}(t) - Y_{i,j}(t)\exp\left(\widehat{\boldsymbol{\beta}}_c'\boldsymbol{X}_{i,j}\right)\frac{\sum_{l=1}^{J}\sum_{k=1}^{n^{(l)}}dN_{k,l}(t)}{S_0(t;\widehat{\boldsymbol{\beta}}_c)}\right\}\right],$$

with $S_0^c(t;\widehat{\boldsymbol{\beta}}_c) = \sum_{l=1}^{J}\sum_{k=1}^{n^{(l)}} Y_{k,l}(t)\exp\left(\widehat{\boldsymbol{\beta}}_c'\boldsymbol{X}_{k,l}\right)$,

$S_1^c(t;\widehat{\boldsymbol{\beta}}_c) = \sum_{l=1}^{J}\sum_{k=1}^{n^{(l)}} Y_{k,l}(t)\exp\left(\widehat{\boldsymbol{\beta}}_c'\boldsymbol{X}_{k,l}\right)\boldsymbol{X}_{k,l}$,

and $S_2^c(t;\widehat{\boldsymbol{\beta}}_c) = \sum_{l=1}^{J}\sum_{k=1}^{n^{(l)}} Y_{k,l}(t)\exp\left(\widehat{\boldsymbol{\beta}}_c'\boldsymbol{X}_{k,l}\right)\boldsymbol{X}_{k,l}\,\boldsymbol{X}_{k,l}'$.

Then, from chapter 4 in Pfeiffer and Gail (2017),

$$\Delta_{i,j}\{d\widehat{\Lambda}_{0,c}(t)\} = \{S_0^c(t;\widehat{\boldsymbol{\beta}}_c)\}^{-1}\left[dN_{i,j}(t) - d\widehat{\Lambda}_{0,c}(t)\,Y_{i,j}(t)\exp\left(\widehat{\boldsymbol{\beta}}_c'\boldsymbol{X}_{i,j}\right) - \right.$$

$$\left. d\widehat{\Lambda}_{0,c}(t)\,S_1^c(t;\widehat{\boldsymbol{\beta}}_c)'\Delta_{i,j}(\widehat{\boldsymbol{\beta}}_c)\right].$$

Note, $\Delta_{i,j}\{d\widehat{\Lambda}_{0,c}(t)\}$ is a linear combination of the increments $dN_{i,j}(t)$ and $d\widehat{\Lambda}_{0,c}(t)$; thus $\Delta_{i,j}\left\{\int_{\tau_1}^{\tau_2} d\widehat{\Lambda}_{0,c}(t)\right\} = \int_{\tau_1}^{\tau_2} \Delta_{i,j}\{d\widehat{\Lambda}_{0,c}(t)\}$. Finally, following Graubard and Fears (2005),

$$\Delta_{i,j}\{\widehat{\pi}_c(\tau_1,\tau_2;\boldsymbol{x})\} = \left\{\frac{\partial \widehat{\pi}_c(\tau_1,\tau_2;\boldsymbol{x})}{\partial \boldsymbol{\beta}}\bigg|_{\boldsymbol{\beta}=\widehat{\boldsymbol{\beta}}_c}\right\}\Delta_{i,j}(\widehat{\boldsymbol{\beta}}_c) + \left[\frac{\partial \widehat{\pi}_c(\tau_1,\tau_2;\boldsymbol{x})}{\partial\left\{\int_{\tau_1}^{\tau_2} d\Lambda_0(t)\right\}}\bigg|_{\Lambda_0(t)=\widehat{\Lambda}_{0,c}(t)}\right]\Delta_{i,j}\left\{\int_{\tau_1}^{\tau_2} d\widehat{\Lambda}_{0,c}(t)\right\},$$

with

$$\frac{\partial \widehat{\pi}_c(\tau_1,\tau_2;\boldsymbol{x})}{\partial \boldsymbol{\beta}}\bigg|_{\boldsymbol{\beta}=\widehat{\boldsymbol{\beta}}_c} = \left\{\int_{\tau_1}^{\tau_2} d\widehat{\Lambda}_{0,c}(t)\exp(\widehat{\boldsymbol{\beta}}_c'\boldsymbol{x})\right\}\exp\left\{-\int_{\tau_1}^{\tau_2} d\widehat{\Lambda}_{0,c}(t)\exp(\widehat{\boldsymbol{\beta}}_c'\boldsymbol{x})\right\}\boldsymbol{x}',$$

$$= \left\{\int_{\tau_1}^{\tau_2} d\widehat{\Lambda}_{0,c}(t)\exp(\widehat{\boldsymbol{\beta}}_c'\boldsymbol{x})\right\}\{1 - \widehat{\pi}_c(\tau_1,\tau_2;\boldsymbol{x})\}\boldsymbol{x}',$$

and

$$\frac{\partial \widehat{\pi}_c(\tau_1,\tau_2;\boldsymbol{x})}{\partial\left\{\int_{\tau_1}^{\tau_2} d\Lambda_0(t)\right\}}\bigg|_{\Lambda_0(t)=\widehat{\Lambda}_{0,c}(t)} = \exp(\widehat{\boldsymbol{\beta}}_c'\boldsymbol{x})\exp\left\{-\int_{\tau_1}^{\tau_2} d\widehat{\Lambda}_{0,c}(t)\exp(\widehat{\boldsymbol{\beta}}_c'\boldsymbol{x})\right\},$$

$$= \exp(\widehat{\boldsymbol{\beta}}_c'\boldsymbol{x})\{1 - \widehat{\pi}_c(\tau_1,\tau_2;\boldsymbol{x})\}.$$

### B.3 Variance estimation from influence functions

Reid and Crépeau (1985) estimated $\text{var}(\widehat{\boldsymbol{\beta}}_c)$ by $\sum_{j=1}^{J} \sum_{i=1}^{n^{(j)}} \boldsymbol{\Delta}_{i,j}(\widehat{\boldsymbol{\beta}}_c) \boldsymbol{\Delta}_{i,j}(\widehat{\boldsymbol{\beta}}_c)'$. Note the relation to the "robust variance" proposed by Barlow (1994) for the case-cohort design. Using the whole cohort, we could similarly estimate $\text{var}(\widehat{\boldsymbol{\theta}}_c)$, $\widehat{\boldsymbol{\theta}}_c \in \{d\widehat{\Lambda}_{0,c}(t), \widehat{\Lambda}_{0,c}(t), \widehat{\pi}_c(\tau_1, \tau_2; \boldsymbol{x})\}$, by $\sum_{j=1}^{J} \sum_{i=1}^{n^{(j)}} \boldsymbol{\Delta}_{i,j}(\widehat{\boldsymbol{\theta}}_c) \boldsymbol{\Delta}_{i,j}(\widehat{\boldsymbol{\theta}}_c)'$.

# Web Appendix C.  ESTIMATION BASED ON DESIGN WEIGHTS FOR THE STRATIFIED CASE-COHORT DESIGN

## C.1 Derivation of the influence functions

Let $\Delta_{i,j}(\widehat{\boldsymbol{\theta}})$ denote the influence of subject $i$ in stratum $j$ on $\widehat{\boldsymbol{\theta}}$, $i \in \{1, \dots, n^{(j)}\}$, $j \in \{1, \dots, J\}$, $\widehat{\boldsymbol{\theta}} \in \{\widehat{\boldsymbol{\beta}}, d\widehat{\Lambda}_0(t), \widehat{\Lambda}_0(t), \widehat{\pi}(\tau_1, \tau_2; \boldsymbol{x})\}$. We can rewrite the estimating equation $\sum_{j=1}^{J} \sum_{i=1}^{n^{(j)}} \int_t \left\{ \boldsymbol{X}_{i,j} - \frac{S_1(t;\widehat{\boldsymbol{\beta}})}{S_0(t;\widehat{\boldsymbol{\beta}})} \right\} dN_{i,j}(t) = 0$ as

$$\boldsymbol{G}_1 - \int_t \frac{S_1(t;\widehat{\boldsymbol{\beta}})}{S_0(t;\widehat{\boldsymbol{\beta}})} \left\{ \sum_{l=1}^{J} \sum_{k=1}^{n^{(l)}} \xi_{k,l}\, w_{k,l}\, dN_{k,l}(t) \right\} = 0,$$

with $\boldsymbol{G}_1 = \sum_{j=1}^{J} \sum_{i=1}^{n^{(j)}} \int_t \xi_{i,j}\, w_{i,j}\, \boldsymbol{X}_{i,j}\, dN_{i,j}(t)$, because $\xi_{i,j}\, w_{i,j} = 1$ for any subject $i$ in stratum $j$ such that $\int_t dN_{i,j}(t) = 1$ (i.e., all the cases are included in the phase-two sample and have unit design sampling weights). Following Graubard and Fears (2005), we have

$$\Delta_{i,j}\{\boldsymbol{G}_1\} + \int_t \left[ -\frac{\Delta_{i,j}\{\sum_{l=1}^{J} \sum_{k=1}^{n^{(l)}} \xi_{k,l}\, w_{k,l}\, dN_{k,l}(t)\} \times S_1(t;\widehat{\boldsymbol{\beta}})}{S_0(t;\widehat{\boldsymbol{\beta}})} - \right. \tag{1}$$

$$\left. \frac{\{\sum_{l=1}^{J} \sum_{k=1}^{n^{(l)}} \xi_{k,l}\, w_{k,l}\, dN_{k,l}(t)\} \times \Delta_{i,j}\{S_1(t;\widehat{\boldsymbol{\beta}})\}}{S_0(t;\widehat{\boldsymbol{\beta}})} + \frac{\{\sum_{l=1}^{J} \sum_{k=1}^{n^{(l)}} \xi_{k,l}\, w_{k,l}\, dN_{k,l}(t)\} \times S_1(t;\widehat{\boldsymbol{\beta}}) \times \Delta_{i,j}\{S_0(t;\widehat{\boldsymbol{\beta}})\}}{S_0(t;\widehat{\boldsymbol{\beta}})^2} \right] = 0,$$

with

$$\Delta_{i,j}\{\boldsymbol{G}_1\} = \int_t \xi_{i,j}\, w_{i,j}\, \boldsymbol{X}_{i,j}\, dN_{i,j}(t),$$

$$\Delta_{i,j}\left\{\sum_{l=1}^{J} \sum_{k=1}^{n^{(l)}} \xi_{k,l}\, w_{k,l}\, dN_{k,l}(t)\right\} = \xi_{i,j}\, w_{i,j}\, dN_{i,j}(t),$$

$$\Delta_{i,j}\{S_1(t;\widehat{\boldsymbol{\beta}})\} = \xi_{i,j}\, w_{i,j}\, Y_{i,j}(t)\, \exp(\widehat{\boldsymbol{\beta}}'\boldsymbol{X}_{i,j})\, \boldsymbol{X}_{i,j} + \left\{ \frac{\partial S_1(t,\boldsymbol{\beta})}{\partial \boldsymbol{\beta}} \bigg|_{\boldsymbol{\beta}=\widetilde{\boldsymbol{\beta}}} \right\} \Delta_{i,j}(\widehat{\boldsymbol{\beta}}),$$

$$\frac{\partial S_1(t,\boldsymbol{\beta})}{\partial \boldsymbol{\beta}} \bigg|_{\boldsymbol{\beta}=\widehat{\boldsymbol{\beta}}} = \sum_{l=1}^{J} \sum_{k=1}^{n^{(l)}} \xi_{k,l}\, w_{k,l}\, Y_{k,l}(t)\, \exp(\widehat{\boldsymbol{\beta}}'\boldsymbol{X}_{k,l})\, \boldsymbol{X}_{k,l}\, \boldsymbol{X}_{k,l}' \equiv S_2(t;\widehat{\boldsymbol{\beta}}),$$

$$\Delta_{i,j}\{S_0(t;\widehat{\boldsymbol{\beta}})\} = \xi_{i,j}\, w_{i,j}\, Y_{i,j}(t) \exp(\widehat{\boldsymbol{\beta}}'\boldsymbol{X}_{i,j}) + \left\{\frac{\partial S_0(t;\boldsymbol{\beta})}{\partial \boldsymbol{\beta}}\bigg|_{\boldsymbol{\beta}=\widehat{\boldsymbol{\beta}}}\right\}\Delta_{i,j}(\widehat{\boldsymbol{\beta}}),$$

and $\frac{\partial S_0(t;\boldsymbol{\beta})}{\partial \boldsymbol{\beta}}\bigg|_{\boldsymbol{\beta}=\widehat{\boldsymbol{\beta}}} = \sum_{l=1}^{J}\sum_{k=1}^{n^{(l)}} \xi_{k,l}\, w_{k,l}\, Y_{k,l}(t) \exp(\widehat{\boldsymbol{\beta}}'\boldsymbol{X}_{k,l})\, \boldsymbol{X}_{k,l}{}' = \boldsymbol{S}_1(t;\widehat{\boldsymbol{\beta}})'.$

Note, the term between square brackets in Equation (1) is a linear combination of $dN_{i,j}(t)$. Hence

$$\Delta_{i,j}(\widehat{\boldsymbol{\beta}}) = \left[\sum_{l=1}^{J}\sum_{k=1}^{n^{(l)}} \int_t \xi_{k,l}\, w_{k,l} \left\{\frac{S_2(t;\widehat{\boldsymbol{\beta}})}{S_0(t;\widehat{\boldsymbol{\beta}})} - \frac{S_1(t;\widehat{\boldsymbol{\beta}})S_1(t;\widehat{\boldsymbol{\beta}})'}{S_0(t;\widehat{\boldsymbol{\beta}})^2}\right\} dN_{k,l}(t)\right]^{-1} \times \left[\int_t \left\{\boldsymbol{X}_{i,j} - \frac{S_1(t;\widehat{\boldsymbol{\beta}})}{S_0(t;\widehat{\boldsymbol{\beta}})}\right\} \times \right.$$

$$\left.\left\{\xi_{i,j}\, w_{i,j}\, dN_{i,j}(t) - \xi_{i,j}\, w_{i,j}\, Y_{i,j}(t) \exp(\widehat{\boldsymbol{\beta}}'\boldsymbol{X}_{i,j})\frac{\sum_{l=1}^{J}\sum_{k=1}^{n^{(l)}} \xi_{k,l}\, w_{k,l}\, dN_{k,l}(t)}{S_0(t;\widehat{\boldsymbol{\beta}})}\right\}\right],$$

and we can write $\Delta_{i,j}(\widehat{\boldsymbol{\beta}}) = \xi_{i,j}\, w_{i,j}\, \boldsymbol{IF}^{(2)}_{i,j}(\widehat{\boldsymbol{\beta}})$, with

$$\boldsymbol{IF}^{(2)}_{i,j}(\widehat{\boldsymbol{\beta}}) = \left[\sum_{l=1}^{J}\sum_{k=1}^{n^{(l)}} \int_t \xi_{k,l}\, w_{k,l}\left\{\frac{S_2(t;\widehat{\boldsymbol{\beta}})}{S_0(t;\widehat{\boldsymbol{\beta}})} - \frac{S_1(t;\widehat{\boldsymbol{\beta}})S_1(t;\widehat{\boldsymbol{\beta}})'}{S_0(t;\widehat{\boldsymbol{\beta}})^2}\right\} dN_{k,l}(t)\right]^{-1} \left[\int_t \left\{\boldsymbol{X}_{i,j} - \frac{S_1(t;\widehat{\boldsymbol{\beta}})}{S_0(t;\widehat{\boldsymbol{\beta}})}\right\} \times \right.$$

$$\left.\left\{dN_{i,j}(t) - Y_{i,j}(t) \exp(\widehat{\boldsymbol{\beta}}'\boldsymbol{X}_{i,j})\frac{\sum_{l=1}^{J}\sum_{k=1}^{n^{(l)}} \xi_{k,l}\, w_{k,l}\, dN_{k,l}(t)}{S_0(t;\widehat{\boldsymbol{\beta}})}\right\}\right].$$

Then we know $\sum_{j=1}^{J}\sum_{i=1}^{n^{(j)}}\{dN_{i,j}(t)\} - d\widehat{\Lambda}_0(t) \times S_0(t;\widehat{\boldsymbol{\beta}}) = 0$, as we estimate the baseline hazard non-parametrically by $d\widehat{\Lambda}_0(t) = \frac{\sum_{j=1}^{J}\sum_{i=1}^{n^{(j)}} dN_{i,j}(t)}{S_0(t;\widehat{\boldsymbol{\beta}})}$. Thus, following similar arguments as for

$\Delta_{i,j}(\widehat{\boldsymbol{\beta}})$, we have $dN_{i,j}(t) - d\widehat{\Lambda}_0(t)\Delta_{i,j}\{S_0(t;\widehat{\boldsymbol{\beta}})\} - S_0(t;\widehat{\boldsymbol{\beta}})\Delta_{i,j}\{d\widehat{\Lambda}_0(t)\} = 0$. In other words,

$$dN_{i,j}(t) - d\widehat{\Lambda}_0(t)\, \xi_{i,j}\, w_{i,j}Y_{i,j}(t) \exp(\widehat{\boldsymbol{\beta}}'\boldsymbol{X}_{i,j}) - d\widehat{\Lambda}_0(t)\boldsymbol{S}_1(t;\widehat{\boldsymbol{\beta}})'\Delta_{i,j}(\widehat{\boldsymbol{\beta}}) -$$

$$S_0(t;\widehat{\boldsymbol{\beta}})\Delta_{i,j}\{d\widehat{\Lambda}_0(t)\} = 0,$$

and as a result

$$\Delta_{i,j}\{d\widehat{\Lambda}_0(t)\} = \{S_0(t;\widehat{\boldsymbol{\beta}})\}^{-1}\big[dN_{i,j}(t) - d\widehat{\Lambda}_0(t)\, \xi_{i,j}\, w_{i,j}\, Y_{i,j}(t) \exp(\widehat{\boldsymbol{\beta}}'\boldsymbol{X}_{i,j}) -$$

$$d\widehat{\Lambda}_0(t)\boldsymbol{S}_1(t;\widehat{\boldsymbol{\beta}})'\Delta_{i,j}(\widehat{\boldsymbol{\beta}})\big].$$

We can thus write $\Delta_{i,j}\{d\widehat{\Lambda}_0(t)\} = \xi_{i,j} w_{i,j} IF^{(2)}_{i,j}\{d\widehat{\Lambda}_0(t)\}$, with $IF^{(2)}_{i,j}\{d\widehat{\Lambda}_0(t)\} = \{S_0(t;\widehat{\boldsymbol{\beta}})\}^{-1}\{dN_{i,j}(t) - d\widehat{\Lambda}_0(t) Y_{i,j}(t) \exp(\widehat{\boldsymbol{\beta}}'\boldsymbol{X}_{i,j}) - d\widehat{\Lambda}_0(t) \boldsymbol{S}_1(t;\widehat{\boldsymbol{\beta}})' IF^{(2)}_{i,j}(\widehat{\boldsymbol{\beta}})\}$, as $\xi_{i,j} w_{i,j} = 1$ if $dN_{i,j}(t) = 1$.

Both $IF^{(2)}_{i,j}\{d\widehat{\Lambda}_0(t)\}$ and $\Delta_{i,j}\{d\widehat{\Lambda}_0(t)\}$ are linear combination of the increments $dN_{i,j}(t)$ and $d\widehat{\Lambda}_0(t)$. Thus $\Delta_{i,j}\left\{\int_{\tau_1}^{\tau_2} d\widehat{\Lambda}_0(t)\right\} = \int_{\tau_1}^{\tau_2} \Delta_{i,j}\{d\widehat{\Lambda}_0(t)\}$, that we can rewrite $\xi_{i,j} w_{i,j} IF^{(2)}_{i,j}\left\{\int_{\tau_1}^{\tau_2} d\widehat{\Lambda}_0(t)\right\}$, with $IF^{(2)}_{i,j}\left\{\int_{\tau_1}^{\tau_2} d\widehat{\Lambda}_0(t)\right\} = \int_{\tau_1}^{\tau_2} IF^{(2)}_{i,j}\{d\widehat{\Lambda}_0(t)\}$. Finally,

$$\Delta_{i,j}\{\widehat{\pi}(\tau_1, \tau_2; \boldsymbol{x})\} = \left\{\frac{\partial \widehat{\pi}(\tau_1, \tau_2; \boldsymbol{x})}{\partial \boldsymbol{\beta}}\bigg|_{\boldsymbol{\beta}=\widehat{\boldsymbol{\beta}}}\right\} \Delta_{i,j}(\widehat{\boldsymbol{\beta}}) + \left[\frac{\partial \widehat{\pi}(\tau_1, \tau_2; \boldsymbol{x})}{\partial \{\int_{\tau_1}^{\tau_2} d\Lambda_0(t)\}}\bigg|_{d\Lambda_0(t)=d\widehat{\Lambda}_0(t)}\right] \Delta_{i,j}\left\{\int_{\tau_1}^{\tau_2} d\widehat{\Lambda}_0(t)\right\},$$

and we can rewrite $\Delta_{i,j}\{\widehat{\pi}(\tau_1, \tau_2; \boldsymbol{x})\} = \xi_{i,j} w_{i,j} IF^{(2)}_{i,j}\{\widehat{\pi}(\tau_1, \tau_2; \boldsymbol{x})\}$ with

$$\frac{\partial \widehat{\pi}(\tau_1, \tau_2; \boldsymbol{x})}{\partial \boldsymbol{\beta}}\bigg|_{\boldsymbol{\beta}=\widehat{\boldsymbol{\beta}}} = \left\{\int_{\tau_1}^{\tau_2} d\widehat{\Lambda}_0(t) \exp(\widehat{\boldsymbol{\beta}}'\boldsymbol{x})\right\} \exp\left\{-\int_{\tau_1}^{\tau_2} d\widehat{\Lambda}_0(t) \exp(\widehat{\boldsymbol{\beta}}'\boldsymbol{x})\right\} \boldsymbol{x}',$$

$$= \left\{\int_{\tau_1}^{\tau_2} d\widehat{\Lambda}_0(t) \exp(\widehat{\boldsymbol{\beta}}'\boldsymbol{x})\right\}\{1 - \widehat{\pi}(\tau_1, \tau_2; \boldsymbol{x})\} \boldsymbol{x}', \text{ and}$$

$$\frac{\partial \widehat{\pi}(\tau_1, \tau_2; \boldsymbol{x})}{\partial \int_{\tau_1}^{\tau_2} d\Lambda_0(t)}\bigg|_{d\Lambda_0(t)=d\widehat{\Lambda}_0(t)} = \exp(\widehat{\boldsymbol{\beta}}'\boldsymbol{x}) \exp\left\{-\int_{\tau_1}^{\tau_2} d\widehat{\Lambda}_0(t) \exp(\widehat{\boldsymbol{\beta}}'\boldsymbol{x})\right\},$$

$$= \exp(\widehat{\boldsymbol{\beta}}'\boldsymbol{x})\{1 - \widehat{\pi}(\tau_1, \tau_2; \boldsymbol{x})\}.$$

Observe that for any $\widehat{\boldsymbol{\theta}} \in \{\widehat{\boldsymbol{\beta}}, d\widehat{\Lambda}_0(t), \widehat{\Lambda}_0(t), \widehat{\pi}(\tau_1, \tau_2; \boldsymbol{x})\}$, if subject $i$ in stratum $j$ is not in the stratified case-cohort, then $\boldsymbol{\Delta}_{i,j}(\widehat{\boldsymbol{\theta}})$ is zero, $i \in \{1, ..., n^{(j)}\}, j \in \{1, ..., J\}$.

### C.2 Comparison with the variance of Samuelsen, Ånestad and Skrondal (2007)

Although they do not mention influence functions, the variance estimate for $\widehat{\boldsymbol{\beta}}$ from Equation (14) in Section 3.3 in the Main Document is very close to that of Samuelsen et al. (2007). There are two subtle differences. The first is that Samuelsen et al. (2007), following Borgan et al. (2000), redefined the strata by excluding the cases; we do not, and the non-cases stratum-specific design weights thus depend on the number of individuals sampled in each stratum, not the actual number of non-cases sampled in each stratum. Then, the phase-two component proposed by Samuelsen et al. (2007), with a slight modification of their notation so that it is closer to ours, is

$$\sum_{j=1}^{J} \frac{\widetilde{m}^{(j)}\left(1-\frac{\widetilde{m}^{(j)}}{\widetilde{n}^{(j)}}\right)}{\widetilde{m}^{(j)}-1} \sum_{i=1}^{\widetilde{n}^{(j)}} (\boldsymbol{D}_{i,j} - \overline{\boldsymbol{D}}_j)(\boldsymbol{D}_{i,j} - \overline{\boldsymbol{D}}_j)',$$

where $\widetilde{n}^{(j)}$ is the size of stratum $j$ after having removed all of the cases, and $\widetilde{m}^{(j)}$ is the number of non-cases sampled among the $m^{(j)}$ individuals sampled in stratum $j$. $\boldsymbol{D}_{i,j}$ is the analogue of our $\boldsymbol{\Delta}_{i,j}(\widehat{\boldsymbol{\beta}})$, and $\overline{\boldsymbol{D}}_j = \frac{1}{\widetilde{m}^{(j)}} \sum_{i=1}^{\widetilde{n}^{(j)}} \boldsymbol{D}_{i,j}$. This quantity equals

$$\sum_{j=1}^{J} \frac{\widetilde{m}^{(j)}\left(1-\frac{\widetilde{m}^{(j)}}{\widetilde{n}^{(j)}}\right)}{\widetilde{m}^{(j)}-1} \left( \sum_{i=1}^{\widetilde{n}^{(j)}} \boldsymbol{D}_{i,j}\boldsymbol{D}_{i,j}' - \frac{1}{\widetilde{m}^{(j)}} \sum_{i=1}^{\widetilde{n}^{(j)}} \sum_{k=1}^{\widetilde{n}^{(j)}} \boldsymbol{D}_{i,j}\boldsymbol{D}_{k,j}' \right)$$

$$= \sum_{j=1}^{J} \frac{\widetilde{m}^{(j)}\left(1-\frac{\widetilde{m}^{(j)}}{\widetilde{n}^{(j)}}\right)}{\widetilde{m}^{(j)}-1} \left( \frac{\widetilde{m}^{(j)}-1}{\widetilde{m}^{(j)}} \sum_{i=1}^{\widetilde{n}^{(j)}} \boldsymbol{D}_{i,j}\boldsymbol{D}_{i,j}' - \frac{1}{\widetilde{m}^{(j)}} \sum_{i=1}^{\widetilde{n}^{(j)}} \sum_{\substack{k=1,\\ k\neq i}}^{\widetilde{n}^{(j)}} \boldsymbol{D}_{i,j}\boldsymbol{D}_{k,j}' \right),$$

$$= \sum_{j=1}^{J} \left\{ \left(1 - \frac{\widetilde{m}^{(j)}}{\widetilde{n}^{(j)}}\right) \sum_{i=1}^{\widetilde{n}^{(j)}} \boldsymbol{D}_{i,j}\boldsymbol{D}_{i,j}' - \frac{1}{\widetilde{m}^{(j)}-1} \left(1 - \frac{\widetilde{m}^{(j)}}{\widetilde{n}^{(j)}}\right) \sum_{i=1}^{\widetilde{n}^{(j)}} \sum_{\substack{k=1,\\ k\neq i}}^{\widetilde{n}^{(j)}} \boldsymbol{D}_{i,j}\boldsymbol{D}_{k,j}' \right\}.$$

Note that the sum in the phase-two component of the variance reduces to a sum only over the sampled non-cases, but $\left(1 - \frac{\widetilde{m}^{(j)}}{\widetilde{n}^{(j)}}\right)$ and $\frac{1}{\widetilde{m}^{(j)}-1}\left(1 - \frac{\widetilde{m}^{(j)}}{\widetilde{n}^{(j)}}\right)$ would be $\left(1 - \frac{m^{(j)}}{n^{(j)}}\right)$ and $\frac{1}{m^{(j)}-1}\left(1 - \frac{m^{(j)}}{n^{(j)}}\right)$, respectively, in our Equation (3) in the Main Document.

Second, we use the influence functions to estimate the phase-one (superpopulation) component of the variance of $\widehat{\boldsymbol{\beta}}$, whereas Samuelsen et al. (2007) use the inverse of the observed information matrix.

# Web Appendix D. ESTIMATION USING CALIBRATED WEIGHTS FOR THE STRATIFIED CASE-COHORT DESIGN

## D.1 Estimating equation for calibration

Let $\boldsymbol{A}_{i,j}$ be the vector of $q$ auxiliary variables for individual $i$ in stratum $j$, $i \in \{1, \ldots, n^{(j)}\}$, $j \in \{1, \ldots, J\}$. Following Breslow et al. (2009a) and Shin et al. (2020), we obtained the auxiliary variables from (*i*) a variable that is identically equal to 1 for every individual in the cohort; (*ii*) the influences for the log-relative hazard parameters from the Cox model with imputed cohort data; and (*iii*) the products of total follow-up time (on the time interval for which the pure risk is to be estimated) and relative hazard for the imputed cohort data and with the log-relative hazard parameters estimated from the Cox model with stratified case-cohort data and weights calibrated with (*ii*). The imputed cohort data is obtained from using phase-one data to impute covariate values (for covariates measured only in phase-two) for all members of the cohort. See Breslow et al. (2009a) and Shin et al. (2020) for details on the construction of the auxiliary variables.

We want to use the calibrated weights $\left(w_{i,j}^*\right)_{i \in \{1,\ldots,n^{(j)}\}, j \in \{1,\ldots,J\}}$ that are as close as possible to the design weights $\left(w_{i,j}\right)_{i \in \{1,\ldots,n^{(j)}\}, j \in \{1,\ldots,J\}}$, and such that we correctly estimate the totals of the auxiliary variables (i.e., such that $\sum_{j=1}^{J} \sum_{i=1}^{n^{(j)}} \xi_{i,j} w_{i,j}^* \boldsymbol{A}_{i,j} = \sum_{j=1}^{J} \sum_{i=1}^{n^{(j)}} \boldsymbol{A}_{i,j}$). In other words, we want to minimize $\sum_{j=1}^{J} \sum_{i=1}^{n^{(j)}} \xi_{i,j} \, \delta(w_{i,j}^*, w_{i,j})$, for $\delta$ a given distance, under the constraint $\sum_{j=1}^{J} \sum_{i=1}^{n^{(j)}} (\xi_{i,j} w_{i,j}^* \boldsymbol{A}_{i,j} - \boldsymbol{A}_{i,j}) = 0$. To solve this constrained optimization problem, we consider $\boldsymbol{\eta}$, a vector of $q$ Lagrange multipliers, and look for a stationary point of the Lagrangian function $L: \left(w_{1,1}^*, \ldots, w_{n^{(J)},J}^*, \eta\right) \mapsto \sum_{j=1}^{J} \sum_{i=1}^{n^{(J)}} \{\xi_{i,j}\ \delta(w_{i,j}^*, w_{i,j}) + \boldsymbol{\eta}'(\xi_{i,j}\ w_{i,j}^* \boldsymbol{A}_{i,j} - \boldsymbol{A}_{i,j})\}$. To do so, we solve the $n+1$ following equations, obtained from the gradient of $L$:

$$\begin{cases} \xi_{1,1} \left\{ \dfrac{\partial \delta(w_{1,1}^*, w_{1,1})}{\partial w_{1,1}^*} + \boldsymbol{\eta}' \boldsymbol{A}_{1,1} \right\} = 0, \\ \quad \vdots \\ \xi_{n^{(J)},J} \left\{ \dfrac{\partial \delta(w_{n^{(J)},J}^*, w_{n^{(J)},J})}{\partial w_{n^{(J)},J}^*} + \boldsymbol{\eta}' \boldsymbol{A}_{n^{(J)},J} \right\} = 0, \\ \sum_{j=1}^{J} \sum_{i=1}^{n^{(j)}} (\xi_{i,j}\ w_{i,j}^* \boldsymbol{A}_{i,j} - \boldsymbol{A}_{i,j}) = 0. \end{cases}$$

We use $\delta: (a,b) \mapsto a \log\left(\dfrac{a}{b}\right) + b - a$ as distance measure, which leads to the raking procedure (Deville and Sarndal, 1992; Breslow et al., 2009) and to solving the estimating equation

$$\sum_{j=1}^{J} \sum_{i=1}^{n^{(j)}} \{\xi_{i,j}\ w_{i,j} \exp(\boldsymbol{\eta}' \boldsymbol{A}_{i,j}) \boldsymbol{A}_{i,j} - \boldsymbol{A}_{i,j}\} = 0.$$

From the estimate $\widehat{\boldsymbol{\eta}}$, we then obtain the calibrated weights $w_{i,j}^* = w_{i,j} \exp(\widehat{\boldsymbol{\eta}}' \boldsymbol{A}_{i,j})$, $i \in \{1, \ldots, n^{(j)}\}, j \in \{1, \ldots, J\}$.

**D.2 Derivation of the influence functions**

Let $\boldsymbol{\Delta}_{i,j}(\widehat{\boldsymbol{\theta}}^*)$ denote the influence of subject $i$ in stratum $j$ on $\widehat{\boldsymbol{\theta}}^*$, $i \in \{1, \ldots, n^{(j)}\}, j \in \{1, \ldots, J\}$, $\widehat{\boldsymbol{\theta}}^* \in \{\widehat{\boldsymbol{\eta}}, \widehat{\boldsymbol{\beta}}^*, d\widehat{\Lambda}_0^*(t), \widehat{\Lambda}_0^*(t), \widehat{\pi}^*(\tau_1, \tau_2; x)\}$. As $\sum_{j=1}^{J} \sum_{i=1}^{n^{(j)}} \{\xi_{i,j}\ w_{i,j} \exp(\widehat{\boldsymbol{\eta}}' \boldsymbol{A}_{i,j}) \boldsymbol{A}_{i,j} - \boldsymbol{A}_{i,j}\} = 0$, then, following Graubard and Fears (2005)

$$\boldsymbol{A}_{i,j} - \xi_{i,j}\ w_{i,j} \exp(\widehat{\boldsymbol{\eta}}' \boldsymbol{A}_{i,j})\ \boldsymbol{A}_{i,j} - \left\{ \sum_{l=1}^{J} \sum_{k=1}^{n^{(l)}} \xi_{k,l}\ w_{k,l} \exp(\widehat{\boldsymbol{\eta}}' \boldsymbol{A}_{k,l})\ \boldsymbol{A}_{k,l} \boldsymbol{A}_{k,l}' \right\} \boldsymbol{\Delta}_{i,j}(\widehat{\boldsymbol{\eta}}) = 0,$$

and as a result

$$\boldsymbol{\Delta}_{i,j}(\widehat{\boldsymbol{\eta}}) = \left\{ \sum_{l=1}^{J} \sum_{k=1}^{n^{(l)}} \xi_{k,l}\ w_{k,l} \exp(\widehat{\boldsymbol{\eta}}' \boldsymbol{A}_{k,l})\ \boldsymbol{A}_{k,l} \boldsymbol{A}_{k,l}' \right\}^{-1} \{\boldsymbol{A}_{i,j} - \xi_{i,j}\ w_{i,j} \exp(\widehat{\boldsymbol{\eta}}' \boldsymbol{A}_{i,j})\ \boldsymbol{A}_{i,j}\}.$$

We can then write

$$\boldsymbol{\Delta}_{i,j}(\widehat{\boldsymbol{\eta}}) = \boldsymbol{IF}_{i,j}^{(1)}(\widehat{\boldsymbol{\eta}}) + \xi_{i,j}\ w_{i,j}\ \boldsymbol{IF}_{i,j}^{(2)}(\widehat{\boldsymbol{\eta}}),$$

with $IF_{i,j}^{(1)}(\widehat{\boldsymbol{\eta}}) = \left\{\sum_{l=1}^{J}\sum_{k=1}^{n^{(j)}} \xi_{k,l}\, w_{k,l}\, \exp(\widehat{\boldsymbol{\eta}}' \boldsymbol{A}_{k,l})\, \boldsymbol{A}_{k,l}\boldsymbol{A}_{k,l}'\right\}^{-1} \boldsymbol{A}_{i,j}$,

and $IF_{i,j}^{(2)}(\widehat{\boldsymbol{\eta}}) = -\exp(\widehat{\boldsymbol{\eta}}' \boldsymbol{A}_{i,j})\left\{\sum_{l=1}^{J}\sum_{k=1}^{n^{(j)}} \xi_{k,l}\, w_{k,l}\, \exp(\widehat{\boldsymbol{\eta}}' \boldsymbol{A}_{k,l})\, \boldsymbol{A}_{k,l}\boldsymbol{A}_{k,l}'\right\}^{-1} \boldsymbol{A}_{i,j}$.

Then we can write the estimating equation

$$\sum_{j=1}^{J}\sum_{i=1}^{n^{(j)}} \int_t \xi_{i,j}\, w_{i,j}\, \exp(\widehat{\boldsymbol{\eta}}' \boldsymbol{A}_{i,j})\left\{\boldsymbol{X}_{i,j} - \frac{S_1^*(t;\widehat{\boldsymbol{\eta}},\widehat{\boldsymbol{\beta}}^*)}{S_0^*(t;\widehat{\boldsymbol{\eta}},\widehat{\boldsymbol{\beta}}^*)}\right\} dN_{i,j}(t) = 0,$$

as

$$\boldsymbol{G}_1^*(\widehat{\boldsymbol{\eta}}) - \int_t \frac{S_1^*(t;\widehat{\boldsymbol{\eta}},\widehat{\boldsymbol{\beta}}^*)}{S_0^*(t;\widehat{\boldsymbol{\eta}},\widehat{\boldsymbol{\beta}}^*)}\left\{\sum_{l=1}^{J}\sum_{k=1}^{n^{(l)}} \xi_{k,l}\, w_{k,l}\, \exp(\widehat{\boldsymbol{\eta}}' \boldsymbol{A}_{k,l})\, dN_{k,l}(t)\right\} = 0,$$

with $\boldsymbol{G}_1^*(\widehat{\boldsymbol{\eta}}) = \sum_{j=1}^{J}\sum_{i=1}^{n^{(j)}} \int_t \xi_{i,j}\, w_{i,j}\, \exp(\widehat{\boldsymbol{\eta}}' \boldsymbol{A}_{i,j})\, dN_{i,j}(t)\, \boldsymbol{X}_{i,j}$.

Following similar arguments as for $\boldsymbol{\Delta}_{i,j}(\widehat{\boldsymbol{\eta}})$, we have

$$\boldsymbol{\Delta}_{i,j}\{\boldsymbol{G}_1^*(\widehat{\boldsymbol{\eta}})\} + \int_t \left[-\frac{\Delta_{i,j}\left\{\sum_{l=1}^{J}\sum_{k=1}^{n^{(l)}} \xi_{k,l}\, w_{k,l}\, \exp(\widehat{\boldsymbol{\eta}}' \boldsymbol{A}_{k,l})\, dN_{k,l}(t)\right\}\times S_1^*(t;\widehat{\boldsymbol{\eta}},\widehat{\boldsymbol{\beta}}^*)}{S_0^*(t;\widehat{\boldsymbol{\eta}},\widehat{\boldsymbol{\beta}}^*)} - \right.$$

$$\frac{\left\{\sum_{l=1}^{J}\sum_{k=1}^{n^{(l)}} \xi_{k,l}\, w_{k,l}\, \exp(\widehat{\boldsymbol{\eta}}' \boldsymbol{A}_{k,l})\, dN_{k,l}(t)\right\}\times \Delta_{i,j}\{S_1^*(t;\widehat{\boldsymbol{\eta}},\widehat{\boldsymbol{\beta}}^*)\}}{S_0^*(t;\widehat{\boldsymbol{\eta}},\widehat{\boldsymbol{\beta}}^*)} +$$

$$\left.\frac{\left\{\sum_{l=1}^{J}\sum_{k=1}^{n^{(l)}} \xi_{k,l}\, w_{k,l}\, \exp(\widehat{\boldsymbol{\eta}}' \boldsymbol{A}_{k,l})\, dN_{k,l}(t)\right\}\times S_1^*(t;\widehat{\boldsymbol{\eta}},\widehat{\boldsymbol{\beta}}^*)\times \Delta_{i,j}\{S_0^*(t;\widehat{\boldsymbol{\eta}},\widehat{\boldsymbol{\beta}}^*)\}}{S_0^*(t;\widehat{\boldsymbol{\eta}},\widehat{\boldsymbol{\beta}}^*)^2}\right] = 0, \quad (2)$$

with

$$\boldsymbol{\Delta}_{i,j}\{\boldsymbol{G}_1^*(\widehat{\boldsymbol{\eta}})\} = \int_t \xi_{i,j}\, w_{i,j}\, \exp(\widehat{\boldsymbol{\eta}}' \boldsymbol{A}_{i,j})\, dN_{i,j}(t)\, \boldsymbol{X}_{i,j} + \left\{\frac{\partial \boldsymbol{G}_1^*(\boldsymbol{\eta})}{\partial \boldsymbol{\eta}}\bigg|_{\boldsymbol{\eta}=\widehat{\boldsymbol{\eta}}}\right\} \boldsymbol{\Delta}_{i,j}(\widehat{\boldsymbol{\eta}}),$$

$$\Delta_{i,j}\left\{\sum_{l=1}^{J}\sum_{k=1}^{n^{(l)}} \xi_{k,l}\, w_{k,l}\, \exp(\widehat{\boldsymbol{\eta}}' \boldsymbol{A}_{k,l})\, dN_{k,l}(t)\right\} = \xi_{i,j}\, w_{i,j}\, \exp(\widehat{\boldsymbol{\eta}}' \boldsymbol{A}_{i,j})\, dN_{i,j}(t) +$$

$$\left[\frac{\partial\left\{\sum_{l=1}^{J}\sum_{k=1}^{n^{(l)}} \xi_{k,l}\, w_{k,l}\, \exp(\boldsymbol{\eta}' \boldsymbol{A}_{k,l})\, dN_{k,l}(t)\right\}}{\partial \boldsymbol{\eta}}\bigg|_{\boldsymbol{\eta}=\widehat{\boldsymbol{\eta}}}\right] \boldsymbol{\Delta}_{i,j}(\widehat{\boldsymbol{\eta}}),$$

$$\frac{\partial \boldsymbol{G}_1^*(\boldsymbol{\eta})}{\partial \boldsymbol{\eta}}\bigg|_{\boldsymbol{\eta}=\widehat{\boldsymbol{\eta}}} = \sum_{l=1}^{J}\sum_{k=1}^{n^{(l)}} \int_t \xi_{k,l}\, w_{k,l}\, \exp(\widehat{\boldsymbol{\eta}}' \boldsymbol{A}_{k,l})\, \boldsymbol{X}_{k,l}\boldsymbol{A}_{k,l}'\, dN_{k,l}(t),$$

$$\frac{\partial\left\{\sum_{l=1}^{J}\sum_{k=1}^{n^{(l)}}\xi_{k,l}\,w_{k,l}\exp(\boldsymbol{\eta}'\boldsymbol{A}_{k,l})\,dN_{k,l}(t)\right\}}{\partial\boldsymbol{\eta}}\bigg|_{\boldsymbol{\eta}=\widehat{\boldsymbol{\eta}}}=\sum_{l=1}^{J}\sum_{k=1}^{n^{(l)}}\xi_{k,l}\,w_{k,l}\exp(\widehat{\boldsymbol{\eta}}'\boldsymbol{A}_{k,l})\,\boldsymbol{A}_{k,l}'\,dN_{k,l}(t),$$

$$\Delta_{i,j}\{S_1^*(t;\widehat{\boldsymbol{\eta}},\widehat{\boldsymbol{\beta}}^*)\}=\xi_{i,j}\,w_{i,j}\exp(\widehat{\boldsymbol{\eta}}'\boldsymbol{A}_{i,j})\,Y_{i,j}(t)\exp\left(\widehat{\boldsymbol{\beta}}^{*'}\boldsymbol{X}_{i,j}\right)\boldsymbol{X}_{i,j}+\left\{\frac{\partial S_1^*(t;\widehat{\boldsymbol{\eta}},\boldsymbol{\beta})}{\partial\boldsymbol{\beta}}\bigg|_{\boldsymbol{\beta}=\widehat{\boldsymbol{\beta}}^*}\right\}\Delta_{i,j}(\widetilde{\boldsymbol{\beta}})+$$

$$\left\{\frac{\partial S_1^*(t;\boldsymbol{\eta},\widehat{\boldsymbol{\beta}}^*)}{\partial\boldsymbol{\eta}}\bigg|_{\boldsymbol{\eta}=\widehat{\boldsymbol{\eta}}}\right\}\Delta_{i,j}(\widehat{\boldsymbol{\eta}}),$$

$$\frac{\partial S_1^*(t;\widehat{\boldsymbol{\eta}},\boldsymbol{\beta})}{\partial\boldsymbol{\beta}}\bigg|_{\boldsymbol{\beta}=\widehat{\boldsymbol{\beta}}^*}=\sum_{l=1}^{J}\sum_{k=1}^{n^{(l)}}\xi_{k,l}\,w_{k,l}\exp(\widehat{\boldsymbol{\eta}}'\boldsymbol{A}_{k,l})\,Y_{k,l}(t)\exp\left(\widehat{\boldsymbol{\beta}}^{*'}\boldsymbol{X}_{k,l}\right)\boldsymbol{X}_{k,l}\,\boldsymbol{X}_{k,l}'\equiv S_2^*(t;\widehat{\boldsymbol{\eta}},\widehat{\boldsymbol{\beta}}^*),$$

$$\frac{\partial S_1^*(t;\boldsymbol{\eta},\widehat{\boldsymbol{\beta}}^*)}{\partial\boldsymbol{\eta}}\bigg|_{\boldsymbol{\eta}=\widehat{\boldsymbol{\eta}}}=\sum_{l=1}^{J}\sum_{k=1}^{n^{(l)}}\xi_{k,l}\,w_{k,l}\exp(\widehat{\boldsymbol{\eta}}'\boldsymbol{A}_{k,l})\,Y_{k,l}(t)\exp\left(\widehat{\boldsymbol{\beta}}^{*'}\boldsymbol{X}_{k,l}\right)\boldsymbol{X}_{k,l}\,\boldsymbol{A}_{k,l}',$$

$$\Delta_{i,j}\{S_0^*(t;\widehat{\boldsymbol{\eta}},\widehat{\boldsymbol{\beta}}^*)\}=\xi_{i,j}\,w_{i,j}\exp(\widehat{\boldsymbol{\eta}}'\boldsymbol{A}_{i,j})\,Y_{i,j}(t)\exp\left(\widehat{\boldsymbol{\beta}}^{*'}\boldsymbol{X}_{i,j}\right)+\left\{\frac{\partial S_0^*(t;\widehat{\boldsymbol{\eta}},\boldsymbol{\beta})}{\partial\boldsymbol{\beta}}\bigg|_{\boldsymbol{\beta}=\widehat{\boldsymbol{\beta}}^*}\right\}\Delta_{i,j}(\widehat{\boldsymbol{\beta}}^*)+$$

$$\left\{\frac{\partial S_0^*(t;\boldsymbol{\eta},\widehat{\boldsymbol{\beta}}^*)}{\partial\boldsymbol{\eta}}\bigg|_{\boldsymbol{\eta}=\widehat{\boldsymbol{\eta}}}\right\}\Delta_{i,j}(\widehat{\boldsymbol{\eta}}),$$

$$\frac{\partial S_0^*(t;\widehat{\boldsymbol{\eta}},\boldsymbol{\beta})}{\partial\boldsymbol{\beta}}\bigg|_{\boldsymbol{\beta}=\widehat{\boldsymbol{\beta}}^*}=\sum_{l=1}^{J}\sum_{k=1}^{n^{(l)}}\xi_{k,l}\,w_{k,l}\exp(\widehat{\boldsymbol{\eta}}'\boldsymbol{A}_{k,l})\,Y_{k,l}(t)\exp\left(\widehat{\boldsymbol{\beta}}^{*'}\boldsymbol{X}_{k,l}\right)\boldsymbol{X}_{k,l}'=S_1^*(t;\widehat{\boldsymbol{\eta}},\widehat{\boldsymbol{\beta}}^*)',$$

and $\frac{\partial S_0^*(t;\boldsymbol{\eta},\widehat{\boldsymbol{\beta}}^*)}{\partial\boldsymbol{\eta}}\bigg|_{\boldsymbol{\eta}=\widehat{\boldsymbol{\eta}}}=\sum_{l=1}^{J}\sum_{k=1}^{n^{(l)}}\xi_{k,l}\,w_{k,l}\exp(\widehat{\boldsymbol{\eta}}'\boldsymbol{A}_{k,l})\,Y_{k,l}(t)\exp\left(\widehat{\boldsymbol{\beta}}^{*'}\boldsymbol{X}_{k,l}\right)\boldsymbol{A}_{k,l}'.$

Note, $\sum_{l=1}^{J}\sum_{k=1}^{n^{(l)}}\xi_{k,l}\,w_{k,l}\exp(\widehat{\boldsymbol{\eta}}'\boldsymbol{A}_{k,l})\,dN_{k,l}(t)$, $\frac{\partial\left\{\sum_{l=1}^{J}\sum_{k=1}^{n^{(l)}}\xi_{k,l}\,w_{k,l}\exp(\boldsymbol{\eta}'\boldsymbol{A}_{k,l})\,dN_{k,l}(t)\right\}}{\partial\boldsymbol{\eta}}\bigg|_{\boldsymbol{\eta}=\widehat{\boldsymbol{\eta}}}$ and thus $\Delta_{i,j}\left\{\sum_{l=1}^{J}\sum_{k=1}^{n^{(l)}}\xi_{k,l}\,w_{k,l}\exp(\widehat{\boldsymbol{\eta}}'\boldsymbol{A}_{k,l})\,dN_{k,l}(t)\right\}$ and the component between square brackets in Equation (2) above are linear combinations of the increments $dN_{i,j}$. Thus

$$\Delta_{i,j}(\widetilde{\boldsymbol{\beta}})=\left[\sum_{l=1}^{J}\sum_{k=1}^{n^{(l)}}\int_t\xi_{k,l}\,w_{k,l}\exp(\widehat{\boldsymbol{\eta}}'\boldsymbol{A}_{k,l})\left\{\frac{S_2^*(t;\widehat{\boldsymbol{\eta}},\widehat{\boldsymbol{\beta}}^*)}{S_0^*(t;\widehat{\boldsymbol{\eta}},\widehat{\boldsymbol{\beta}}^*)}-\frac{S_1^*(t;\widehat{\boldsymbol{\eta}},\widehat{\boldsymbol{\beta}}^*)S_1^*(t;\widehat{\boldsymbol{\eta}},\widehat{\boldsymbol{\beta}}^*)'}{S_0^*(t;\widehat{\boldsymbol{\eta}},\widehat{\boldsymbol{\beta}}^*)^2}\right\}dN_{k,l}(t)\right]^{-1}\times$$

$$\left[\int_t\left\{\boldsymbol{X}_{i,j}-\frac{S_1^*(t;\widehat{\boldsymbol{\eta}},\widehat{\boldsymbol{\beta}}^*)}{S_0^*(t;\widehat{\boldsymbol{\eta}},\widehat{\boldsymbol{\beta}}^*)}\right\}\times\left\{\xi_{i,j}\,w_{i,j}\exp(\widehat{\boldsymbol{\eta}}'\boldsymbol{A}_{i,j})\,dN_{i,j}(t)-\right.\right.$$

$$\left.\left.\xi_{i,j}\,w_{i,j}\exp(\widehat{\boldsymbol{\eta}}'\boldsymbol{A}_{i,j})\,Y_{i,j}(t)\exp\left(\widehat{\boldsymbol{\beta}}^{*'}\boldsymbol{X}_{i,j}\right)\frac{\sum_{l=1}^{J}\sum_{k=1}^{n^{(l)}}\xi_{k,l}\,w_{k,l}\exp(\widehat{\boldsymbol{\eta}}'\boldsymbol{A}_{k,l})\,dN_{k,l}(t)}{S_0^*(t;\widehat{\boldsymbol{\eta}},\widehat{\boldsymbol{\beta}}^*)}\right\}\right]+$$

$$\left[\sum_{l=1}^{J}\sum_{k=1}^{n^{(l)}}\int_{t}\xi_{k,l}\,w_{k,l}\exp(\widehat{\boldsymbol{\eta}}'\boldsymbol{A}_{k,l})\left\{\frac{S_{2}^{*}(t;\widehat{\boldsymbol{\eta}},\widehat{\boldsymbol{\beta}}^{*})}{S_{0}^{*}(t;\widehat{\boldsymbol{\eta}},\widehat{\boldsymbol{\beta}}^{*})}-\frac{S_{1}^{*}(t;\widehat{\boldsymbol{\eta}},\widehat{\boldsymbol{\beta}}^{*})S_{1}^{*}(t;\widehat{\boldsymbol{\eta}},\widehat{\boldsymbol{\beta}}^{*})'}{S_{0}^{*}(t;\widehat{\boldsymbol{\eta}},\widehat{\boldsymbol{\beta}}^{*})^{2}}\right\}dN_{k,l}(t)\right]^{-1}\times\left(\left\{\frac{\partial G_{1}^{*}(\boldsymbol{\eta})}{\partial\boldsymbol{\eta}}\bigg|_{\boldsymbol{\eta}=\widehat{\boldsymbol{\eta}}}\right\}+\right.$$

$$-\int_{t}\frac{S_{1}^{*}(t;\widehat{\boldsymbol{\eta}},\widehat{\boldsymbol{\beta}}^{*})}{S_{0}^{*}(t;\widehat{\boldsymbol{\eta}},\widehat{\boldsymbol{\beta}}^{*})}\times\left[\frac{\partial\left\{\sum_{l=1}^{J}\sum_{k=1}^{n^{(l)}}\xi_{k,l}\,w_{k,l}\exp(\boldsymbol{\eta}'\boldsymbol{A}_{k,l})\,dN_{k,l}(t)\right\}}{\partial\boldsymbol{\eta}}\bigg|_{\boldsymbol{\eta}=\widehat{\boldsymbol{\eta}}}\right]-\int_{t}\frac{\left\{\sum_{l=1}^{J}\sum_{k=1}^{n^{(l)}}\xi_{k,l}\,w_{k,l}\exp(\widehat{\boldsymbol{\eta}}'\boldsymbol{A}_{k,l})\,dN_{k,l}(t)\right\}}{S_{0}^{*}(t;\widehat{\boldsymbol{\eta}},\widehat{\boldsymbol{\beta}}^{*})}\times$$

$$\left\{\frac{\partial S_{1}^{*}(t;\boldsymbol{\eta},\widehat{\boldsymbol{\beta}}^{*})}{\partial\boldsymbol{\eta}}\bigg|_{\boldsymbol{\eta}=\widehat{\boldsymbol{\eta}}}\right\}+\int_{t}\frac{\left\{\sum_{l=1}^{J}\sum_{k=1}^{n^{(l)}}\xi_{k,l}\,w_{k,l}\exp(\widehat{\boldsymbol{\eta}}'\boldsymbol{A}_{k,l})\,dN_{k,l}(t)\right\}\tilde{S}_{1}(t;\widetilde{\boldsymbol{\gamma}},\widetilde{\boldsymbol{\beta}})}{S_{0}^{*}(t;\widehat{\boldsymbol{\eta}},\widehat{\boldsymbol{\beta}}^{*})^{2}}\times\left\{\frac{\partial S_{0}^{*}(t;\boldsymbol{\eta},\widehat{\boldsymbol{\beta}}^{*})}{\partial\boldsymbol{\eta}}\bigg|_{\boldsymbol{\eta}=\widehat{\boldsymbol{\eta}}}\right\}\right)\boldsymbol{\Delta}_{i,j}(\widetilde{\boldsymbol{\gamma}}),$$

that we can write as $\boldsymbol{\Delta}_{i,j}(\widehat{\boldsymbol{\beta}}^{*})=\boldsymbol{IF}_{i,j}^{(1)}(\widehat{\boldsymbol{\beta}}^{*})+\xi_{i,j}\,w_{i,j}\,\boldsymbol{IF}_{i,j}^{(2)}(\widehat{\boldsymbol{\beta}}^{*})$, with

$$\boldsymbol{IF}_{i,j}^{(1)}(\widehat{\boldsymbol{\beta}}^{*})=\left\{\sum_{l=1}^{J}\sum_{k=1}^{n^{(j)}}\xi_{k,l}\,w_{k,l}\exp(\widehat{\boldsymbol{\eta}}'\boldsymbol{A}_{k,l})\boldsymbol{Z}_{k,l}\boldsymbol{A}_{k,l}'\right\}\times\boldsymbol{IF}_{i,j}^{(1)}(\widehat{\boldsymbol{\eta}}),\text{ and}$$

$$\boldsymbol{IF}_{i,j}^{(2)}(\widehat{\boldsymbol{\beta}}^{*})=\exp(\widehat{\boldsymbol{\eta}}'\boldsymbol{A}_{i,j})\,\boldsymbol{Z}_{i,j}+\left\{\sum_{l=1}^{J}\sum_{k=1}^{n^{(j)}}\xi_{k,l}\,w_{k,l}\exp(\widehat{\boldsymbol{\eta}}'\boldsymbol{A}_{k,l})\,\boldsymbol{Z}_{k,l}\boldsymbol{A}_{k,l}'\right\}\times\boldsymbol{IF}_{i,j}^{(2)}(\widehat{\boldsymbol{\eta}}),$$

Where
$$\boldsymbol{Z}_{i,j}=\left[\sum_{l=1}^{J}\sum_{k=1}^{n^{(l)}}\int_{t}\xi_{k,l}\,w_{k,l}\exp(\widehat{\boldsymbol{\eta}}'\boldsymbol{A}_{k,l})\left\{\frac{S_{2}^{*}(t;\widehat{\boldsymbol{\eta}},\widehat{\boldsymbol{\beta}}^{*})}{S_{0}^{*}(t;\widehat{\boldsymbol{\eta}},\widehat{\boldsymbol{\beta}}^{*})}-\right.\right.$$

$$\left.\left.\frac{S_{1}^{*}(t;\widehat{\boldsymbol{\eta}},\widehat{\boldsymbol{\beta}}^{*})S_{1}^{*}(t;\widehat{\boldsymbol{\eta}},\widehat{\boldsymbol{\beta}}^{*})'}{S_{0}^{*}(t;\widehat{\boldsymbol{\eta}},\widehat{\boldsymbol{\beta}}^{*})^{2}}\right\}dN_{k,l}(t)\right]^{-1}\left[\int_{t}\left\{\boldsymbol{X}_{i,j}-\frac{S_{1}^{*}(t;\widehat{\boldsymbol{\eta}},\widehat{\boldsymbol{\beta}}^{*})}{S_{0}^{*}(t;\widehat{\boldsymbol{\eta}},\widehat{\boldsymbol{\beta}}^{*})}\right\}\times\left\{\xi_{i,j}\,w_{i,j}\exp(\widehat{\boldsymbol{\eta}}'\boldsymbol{A}_{i,j})\,dN_{i,j}(t)-\right.\right.$$

$$\left.\left.\xi_{i,j}\,w_{i,j}\exp(\widehat{\boldsymbol{\eta}}'\boldsymbol{A}_{i,j})\,Y_{i,j}(t)\exp(\widehat{\boldsymbol{\beta}}^{*'}\boldsymbol{X}_{i,j})\frac{\sum_{l=1}^{J}\sum_{k=1}^{n^{(l)}}\xi_{k,l}\,w_{k,l}\exp(\widehat{\boldsymbol{\eta}}'\boldsymbol{A}_{k,l})\,dN_{k,l}(t)}{S_{0}^{*}(t;\widehat{\boldsymbol{\eta}},\widehat{\boldsymbol{\beta}}^{*})}\right\}\right].$$

The estimating equation $\sum_{j=1}^{J}\sum_{i=1}^{n^{(j)}}\{dN_{i,j}(t)\}-d\widehat{\Lambda}_{0}^{*}(t)\times S_{0}^{*}(t;\widehat{\boldsymbol{\eta}},\widehat{\boldsymbol{\beta}}^{*})=0$ leads to the Breslow estimate $d\widehat{\Lambda}_{0}^{*}(t)=\frac{\sum_{j=1}^{J}\sum_{i=1}^{n^{(j)}}dN_{i,j}(t)}{S_{0}^{*}(t;\widehat{\boldsymbol{\eta}},\widehat{\boldsymbol{\beta}}^{*})}$. Thus

$$dN_{i,j}(t)-d\widehat{\Lambda}_{0}^{*}(t)\,\boldsymbol{\Delta}_{i,j}\{S_{0}^{*}(t;\widehat{\boldsymbol{\eta}},\widehat{\boldsymbol{\beta}}^{*})\}-S_{0}^{*}(t;\widehat{\boldsymbol{\eta}},\widehat{\boldsymbol{\beta}}^{*})\,\boldsymbol{\Delta}_{i,j}\{d\widehat{\Lambda}_{0}^{*}(t)\}=0,$$

or, in other words,

$$dN_{i,j}(t)-d\widehat{\Lambda}_{0}^{*}(t)\,\xi_{i,j}\,w_{i,j}\exp(\widehat{\boldsymbol{\eta}}'\boldsymbol{A}_{i,j})\,Y_{i,j}(t)\exp(\widehat{\boldsymbol{\beta}}^{*'}\boldsymbol{X}_{i,j})-d\widehat{\Lambda}_{0}^{*}(t)\left\{\frac{\partial S_{0}^{*}(t;\widehat{\boldsymbol{\eta}},\boldsymbol{\beta})}{\partial\boldsymbol{\beta}}\bigg|_{\boldsymbol{\beta}=\widehat{\boldsymbol{\beta}}^{*}}\right\}'\boldsymbol{\Delta}_{i,j}(\widehat{\boldsymbol{\beta}}^{*})$$

$$-d\widehat{\Lambda}_{0}^{*}(t)\left\{\frac{\partial S_{0}^{*}(t;\boldsymbol{\eta},\widehat{\boldsymbol{\beta}}^{*})}{\partial\boldsymbol{\eta}}\bigg|_{\boldsymbol{\eta}=\widehat{\boldsymbol{\eta}}}\right\}'\boldsymbol{\Delta}_{i,j}(\widehat{\boldsymbol{\eta}})-S_{0}^{*}(t;\widehat{\boldsymbol{\eta}},\widehat{\boldsymbol{\beta}}^{*})\,\boldsymbol{\Delta}_{i,j}\{d\widehat{\Lambda}_{0}^{*}(t)\}=0.$$

As a result

$$\Delta_{i,j}\{d\widehat{\Lambda}_0^*(t)\} = \{S_0^*(t;\widehat{\boldsymbol{\eta}},\widehat{\boldsymbol{\beta}}^*)\}^{-1}\Big[dN_{i,j}(t) - d\widehat{\Lambda}_0^*(t)\,\xi_{i,j}\,w_{i,j}\exp(\widehat{\boldsymbol{\eta}}'\boldsymbol{A}_{i,j})\,Y_{i,j}(t)\exp\left(\widehat{\boldsymbol{\beta}}^{*'}\boldsymbol{X}_{i,j}\right) -$$

$$d\widehat{\Lambda}_0^*(t)\,\boldsymbol{S}_1^*(t;\widehat{\boldsymbol{\eta}},\widehat{\boldsymbol{\beta}}^*)'\Delta_{i,j}(\widehat{\boldsymbol{\beta}}^*) -$$

$$d\widehat{\Lambda}_0^*(t)\left\{\sum_{l=1}^{J}\sum_{k=1}^{n^{(l)}}\xi_{k,l}\,w_{k,l}\exp(\widehat{\boldsymbol{\eta}}'\boldsymbol{A}_{k,l})\,Y_{k,l}(t)\exp\left(\widehat{\boldsymbol{\beta}}^{*'}\boldsymbol{X}_{k,l}\right)\boldsymbol{A}_{k,l}\right\}'\Delta_{i,j}(\widehat{\boldsymbol{\eta}})\Big],$$

that we can rewrite $\Delta_{i,j}\{d\widehat{\Lambda}_0^*(t)\} = IF_{i,j}^{(1)}\left(d\widehat{\Lambda}_0^*(t)\right) + \xi_{i,j}\,w_{i,j}\,IF_{i,j}^{(2)}\left(d\widehat{\Lambda}_0^*(t)\right)$,

with $IF_{i,j}^{(1)}\{d\widehat{\Lambda}_0^*(t)\} = \left\{\sum_{l=1}^{J}\sum_{k=1}^{n^{(l)}}\xi_{k,l}\,w_{k,l}\exp(\widehat{\boldsymbol{\eta}}'\boldsymbol{A}_{k,l})\,H_{k,l}(t)\,\boldsymbol{A}_{k,l}'\right\} \times \boldsymbol{IF}_{i,j}^{(1)}(\widehat{\boldsymbol{\eta}})$,

$IF_{i,j}^{(2)}\{d\widehat{\Lambda}_0^*(t)\} = \{S_0^*(t;\widehat{\boldsymbol{\eta}},\widehat{\boldsymbol{\beta}}^*)\}^{-1}dN_{i,j}(t) +$

$\exp(\widehat{\boldsymbol{\eta}}'\boldsymbol{A}_{i,j})\,H_{i,j}(t) + \left\{\sum_{l=1}^{J}\sum_{k=1}^{n^{(l)}}\xi_{k,l}\,w_{k,l}\exp(\widehat{\boldsymbol{\eta}}'\boldsymbol{A}_{k,l})\,H_{k,l}(t)\,\boldsymbol{A}_{k,l}'\right\} \times \boldsymbol{IF}_{i,j}^{(2)}(\widehat{\boldsymbol{\eta}})$,

where $\xi_{i,j}w_{i,j} = 1$ if $dN_{i,j}(t) = 1$, and $H_{i,j}(t) = -\{S_0^*(t;\widehat{\boldsymbol{\eta}},\widehat{\boldsymbol{\beta}}^*)\}^{-1}d\widehat{\Lambda}_0^*(t)\{\boldsymbol{S}_1^*(t;\widehat{\boldsymbol{\eta}},\widehat{\boldsymbol{\beta}}^*)'\boldsymbol{Z}_{i,j} +$

$K_{i,j}(t)\}$ and $K_{i,j}(t) = Y_{i,j}(t)\exp(\widehat{\boldsymbol{\beta}}^{*'}\boldsymbol{X}_{i,j})$.

Note, $H_{i,j}(t)$, and thus $IF_{i,j}^{(1)}\{d\widehat{\Lambda}_0^*(t)\}$ and $IF_{i,j}^{(2)}\{d\widehat{\Lambda}_0^*(t)\}$, are linear combinations of the increments $dN_{i,j}(t)$ and $d\widehat{\Lambda}_0^*(t)$. Thus $\Delta_{i,j}\left\{\int_{\tau_1}^{\tau_2}d\widehat{\Lambda}_0^*(t)\right\} = \int_{\tau_1}^{\tau_2}\Delta_{i,j}\{d\widehat{\Lambda}_0^*(t)\}$, that we can rewrite as

$IF_{i,j}^{(1)}\left\{\int_{\tau_1}^{\tau_2}d\widehat{\Lambda}_0^*(t)\right\} + \xi_{i,j}\,w_{i,j}\,IF_{i,j}^{(2)}\left\{\int_{\tau_1}^{\tau_2}d\widehat{\Lambda}_0^*(t)\right\}$, with $IF_{i,j}^{(1)}\{\widehat{\Lambda}_0^*(\tau_2)\} = \int_{\tau_1}^{\tau_2}IF_{i,j}^{(1)}\{d\widehat{\Lambda}_0^*(t)\}$, and

$IF_{i,j}^{(2)}\{\widehat{\Lambda}_0^*(\tau_2)\} = \int_{\tau_1}^{\tau_2}IF_{i,j}^{(2)}\{d\widehat{\Lambda}_0^*(t)\}$.

Finally

$$\Delta_{i,j}\{\widehat{\pi}^*(\tau_1,\tau_2;\boldsymbol{x})\} = \left\{\frac{\partial\widehat{\pi}^*(\tau_1,\tau_2;\boldsymbol{x})}{\partial\boldsymbol{\beta}}\bigg|_{\boldsymbol{\beta}=\widehat{\boldsymbol{\beta}}^*}\right\}\Delta_{i,j}(\widehat{\boldsymbol{\beta}}^*) + \left[\frac{\partial\widehat{\pi}^*(\tau_1,\tau_2;\boldsymbol{x})}{\partial\{\int_{\tau_1}^{\tau_2}d\Lambda_0(t)\}}\bigg|_{d\Lambda_0(t)=d\widehat{\Lambda}_0^*(t)}\right]\Delta_{i,j}\left\{\int_{\tau_1}^{\tau_2}d\widehat{\Lambda}_0^*(t)\right\},$$

with $\dfrac{\partial\widehat{\pi}^*(\tau_1,\tau_2;\boldsymbol{x})}{\partial\boldsymbol{\beta}}\bigg|_{\boldsymbol{\beta}=\widehat{\boldsymbol{\beta}}^*} = \left\{\int_{\tau_1}^{\tau_2}d\widehat{\Lambda}_0^*(t)\exp(\widehat{\boldsymbol{\beta}}^{*'}\boldsymbol{x})\right\}\{1 - \widehat{\pi}^*(\tau_1,\tau_2;\boldsymbol{x})\}\boldsymbol{x}'$,

and $\dfrac{\partial\widehat{\pi}^*(\tau_1,\tau_2;\boldsymbol{x})}{\partial\int_{\tau_1}^{\tau_2}d\Lambda_0(t)}\bigg|_{d\Lambda_0(t)=d\widehat{\Lambda}_0^*(t)} = \exp(\widehat{\boldsymbol{\beta}}'\boldsymbol{x})\{1 - \widehat{\pi}^*(\tau_1,\tau_2;\boldsymbol{x})\}$,

and we can rewrite $\Delta_{i,j}\{\hat{\pi}^*(\tau_1,\tau_2;\boldsymbol{x})\}$ as $IF_{i,j}^{(1)}\{\hat{\pi}^*(\tau_1,\tau_2;\boldsymbol{x})\} + \xi_{i,j}\, w_{i,j}\, IF_{i,j}^{(2)}\{\hat{\pi}^*(\tau_1,\tau_2;\boldsymbol{x})\}$.

### D.3 Comment on the choice of the auxiliary variables

As mentioned in Section 4.3 in the Main Document, for any $\widehat{\boldsymbol{\theta}}^* \in \{\widehat{\boldsymbol{\eta}}, \widehat{\boldsymbol{\beta}}^*, d\widehat{\Lambda}_0^*(t), \widehat{\Lambda}_0^*(t), \hat{\pi}^*(\tau_1,\tau_2;\boldsymbol{x})\}$, we estimate $\text{var}(\widehat{\boldsymbol{\theta}}^*)$ by

$$\frac{n}{n-1}\sum_{j=1}^{J}\sum_{i=1}^{n^{(j)}}\left\{\boldsymbol{IF}_{i,j}^{(1)}(\widehat{\boldsymbol{\theta}}^*)\boldsymbol{IF}_{i,j}^{(1)}(\widehat{\boldsymbol{\theta}}^*)' + 2\,\xi_{i,j}\,w_{i,j}\,\boldsymbol{IF}_{i,j}^{(2)}(\widehat{\boldsymbol{\theta}}^*)\boldsymbol{IF}_{i,j}^{(2)}(\widehat{\boldsymbol{\theta}}^*)' + \right.$$

$$\left. \xi_{i,j}\,w_{i,j}\,\boldsymbol{IF}_{i,j}^{(2)}(\widehat{\boldsymbol{\theta}}^*)\boldsymbol{IF}_{i,j}^{(2)}(\widehat{\boldsymbol{\theta}}^*)'\right\} + $$

$$\sum_{j=1}^{J}\sum_{i=1}^{n^{(j)}}\sum_{k=1}^{n^{(j)}} w_{i,k,j}\,\sigma_{i,k,j}\,w_{i,j}\,w_{k,j}\xi_{i,j}\,\xi_{k,j}\boldsymbol{IF}_{i,j}^{(2)}(\widehat{\boldsymbol{\theta}}^*)\boldsymbol{IF}_{k,j}^{(2)}(\widehat{\boldsymbol{\theta}}^*)',$$

with $\sigma_{i,k,j} = \frac{m^{(j)}}{n^{(j)}}\frac{m^{(j)}-1}{n^{(j)}-1} - \left(\frac{m^{(j)}}{n^{(j)}}\right)^2$ if individuals $i$ and $k$ in stratum $j$, $k \neq i$, are both non-cases, and $\sigma_{i,k,j} = 0$ otherwise, and with $\sigma_{i,i,j} = \frac{m^{(j)}}{n^{(j)}}\left(1 - \frac{m^{(j)}}{n^{(j)}}\right)$ if individual $i$ in stratum $j$ is a non-case and $\sigma_{i,i,j} = 0$ otherwise, $i,k \in \{1,\ldots,n^{(j)}\}, j \in \{1,\ldots,J\}$.

Observe that $\boldsymbol{IF}_{i,j}^{(1)}(\widehat{\boldsymbol{\beta}}^*)$ given in Web Appendix D.2 is the predicted value for the $i$-th individual in the $j$-th stratum from the weighted linear regression model of the $\xi_{i,j}\boldsymbol{Z}_{i,j}$ on the $\xi_{i,j}\boldsymbol{A}_{i,j}$, using the calibrated weights $w_{i,j}^*$, $i \in \{1,\ldots,n^{(j)}\}, j \in \{1,\ldots,J\}$. Then, $\xi_{i,j}\,w_{i,j}\,\boldsymbol{IF}_{i,j}^{(2)}(\widehat{\boldsymbol{\beta}}^*)$ is the $\sum_{l=1}^{j-1} n^{(l)} + i$-th weighted residual from this weighted linear regression, and it is zero if individual $i$ in stratum $j$ is not in the stratified case-cohort (i.e., not a case or not a sampled non-case), $i \in \{1,\ldots,n^{(j)}\}, j \in \{1,\ldots,J\}$. Similarly, $IF_{i,j}^{(1)}\{d\widehat{\Lambda}_0^*(t)\}$ is the predicted value for the $i$-th individual in the $j$-th stratum from the weighted linear regression of the $\xi_{i,j}H_{i,j}(t)$ on the $\xi_{i,j}\boldsymbol{A}_{i,j}$, using weights $w_{i,j}^*$, and $IF_{i,j}^{(2)}\{d\widehat{\Lambda}_0^*(t)\}$ is zero if individual $i$ in stratum $j$ is not in the stratified case-cohort, $i \in \{1,\ldots,n^{(j)}\}, j \in \{1,\ldots,J\}$. Overall, for any $\widehat{\boldsymbol{\theta}}^* \in \{\widehat{\boldsymbol{\eta}}, \widehat{\boldsymbol{\beta}}^*, d\widehat{\Lambda}_0^*(t), \widehat{\Lambda}_0^*(t), \hat{\pi}^*(\tau_1,\tau_2;\boldsymbol{x})\}$, $\xi_{i,j}\,w_{i,j}\,\boldsymbol{IF}_{i,j}^{(2)}(\widehat{\boldsymbol{\theta}}^*)$ is zero if individual $i$ in stratum $j$ is not in the stratified case-cohort; such individual affects $\widehat{\boldsymbol{\theta}}^*$ through her/his influence on $\widehat{\boldsymbol{\eta}}$, as he/she is used to calibrate the design weights. In addition, for any individuals $i$ and $k$ in stratum $j$ such that $\sigma_{i,k,j}$ and $\sigma_{i,i,j}$ are non-zero (i.e., non-cases), $\xi_{i,j} w_{i,j} \boldsymbol{IF}^{(2)}_{i,j}(\widehat{\boldsymbol{\theta}}^*)$ and $\xi_{k,j} w_{k,j} \boldsymbol{IF}^{(2)}_{k,j}(\widehat{\boldsymbol{\theta}}^*)$ are weighted residuals from a weighted linear regression.

Then, one can expect the phase-two component of the variance of $\widehat{\boldsymbol{\theta}}^*$ to be smaller, and subsequently the overall variance to be smaller, when these weighted residuals are close to zero. This should be case the case for $\widehat{\boldsymbol{\theta}}^* \in \{\widehat{\boldsymbol{\beta}}^*, \mathrm{d}\widehat{\Lambda}_0^*(t), \widehat{\Lambda}_0^*(t), \widehat{\pi}^*(\tau_1, \tau_2; \boldsymbol{x})\}$ when $\boldsymbol{A}$ contains the influences for the log-relative hazard and the total follow-up times on the pure risk time interval multiplied by the relative hazards (as detailed in Section 4.1 in the Main Document and in Web Appendix D.1). Indeed, observe that $\boldsymbol{Z}_{i,j}$ given in Web Appendix D.2 is the "direct" influence of individual $i$ in stratum $j$ on $\widehat{\boldsymbol{\beta}}^*$ (i.e., not through his/her influence on the calibrated weights), $i \in \{1, \ldots, n^{(j)}\}, j \in \{1, \ldots, J\}$. Its form is similar to that of the influence when using design weights; see Web Appendix C.1. On the other hand, $\int_{\tau_1}^{\tau_2} K_{i,j}(t)\mathrm{d}t$ corresponds to the total follow-up time in the interval $(\tau_1, \tau_2]$ multiplied by the estimated relative hazard of individual $i$ in stratum $j$, $i \in \{1, \ldots, n^{(j)}\}, j \in \{1, \ldots, J\}$. Although this point was not mentioned by Shin et al. (2020), it supports their choice of auxiliary variables over that of Breslow and Lumley (2013).

# Web Appendix E. SIMULATIONS

## E.1 Fixed parameter values used for the simulations in Section 7 of the Main Document

|  | $X_1 < -2$ | $-2 \leq X_1 < 1$ | $1 \leq X_1$ |
|---|---|---|---|
| $p_{0\|X_1}$ | 0.7 | 0.45 | 0.4 |
| $p_{1\|X_1}$ | 0.05 | 0.2 | 0.3 |
| $p_{2\|X_1}$ | 0.25 | 0.35 | 0.3 |

**WEB TABLE 1-** Parameter values for $p_{0|X_1} = P(X_2 = 0|X_1)$, $p_{1|X_1} = P(X_2 = 1|X_1)$ and $p_{2|X_1} = P(X_2 = 2|X_1)$, used for the simulation of $X_2$.

| $\alpha_1$ | $\alpha_2$ | $\beta_1$ | $\beta_2$ | $\beta_3$ |
|---|---|---|---|---|
| 0.05 | -0.35 | -0.2 | 0.25 | -0.3 |

**WEB TABLE 2-** Parameters values used for the simulation of $X_3$ and T.

## E.2 Simulations results for all investigated scenarios

Recall that estimation was performed using the stratified case-cohort with design weights (SCC); the stratified case-cohort with calibrated weights (SCC.Calib); the unstratified case-cohort with design weights (USCC); and the unstratified case-cohort with calibrated weights (USCC.Calib). Then, for SCC, we used the variance estimate with superpopulation and phase-two variance components ($\hat{V}$) given in Equation (14) in Section 3.3 in the Main Document; and the robust variance estimate ($\hat{V}_{\text{Robust}}$) given in Equation (15) in Section 3.3 in the Main Document. For SCC.Calib, we use the variance estimate with superpopulation and phase-two variance components ($\hat{V}$) given in Equation (18) in Section 4.3 in the Main Document; and the robust variance estimate ($\hat{V}_{\text{Robust}}$) given in Equation (19) in Section 4.3 in the Main Document. For USCC and USCC.Calib, we used the variance estimates obtained from the simplified versions of these equations when $J = 1$. We also estimated these parameters using the data from the whole cohort, to serve as a point of reference (Cohort).

Simulations results for all 12 scenarios (with $n \in \{5 \times 10^3, 10^4\}$, $p_Y \in \{0.02, 0.05, 0.1\}$ and $N \in \{2, 4\}$) are displayed in **WEB TABLE 3** to **WEB TABLE 20**. More precisely, Web Tables **WEB TABLE 3** to **WEB TABLE 8** display the coverages of 95% CIs, **WEB TABLE 9** to **WEB TABLE 14** display the mean of estimated variances, and **WEB TABLE 15** to **WEB TABLE 20** display the empirical variances and ratios of empirical variances with that from using the whole cohort, for $\beta_1$, $\beta_2$, $\beta_3$ and $\log\{\pi(\tau_1, \tau_2; \boldsymbol{x})\}$, $(\tau_1, \tau_2] = (0,8]$ and $\boldsymbol{x} \in \{(-1, 1, -0.6)', (1, -1, 0.6)', (1, 1, 0.6)'\}$, respectively. Interpretation of the results and conclusions are identical to that given in Section 7.2 in the Main Document.

### E.3 Comment on the robust variance estimation

Barlow (1994) also mentioned stratified sampling of the subcohort, and he stated that his robust formula could be used for variance estimation, but he did not specify the formula. Similarly, other authors who mentioned the robust variance with stratified case-cohort data did not provide the formula, even when comparing variance estimates (Samuelsen et al., 2007; Gray, 2009). On the other hand, Jiao (2002) stated that the "naïve" robust variance estimate (i.e., the sum of squared influences) of the relative hazard, as proposed by Barlow (1994), is not valid when the sampling of the subcohort is stratified. She proposed an extension of the robust variance for the stratified case-cohort design. But because no difference was observed between the two robust variance estimates in our simulations (results not shown), we used Barlow's initial robust variance formula. Jiao (2002) conjectured that Barlow (1994) could simply be referring the sum of the squared influences, and we believe that many practitioners use the sum of squared influences, as in Barlow (1994).

| Cohort | SCC | | SCC.Calib | | USCC | | USCC.Calib | | $n$ | $K$ | $p_Y$ | $\beta_1$ |
|---|---|---|---|---|---|---|---|---|---|---|---|---|
| | $\hat{V}_{\text{Robust}}$ | $\hat{V}$ | $\hat{V}_{\text{Robust}}$ | $\hat{V}$ | $\hat{V}_{\text{Robust}}$ | $\hat{V}$ | $\hat{V}_{\text{Robust}}$ | $\hat{V}$ | | | | |
| 0.944 | 0.9682* | 0.9516 | 0.9528 | 0.95 | 0.9558 | 0.956* | 0.9524 | 0.9524 | 5000 | 2 | 0.02 | -0.2 |
| 0.9512 | 0.961* | 0.9524 | 0.953 | 0.9522 | 0.9512 | 0.9514 | 0.9462 | 0.9466 | 5000 | 4 | 0.02 | -0.2 |
| 0.9476 | 0.9668* | 0.9524 | 0.9538 | 0.95 | 0.947 | 0.947 | 0.9452 | 0.945 | 10000 | 2 | 0.02 | -0.2 |
| 0.948 | 0.9596* | 0.9514 | 0.9514 | 0.95 | 0.9502 | 0.9502 | 0.951 | 0.951 | 10000 | 4 | 0.02 | -0.2 |
| 0.9476 | 0.968* | 0.9548 | 0.9558 | 0.953 | 0.9528 | 0.953 | 0.948 | 0.948 | 5000 | 2 | 0.05 | -0.2 |
| 0.954 | 0.9622* | 0.9546 | 0.9524 | 0.951 | 0.9528 | 0.953 | 0.9538 | 0.9538 | 5000 | 4 | 0.05 | -0.2 |
| 0.9552 | 0.9654* | 0.953 | 0.9558 | 0.954 | 0.9536 | 0.9536 | 0.9556 | 0.9558 | 10000 | 2 | 0.05 | -0.2 |
| 0.9482 | 0.9566* | 0.9516 | 0.9526 | 0.9512 | 0.9468 | 0.9468 | 0.949 | 0.949 | 10000 | 4 | 0.05 | -0.2 |
| 0.947 | 0.9598* | 0.9448 | 0.948 | 0.9472 | 0.9522 | 0.9522 | 0.948 | 0.948 | 5000 | 2 | 0.1 | -0.2 |
| 0.9544 | 0.9584* | 0.9538 | 0.957* | 0.957* | 0.955 | 0.955 | 0.9568* | 0.9568* | 5000 | 4 | 0.1 | -0.2 |
| 0.9488 | 0.9618* | 0.9506 | 0.9504 | 0.9486 | 0.9474 | 0.9476 | 0.9496 | 0.9496 | 10000 | 2 | 0.1 | -0.2 |
| 0.9526 | 0.9552 | 0.9496 | 0.9532 | 0.9514 | 0.9478 | 0.9478 | 0.9486 | 0.9486 | 10000 | 4 | 0.1 | -0.2 |

**WEB TABLE 3-** Coverage of 95% CIs for log-relative hazard parameter $\beta_1$ for different sampling designs and methods of analysis and variance estimation in 5,000 simulated cohorts. * indicates coverage outside the expected interval [0.9440; 0.9560].

| Cohort | SCC | | SCC.Calib | | USCC | | USCC.Calib | | $n$ | $K$ | $p_Y$ | $\beta_2$ |
| | $\hat{V}_{Robust}$ | $\hat{V}$ | $\hat{V}_{Robust}$ | $\hat{V}$ | $\hat{V}_{Robust}$ | $\hat{V}$ | $\hat{V}_{Robust}$ | $\hat{V}$ | | | | |
|---|---|---|---|---|---|---|---|---|---|---|---|---|
| 0.9532 | 0.9714* | 0.9568* | 0.9552 | 0.9548 | 0.9536 | 0.9536 | 0.9498 | 0.9494 | 5000 | 2 | 0.02 | 0.25 |
| 0.9514 | 0.9666* | 0.955 | 0.9542 | 0.9538 | 0.9564* | 0.9564* | 0.95 | 0.95 | 5000 | 4 | 0.02 | 0.25 |
| 0.9486 | 0.9716* | 0.9522 | 0.9514 | 0.9512 | 0.9554 | 0.9554 | 0.9506 | 0.9506 | 10000 | 2 | 0.02 | 0.25 |
| 0.9488 | 0.9606* | 0.9504 | 0.9468 | 0.9468 | 0.9488 | 0.9492 | 0.948 | 0.948 | 10000 | 4 | 0.02 | 0.25 |
| 0.9488 | 0.967* | 0.9488 | 0.9498 | 0.9492 | 0.953 | 0.953 | 0.9508 | 0.9508 | 5000 | 2 | 0.05 | 0.25 |
| 0.9524 | 0.9608* | 0.9502 | 0.9512 | 0.9508 | 0.9542 | 0.9542 | 0.9528 | 0.9528 | 5000 | 4 | 0.05 | 0.25 |
| 0.9506 | 0.9676* | 0.95 | 0.952 | 0.952 | 0.9506 | 0.9506 | 0.9526 | 0.9526 | 10000 | 2 | 0.05 | 0.25 |
| 0.951 | 0.9562* | 0.9488 | 0.9488 | 0.9484 | 0.9504 | 0.9504 | 0.9514 | 0.9514 | 10000 | 4 | 0.05 | 0.25 |
| 0.9512 | 0.9656* | 0.95 | 0.952 | 0.9518 | 0.947 | 0.947 | 0.9516 | 0.9516 | 5000 | 2 | 0.1 | 0.25 |
| 0.9496 | 0.9592* | 0.952 | 0.951 | 0.951 | 0.9504 | 0.9506 | 0.9532 | 0.9532 | 5000 | 4 | 0.1 | 0.25 |
| 0.9504 | 0.9636* | 0.9504 | 0.952 | 0.9518 | 0.9524 | 0.9524 | 0.9534 | 0.9534 | 10000 | 2 | 0.1 | 0.25 |
| 0.952 | 0.9588* | 0.953 | 0.9514 | 0.9512 | 0.9496 | 0.9496 | 0.9512 | 0.9512 | 10000 | 4 | 0.1 | 0.25 |

**WEB TABLE 4-** Coverage of 95% CIs for log-relative hazard parameter $\beta_2$ for different sampling designs and methods of analysis and variance estimation in 5,000 simulated cohorts. * indicates coverage outside the expected interval [0.9440; 0.9560].

| Cohort | SCC | | SCC.Calib | | USCC | | USCC.Calib | | $n$ | $K$ | $p_Y$ | $\beta_3$ |
| --- | --- | --- | --- | --- | --- | --- | --- | --- | --- | --- | --- | --- |
| | $\hat{V}_{\text{Robust}}$ | $\hat{V}$ | $\hat{V}_{\text{Robust}}$ | $\hat{V}$ | $\hat{V}_{\text{Robust}}$ | $\hat{V}$ | $\hat{V}_{\text{Robust}}$ | $\hat{V}$ | | | | |
| 0.9442 | 0.9546 | 0.9548 | 0.9508 | 0.9506 | 0.9482 | 0.9486 | 0.9468 | 0.9468 | 5000 | 2 | 0.02 | -0.3 |
| 0.9464 | 0.9478 | 0.9478 | 0.9478 | 0.9474 | 0.9566* | 0.9566* | 0.9544 | 0.9544 | 5000 | 4 | 0.02 | -0.3 |
| 0.9422* | 0.9492 | 0.9494 | 0.9474 | 0.9474 | 0.9454 | 0.9454 | 0.9476 | 0.9478 | 10000 | 2 | 0.02 | -0.3 |
| 0.9506 | 0.9516 | 0.9516 | 0.9522 | 0.9522 | 0.9524 | 0.9524 | 0.9502 | 0.9502 | 10000 | 4 | 0.02 | -0.3 |
| 0.9484 | 0.9462 | 0.9466 | 0.9458 | 0.946 | 0.9522 | 0.9522 | 0.9506 | 0.9506 | 5000 | 2 | 0.05 | -0.3 |
| 0.9464 | 0.95 | 0.95 | 0.9482 | 0.948 | 0.948 | 0.948 | 0.9444 | 0.9446 | 5000 | 4 | 0.05 | -0.3 |
| 0.9504 | 0.9466 | 0.9468 | 0.9498 | 0.9498 | 0.9456 | 0.9458 | 0.95 | 0.95 | 10000 | 2 | 0.05 | -0.3 |
| 0.9498 | 0.9522 | 0.9524 | 0.9516 | 0.9516 | 0.9512 | 0.9512 | 0.9524 | 0.9524 | 10000 | 4 | 0.05 | -0.3 |
| 0.9494 | 0.951 | 0.951 | 0.951 | 0.951 | 0.949 | 0.949 | 0.9514 | 0.9514 | 5000 | 2 | 0.1 | -0.3 |
| 0.9508 | 0.9498 | 0.9498 | 0.951 | 0.951 | 0.952 | 0.952 | 0.9492 | 0.9492 | 5000 | 4 | 0.1 | -0.3 |
| 0.9532 | 0.9518 | 0.9518 | 0.9516 | 0.9516 | 0.9516 | 0.9516 | 0.9532 | 0.9532 | 10000 | 2 | 0.1 | -0.3 |
| 0.9482 | 0.9536 | 0.9536 | 0.9512 | 0.9512 | 0.949 | 0.949 | 0.95 | 0.9502 | 10000 | 4 | 0.1 | -0.3 |

**WEB TABLE 5-** Coverage of 95% CIs for log-relative hazard parameter $\beta_3$ for different sampling designs and methods of analysis and variance estimation in 5,000 simulated cohorts. * indicates coverage outside the expected interval [0.9440; 0.9560].

| Cohort | SCC | | SCC.Calib | | USCC | | USCC.Calib | | $n$ | $K$ | $p_Y$ | $\log\{\pi(\tau_1,\tau_2;x)\}$ |
| | $\hat{V}_{\text{Robust}}$ | $\hat{V}$ | $\hat{V}_{\text{Robust}}$ | $\hat{V}$ | $\hat{V}_{\text{Robust}}$ | $\hat{V}$ | $\hat{V}_{\text{Robust}}$ | $\hat{V}$ | | | | |
|---|---|---|---|---|---|---|---|---|---|---|---|---|
| 0.9466 | 0.9722* | 0.95 | 0.9558 | 0.9516 | 0.9634* | 0.9494 | 0.9526 | 0.9526 | 5000 | 2 | 0.02 | -3.948 |
| 0.9484 | 0.9602* | 0.9492 | 0.9508 | 0.9506 | 0.9558 | 0.9456 | 0.95 | 0.95 | 5000 | 4 | 0.02 | -3.948 |
| 0.956* | 0.973* | 0.9568* | 0.96 | 0.9546 | 0.9656* | 0.9526 | 0.955 | 0.955 | 10000 | 2 | 0.02 | -3.948 |
| 0.951 | 0.9656* | 0.9538 | 0.9538 | 0.952 | 0.9602* | 0.9526 | 0.9512 | 0.9512 | 10000 | 4 | 0.02 | -3.948 |
| 0.9488 | 0.9694* | 0.9494 | 0.9528 | 0.9494 | 0.958* | 0.948 | 0.9506 | 0.9506 | 5000 | 2 | 0.05 | -3.046 |
| 0.9546 | 0.9634* | 0.955 | 0.956* | 0.954 | 0.9596* | 0.9538 | 0.9524 | 0.9524 | 5000 | 4 | 0.05 | -3.046 |
| 0.9556 | 0.971* | 0.9556 | 0.9602* | 0.9562* | 0.9656* | 0.9554 | 0.9542 | 0.9542 | 10000 | 2 | 0.05 | -3.046 |
| 0.9534 | 0.9604* | 0.9518 | 0.9554 | 0.953 | 0.962* | 0.9538 | 0.9534 | 0.9534 | 10000 | 4 | 0.05 | -3.046 |
| 0.9506 | 0.9658* | 0.951 | 0.954 | 0.951 | 0.9622* | 0.954 | 0.9522 | 0.9522 | 5000 | 2 | 0.1 | -2.377 |
| 0.9496 | 0.9566* | 0.9512 | 0.951 | 0.9504 | 0.9544 | 0.9514 | 0.9506 | 0.9508 | 5000 | 4 | 0.1 | -2.377 |
| 0.9484 | 0.9634* | 0.9476 | 0.9534 | 0.9504 | 0.957* | 0.9484 | 0.9444 | 0.9444 | 10000 | 2 | 0.1 | -2.377 |
| 0.9538 | 0.9586* | 0.9524 | 0.9548 | 0.954 | 0.9594* | 0.9562* | 0.9558 | 0.9558 | 10000 | 4 | 0.1 | -2.377 |

**WEB TABLE 6-** Coverage of 95% CIs for log-pure risk $\log\{\pi(\tau_1,\tau_2;x)\}$, with $(\tau_1,\tau_2]=(0,8]$ and $x=(-1,1,-0.6)'$, for different sampling designs and methods of analysis and variance estimation in 5,000 simulated cohorts. * indicates coverage outside the expected interval [0.9440; 0.9560].

| Cohort | SCC | | SCC.Calib | | USCC | | USCC.Calib | | $n$ | $K$ | $p_Y$ | $\log\{\pi(\tau_1,\tau_2;x)\}$ |
| --- | --- | --- | --- | --- | --- | --- | --- | --- | --- | --- | --- | --- |
| | $\hat{V}_{Robust}$ | $\hat{V}$ | $\hat{V}_{Robust}$ | $\hat{V}$ | $\hat{V}_{Robust}$ | $\hat{V}$ | $\hat{V}_{Robust}$ | $\hat{V}$ | | | | |
| 0.9476 | 0.9754* | 0.9504 | 0.95 | 0.9492 | 0.9552 | 0.9526 | 0.9516 | 0.9516 | 5000 | 2 | 0.02 | -5.201 |
| 0.946 | 0.9642* | 0.948 | 0.9472 | 0.9472 | 0.953 | 0.9514 | 0.9454 | 0.9454 | 5000 | 4 | 0.02 | -5.201 |
| 0.952 | 0.9722* | 0.9518 | 0.9522 | 0.9516 | 0.954 | 0.9522 | 0.9472 | 0.947 | 10000 | 2 | 0.02 | -5.201 |
| 0.946 | 0.9582* | 0.9482 | 0.9464 | 0.946 | 0.9528 | 0.9508 | 0.9456 | 0.9456 | 10000 | 4 | 0.02 | -5.201 |
| 0.9456 | 0.968* | 0.9462 | 0.9474 | 0.9472 | 0.9512 | 0.9486 | 0.946 | 0.946 | 5000 | 2 | 0.05 | -4.288 |
| 0.9488 | 0.962* | 0.9508 | 0.9498 | 0.9494 | 0.9516 | 0.95 | 0.9478 | 0.9478 | 5000 | 4 | 0.05 | -4.288 |
| 0.9508 | 0.966* | 0.949 | 0.9464 | 0.9462 | 0.9558 | 0.9522 | 0.9494 | 0.9494 | 10000 | 2 | 0.05 | -4.288 |
| 0.9526 | 0.9576* | 0.9508 | 0.9524 | 0.9524 | 0.9526 | 0.9512 | 0.9544 | 0.9544 | 10000 | 4 | 0.05 | -4.288 |
| 0.9442 | 0.9666* | 0.9492 | 0.9484 | 0.948 | 0.9518 | 0.9494 | 0.9514 | 0.9514 | 5000 | 2 | 0.1 | -3.602 |
| 0.9548 | 0.9616* | 0.9546 | 0.9544 | 0.9544 | 0.954 | 0.9534 | 0.9528 | 0.9528 | 5000 | 4 | 0.1 | -3.602 |
| 0.9526 | 0.9698* | 0.9568* | 0.9536 | 0.9532 | 0.9544 | 0.9522 | 0.9508 | 0.9508 | 10000 | 2 | 0.1 | -3.602 |
| 0.9514 | 0.959* | 0.9518 | 0.9512 | 0.951 | 0.9524 | 0.9514 | 0.9522 | 0.9522 | 10000 | 4 | 0.1 | -3.602 |

**WEB TABLE 7-** Coverage of 95% CIs for log-pure risk $\log\{\pi(\tau_1,\tau_2;x)\}$, with $(\tau_1,\tau_2] = (0,8]$ and $x = (1,-1,0.6)'$, for different sampling designs and methods of analysis and variance estimation in 5,000 simulated cohorts. * indicates coverage outside the expected interval [0.9440; 0.9560].

| Cohort | SCC | | SCC.Calib | | USCC | | USCC.Calib | | $n$ | $K$ | $p_Y$ | $\log\{\pi(\tau_1,\tau_2;x)\}$ |
|---|---|---|---|---|---|---|---|---|---|---|---|---|
| | $\widehat{V}_{Robust}$ | $\widehat{V}$ | $\widehat{V}_{Robust}$ | $\widehat{V}$ | $\widehat{V}_{Robust}$ | $\widehat{V}$ | $\widehat{V}_{Robust}$ | $\widehat{V}$ | | | | |
| 0.955 | 0.9712* | 0.9582* | 0.9562* | 0.9546 | 0.9658* | 0.9596 | 0.9584* | 0.9586 | 5000 | 2 | 0.02 | -4.702 |
| 0.9468 | 0.9578* | 0.9494 | 0.9494 | 0.949 | 0.9534 | 0.95 | 0.9482 | 0.9482 | 5000 | 4 | 0.02 | -4.702 |
| 0.952 | 0.9666* | 0.9542 | 0.9532 | 0.9516 | 0.9574* | 0.95 | 0.9502 | 0.9502 | 10000 | 2 | 0.02 | -4.702 |
| 0.9492 | 0.9584* | 0.9504 | 0.9516 | 0.9512 | 0.956* | 0.9522 | 0.9482 | 0.9482 | 10000 | 4 | 0.02 | -4.702 |
| 0.9488 | 0.962* | 0.9488 | 0.9498 | 0.949 | 0.9572* | 0.949 | 0.9506 | 0.9506 | 5000 | 2 | 0.05 | -3.793 |
| 0.95 | 0.9602* | 0.953 | 0.9526 | 0.9516 | 0.9526 | 0.9502 | 0.951 | 0.951 | 5000 | 4 | 0.05 | -3.793 |
| 0.9462 | 0.9606* | 0.9468 | 0.9494 | 0.948 | 0.953 | 0.9444 | 0.946 | 0.946 | 10000 | 2 | 0.05 | -3.793 |
| 0.9496 | 0.9562* | 0.9494 | 0.9538 | 0.9536 | 0.9518 | 0.95 | 0.9516 | 0.9516 | 10000 | 4 | 0.05 | -3.793 |
| 0.9452 | 0.9608* | 0.9486 | 0.9484 | 0.9484 | 0.9514 | 0.9466 | 0.9458 | 0.9458 | 5000 | 2 | 0.1 | -3.111 |
| 0.9516 | 0.957* | 0.9522 | 0.9528 | 0.9524 | 0.9544 | 0.9506 | 0.9508 | 0.951 | 5000 | 4 | 0.1 | -3.111 |
| 0.9502 | 0.9606* | 0.9512 | 0.951 | 0.9504 | 0.9522 | 0.9466 | 0.9474 | 0.9474 | 10000 | 2 | 0.1 | -3.111 |
| 0.9528 | 0.956* | 0.9512 | 0.9518 | 0.9516 | 0.9526 | 0.9508 | 0.9524 | 0.9524 | 10000 | 4 | 0.1 | -3.111 |

**WEB TABLE 8-** Coverage of 95% CIs for log-pure risk $\log\{\pi(\tau_1,\tau_2;x)\}$, with $(\tau_1,\tau_2] = (0,8]$ and $x = (1,1,0.6)'$, for different sampling designs and methods of analysis and variance estimation in 5,000 simulated cohorts. * indicates coverage outside the expected interval [0.9440; 0.9560].

| Cohort | SCC | | SCC.Calib | | USCC | | USCC.Calib | | $n$ | $K$ | $p_Y$ | $\beta_1$ |
| --- | --- | --- | --- | --- | --- | --- | --- | --- | --- | --- | --- | --- |
| | $\hat{V}_{\text{Robust}}$ | $\hat{V}$ | $\hat{V}_{\text{Robust}}$ | $\hat{V}$ | $\hat{V}_{\text{Robust}}$ | $\hat{V}$ | $\hat{V}_{\text{Robust}}$ | $\hat{V}$ | | | | |
| 0.0138 | 0.0212 | 0.0182 | 0.0177 | 0.0172 | 0.0222 | 0.0223 | 0.0184 | 0.0184 | 5000 | 2 | 0.02 | -0.2 |
| 0.0139 | 0.0174 | 0.016 | 0.0157 | 0.0155 | 0.0177 | 0.0178 | 0.0159 | 0.0159 | 5000 | 4 | 0.02 | -0.2 |
| 0.0069 | 0.0102 | 0.0087 | 0.0085 | 0.0083 | 0.0106 | 0.0106 | 0.0087 | 0.0087 | 10000 | 2 | 0.02 | -0.2 |
| 0.0069 | 0.0085 | 0.0078 | 0.0076 | 0.0075 | 0.0086 | 0.0087 | 0.0077 | 0.0077 | 10000 | 4 | 0.02 | -0.2 |
| 0.0056 | 0.0079 | 0.0069 | 0.0067 | 0.0065 | 0.0082 | 0.0082 | 0.0068 | 0.0068 | 5000 | 2 | 0.05 | -0.2 |
| 0.0056 | 0.0066 | 0.0061 | 0.0061 | 0.006 | 0.0067 | 0.0067 | 0.0061 | 0.0061 | 5000 | 4 | 0.05 | -0.2 |
| 0.0028 | 0.0039 | 0.0034 | 0.0033 | 0.0032 | 0.004 | 0.004 | 0.0033 | 0.0033 | 10000 | 2 | 0.05 | -0.2 |
| 0.0028 | 0.0033 | 0.0031 | 0.003 | 0.003 | 0.0033 | 0.0033 | 0.003 | 0.003 | 10000 | 4 | 0.05 | -0.2 |
| 0.0029 | 0.0037 | 0.0033 | 0.0032 | 0.0032 | 0.0039 | 0.0039 | 0.0033 | 0.0033 | 5000 | 2 | 0.1 | -0.2 |
| 0.0029 | 0.0032 | 0.003 | 0.003 | 0.003 | 0.0032 | 0.0032 | 0.003 | 0.003 | 5000 | 4 | 0.1 | -0.2 |
| 0.0014 | 0.0019 | 0.0017 | 0.0016 | 0.0016 | 0.0019 | 0.0019 | 0.0016 | 0.0016 | 10000 | 2 | 0.1 | -0.2 |
| 0.0014 | 0.0016 | 0.0015 | 0.0015 | 0.0015 | 0.0016 | 0.0016 | 0.0015 | 0.0015 | 10000 | 4 | 0.1 | -0.2 |

**WEB TABLE 9-** Mean of estimated variances of log-relative hazard parameter $\beta_1$ for different sampling designs and methods of analysis and variance estimation in 5,000 simulated cohorts.

| Cohort | SCC | | SCC.Calib | | USCC | | USCC.Calib | | $n$ | $K$ | $p_Y$ | $\beta_2$ |
| --- | --- | --- | --- | --- | --- | --- | --- | --- | --- | --- | --- | --- |
| | $\hat{V}_{\text{Robust}}$ | $\hat{V}$ | $\hat{V}_{\text{Robust}}$ | $\hat{V}$ | $\hat{V}_{\text{Robust}}$ | $\hat{V}$ | $\hat{V}_{\text{Robust}}$ | $\hat{V}$ | | | | |
| 0.0197 | 0.0286 | 0.0237 | 0.0226 | 0.0225 | 0.029 | 0.0291 | 0.0227 | 0.0227 | 5000 | 2 | 0.02 | 0.25 |
| 0.0198 | 0.024 | 0.0217 | 0.0212 | 0.0211 | 0.0243 | 0.0243 | 0.0213 | 0.0213 | 5000 | 4 | 0.02 | 0.25 |
| 0.0097 | 0.0139 | 0.0114 | 0.0109 | 0.0109 | 0.014 | 0.014 | 0.0109 | 0.0109 | 10000 | 2 | 0.02 | 0.25 |
| 0.0097 | 0.0117 | 0.0106 | 0.0103 | 0.0103 | 0.0118 | 0.0118 | 0.0103 | 0.0103 | 10000 | 4 | 0.02 | 0.25 |
| 0.0079 | 0.0108 | 0.0091 | 0.0087 | 0.0087 | 0.0109 | 0.0109 | 0.0087 | 0.0087 | 5000 | 2 | 0.05 | 0.25 |
| 0.0079 | 0.0092 | 0.0084 | 0.0082 | 0.0082 | 0.0092 | 0.0092 | 0.0082 | 0.0082 | 5000 | 4 | 0.05 | 0.25 |
| 0.0039 | 0.0054 | 0.0045 | 0.0043 | 0.0043 | 0.0054 | 0.0054 | 0.0043 | 0.0043 | 10000 | 2 | 0.05 | 0.25 |
| 0.0039 | 0.0046 | 0.0042 | 0.0041 | 0.0041 | 0.0046 | 0.0046 | 0.0041 | 0.0041 | 10000 | 4 | 0.05 | 0.25 |
| 0.004 | 0.0052 | 0.0045 | 0.0043 | 0.0043 | 0.0052 | 0.0052 | 0.0043 | 0.0043 | 5000 | 2 | 0.1 | 0.25 |
| 0.004 | 0.0044 | 0.0042 | 0.0041 | 0.0041 | 0.0044 | 0.0044 | 0.0041 | 0.0041 | 5000 | 4 | 0.1 | 0.25 |
| 0.002 | 0.0026 | 0.0022 | 0.0021 | 0.0021 | 0.0026 | 0.0026 | 0.0021 | 0.0021 | 10000 | 2 | 0.1 | 0.25 |
| 0.002 | 0.0022 | 0.0021 | 0.0021 | 0.0021 | 0.0022 | 0.0022 | 0.002 | 0.002 | 10000 | 4 | 0.1 | 0.25 |

**WEB TABLE 10-** Mean of estimated variances of log-relative hazard parameter $\beta_2$ for different sampling designs and methods of analysis and variance estimation in 5,000 simulated cohorts.

| Cohort | SCC | | SCC.Calib | | USCC | | USCC.Calib | | $n$ | $K$ | $p_Y$ | $\beta_3$ |
| --- | --- | --- | --- | --- | --- | --- | --- | --- | --- | --- | --- | --- |
| | $\widehat{V}_{\text{Robust}}$ | $\widehat{V}$ | $\widehat{V}_{\text{Robust}}$ | $\widehat{V}$ | $\widehat{V}_{\text{Robust}}$ | $\widehat{V}$ | $\widehat{V}_{\text{Robust}}$ | $\widehat{V}$ | | | | |
| 0.0136 | 0.0214 | 0.0215 | 0.0179 | 0.0179 | 0.0223 | 0.0223 | 0.0183 | 0.0183 | 5000 | 2 | 0.02 | -0.3 |
| 0.0136 | 0.0173 | 0.0173 | 0.0157 | 0.0156 | 0.0178 | 0.0178 | 0.0159 | 0.0159 | 5000 | 4 | 0.02 | -0.3 |
| 0.0068 | 0.0102 | 0.0102 | 0.0085 | 0.0085 | 0.0107 | 0.0107 | 0.0087 | 0.0087 | 10000 | 2 | 0.02 | -0.3 |
| 0.0068 | 0.0084 | 0.0084 | 0.0076 | 0.0076 | 0.0086 | 0.0086 | 0.0077 | 0.0077 | 10000 | 4 | 0.02 | -0.3 |
| 0.0055 | 0.0079 | 0.0079 | 0.0067 | 0.0067 | 0.0082 | 0.0082 | 0.0068 | 0.0068 | 5000 | 2 | 0.05 | -0.3 |
| 0.0055 | 0.0066 | 0.0066 | 0.006 | 0.006 | 0.0067 | 0.0067 | 0.0061 | 0.0061 | 5000 | 4 | 0.05 | -0.3 |
| 0.0028 | 0.0039 | 0.0039 | 0.0033 | 0.0033 | 0.004 | 0.004 | 0.0033 | 0.0033 | 10000 | 2 | 0.05 | -0.3 |
| 0.0028 | 0.0032 | 0.0032 | 0.003 | 0.003 | 0.0033 | 0.0033 | 0.003 | 0.003 | 10000 | 4 | 0.05 | -0.3 |
| 0.0028 | 0.0037 | 0.0037 | 0.0032 | 0.0032 | 0.0038 | 0.0038 | 0.0033 | 0.0033 | 5000 | 2 | 0.1 | -0.3 |
| 0.0028 | 0.0031 | 0.0031 | 0.003 | 0.003 | 0.0032 | 0.0032 | 0.003 | 0.003 | 5000 | 4 | 0.1 | -0.3 |
| 0.0014 | 0.0018 | 0.0018 | 0.0016 | 0.0016 | 0.0019 | 0.0019 | 0.0016 | 0.0016 | 10000 | 2 | 0.1 | -0.3 |
| 0.0014 | 0.0016 | 0.0016 | 0.0015 | 0.0015 | 0.0016 | 0.0016 | 0.0015 | 0.0015 | 10000 | 4 | 0.1 | -0.3 |

**WEB TABLE 11-** Mean of estimated variances of log-relative hazard parameter $\beta_3$ from using different sampling designs, methods of analysis and variance estimation, over 5,000 simulated cohorts.

| Cohort | SCC | | SCC.Calib | | USCC | | USCC.Calib | | $n$ | $K$ | $p_Y$ | $\log\{\pi(\tau_1,\tau_2;\boldsymbol{x})\}$ |
|---|---|---|---|---|---|---|---|---|---|---|---|---|
| | $\hat{V}_{\text{Robust}}$ | $\hat{V}$ | $\hat{V}_{\text{Robust}}$ | $\hat{V}$ | $\hat{V}_{\text{Robust}}$ | $\hat{V}$ | $\hat{V}_{\text{Robust}}$ | $\hat{V}$ | | | | |
| 0.0248 | 0.0354 | 0.029 | 0.0296 | 0.0285 | 0.0374 | 0.0329 | 0.0306 | 0.0306 | 5000 | 2 | 0.02 | -3.948 |
| 0.0248 | 0.0298 | 0.0268 | 0.0271 | 0.0266 | 0.0307 | 0.0286 | 0.0275 | 0.0275 | 5000 | 4 | 0.02 | -3.948 |
| 0.0122 | 0.0172 | 0.014 | 0.0142 | 0.0137 | 0.0181 | 0.0159 | 0.0145 | 0.0145 | 10000 | 2 | 0.02 | -3.948 |
| 0.0122 | 0.0145 | 0.013 | 0.0131 | 0.0129 | 0.015 | 0.0139 | 0.0133 | 0.0133 | 10000 | 4 | 0.02 | -3.948 |
| 0.0096 | 0.013 | 0.0109 | 0.011 | 0.0106 | 0.0137 | 0.0122 | 0.0112 | 0.0112 | 5000 | 2 | 0.05 | -3.046 |
| 0.0096 | 0.0111 | 0.0102 | 0.0102 | 0.0101 | 0.0114 | 0.0108 | 0.0103 | 0.0103 | 5000 | 4 | 0.05 | -3.046 |
| 0.0048 | 0.0064 | 0.0054 | 0.0054 | 0.0052 | 0.0068 | 0.006 | 0.0055 | 0.0055 | 10000 | 2 | 0.05 | -3.046 |
| 0.0048 | 0.0055 | 0.005 | 0.0051 | 0.005 | 0.0057 | 0.0053 | 0.0051 | 0.0051 | 10000 | 4 | 0.05 | -3.046 |
| 0.0047 | 0.0059 | 0.0052 | 0.0051 | 0.005 | 0.0062 | 0.0057 | 0.0052 | 0.0052 | 5000 | 2 | 0.1 | -2.377 |
| 0.0047 | 0.0051 | 0.0048 | 0.0048 | 0.0048 | 0.0052 | 0.005 | 0.0049 | 0.0049 | 5000 | 4 | 0.1 | -2.377 |
| 0.0023 | 0.0029 | 0.0026 | 0.0026 | 0.0025 | 0.0031 | 0.0028 | 0.0026 | 0.0026 | 10000 | 2 | 0.1 | -2.377 |
| 0.0023 | 0.0025 | 0.0024 | 0.0024 | 0.0024 | 0.0026 | 0.0025 | 0.0024 | 0.0024 | 10000 | 4 | 0.1 | -2.377 |

**WEB TABLE 12-** Mean of estimated variances of log-pure risk $\log\{\pi(\tau_1,\tau_2;\boldsymbol{x})\}$, with $(\tau_1,\tau_2]=(0,8]$ and $\boldsymbol{x}=(-1,1,-0.6)'$, from using different sampling designs, methods of analysis and variance estimation, over 5,000 simulated cohorts.

| Cohort | SCC | | SCC.Calib | | USCC | | USCC.Calib | | $n$ | $K$ | $p_Y$ | $\log\{\pi(\tau_1,\tau_2;\boldsymbol{x})\}$ |
|---|---|---|---|---|---|---|---|---|---|---|---|---|
| | $\hat{V}_{\text{Robust}}$ | $\hat{V}$ | $\hat{V}_{\text{Robust}}$ | $\hat{V}$ | $\hat{V}_{\text{Robust}}$ | $\hat{V}$ | $\hat{V}_{\text{Robust}}$ | $\hat{V}$ | | | | |
| 0.1249 | 0.1773 | 0.1441 | 0.141 | 0.1401 | 0.1788 | 0.1744 | 0.1453 | 0.1452 | 5000 | 2 | 0.02 | -5.201 |
| 0.1258 | 0.1507 | 0.1349 | 0.1335 | 0.1331 | 0.1511 | 0.1489 | 0.1348 | 0.1348 | 5000 | 4 | 0.02 | -5.201 |
| 0.062 | 0.0861 | 0.0697 | 0.0684 | 0.0681 | 0.086 | 0.0837 | 0.0696 | 0.0696 | 10000 | 2 | 0.02 | -5.201 |
| 0.0618 | 0.0733 | 0.0654 | 0.0648 | 0.0647 | 0.0731 | 0.072 | 0.0653 | 0.0653 | 10000 | 4 | 0.02 | -5.201 |
| 0.0496 | 0.0666 | 0.0549 | 0.0538 | 0.0536 | 0.0661 | 0.0645 | 0.0546 | 0.0546 | 5000 | 2 | 0.05 | -4.288 |
| 0.0496 | 0.0571 | 0.0518 | 0.0513 | 0.0513 | 0.0569 | 0.0562 | 0.0518 | 0.0518 | 5000 | 4 | 0.05 | -4.288 |
| 0.0247 | 0.0329 | 0.0271 | 0.0265 | 0.0265 | 0.0325 | 0.0317 | 0.0268 | 0.0268 | 10000 | 2 | 0.05 | -4.288 |
| 0.0247 | 0.0284 | 0.0257 | 0.0255 | 0.0255 | 0.0282 | 0.0278 | 0.0257 | 0.0257 | 10000 | 4 | 0.05 | -4.288 |
| 0.0248 | 0.0316 | 0.0268 | 0.0263 | 0.0262 | 0.0311 | 0.0304 | 0.0266 | 0.0266 | 5000 | 2 | 0.1 | -3.602 |
| 0.0248 | 0.0273 | 0.0255 | 0.0253 | 0.0253 | 0.0271 | 0.0268 | 0.0254 | 0.0254 | 5000 | 4 | 0.1 | -3.602 |
| 0.0124 | 0.0157 | 0.0133 | 0.013 | 0.013 | 0.0154 | 0.0151 | 0.0132 | 0.0132 | 10000 | 2 | 0.1 | -3.602 |
| 0.0124 | 0.0136 | 0.0127 | 0.0126 | 0.0126 | 0.0134 | 0.0133 | 0.0126 | 0.0126 | 10000 | 4 | 0.1 | -3.602 |

**WEB TABLE 13-** Mean of estimated variances of log-pure risk $\log\{\pi(\tau_1,\tau_2;\boldsymbol{x})\}$, with $(\tau_1,\tau_2]=(0,8]$ and $\boldsymbol{x}=(1,-1,0.6)'$, for different sampling designs and methods of analysis and variance estimation in 5,000 simulated cohorts.

| Cohort | SCC | | SCC.Calib | | USCC | | USCC.Calib | | $n$ | $K$ | $p_Y$ | $\log\{\pi(\tau_1,\tau_2;\boldsymbol{x})\}$ |
|---|---|---|---|---|---|---|---|---|---|---|---|---|
| | $\hat{V}_{\text{Robust}}$ | $\hat{V}$ | $\hat{V}_{\text{Robust}}$ | $\hat{V}$ | $\hat{V}_{\text{Robust}}$ | $\hat{V}$ | $\hat{V}_{\text{Robust}}$ | $\hat{V}$ | | | | |
| 0.0557 | 0.0789 | 0.0698 | 0.0672 | 0.0665 | 0.0806 | 0.0761 | 0.0689 | 0.0689 | 5000 | 2 | 0.02 | -4.702 |
| 0.0559 | 0.067 | 0.0626 | 0.0613 | 0.061 | 0.0679 | 0.0657 | 0.0621 | 0.0621 | 5000 | 4 | 0.02 | -4.702 |
| 0.0277 | 0.0379 | 0.0333 | 0.0319 | 0.0316 | 0.0386 | 0.0363 | 0.0328 | 0.0328 | 10000 | 2 | 0.02 | -4.702 |
| 0.0277 | 0.0325 | 0.0303 | 0.0297 | 0.0295 | 0.0328 | 0.0317 | 0.03 | 0.03 | 10000 | 4 | 0.02 | -4.702 |
| 0.0221 | 0.0291 | 0.0259 | 0.0249 | 0.0247 | 0.0294 | 0.0278 | 0.0253 | 0.0253 | 5000 | 2 | 0.05 | -3.793 |
| 0.0221 | 0.0251 | 0.0237 | 0.0232 | 0.0231 | 0.0253 | 0.0246 | 0.0235 | 0.0235 | 5000 | 4 | 0.05 | -3.793 |
| 0.011 | 0.0143 | 0.0127 | 0.0122 | 0.0121 | 0.0145 | 0.0136 | 0.0124 | 0.0124 | 10000 | 2 | 0.05 | -3.793 |
| 0.011 | 0.0125 | 0.0117 | 0.0115 | 0.0115 | 0.0125 | 0.0122 | 0.0116 | 0.0116 | 10000 | 4 | 0.05 | -3.793 |
| 0.011 | 0.0136 | 0.0123 | 0.0119 | 0.0118 | 0.0137 | 0.013 | 0.0121 | 0.0121 | 5000 | 2 | 0.1 | -3.111 |
| 0.011 | 0.0119 | 0.0114 | 0.0113 | 0.0113 | 0.0119 | 0.0117 | 0.0114 | 0.0114 | 5000 | 4 | 0.1 | -3.111 |
| 0.0055 | 0.0068 | 0.0061 | 0.0059 | 0.0059 | 0.0068 | 0.0065 | 0.006 | 0.006 | 10000 | 2 | 0.1 | -3.111 |
| 0.0055 | 0.0059 | 0.0057 | 0.0056 | 0.0056 | 0.0059 | 0.0058 | 0.0056 | 0.0056 | 10000 | 4 | 0.1 | -3.111 |

**WEB TABLE 14-** Mean of estimated variances of log-pure risk $\log\{\pi(\tau_1,\tau_2;\boldsymbol{x})\}$, with $(\tau_1,\tau_2]=(0,8]$ and $\boldsymbol{x}=(1,1,0.6)'$, for different sampling designs and methods of analysis and variance estimation in 5,000 simulated cohorts.

| Empirical variance | | | | | Ratio of empirical variance with the whole cohort to empirical variance with | | | | | | | |
|---|---|---|---|---|---|---|---|---|---|---|---|---|
| Cohort | SCC | SCC.Calib | USCC | USCC.Calib | SCC | SCC.Calib | USCC | USCC.Calib | $n$ | $K$ | $p_Y$ | $\beta_1$ |
| 0.0138 | 0.0181 | 0.017 | 0.0223 | 0.0177 | 0.763 | 0.8141 | 0.62 | 0.7794 | 5000 | 2 | 0.02 | -0.2 |
| 0.0138 | 0.0159 | 0.0152 | 0.0177 | 0.0158 | 0.8678 | 0.9054 | 0.7769 | 0.8711 | 5000 | 4 | 0.02 | -0.2 |
| 0.007 | 0.0085 | 0.0082 | 0.0108 | 0.0087 | 0.8161 | 0.8521 | 0.6408 | 0.8013 | 10000 | 2 | 0.02 | -0.2 |
| 0.0069 | 0.0076 | 0.0074 | 0.0086 | 0.0076 | 0.9055 | 0.9322 | 0.8017 | 0.9039 | 10000 | 4 | 0.02 | -0.2 |
| 0.0056 | 0.0068 | 0.0064 | 0.0082 | 0.0067 | 0.8327 | 0.8769 | 0.6897 | 0.8401 | 5000 | 2 | 0.05 | -0.2 |
| 0.0055 | 0.006 | 0.0059 | 0.0066 | 0.006 | 0.917 | 0.9363 | 0.8331 | 0.9207 | 5000 | 4 | 0.05 | -0.2 |
| 0.0027 | 0.0033 | 0.0031 | 0.004 | 0.0033 | 0.8367 | 0.8832 | 0.6895 | 0.839 | 10000 | 2 | 0.05 | -0.2 |
| 0.0028 | 0.003 | 0.0029 | 0.0033 | 0.003 | 0.9275 | 0.9547 | 0.8424 | 0.9402 | 10000 | 4 | 0.05 | -0.2 |
| 0.003 | 0.0034 | 0.0033 | 0.0039 | 0.0033 | 0.8631 | 0.9076 | 0.7548 | 0.8862 | 5000 | 2 | 0.1 | -0.2 |
| 0.0028 | 0.003 | 0.0029 | 0.0032 | 0.003 | 0.9535 | 0.9759 | 0.8931 | 0.9574 | 5000 | 4 | 0.1 | -0.2 |
| 0.0014 | 0.0016 | 0.0016 | 0.0019 | 0.0016 | 0.8596 | 0.9023 | 0.7398 | 0.8725 | 10000 | 2 | 0.1 | -0.2 |
| 0.0014 | 0.0015 | 0.0015 | 0.0016 | 0.0015 | 0.9603 | 0.973 | 0.8908 | 0.9485 | 10000 | 4 | 0.1 | -0.2 |

**WEB TABLE 15-** Empirical variance and ratio of empirical variance (to that with the whole cohort) of log-relative hazard parameter $\beta_1$, for different sampling designs and methods of analysis and variance estimation in 5,000 simulated cohorts.

| Empirical variance | | | | | Ratio of empirical variance with the whole cohort to empirical variance with | | | | | | | |
|---|---|---|---|---|---|---|---|---|---|---|---|---|
| Cohort | SCC | SCC.Calib | USCC | USCC.Calib | SCC | SCC.Calib | USCC | USCC.Calib | $n$ | $K$ | $p_Y$ | $\beta_2$ |
| 0.0197 | 0.0235 | 0.0223 | 0.0287 | 0.0224 | 0.8415 | 0.8867 | 0.6885 | 0.883 | 5000 | 2 | 0.02 | 0.25 |
| 0.0195 | 0.0213 | 0.0205 | 0.0239 | 0.0211 | 0.9165 | 0.9497 | 0.816 | 0.9248 | 5000 | 4 | 0.02 | 0.25 |
| 0.01 | 0.0115 | 0.0109 | 0.014 | 0.0112 | 0.871 | 0.9155 | 0.7143 | 0.8955 | 10000 | 2 | 0.02 | 0.25 |
| 0.0099 | 0.0108 | 0.0105 | 0.012 | 0.0105 | 0.9173 | 0.9499 | 0.8279 | 0.9447 | 10000 | 4 | 0.02 | 0.25 |
| 0.008 | 0.0092 | 0.0088 | 0.0108 | 0.0088 | 0.868 | 0.91 | 0.7415 | 0.909 | 5000 | 2 | 0.05 | 0.25 |
| 0.008 | 0.0085 | 0.0083 | 0.0093 | 0.0083 | 0.9404 | 0.9647 | 0.8621 | 0.9594 | 5000 | 4 | 0.05 | 0.25 |
| 0.0039 | 0.0045 | 0.0042 | 0.0054 | 0.0042 | 0.863 | 0.9241 | 0.725 | 0.9221 | 10000 | 2 | 0.05 | 0.25 |
| 0.0038 | 0.0042 | 0.0041 | 0.0044 | 0.004 | 0.9263 | 0.9501 | 0.8763 | 0.9694 | 10000 | 4 | 0.05 | 0.25 |
| 0.004 | 0.0045 | 0.0043 | 0.0052 | 0.0043 | 0.8895 | 0.9347 | 0.7727 | 0.947 | 5000 | 2 | 0.1 | 0.25 |
| 0.0041 | 0.0042 | 0.0042 | 0.0045 | 0.0042 | 0.9647 | 0.9732 | 0.9184 | 0.9858 | 5000 | 4 | 0.1 | 0.25 |
| 0.002 | 0.0022 | 0.0021 | 0.0025 | 0.0021 | 0.8994 | 0.9384 | 0.7814 | 0.9437 | 10000 | 2 | 0.1 | 0.25 |
| 0.002 | 0.0021 | 0.002 | 0.0022 | 0.002 | 0.9564 | 0.9701 | 0.8997 | 0.9801 | 10000 | 4 | 0.1 | 0.25 |

**WEB TABLE 16-** Empirical variance and ratio of empirical variance (to that with the whole cohort) of log-relative hazard parameter $\beta_2$, for different sampling designs and methods of analysis and variance estimation in 5,000 simulated cohorts.

| Empirical variance | | | | | Ratio of empirical variance with the whole cohort to empirical variance with | | | | | | | |
|---|---|---|---|---|---|---|---|---|---|---|---|---|
| Cohort | SCC | SCC.Calib | USCC | USCC.Calib | SCC | SCC.Calib | USCC | USCC.Calib | $n$ | $K$ | $p_Y$ | $\beta_3$ |
| 0.014 | 0.0217 | 0.0179 | 0.0223 | 0.0182 | 0.6443 | 0.7801 | 0.6254 | 0.7671 | 5000 | 2 | 0.02 | -0.3 |
| 0.0135 | 0.0176 | 0.0155 | 0.0175 | 0.0154 | 0.7652 | 0.8709 | 0.7707 | 0.8774 | 5000 | 4 | 0.02 | -0.3 |
| 0.0069 | 0.0103 | 0.0084 | 0.0109 | 0.0088 | 0.6734 | 0.822 | 0.6335 | 0.7884 | 10000 | 2 | 0.02 | -0.3 |
| 0.0067 | 0.0085 | 0.0075 | 0.0085 | 0.0076 | 0.7921 | 0.8962 | 0.7916 | 0.8881 | 10000 | 4 | 0.02 | -0.3 |
| 0.0056 | 0.0081 | 0.0067 | 0.0082 | 0.0068 | 0.6875 | 0.8256 | 0.6781 | 0.818 | 5000 | 2 | 0.05 | -0.3 |
| 0.0057 | 0.0066 | 0.0061 | 0.0069 | 0.0062 | 0.8593 | 0.9308 | 0.8276 | 0.9152 | 5000 | 4 | 0.05 | -0.3 |
| 0.0028 | 0.004 | 0.0033 | 0.0041 | 0.0034 | 0.696 | 0.8445 | 0.6862 | 0.8338 | 10000 | 2 | 0.05 | -0.3 |
| 0.0027 | 0.0032 | 0.0029 | 0.0032 | 0.0029 | 0.8468 | 0.9291 | 0.8443 | 0.9269 | 10000 | 4 | 0.05 | -0.3 |
| 0.0029 | 0.0037 | 0.0032 | 0.0039 | 0.0032 | 0.7675 | 0.8895 | 0.7349 | 0.881 | 5000 | 2 | 0.1 | -0.3 |
| 0.0028 | 0.0031 | 0.0029 | 0.0032 | 0.003 | 0.9157 | 0.9606 | 0.8901 | 0.9391 | 5000 | 4 | 0.1 | -0.3 |
| 0.0014 | 0.0018 | 0.0016 | 0.0019 | 0.0016 | 0.7728 | 0.8803 | 0.7393 | 0.8669 | 10000 | 2 | 0.1 | -0.3 |
| 0.0014 | 0.0016 | 0.0015 | 0.0016 | 0.0015 | 0.9 | 0.9536 | 0.8925 | 0.9475 | 10000 | 4 | 0.1 | -0.3 |

**WEB TABLE 17-** Empirical variance and ratio of empirical variance (to that with the whole cohort) of log-relative hazard parameter $\beta_3$, for different sampling designs and methods of analysis and variance estimation in 5,000 simulated cohorts.

| Empirical variance | | | | | Ratio of empirical variance with the whole cohort to empirical variance with | | | | $n$ | $K$ | $p_Y$ | $\log\{\pi(\tau_1,\tau_2;\boldsymbol{x})\}$ |
|---|---|---|---|---|---|---|---|---|---|---|---|---|
| Cohort | SCC | SCC.Calib | USCC | USCC.Calib | SCC | SCC.Calib | USCC | USCC.Calib | | | | |
| 0.0254 | 0.0295 | 0.0286 | 0.0332 | 0.03 | 0.8606 | 0.8887 | 0.764 | 0.8475 | 5000 | 2 | 0.02 | -3.948 |
| 0.0255 | 0.0275 | 0.0269 | 0.0293 | 0.0278 | 0.9255 | 0.9458 | 0.8689 | 0.9146 | 5000 | 4 | 0.02 | -3.948 |
| 0.0119 | 0.0136 | 0.0133 | 0.0158 | 0.014 | 0.87 | 0.895 | 0.7522 | 0.8482 | 10000 | 2 | 0.02 | -3.948 |
| 0.0119 | 0.0127 | 0.0124 | 0.0137 | 0.0129 | 0.9384 | 0.9558 | 0.8689 | 0.9234 | 10000 | 4 | 0.02 | -3.948 |
| 0.0097 | 0.0108 | 0.0105 | 0.0124 | 0.0112 | 0.8983 | 0.9279 | 0.7826 | 0.8691 | 5000 | 2 | 0.05 | -3.046 |
| 0.0094 | 0.0099 | 0.0098 | 0.0105 | 0.0101 | 0.952 | 0.9621 | 0.899 | 0.9363 | 5000 | 4 | 0.05 | -3.046 |
| 0.0046 | 0.0052 | 0.005 | 0.0058 | 0.0053 | 0.8931 | 0.9251 | 0.796 | 0.8749 | 10000 | 2 | 0.05 | -3.046 |
| 0.0047 | 0.0049 | 0.0048 | 0.0052 | 0.0049 | 0.9441 | 0.9617 | 0.9008 | 0.9519 | 10000 | 4 | 0.05 | -3.046 |
| 0.0046 | 0.0052 | 0.005 | 0.0056 | 0.0051 | 0.9003 | 0.9276 | 0.8338 | 0.9061 | 5000 | 2 | 0.1 | -2.377 |
| 0.0047 | 0.0048 | 0.0048 | 0.005 | 0.0049 | 0.9741 | 0.9824 | 0.9289 | 0.9628 | 5000 | 4 | 0.1 | -2.377 |
| 0.0023 | 0.0026 | 0.0025 | 0.0028 | 0.0026 | 0.9076 | 0.9394 | 0.8249 | 0.8978 | 10000 | 2 | 0.1 | -2.377 |
| 0.0023 | 0.0023 | 0.0023 | 0.0024 | 0.0024 | 0.9735 | 0.9836 | 0.9366 | 0.9703 | 10000 | 4 | 0.1 | -2.377 |

**WEB TABLE 18-** Empirical variance and ratio of empirical variance (to that with the whole cohort) of log-pure risk $\log\{\pi(\tau_1,\tau_2;\boldsymbol{x})\}$, with $(\tau_1,\tau_2] = (0,8]$ and $\boldsymbol{x} = (-1, 1, -0.6)'$, for different sampling designs and methods of analysis and variance estimation in 5,000 simulated cohorts.

| Empirical variance | | | | | Ratio of empirical variance with the whole cohort to empirical variance with | | | | $n$ | $K$ | $p_Y$ | $\log\{\pi(\tau_1,\tau_2;\boldsymbol{x})\}$ |
|---|---|---|---|---|---|---|---|---|---|---|---|---|
| Cohort | SCC | SCC.Calib | USCC | USCC.Calib | SCC | SCC.Calib | USCC | USCC.Calib | | | | |
| 0.1248 | 0.1423 | 0.1423 | 0.1388 | 0.1725 | 0.8772 | 0.8991 | 0.7233 | 0.8799 | 5000 | 2 | 0.02 | -5.201 |
| 0.1294 | 0.1382 | 0.1382 | 0.1343 | 0.1523 | 0.9363 | 0.9635 | 0.85 | 0.9352 | 5000 | 4 | 0.02 | -5.201 |
| 0.0618 | 0.0688 | 0.0688 | 0.0662 | 0.0823 | 0.8983 | 0.934 | 0.7518 | 0.8891 | 10000 | 2 | 0.02 | -5.201 |
| 0.062 | 0.0657 | 0.0657 | 0.0646 | 0.0721 | 0.9438 | 0.96 | 0.8604 | 0.9481 | 10000 | 4 | 0.02 | -5.201 |
| 0.0514 | 0.0568 | 0.0568 | 0.0551 | 0.0647 | 0.9047 | 0.9337 | 0.795 | 0.919 | 5000 | 2 | 0.05 | -4.288 |
| 0.0499 | 0.052 | 0.052 | 0.0513 | 0.0562 | 0.9599 | 0.9719 | 0.8882 | 0.9577 | 5000 | 4 | 0.05 | -4.288 |
| 0.0241 | 0.0266 | 0.0266 | 0.026 | 0.0313 | 0.9064 | 0.9295 | 0.7713 | 0.9185 | 10000 | 2 | 0.05 | -4.288 |
| 0.0245 | 0.0255 | 0.0255 | 0.0252 | 0.0275 | 0.9636 | 0.9742 | 0.8908 | 0.9717 | 10000 | 4 | 0.05 | -4.288 |
| 0.025 | 0.027 | 0.027 | 0.0261 | 0.0305 | 0.9256 | 0.9544 | 0.8188 | 0.9382 | 5000 | 2 | 0.1 | -3.602 |
| 0.0243 | 0.0248 | 0.0248 | 0.0247 | 0.0259 | 0.9789 | 0.9813 | 0.9362 | 0.9815 | 5000 | 4 | 0.1 | -3.602 |
| 0.0125 | 0.0133 | 0.0133 | 0.013 | 0.0151 | 0.9411 | 0.9608 | 0.8257 | 0.9424 | 10000 | 2 | 0.1 | -3.602 |
| 0.0122 | 0.0125 | 0.0125 | 0.0124 | 0.0132 | 0.9762 | 0.982 | 0.92 | 0.9778 | 10000 | 4 | 0.1 | -3.602 |

**WEB TABLE 19-** Empirical variance and ratio of empirical variance (to that with the whole cohort) of log-pure risk $log\{\pi(\tau_1,\tau_2;\boldsymbol{x})\}$, with $(\tau_1,\tau_2]=(0,8]$ and $\boldsymbol{x}=(1,-1,0.6)'$, for different sampling designs and methods of analysis and variance estimation in 5,000 simulated cohorts.

| Empirical variance | | | | | Ratio of empirical variance with the whole cohort to empirical variance with | | | | $n$ | $K$ | $p_Y$ | $\log\{\pi(\tau_1,\tau_2;\boldsymbol{x})\}$ |
|---|---|---|---|---|---|---|---|---|---|---|---|---|
| Cohort | SCC | SCC.Calib | USCC | USCC.Calib | SCC | SCC.Calib | USCC | USCC.Calib | | | | |
| 0.0537 | 0.0668 | 0.0631 | 0.0721 | 0.0645 | 0.8032 | 0.85 | 0.7448 | 0.8327 | 5000 | 2 | 0.02 | -4.702 |
| 0.058 | 0.0648 | 0.0622 | 0.0678 | 0.0634 | 0.8962 | 0.9327 | 0.8557 | 0.9156 | 5000 | 4 | 0.02 | -4.702 |
| 0.0274 | 0.0326 | 0.0309 | 0.0361 | 0.0325 | 0.8421 | 0.8865 | 0.7589 | 0.8439 | 10000 | 2 | 0.02 | -4.702 |
| 0.0277 | 0.0303 | 0.0292 | 0.0315 | 0.03 | 0.914 | 0.9481 | 0.8793 | 0.9248 | 10000 | 4 | 0.02 | -4.702 |
| 0.0226 | 0.0265 | 0.0252 | 0.028 | 0.0253 | 0.8506 | 0.8934 | 0.8069 | 0.8907 | 5000 | 2 | 0.05 | -3.793 |
| 0.0218 | 0.0233 | 0.0227 | 0.0246 | 0.0232 | 0.935 | 0.9592 | 0.8868 | 0.9397 | 5000 | 4 | 0.05 | -3.793 |
| 0.0111 | 0.0128 | 0.0121 | 0.014 | 0.0126 | 0.8618 | 0.9145 | 0.7903 | 0.8766 | 10000 | 2 | 0.05 | -3.793 |
| 0.0111 | 0.0118 | 0.0114 | 0.0123 | 0.0116 | 0.9405 | 0.9741 | 0.9017 | 0.9548 | 10000 | 4 | 0.05 | -3.793 |
| 0.0114 | 0.0126 | 0.0122 | 0.0135 | 0.0123 | 0.9051 | 0.938 | 0.8434 | 0.928 | 5000 | 2 | 0.1 | -3.111 |
| 0.0109 | 0.0113 | 0.0112 | 0.0116 | 0.0114 | 0.9625 | 0.9744 | 0.9405 | 0.9582 | 5000 | 4 | 0.1 | -3.111 |
| 0.0055 | 0.0061 | 0.0058 | 0.0066 | 0.006 | 0.9004 | 0.9369 | 0.8318 | 0.9183 | 10000 | 2 | 0.1 | -3.111 |
| 0.0054 | 0.0056 | 0.0055 | 0.0058 | 0.0057 | 0.963 | 0.9815 | 0.9306 | 0.9582 | 10000 | 4 | 0.1 | -3.111 |

**WEB TABLE 20-** Empirical variance and ratio of empirical variance (to that with the whole cohort) of log-pure risk $log\{\pi(\tau_1,\tau_2;\boldsymbol{x})\}$, with $(\tau_1,\tau_2]=(0,8]$ and $\boldsymbol{x}=(1,1,0.6)'$, for different sampling designs and methods of analysis and variance estimation in 5,000 simulated cohorts.

### E.4 Additional simulations, with design weights poststratified on the number of non-cases

In the simulation in Section 7 in the Main Document and in Web Appendix E.2, the results for SCC and USCC were obtained from using the design weights. In the literature, several authors computed the design weights after having redefined the strata by excluding the cases (Borgan et al., 2000; Samuelsen et al., 2007); this amounts to poststratification of the weights on the number of non-cases. In this Web Appendix, we performed the same simulations as before, except that in addition to estimation with SCC and USCC, estimation was also performed using the stratified case-cohort with post-stratified weights (SCC.Poststrat); and the unstratified case-cohort with post-stratified weights (USCC.Poststrat).

The ratio of the empirical variance with SCC to the empirical variance with SCC.Poststrat and the ratio of the empirical variance with USCC to the empirical variance with USCC.Poststrat, are displayed for $\beta_1, \beta_2$ and $\beta_3$ (**WEB TABLE 21**) and for $\pi(\tau_1, \tau_2; x)$, with $(\tau_1, \tau_2] = (0,8]$ and $x \in \{(-1, 1, -0.6)', (1, -1, 0.6)', (1, 1, 0.6)'\}$ (**WEB TABLE 22**). These ratios measure the efficiency gain from using the post-stratified weights instead of the design weights, with the stratified case-cohort and with the unstratified case-cohort designs, respectively. The ratios are all close to one, indicating almost no improvement with poststratification of the weights on the number of non-cases.

| $\beta_1$ | | $\beta_2$ | | $\beta_3$ | | $n$ | $K$ | $p_Y$ |
|---|---|---|---|---|---|---|---|---|
| SCC / SCC.Poststrat | USCC / USCC.Poststrat | SCC / SCC.Poststrat | USCC / USCC.Poststrat | SCC / SCC.Poststrat | USCC / USCC.Poststrat | | | |
| 1.002 | 1 | 1.001 | 1 | 1 | 1 | 5000 | 2 | 0.02 |
| 1.003 | 1 | 1.003 | 1 | 1 | 1 | 5000 | 4 | 0.02 |
| 1.001 | 1 | 1.004 | 1 | 1 | 1 | 10000 | 2 | 0.02 |
| 1 | 1 | 1.002 | 1 | 1 | 1 | 10000 | 4 | 0.02 |
| 1.005 | 1 | 1.004 | 1 | 1 | 1 | 5000 | 2 | 0.05 |
| 1.001 | 1 | 1 | 1 | 1 | 1 | 5000 | 4 | 0.05 |
| 1.006 | 1 | 1.006 | 1 | 1 | 1 | 10000 | 2 | 0.05 |
| 1.004 | 1 | 1.004 | 1 | 1 | 1 | 10000 | 4 | 0.05 |
| 1.015 | 1 | 1.011 | 1 | 1 | 1 | 5000 | 2 | 0.1 |
| 1 | 1 | 1.006 | 1 | 1 | 1 | 5000 | 4 | 0.1 |
| 1.012 | 1 | 1.01 | 1 | 1 | 1 | 10000 | 2 | 0.1 |
| 1.004 | 1 | 1.003 | 1 | 1 | 1 | 10000 | 4 | 0.1 |

**WEB TABLE 21-** Ratio of the empirical variance with SCC to the empirical variance with SCC. Poststrat and ratio of the empirical variance with USCC to the empirical variance with USCC.Poststrat, for the log-relative hazard parameters $\beta_1$, $\beta_2$ and $\beta_3$, from 5,000 simulated cohorts. The ratio is a measure of efficiency gain from using the poststratified weights.

| $x = (-1, 1, -0.6)'$ | | $x = (1, -1, 0.6)'$ | | $x = (1, 1, 0.6)'$ | | $n$ | $K$ | $p_Y$ |
|---|---|---|---|---|---|---|---|---|
| SCC / SCC.Poststrat | USCC / USCC.Poststrat | SCC / SCC.Poststrat | USCC / USCC.Poststrat | SCC / SCC.Poststrat | USCC / USCC.Poststrat | | | |
| 1.005 | 1.005 | 1.006 | 1.001 | 1.003 | 1 | 5000 | 2 | 0.02 |
| 1.001 | 1.002 | 1.005 | 0.999 | 1.002 | 1 | 5000 | 4 | 0.02 |
| 1.002 | 1.003 | 1.006 | 1.001 | 1 | 1.001 | 10000 | 2 | 0.02 |
| 1.002 | 1.003 | 1.002 | 1.001 | 1.001 | 1 | 10000 | 4 | 0.02 |
| 1.009 | 1.005 | 1.007 | 1.002 | 1.008 | 1.002 | 5000 | 2 | 0.05 |
| 1.008 | 1.005 | 1 | 1.001 | 1.002 | 1.001 | 5000 | 4 | 0.05 |
| 1.005 | 1.005 | 1.008 | 1.001 | 1.007 | 1.003 | 10000 | 2 | 0.05 |
| 1.004 | 1 | 1.005 | 1.001 | 1.005 | 1.002 | 10000 | 4 | 0.05 |
| 1.018 | 1.011 | 1.012 | 1.003 | 1.011 | 1.006 | 5000 | 2 | 0.1 |
| 1.005 | 1.001 | 1.006 | 1 | 1.002 | 1.002 | 5000 | 4 | 0.1 |
| 1.019 | 1.007 | 1.01 | 1.001 | 1.006 | 1.001 | 10000 | 2 | 0.1 |
| 1.002 | 1.003 | 1.004 | 1 | 1.004 | 1.002 | 10000 | 4 | 0.1 |

**WEB TABLE 22-** Ratio of the empirical variance with SCC to the empirical variance with SCC. Poststrat and ratio of the empirical variance with USCC to the empirical variance with USCC.Poststrat, for pure risks $\pi(\tau_1, \tau_2; x)$, with $(\tau_1, \tau_2] = (0,8]$ and $x \in \{(-1, 1, -0.6)', (1, -1, 0.6)', (1, 1, 0.6)'\}$, from 5,000 simulated cohorts. The ratio is a measure of efficiency gain from using the poststratified weights.

### E.5 Additional simulations, with weights calibrated using weaker proxies

In the simulations in Section 7 in the Main Document and in Web Appendix E.2, the results for SCC.Calib were obtained by using proxies $\widetilde{X} = (\widetilde{X}_1, \widetilde{X}_2, \widetilde{X}_3)'$, with $\widetilde{X}_1 = X_1 + \varepsilon_1$, $\varepsilon_1 \sim \mathcal{N}(0, 0.75^2)$, and thus such that $\text{corr}(\widetilde{X}_1, X_1) = 0.8$. In this Section, we performed the same simulations, except that we used $\varepsilon_1 \sim \mathcal{N}(0, 1.2^2)$, so that $\text{corr}(\widetilde{X}_1, X_1) \approx 0.64$.

**WEB TABLE 23** and **WEB TABLE 24** display the estimation results with SCC.Calib for the log-relative hazard $\beta_1$ and the pure risk with profile $x = (-1, 1, -0.6)'$, respectively. Because of the weaker proxy for $X_1$, the robust variance formula overestimated the variance and yielded supra-nominal confidence interval coverage in many scenarios. In addition, the efficiency gain is more modest than with the stronger proxy; see empirical variances in **WEB TABLE 15** and **WEB TABLE 18** in Web Appendix E.2.

| | SCC.Calib | | | | | | | |
|---|---|---|---|---|---|---|---|---|
| Empirical variance | Mean of estimated variance | | Coverage of 95% CIs | | $n$ | $K$ | $p_Y$ | $\beta_1$ |
| | $\hat{V}_{Robust}$ | $\hat{V}$ | $\hat{V}_{Robust}$ | $\hat{V}$ | | | | |
| 0.0184 | 0.0193 | 0.0182 | 0.9582* | 0.9524 | 5000 | 2 | 0.02 | -0.2 |
| 0.0161 | 0.0165 | 0.016 | 0.954 | 0.9506 | 5000 | 4 | 0.02 | -0.2 |
| 0.0085 | 0.0091 | 0.0086 | 0.9584* | 0.9528 | 10000 | 2 | 0.02 | -0.2 |
| 0.0076 | 0.008 | 0.0077 | 0.9562* | 0.9532 | 10000 | 4 | 0.02 | -0.2 |
| 0.0069 | 0.0072 | 0.0068 | 0.954 | 0.9478 | 5000 | 2 | 0.05 | -0.2 |
| 0.0061 | 0.0063 | 0.0061 | 0.9516 | 0.9486 | 5000 | 4 | 0.05 | -0.2 |
| 0.0033 | 0.0035 | 0.0033 | 0.9586* | 0.9526 | 10000 | 2 | 0.05 | -0.2 |
| 0.0031 | 0.0031 | 0.003 | 0.9526 | 0.9494 | 10000 | 4 | 0.05 | -0.2 |
| 0.0033 | 0.0034 | 0.0033 | 0.9582* | 0.9536 | 5000 | 2 | 0.1 | -0.2 |
| 0.0031 | 0.0031 | 0.003 | 0.9482 | 0.9466 | 5000 | 4 | 0.1 | -0.2 |
| 0.0016 | 0.0017 | 0.0016 | 0.959* | 0.9528 | 10000 | 2 | 0.1 | -0.2 |
| 0.0015 | 0.0015 | 0.0015 | 0.952 | 0.9496 | 10000 | 4 | 0.1 | -0.2 |

**WEB TABLE 23-** Estimation results with SCC.Calib (with the weaker proxy) for the log-relative hazard $\beta_1$, from using different variance estimation methods in 5,000 simulated cohorts. * indicates coverage outside the expected interval [0.9440; 0.9560].

|  | SCC.Calib | | | | | | | |
|---|---|---|---|---|---|---|---|---|
| Empirical variance | Mean of estimated variance | | Coverage of 95% CIs | | $n$ | $K$ | $p_Y$ | $\log\{\pi(\tau_1, \tau_2; \boldsymbol{x})\}$ |
|  | $\hat{V}_{Robust}$ | $\hat{V}$ | $\hat{V}_{Robust}$ | $\hat{V}$ | | | | |
| 0.0292 | 0.031 | 0.0292 | 0.9576* | 0.9514 | 5000 | 2 | 0.02 | -3.948 |
| 0.027 | 0.0278 | 0.027 | 0.9526 | 0.949 | 5000 | 4 | 0.02 | -3.948 |
| 0.014 | 0.0148 | 0.014 | 0.959* | 0.9524 | 10000 | 2 | 0.02 | -3.948 |
| 0.0128 | 0.0135 | 0.0131 | 0.9584* | 0.9536 | 10000 | 4 | 0.02 | -3.948 |
| 0.0106 | 0.0114 | 0.0108 | 0.9618* | 0.9562* | 5000 | 2 | 0.05 | -3.046 |
| 0.0105 | 0.0104 | 0.0102 | 0.9478 | 0.945 | 5000 | 4 | 0.05 | -3.046 |
| 0.0053 | 0.0056 | 0.0053 | 0.954 | 0.9494 | 10000 | 2 | 0.05 | -3.046 |
| 0.005 | 0.0051 | 0.005 | 0.9546 | 0.9516 | 10000 | 4 | 0.05 | -3.046 |
| 0.005 | 0.0053 | 0.0051 | 0.9532 | 0.9486 | 5000 | 2 | 0.1 | -2.377 |
| 0.005 | 0.0049 | 0.0048 | 0.9494 | 0.9472 | 5000 | 4 | 0.1 | -2.377 |
| 0.0024 | 0.0026 | 0.0025 | 0.959* | 0.9548 | 10000 | 2 | 0.1 | -2.377 |
| 0.0025 | 0.0024 | 0.0024 | 0.948 | 0.9462 | 10000 | 4 | 0.1 | -2.377 |

**WEB TABLE 24-** Estimation results with SCC.Calib (with the weaker proxy) for the pure risk with profile $\boldsymbol{x} = (-1, 1, -0.6)'$, from using different variance estimation methods in 5,000 simulated cohorts. * indicates coverage outside the expected interval [0.9440; 0.9560].

# Web Appendix F. GOLESTAN DATA ANALYSIS – RESULTS FROM 2,500 REPLICATIONS

We replicated the data analysis in Section 8 in the Main Document with 2,500 cohorts of size $n = 20{,}000$ sampled with replacement from the Golestan dataset. In this way, we were able to compare the means of the estimated variances with the empirical variances. Recall that we used the age-scale and ran a Cox proportional hazards model predicting mortality from baseline variables: $x_1 =$ indicator of male gender, $x_2 =$ wealth score, $x_3 =$ indicator of former smoker (cigarettes, nass, or opium), $x_4 =$ indicator of current smoker, $x_5 =$ indicator of morbidity, $x_6 = x_1 x_3$ and $x_7 = x_1 x_4$. For stratified sampling, **WEB TABLE 25** and **WEB TABLE 26** display the estimation results for log-relative hazard and covariate specific pure risks, respectively. Results for unstratified sampling are displayed in **WEB TABLE 27** and **WEB TABLE 28.**

When using design weights in the stratified design, the robust variance formula overestimated the variances of log-relative hazards for covariates $x_1$ and $x_2$ and of the pure risks with profiles $\boldsymbol{x} \in \{(0, -0.4, 0, 1, \mathbf{0}_3)', (\mathbf{0}_4, 1, \mathbf{0}_2)', \mathbf{0}_7'\}$. For the log-relative hazard of $x_6$ and for the pure risk with profile $\boldsymbol{x} = (0, 0.4, 0, 1, \mathbf{0}_3)'$, the mean of the robust variance estimates was actually the closest to the empirical variance, but the difference between the two means of variance estimates was negligible. This may be due to a violation of the proportional hazard assumption, or because the available smoking variable is too crude (for example, two individuals smoking 2 and 30 cigarettes per day, respectively, may share the same covariate profile). In addition, in this data example, the robust variance formula led to valid inference for many parameters in the stratified design, possibly because stratification was based on only one of the 7 covariates, namely $x_1$. On the other hand, the robust variance formula slightly overestimated the variance for the pure risks with profiles $\boldsymbol{x} \in$ $\{(\mathbf{0}_4, 1, \mathbf{0}_2)', \mathbf{0}_7'\}$, in the unstratified design with design weights. For the two other pure risks, the mean of the robust variance estimates was the closest to the empirical variance.

Stratification improved efficiency, but calibration of the design weights led to even more efficient estimation. In particular, the precision from using calibrated weights in the unstratified case-cohort design was comparable to that from using the whole cohort, or to that from using the stratified case-cohort with calibrated weights, except for the log-relative hazard of covariate $x_2$.

| Parameter | Cohort | | SCC | | | SCC.Calib | | |
|---|---|---|---|---|---|---|---|---|
| | Emp. variance | $\hat{V}_{\text{Robust}}$ | Emp. variance | $\hat{V}_{\text{Robust}}$ | $\hat{V}$ | Emp. variance | $\hat{V}_{\text{Robust}}$ | $\hat{V}$ |
| $\beta_1$ | 0.002 | 0.002 | 0.0024 | 0.0028 | 0.0024 | 0.002 | 0.002 | 0.002 |
| $\beta_2$ | 0.0071 | 0.0073 | 0.0103 | 0.0107 | 0.0104 | 0.0076 | 0.0077 | 0.0077 |
| $\beta_3$ | 0.0253 | 0.0254 | 0.0434 | 0.0419 | 0.0419 | 0.0253 | 0.0261 | 0.0261 |
| $\beta_4$ | 0.0042 | 0.0042 | 0.0081 | 0.008 | 0.008 | 0.0042 | 0.0042 | 0.0042 |
| $\beta_5$ | 0.0013 | 0.0013 | 0.002 | 0.002 | 0.002 | 0.0013 | 0.0013 | 0.0013 |
| $\beta_6$ | 0.0321 | 0.0319 | 0.0518 | 0.0501 | 0.05 | 0.0321 | 0.0326 | 0.0326 |
| $\beta_7$ | 0.0063 | 0.0063 | 0.0109 | 0.0109 | 0.0109 | 0.0064 | 0.0064 | 0.0064 |

(All rows under Log-relative hazard)

**WEB TABLE 25-** Empirical variance and mean of estimated variances of log-relative hazard parameters from using different methods of analysis and variance estimation with stratified sampling, in 2,500 replications from the Golestan cohort. $\beta_p$ denotes the log-relative hazard parameter of covariate $x_p$, $p \in \{1, ..., 7\}$.

| Parameter | | Cohort | | SCC | | | SCC.Calib | | |
|---|---|---|---|---|---|---|---|---|---|
| | | Emp. variance | $\hat{V}_{Robust}$ | Emp. variance | $\hat{V}_{Robust}$ | $\hat{V}$ | Emp. variance | $\hat{V}_{Robust}$ | $\hat{V}$ |
| $\pi(\tau_1, \tau_2; \boldsymbol{x})$ | $\boldsymbol{x} = (0, -0.4, 0, 1, \boldsymbol{0}_3)'$ | 0.0035 | 0.0036 | 0.0063 | 0.0066 | 0.0064 | 0.0036 | 0.0037 | 0.0037 |
| | $\boldsymbol{x} = (0, 0.4, 0, 1, \boldsymbol{0}_3)'$ | 0.0053 | 0.0052 | 0.0091 | 0.009 | 0.0086 | 0.0054 | 0.0053 | 0.0053 |
| | $\boldsymbol{x} = (\boldsymbol{0}_4, 1, \boldsymbol{0}_2)'$ | 0.0014 | 0.0014 | 0.002 | 0.0022 | 0.0019 | 0.0014 | 0.0014 | 0.0014 |
| | $\boldsymbol{x} = \boldsymbol{0}_7'$ | 0.0012 | 0.0012 | 0.0014 | 0.0018 | 0.0014 | 0.0013 | 0.0013 | 0.0013 |

**WEB TABLE 26-** Empirical variance and mean of estimated variances of pure risk parameters from using different methods of analysis and variance estimation with stratified samples, in 2,500 replications from the Golestan cohort.

| Parameter | Cohort | | USCC | | | USCC.Calib | | |
|---|---|---|---|---|---|---|---|---|
| | Emp. variance | $\hat{V}_{\text{Robust}}$ | Emp. variance | $\hat{V}_{\text{Robust}}$ | $\hat{V}$ | Emp. variance | $\hat{V}_{\text{Robust}}$ | $\hat{V}$ |
| $\beta_1$ | 0.002 | 0.002 | 0.0039 | 0.0039 | 0.0039 | 0.002 | 0.002 | 0.002 |
| $\beta_2$ | 0.0071 | 0.0073 | 0.015 | 0.0146 | 0.0146 | 0.0081 | 0.0083 | 0.0083 |
| $\beta_3$ | 0.0253 | 0.0254 | 0.0566 | 0.0541 | 0.0541 | 0.0254 | 0.0277 | 0.0277 |
| $\beta_4$ | 0.0042 | 0.0042 | 0.0097 | 0.0093 | 0.0093 | 0.0042 | 0.0042 | 0.0042 |
| $\beta_5$ | 0.0013 | 0.0013 | 0.0029 | 0.0028 | 0.0028 | 0.0013 | 0.0013 | 0.0013 |
| $\beta_6$ | 0.0321 | 0.0319 | 0.0707 | 0.0686 | 0.0686 | 0.0322 | 0.0343 | 0.0343 |
| $\beta_7$ | 0.0063 | 0.0063 | 0.0148 | 0.014 | 0.0141 | 0.0064 | 0.0064 | 0.0064 |

Rows labelled under "Log-relative hazard".

**WEB TABLE 27-** Empirical variance and mean of estimated variances of log-relative hazard parameters from using different methods of analysis and variance estimation with unstratified sampling in 2,500 replications from the Golestan cohort. $\beta_p$ denotes the log-relative hazard parameter of covariate $x_p$, $p \in \{1, \ldots, 7\}$.

| Parameter | | Cohort | | USCC | | | USCC.Calib | | |
|---|---|---|---|---|---|---|---|---|---|
| | | Emp. variance | $\hat{V}_{Robust}$ | Emp. variance | $\hat{V}_{Robust}$ | $\hat{V}$ | Emp. variance | $\hat{V}_{Robust}$ | $\hat{V}$ |
| $\pi(\tau_1, \tau_2; \boldsymbol{x})$ | $\boldsymbol{x} = (0, -0.4, 0, 1, \boldsymbol{0}_3)'$ | 0.0035 | 0.0036 | 0.0078 | 0.0077 | 0.0076 | 0.0036 | 0.0037 | 0.0037 |
| | $\boldsymbol{x} = (0, 0.4, 0, 1, \boldsymbol{0}_3)'$ | 0.0053 | 0.0052 | 0.0115 | 0.0109 | 0.0108 | 0.0056 | 0.0055 | 0.0055 |
| | $\boldsymbol{x} = (\boldsymbol{0}_4, 1, \boldsymbol{0}_2)'$ | 0.0014 | 0.0014 | 0.0024 | 0.0025 | 0.0024 | 0.0014 | 0.0015 | 0.0014 |
| | $\boldsymbol{x} = \boldsymbol{0}_7'$ | 0.0012 | 0.0012 | 0.0018 | 0.0019 | 0.0018 | 0.0013 | 0.0013 | 0.0013 |

**WEB TABLE 28-** Empirical variance and mean of estimated variances of pure risk parameters from using different methods of analysis and variance estimation with unstratified sampling in 2,500 replications from the Golestan cohort.

# Web Appendix G. MISSING DATA WHEN THE PHASE-THREE DESIGN SAMPLING PROBABILITIES ARE UNKNOWN

## G.1 Derivation of the influence functions

We let $\boldsymbol{\Delta}_{i,j}(\widetilde{\boldsymbol{\theta}})$ denote the influence of subject $i$ in stratum $j$ on $\widetilde{\boldsymbol{\theta}}$, $i \in \{1, \ldots, n^{(j)}\}, j \in \{1, \ldots, J\}$, $\widetilde{\boldsymbol{\theta}} \in \{\widetilde{\boldsymbol{\gamma}}, \widetilde{\boldsymbol{\beta}}, d\widetilde{\Lambda}_0(t), \widetilde{\Lambda}_0(t), \widetilde{\pi}(\tau_1, \tau_2; \boldsymbol{x})\}$.

As $\sum_{j=1}^{J} \sum_{i=1}^{n^{(j)}} \{\xi_{i,j} \boldsymbol{B}_{i,j} - \exp(\widetilde{\boldsymbol{\gamma}}' \boldsymbol{B}_{i,j}) \xi_{i,j} V_{i,j} \boldsymbol{B}_{i,j}\} = 0$, then, following Graubard and Fears (2005), we know that

$$\xi_{i,j} \boldsymbol{B}_{i,j} - \exp(\widetilde{\boldsymbol{\gamma}}' \boldsymbol{B}_{i,j}) \xi_{i,j} V_{i,j} \boldsymbol{B}_{i,j} - \left\{\sum_{l=1}^{J} \sum_{k=1}^{n^{(l)}} \xi_{k,l} V_{k,l} \exp(\widetilde{\boldsymbol{\gamma}}' \boldsymbol{B}_{k,l}) \boldsymbol{B}_{k,l} \boldsymbol{B}_{k,l}'\right\} \boldsymbol{\Delta}_{i,j}(\widetilde{\boldsymbol{\gamma}}) = 0,$$

and as a result

$$\boldsymbol{\Delta}_{i,j}(\widetilde{\boldsymbol{\gamma}}) = \left\{\sum_{l=1}^{J} \sum_{k=1}^{n^{(l)}} \xi_{k,l} V_{k,l} \exp(\widetilde{\boldsymbol{\gamma}}' \boldsymbol{B}_{k,l}) \boldsymbol{B}_{k,l} \boldsymbol{B}_{k,l}'\right\}^{-1} \{\xi_{i,j} \boldsymbol{B}_{i,j} - \xi_{i,j} V_{i,j} \exp(\widetilde{\boldsymbol{\gamma}}' \boldsymbol{B}_{i,j}) \boldsymbol{B}_{i,j}\}.$$

We can then write

$$\boldsymbol{\Delta}_{i,j}(\widetilde{\boldsymbol{\gamma}}) = \xi_{i,j} \, \boldsymbol{IF}^{(2)}_{i,j}(\widetilde{\boldsymbol{\gamma}}) + \xi_{i,j} V_{i,j} \exp(\widetilde{\boldsymbol{\gamma}}' \boldsymbol{B}_{i,j}) \, \boldsymbol{IF}^{(3)}_{i,j}(\widetilde{\boldsymbol{\gamma}}),$$

with $\boldsymbol{IF}^{(2)}_{i,j}(\widetilde{\boldsymbol{\gamma}}) = \left\{\sum_{l=1}^{J} \sum_{k=1}^{n^{(l)}} \xi_{k,l} V_{k,l} \exp(\widetilde{\boldsymbol{\gamma}}' \boldsymbol{B}_{k,l}) \boldsymbol{B}_{k,l} \boldsymbol{B}_{k,l}'\right\}^{-1} \boldsymbol{B}_{i,j}$,

and $\boldsymbol{IF}^{(3)}_{i,j}(\widetilde{\boldsymbol{\gamma}}) = -\left\{\sum_{l=1}^{J} \sum_{k=1}^{n^{(l)}} \xi_{k,l} V_{k,l} \exp(\widetilde{\boldsymbol{\gamma}}' \boldsymbol{B}_{k,l}) \boldsymbol{B}_{k,l} \boldsymbol{B}_{k,l}'\right\}^{-1} \boldsymbol{B}_{i,j}$.

Then, we know that $\int_t \sum_{j=1}^{J} \sum_{i=1}^{n^{(j)}} V_{i,j} \exp(\widetilde{\boldsymbol{\gamma}}' \boldsymbol{B}_{i,j}) \left\{X_{i,j} - \frac{\tilde{s}_1(t;\widetilde{\boldsymbol{\gamma}},\widetilde{\boldsymbol{\beta}})}{\tilde{s}_0(t;\widetilde{\boldsymbol{\gamma}},\widetilde{\boldsymbol{\beta}})}\right\} dN_{i,j}(t) = 0$. We can rewrite this estimating equation as $\int_t \sum_{j=1}^{J} \sum_{i=1}^{n^{(j)}} \xi_{i,j} V_{i,j} w^{(2)}_{i,j} \exp(\widetilde{\boldsymbol{\gamma}}' \boldsymbol{B}_{i,j}) \left\{X_{i,j} - \frac{\tilde{s}_1(t;\widetilde{\boldsymbol{\gamma}},\widetilde{\boldsymbol{\beta}})}{\tilde{s}_0(t;\widetilde{\boldsymbol{\gamma}},\widetilde{\boldsymbol{\beta}})}\right\} dN_{i,j}(t) = 0$, because $\xi_{i,j} w^{(2)}_{i,j} = 1$ for any subject $i$ in stratum $j$ such that $\int_t dN_{i,j}(t) = 1$ (i.e., all the cases are included in the phase-two sample and have unit phase-two sampling weights). As a result, and following the similar arguments as for $\boldsymbol{\Delta}_{i,j}(\widetilde{\boldsymbol{\gamma}})$, we know

$$\Delta_{i,j}\{\widetilde{G}_1(\widetilde{\gamma})\} + \int_t \left[ -\frac{\Delta_{i,j}\left\{\sum_{l=1}^{J}\sum_{k=1}^{n^{(l)}} \xi_{k,l} V_{k,l} w_{k,l}^{(2)} \exp(\widetilde{\gamma}' B_{k,l}) dN_{k,l}(t)\right\} \times \tilde{S}_1(t;\widetilde{\gamma},\widetilde{\beta})}{\tilde{S}_0(t;\widetilde{\gamma},\widetilde{\beta})} - \right.$$

$$\frac{\left\{\sum_{l=1}^{J}\sum_{k=1}^{n^{(l)}} \xi_{k,l} V_{k,l} w_{k,l}^{(2)} \exp(\widetilde{\gamma}' B_{k,l}) dN_{k,l}(t)\right\} \times \Delta_{i,j}\{\tilde{S}_1(t;\widetilde{\gamma},\widetilde{\beta})\}}{\tilde{S}_0(t;\widetilde{\gamma},\widetilde{\beta})} + \quad (3)$$

$$\left. \frac{\left\{\sum_{l=1}^{J}\sum_{k=1}^{n^{(l)}} \xi_{k,l} V_{k,l} w_{k,l}^{(2)} \exp(\widetilde{\gamma}' B_{k,l}) dN_{k,l}(t)\right\} \times \tilde{S}_1(t;\widetilde{\gamma},\widetilde{\beta}) \times \Delta_{i,j}\{\tilde{S}_0(t;\widetilde{\gamma},\widetilde{\beta})\}}{\{\tilde{S}_0(t;\widetilde{\gamma},\widetilde{\beta})\}^2} \right] = 0,$$

with

$$\widetilde{G}_1(\widetilde{\gamma}) = \int_t \sum_{l=1}^{J}\sum_{k=1}^{n^{(l)}} \xi_{k,l} V_{k,l} w_{k,l}^{(2)} \exp(\widetilde{\gamma}' B_{k,l}) X_{k,l} dN_{k,l}(t),$$

$$\Delta_{i,j}\{\widetilde{G}_1(t,\widetilde{\gamma})\} = \xi_{i,j} V_{i,j} w_{i,j}^{(2)} \exp(\widetilde{\gamma}' B_{i,j}) dN_{i,j}(t) X_{i,j} + \left\{\frac{\partial \widetilde{G}_1(t,\gamma)}{\partial \gamma}\bigg|_{\gamma=\widetilde{\gamma}}\right\} \Delta_{i,j}(\widetilde{\gamma}),$$

$$\Delta_{i,j}\left\{\sum_{l=1}^{J}\sum_{k=1}^{n^{(l)}} \xi_{k,l} V_{k,l} w_{k,l}^{(2)} \exp(\widetilde{\gamma}' B_{k,l}) dN_{k,l}(t)\right\} = \xi_{i,j} V_{i,j} w_{i,j}^{(2)} \exp(\widetilde{\gamma}' B_{i,j}) dN_{i,j}(t) +$$

$$\left[\frac{\partial\left\{\sum_{l=1}^{J}\sum_{k=1}^{n^{(l)}} \xi_{k,l} V_{k,l} w_{k,l}^{(2)} \exp(\gamma' B_{k,l}) dN_{k,l}(t)\right\}}{\partial \gamma}\bigg|_{\gamma=\widetilde{\gamma}}\right] \Delta_{i,j}(\widetilde{\gamma}),$$

$$\frac{\partial \widetilde{G}_1(t,\gamma)}{\partial \gamma}\bigg|_{\gamma=\widetilde{\gamma}} = \sum_{l=1}^{J}\sum_{k=1}^{n^{(l)}} \xi_{k,l} V_{k,l} w_{k,l}^{(2)} \exp(\widetilde{\gamma}' B_{k,l}) dN_{k,l}(t) X_{k,l} B_{k,l}',$$

$$\frac{\partial\left\{\sum_{l=1}^{J}\sum_{k=1}^{n^{(j)}} \xi_{k,l} V_{k,l} w_{k,l}^{(2)} \exp(\gamma' B_{k,l}) dN_{k,l}(t)\right\}}{\partial \gamma}\bigg|_{\gamma=\widetilde{\gamma}} = \sum_{l=1}^{J}\sum_{k=1}^{n^{(l)}} \xi_{k,l} V_{k,l} w_{k,l}^{(2)} \exp(\widetilde{\gamma}' B_{k,l}) dN_{k,l}(t) B_{k,l}',$$

$$\Delta_{i,j}\{\tilde{S}_1(t;\widetilde{\gamma},\widetilde{\beta})\} = \xi_{i,j} V_{i,j} w_{i,j}^{(2)} \exp(\widetilde{\gamma}' B_{i,j}) Y_{i,j}(t) \exp(\widetilde{\beta}' X_{i,j}) X_{i,j} + \left\{\frac{\partial \tilde{S}_1(t;\widetilde{\gamma},\beta)}{\partial \beta}\bigg|_{\beta=\widetilde{\beta}}\right\} \Delta_{i,j}(\widetilde{\beta}) +$$

$$\left\{\frac{\partial \tilde{S}_1(t;\gamma,\widetilde{\beta})}{\partial \gamma}\bigg|_{\gamma=\widetilde{\gamma}}\right\} \Delta_{i,j}(\widetilde{\gamma}),$$

$$\frac{\partial \tilde{S}_1(t;\widetilde{\gamma},\beta)}{\partial \beta}\bigg|_{\beta=\widetilde{\beta}} = \sum_{l=1}^{J}\sum_{k=1}^{n^{(l)}} \xi_{k,l} V_{k,l} w_{k,l}^{(2)} \exp(\widetilde{\gamma}' B_{k,l}) Y_{k,l}(t) \exp(\widetilde{\beta}' X_{k,l}) X_{k,l} X_{k,l}' \equiv \tilde{S}_2(t;\widetilde{\gamma},\widetilde{\beta}),$$

$$\frac{\partial \tilde{S}_1(t;\gamma,\widetilde{\beta})}{\partial \gamma}\bigg|_{\gamma=\widetilde{\gamma}} = \sum_{l=1}^{J}\sum_{k=1}^{n^{(l)}} \xi_{k,l} V_{k,l} w_{k,l}^{(2)} \exp(\widetilde{\gamma}' B_{k,l}) Y_{k,l}(t) \exp(\widetilde{\beta}' X_{k,l}) X_{k,l} B_{k,l}',$$

$$\Delta_{i,j}\{\tilde{S}_0(t;\widetilde{\boldsymbol{\gamma}},\widetilde{\boldsymbol{\beta}})\} = \xi_{i,j}\,V_{i,j}\exp(\widetilde{\boldsymbol{\gamma}}'\boldsymbol{B}_{i,j})\,w^{(2)}_{i,j}\,Y_{i,j}(t)\exp(\widetilde{\boldsymbol{\beta}}'\boldsymbol{X}_{i,j}) + \left\{\frac{\partial\tilde{S}_0(t;\widetilde{\boldsymbol{\gamma}},\boldsymbol{\beta})}{\partial\boldsymbol{\beta}}\bigg|_{\boldsymbol{\beta}=\widetilde{\boldsymbol{\beta}}}\right\}\Delta_{i,j}(\widetilde{\boldsymbol{\beta}}) +$$

$$\left\{\frac{\partial\tilde{S}_0(t;\boldsymbol{\gamma},\widetilde{\boldsymbol{\beta}})}{\partial\boldsymbol{\gamma}}\bigg|_{\boldsymbol{\gamma}=\widetilde{\boldsymbol{\gamma}}}\right\}\Delta_{i,j}(\widetilde{\boldsymbol{\gamma}}),$$

$$\frac{\partial\tilde{S}_0(t;\widetilde{\boldsymbol{\gamma}},\boldsymbol{\beta})}{\partial\boldsymbol{\beta}}\bigg|_{\boldsymbol{\beta}=\widetilde{\boldsymbol{\beta}}} = \sum_{l=1}^{J}\sum_{k=1}^{n^{(l)}}\xi_{k,l}\,V_{k,l}\exp(\widetilde{\boldsymbol{\gamma}}'\boldsymbol{B}_{k,l})\,w^{(2)}_{k,l}\,Y_{k,l}(t)\exp(\widetilde{\boldsymbol{\beta}}'\boldsymbol{X}_{k,l})\,\boldsymbol{X}_{k,l}' = \tilde{S}_1(t;\widetilde{\boldsymbol{\gamma}},\widetilde{\boldsymbol{\beta}})',$$

and $\dfrac{\partial\tilde{S}_0(t;\boldsymbol{\gamma},\widetilde{\boldsymbol{\beta}})}{\partial\boldsymbol{\gamma}}\bigg|_{\boldsymbol{\gamma}=\widetilde{\boldsymbol{\gamma}}} = \sum_{l=1}^{J}\sum_{k=1}^{n^{(l)}}\xi_{k,l}\,V_{k,l}\exp(\widetilde{\boldsymbol{\gamma}}'\boldsymbol{B}_{k,l})\,w^{(2)}_{k,l}\,Y_{k,l}(t)\exp(\widetilde{\boldsymbol{\beta}}'\boldsymbol{X}_{k,l})\,\boldsymbol{B}_{k,l}'.$

Note, $\sum_{l=1}^{J}\sum_{k=1}^{n^{(l)}}\xi_{k,l}\,V_{k,l}\,w^{(2)}_{k,l}\exp(\widetilde{\boldsymbol{\gamma}}'\boldsymbol{B}_{k,l})\,\mathrm{d}N_{k,l}(t)$, $\dfrac{\partial\left\{\sum_{l=1}^{J}\sum_{k=1}^{n^{(J)}}\xi_{k,l}\,V_{k,l}\,w^{(2)}_{k,l}\exp(\boldsymbol{\gamma}'\boldsymbol{B}_{k,l})\mathrm{d}N_{k,l}(t)\right\}}{\partial\boldsymbol{\gamma}}\bigg|_{\boldsymbol{\gamma}=\widetilde{\boldsymbol{\gamma}}}$

and thus $\Delta_{i,j}\left\{\sum_{l=1}^{J}\sum_{k=1}^{n^{(J)}}\xi_{k,l}\,V_{k,l}\,w^{(2)}_{k,l}\exp(\widetilde{\boldsymbol{\gamma}}'\boldsymbol{B}_{k,l})\,\mathrm{d}N_{k,l}(t)\right\}$ and the term between brackets in Equation (3) are linear combinations of $\mathrm{d}N_{i,j}(t)$. Hence

$$\Delta_{i,j}(\widetilde{\boldsymbol{\beta}}) = \left[\sum_{l=1}^{J}\sum_{k=1}^{n^{(l)}}\int_t \xi_{k,l}\,V_{k,l}\,w^{(2)}_{k,l}\exp(\widetilde{\boldsymbol{\gamma}}'\boldsymbol{B}_{k,l})\left\{\frac{\tilde{S}_2(t;\widetilde{\boldsymbol{\gamma}},\widetilde{\boldsymbol{\beta}})}{\tilde{S}_0(t;\widetilde{\boldsymbol{\gamma}},\widetilde{\boldsymbol{\beta}})} - \frac{\tilde{S}_1(t;\widetilde{\boldsymbol{\gamma}},\widetilde{\boldsymbol{\beta}})\tilde{S}_1(t;\widetilde{\boldsymbol{\gamma}},\widetilde{\boldsymbol{\beta}})'}{\tilde{S}_0(t;\widetilde{\boldsymbol{\gamma}},\widetilde{\boldsymbol{\beta}})^2}\right\}\mathrm{d}N_{k,l}(t)\right]^{-1}\times$$

$$\left[\int_t\left\{\boldsymbol{X}_{i,j} - \frac{\tilde{S}_1(t;\widetilde{\boldsymbol{\gamma}},\widetilde{\boldsymbol{\beta}})}{\tilde{S}_0(t;\widetilde{\boldsymbol{\gamma}},\widetilde{\boldsymbol{\beta}})}\right\}\times\left\{\xi_{i,j}\,V_{i,j}\,w^{(2)}_{i,j}\exp(\widetilde{\boldsymbol{\gamma}}'\boldsymbol{B}_{i,j})\,\mathrm{d}N_{i,j}(t) - \right.\right.$$

$$\left.\left.\xi_{i,j}\,V_{i,j}\,w^{(2)}_{i,j}\exp(\widetilde{\boldsymbol{\gamma}}'\boldsymbol{B}_{i,j})\,Y_{i,j}(t)\exp(\widetilde{\boldsymbol{\beta}}'\boldsymbol{X}_{i,j})\times\frac{\sum_{l=1}^{J}\sum_{k=1}^{n^{(J)}}\xi_{k,l}\,V_{k,l}\,w^{(2)}_{k,l}\exp(\widetilde{\boldsymbol{\gamma}}'\boldsymbol{B}_{k,l})\mathrm{d}N_{k,l}(t)}{\tilde{S}_0(t;\widetilde{\boldsymbol{\gamma}},\widetilde{\boldsymbol{\beta}})}\right\}\right] +$$

$$\left[\sum_{l=1}^{J}\sum_{k=1}^{n^{(l)}}\int_t \xi_{k,l}\,V_{k,l}\,w^{(2)}_{k,l}\exp(\widetilde{\boldsymbol{\gamma}}'\boldsymbol{B}_{k,l})\left\{\frac{\tilde{S}_2(t;\widetilde{\boldsymbol{\gamma}},\widetilde{\boldsymbol{\beta}})}{\tilde{S}_0(t;\widetilde{\boldsymbol{\gamma}},\widetilde{\boldsymbol{\beta}})} - \frac{\tilde{S}_1(t;\widetilde{\boldsymbol{\gamma}},\widetilde{\boldsymbol{\beta}})\tilde{S}_1(t;\widetilde{\boldsymbol{\gamma}},\widetilde{\boldsymbol{\beta}})'}{\tilde{S}_0(t;\widetilde{\boldsymbol{\gamma}},\widetilde{\boldsymbol{\beta}})^2}\right\}\mathrm{d}N_{k,l}(t)\right]^{-1}\times$$

$$\left(\frac{\partial\widetilde{G}_1(t,\boldsymbol{\gamma})}{\partial\boldsymbol{\gamma}}\bigg|_{\boldsymbol{\gamma}=\widetilde{\boldsymbol{\gamma}}} - \int_t\frac{\tilde{S}_1(t;\widetilde{\boldsymbol{\gamma}},\widetilde{\boldsymbol{\beta}})}{\tilde{S}_0(t;\widetilde{\boldsymbol{\gamma}},\widetilde{\boldsymbol{\beta}})}\times\left[\frac{\partial\left\{\sum_{l=1}^{J}\sum_{k=1}^{n^{(J)}}\xi_{k,l}\,V_{k,l}\,w^{(2)}_{k,l}\exp(\boldsymbol{\gamma}'\boldsymbol{B}_{k,l})\mathrm{d}N_{k,l}(t)\right\}}{\partial\boldsymbol{\gamma}}\bigg|_{\boldsymbol{\gamma}=\widetilde{\boldsymbol{\gamma}}}\right] - \right.$$

$$\int_t \frac{\sum_{l=1}^{J}\sum_{k=1}^{n^{(J)}}\xi_{k,l}\,V_{k,l}\,w^{(2)}_{k,l}\exp(\boldsymbol{\gamma}'\boldsymbol{B}_{k,l})\mathrm{d}N_{k,l}(t)}{\tilde{S}_0(t;\widetilde{\boldsymbol{\gamma}},\widetilde{\boldsymbol{\beta}})}\times\left\{\frac{\partial\tilde{S}_1(t;\widetilde{\boldsymbol{\gamma}},\boldsymbol{\beta})}{\partial\boldsymbol{\gamma}}\bigg|_{\boldsymbol{\gamma}=\widetilde{\boldsymbol{\gamma}}}\right\} +$$

$$\left.\int_t \frac{\left\{\sum_{l=1}^{J}\sum_{k=1}^{n^{(J)}}\xi_{k,l}\,V_{k,l}\,w^{(2)}_{k,l}\exp(\boldsymbol{\gamma}'\boldsymbol{B}_{k,l})\mathrm{d}N_{k,l}(t)\right\}\tilde{S}_1(t;\widetilde{\boldsymbol{\gamma}},\widetilde{\boldsymbol{\beta}})}{\tilde{S}_0(t;\widetilde{\boldsymbol{\gamma}},\widetilde{\boldsymbol{\beta}})^2}\times\left\{\frac{\partial\tilde{S}_0(t;\boldsymbol{\gamma},\widetilde{\boldsymbol{\beta}})}{\partial\boldsymbol{\gamma}}\bigg|_{\boldsymbol{\gamma}=\widetilde{\boldsymbol{\gamma}}}\right\}\right)\Delta_{i,j}(\widetilde{\boldsymbol{\gamma}}),$$

and we can write

$$\Delta_{i,j}(\widetilde{\boldsymbol{\beta}}) = \xi_{i,j} \boldsymbol{IF}^{(2)}_{i,j}(\widetilde{\boldsymbol{\beta}}) + \xi_{i,j} V_{i,j} \exp(\widetilde{\boldsymbol{\gamma}}' \boldsymbol{B}_{i,j}) \boldsymbol{IF}^{(3)}_{i,j}(\widetilde{\boldsymbol{\beta}}),$$

with $\boldsymbol{IF}^{(2)}_{i,j}(\widetilde{\boldsymbol{\beta}}) = \left\{ \sum_{l=1}^{J} \sum_{k=1}^{n^{(j)}} \xi_{k,l} V_{k,l} \exp(\widetilde{\boldsymbol{\gamma}}' \boldsymbol{B}_{k,l}) \widetilde{\boldsymbol{Z}}_{k,l} \boldsymbol{B}_{k,l}' \right\} \times \boldsymbol{IF}^{(2)}_{i,j}(\widetilde{\boldsymbol{\gamma}})$,

and $\boldsymbol{IF}^{(3)}_{i,j}(\widetilde{\boldsymbol{\beta}}) = \widetilde{\boldsymbol{Z}}_{i,j} + \left\{ \sum_{l=1}^{J} \sum_{k=1}^{n^{(j)}} \xi_{k,l} V_{k,l} \exp(\widetilde{\boldsymbol{\gamma}}' \boldsymbol{B}_{k,l}) \widetilde{\boldsymbol{Z}}_{k,l} \boldsymbol{B}_{k,l}' \right\} \times \boldsymbol{IF}^{(3)}_{i,j}(\widetilde{\boldsymbol{\gamma}})$,

and where $\widetilde{\boldsymbol{Z}}_{i,j} = w^{(2)}_{i,j} \times \left[ \sum_{l=1}^{J} \sum_{k=1}^{n^{(l)}} \int_t \xi_{k,l} V_{k,l} w^{(2)}_{k,l} \exp(\widetilde{\boldsymbol{\gamma}}' \boldsymbol{B}_{k,l}) \left\{ \frac{\tilde{S}_2(t;\widetilde{\boldsymbol{\gamma}},\widetilde{\boldsymbol{\beta}})}{\tilde{S}_0(t;\widetilde{\boldsymbol{\gamma}},\widetilde{\boldsymbol{\beta}})} - \frac{\tilde{S}_1(t;\widetilde{\boldsymbol{\gamma}},\widetilde{\boldsymbol{\beta}}) \tilde{S}_1(t;\widetilde{\boldsymbol{\gamma}},\widetilde{\boldsymbol{\beta}})'}{\tilde{S}_0(t;\widetilde{\boldsymbol{\gamma}},\widetilde{\boldsymbol{\beta}})^2} \right\} dN_{k,l}(t) \right]^{-1} \times \left[ \int_t \left\{ \boldsymbol{X}_{i,j} - \frac{\tilde{S}_1(t;\widetilde{\boldsymbol{\gamma}},\widetilde{\boldsymbol{\beta}})}{\tilde{S}_0(t;\widetilde{\boldsymbol{\gamma}},\widetilde{\boldsymbol{\beta}})} \right\} \times \left\{ dN_{i,j}(t) - Y_{i,j}(t) \exp(\widetilde{\boldsymbol{\beta}}' \boldsymbol{X}_{i,j}) \times \frac{\sum_{l=1}^{J} \sum_{k=1}^{n^{(j)}} \xi_{k,l} V_{k,l} w^{(2)}_{k,l} \exp(\widetilde{\boldsymbol{\gamma}}' \boldsymbol{B}_{k,l}) dN_{k,l}(t)}{\tilde{S}_0(t;\widetilde{\boldsymbol{\gamma}},\widetilde{\boldsymbol{\beta}})} \right\} \right]$.

Then we know $\sum_{j=1}^{J} \sum_{i=1}^{n^{(j)}} \left\{ dN_{i,j}(t) - d\widetilde{\Lambda}_0(t) \xi_{i,j} V_{i,j} w^{(2)}_{i,j} \exp(\widetilde{\boldsymbol{\gamma}}' \boldsymbol{B}_{i,j}) Y_{i,j}(t) \exp(\widetilde{\boldsymbol{\beta}}' \boldsymbol{X}_{i,j}) \right\} = 0$,

as we estimate the baseline hazard non-parametrically by $d\widetilde{\Lambda}_0(t) = \frac{\sum_{j=1}^{J} \sum_{i=1}^{n^{(j)}} dN_{i,j}(t)}{\tilde{S}_0(t;\widetilde{\boldsymbol{\gamma}},\widetilde{\boldsymbol{\beta}})}$, and thus

$$dN_{i,j}(t) - d\widetilde{\Lambda}_0(t) \xi_{i,j} V_{i,j} w^{(2)}_{i,j} \exp(\widetilde{\boldsymbol{\gamma}}' \boldsymbol{B}_{i,j}) Y_{i,j}(t) \exp(\widetilde{\boldsymbol{\beta}}' \boldsymbol{X}_{i,j}) - d\widetilde{\Lambda}_0(t) \tilde{S}_1(t;\widetilde{\boldsymbol{\gamma}},\widetilde{\boldsymbol{\beta}})' \Delta_{i,j}(\widetilde{\boldsymbol{\beta}})$$

$$- d\widetilde{\Lambda}_0(t) \left\{ \sum_{l=1}^{J} \sum_{k=1}^{n^{(j)}} \xi_{k,l} V_{k,l} w^{(2)}_{k,l} \exp(\widetilde{\boldsymbol{\gamma}}' \boldsymbol{B}_{k,l}) Y_{k,l}(t) \exp(\widetilde{\boldsymbol{\beta}}' \boldsymbol{X}_{k,l}) \boldsymbol{B}_{k,l} \right\}' \Delta_{i,j}(\widetilde{\boldsymbol{\gamma}})$$

$$- \tilde{S}_0(t;\widetilde{\boldsymbol{\gamma}},\widetilde{\boldsymbol{\beta}}) \Delta_{i,j}\{d\widetilde{\Lambda}_0(t)\} = 0.$$

As a result,

$$\Delta_{i,j}\{d\widetilde{\Lambda}_0(t)\} = \{\tilde{S}_0(t;\widetilde{\boldsymbol{\gamma}},\widetilde{\boldsymbol{\beta}})\}^{-1} \left[ dN_{i,j}(t) - d\widetilde{\Lambda}_0(t) \xi_{i,j} V_{i,j} w^{(2)}_{i,j} \exp(\widetilde{\boldsymbol{\gamma}}' \boldsymbol{B}_{i,j}) Y_{i,j}(t) \exp(\widetilde{\boldsymbol{\beta}}' \boldsymbol{X}_{i,j}) - \right.$$

$$d\widetilde{\Lambda}_0(t) \tilde{S}_1(t;\widetilde{\boldsymbol{\gamma}},\widetilde{\boldsymbol{\beta}})' \Delta_{i,j}(\widetilde{\boldsymbol{\beta}}) -$$

$$\left. d\widetilde{\Lambda}_0(t) \left\{ \sum_{l=1}^{J} \sum_{k=1}^{n^{(j)}} \xi_{k,l} V_{k,l} w^{(2)}_{k,l} \exp(\widetilde{\boldsymbol{\gamma}}' \boldsymbol{B}_{k,l}) Y_{k,l}(t) \exp(\widetilde{\boldsymbol{\beta}}' \boldsymbol{X}_{k,l}) \boldsymbol{B}_{k,l} \right\}' \Delta_{i,j}(\widetilde{\boldsymbol{\gamma}}) \right],$$

and we can write

$$\Delta_{i,j}\{d\widetilde{\Lambda}_0(t)\} = \xi_{i,j} IF^{(2)}_{i,j}\{d\widetilde{\Lambda}_0(t)\} + \xi_{i,j} V_{i,j} \exp(\widetilde{\boldsymbol{\gamma}}' \boldsymbol{B}_{i,j}) IF^{(3)}_{i,j}\{d\widetilde{\Lambda}_0(t)\},$$

with

$$IF_{i,j}^{(2)}\{d\widetilde{\Lambda}_0(t)\} = \{\widetilde{S}_0(t;\widetilde{\gamma},\widetilde{\beta})\}^{-1} dN_{i,j}(t) + \left\{\sum_{l=1}^{J}\sum_{k=1}^{n^{(j)}} \xi_{k,l} V_{k,l} \exp(\widetilde{\gamma}' B_{k,l}) \widetilde{H}_{k,l}(t) B_{k,l}'\right\} \times$$

$$\boldsymbol{IF}_{i,j}^{(2)}(\widetilde{\gamma}),$$

and $IF_{i,j}^{(3)}\{d\widetilde{\Lambda}_0(t)\} = \widetilde{H}_{i,j}(t) + \left\{\sum_{l=1}^{J}\sum_{k=1}^{n^{(j)}} \xi_{k,l} V_{k,l} \exp(\widetilde{\gamma}' B_{k,l}) \widetilde{H}_{k,l}(t) B_{k,l}'\right\} \times \boldsymbol{IF}_{i,j}^{(3)}(\widetilde{\gamma}),$

and where $\widetilde{H}_{i,j}(t) = -\widetilde{S}_0(t;\widetilde{\gamma},\widetilde{\beta})^{-1} d\widetilde{\Lambda}_0(t) \left\{\widetilde{S}_1(t;\widetilde{\gamma},\widetilde{\beta})' \widetilde{Z}_{i,j} + \widetilde{K}_{i,j}(t)\right\},$

and $\widetilde{K}_{i,j}(t) = w_{i,j}^{(2)} Y_{i,j}(t) \exp(\widetilde{\beta}' X_{i,j}).$

Note, $\widetilde{H}_{i,j}(t)$, and thus $IF_{i,j}^{(2)}\{d\widetilde{\Lambda}_0(t)\}$ and $IF_{i,j}^{(3)}\{d\widetilde{\Lambda}_0(t)\}$, are linear combinations of the increments $dN_{i,j}(t)$ and $d\widetilde{\Lambda}_0(t)$. Thus $\Delta_{i,j}\left\{\int_{\tau_1}^{\tau_2} d\widetilde{\Lambda}_0(t)\right\} = \int_{\tau_1}^{\tau_2} \Delta_{i,j}\{d\widetilde{\Lambda}_0(t)\}$, that we can write

$$\xi_{i,j} IF_{i,j}^{(2)}\left\{\int_{\tau_1}^{\tau_2} d\widetilde{\Lambda}_0(t)\right\} + \xi_{i,j} V_{i,j} \exp(\widetilde{\gamma}' B_{i,j}) IF_{i,j}^{(3)}\left\{\int_{\tau_1}^{\tau_2} d\widetilde{\Lambda}_0(t)\right\},$$

with $IF_{i,j}^{(2)}\left\{\int_{\tau_1}^{\tau_2} d\widetilde{\Lambda}_0(t)\right\} = \sum_{t=0}^{\tau_2} IF_{i,j}^{(2)}\{d\widetilde{\Lambda}_0(t)\},$

and $IF_{i,j}^{(2,3)}\left\{\int_{\tau_1}^{\tau_2} d\widetilde{\Lambda}_0(t)\right\} = \sum_{t=0}^{\tau_2} IF_{i,j}^{(3)}\{d\widetilde{\Lambda}_0(t)\},$ and we finally have

$$\Delta_{i,j}\{\widetilde{\pi}(\tau_1,\tau_2;x)\} = \left\{\frac{\partial \widetilde{\pi}(\tau_1,\tau_2;x)}{\partial \beta}\bigg|_{\beta=\widetilde{\beta}}\right\} \Delta_{i,j}(\widetilde{\beta}) + \left[\frac{\partial \widetilde{\pi}(\tau_1,\tau_2;x)}{\partial \left\{\int_{\tau_1}^{\tau_2} d\Lambda_0(t)\right\}}\bigg|_{d\Lambda_0(t)=d\widetilde{\Lambda}_0(t)}\right] \Delta_{i,j}\left\{\int_{\tau_1}^{\tau_2} d\widetilde{\Lambda}_0(t)\right\},$$

that we can write

$$\xi_{i,j} IF_{i,j}^{(2)}\{\widetilde{\pi}(\tau_1,\tau_2;x)\} + \xi_{i,j} V_{i,j} \exp(\widetilde{\gamma}' B_{i,j}) IF_{i,j}^{(3)}\{\widetilde{\pi}(\tau_1,\tau_2;x)\},$$

with $\frac{\partial \widetilde{\pi}(\tau_1,\tau_2;x)}{\partial \beta}\bigg|_{\beta=\widetilde{\beta}} = \left\{\int_{\tau_1}^{\tau_2} d\widetilde{\Lambda}_0(t) \exp(\widetilde{\beta}'x)\right\}\{1 - \widetilde{\pi}(\tau_1,\tau_2;x)\}x',$

and $\frac{\partial \widetilde{\pi}(\tau_1,\tau_2;x)}{\partial \left\{\int_{\tau_1}^{\tau_2} d\Lambda_0(t)\right\}}\bigg|_{d\Lambda_0(t)=d\widetilde{\Lambda}_0(t)} = \exp(\widetilde{\beta}'x)\{1 - \widetilde{\pi}(\tau_1,\tau_2;x)\}.$

Observe that for any $\widetilde{\theta} \in \{\widetilde{\gamma}, \widetilde{\beta}, d\widetilde{\Lambda}_0(t), \widetilde{\Lambda}_0(t), \widetilde{\pi}(\tau_1,\tau_2;x)\}$, if subject $i$ in stratum $j$ is in the phase-two sample but not in the phase-three sample, then $\xi_{i,j} V_{i,j} \exp(\widetilde{\gamma}' B_{i,j}) \boldsymbol{IF}_{i,j}^{(3)}(\widetilde{\theta})$ is zero. However, $\xi_{i,j}IF_{i,j}^{(2)}(\widetilde{\boldsymbol{\theta}})$ is usually non-zero. In particular, such individual affects $\widetilde{\boldsymbol{\theta}}$ through his/her influence on $\widetilde{\boldsymbol{\gamma}}$, as he/she is used for the estimation of the phase-three sampling weights.

### G.2  Variance decomposition and estimation

As mentioned in Section 5.5. in the Main Document, for any $\widetilde{\boldsymbol{\theta}} \in \{\widetilde{\boldsymbol{\gamma}}, \widetilde{\boldsymbol{\beta}}, d\widetilde{\Lambda}_0(t), \widetilde{\Lambda}_0(t), \widetilde{\pi}(\tau_1, \tau_2; \boldsymbol{x})\}$, using the law of total covariance and the law of total expectation, we can decompose $\text{var}\left\{\sum_{j=1}^{J}\sum_{i=1}^{n^{(j)}}\Delta_{i,j}(\widetilde{\boldsymbol{\theta}})\right\}$ as

$$\text{var}\left(E\left[E\left\{\sum_{j=1}^{J}\sum_{i=1}^{n^{(j)}}\Delta_{i,j}(\widetilde{\boldsymbol{\theta}})\,|C_1,C_2\right\}|C_1\right]\right) + E\left(\text{var}\left[E\left\{\sum_{j=1}^{J}\sum_{i=1}^{n^{(j)}}\Delta_{i,j}(\widetilde{\boldsymbol{\theta}})\,|C_1,C_2\right\}|C_1\right]\right) +$$

$$E\left(E\left[\text{var}\left\{\sum_{j=1}^{J}\sum_{i=1}^{n^{(j)}}\Delta_{i,j}(\widetilde{\boldsymbol{\theta}})\,|C_1,C_2\right\}|C_1\right]\right),$$

where $C_1$ denote the information from the whole cohort, and $C_2$ denote the information from the phase-two sample.

For any $j,l \in \{1,\ldots,J\}$, we know that $E(\xi_{i,j}|C_1) = \frac{1}{w_{i,j}^{(2)}}$, $E(\xi_{i,j}\xi_{k,j}|C_1) = \frac{1}{w_{i,k,j}^{(2)}} = \frac{m^{(j)}(m^{(j)}-1)}{n^{(j)}(n^{(j)}-1)}$ if individuals $i$ and $k$, $i \neq k$, are both non-cases in stratum $j$, and $E(\xi_{i,j}\xi_{k,j}|C_1) = \frac{1}{w_{i,j}^{(2)}} \times \frac{1}{w_{k,j}^{(2)}}$ if $i$ and/or $k$, $i \neq k$, is a case in stratum $j$. In addition, $E(\xi_{i,j}\xi_{k,l}|C_1) = \frac{1}{w_{i,j}^{(2)}} \times \frac{1}{w_{k,l}^{(2)}}$ if individuals $i$ and $k$ are in stratum $j$ and stratum $l$, respectively, with $j \neq l$. Thus, $\text{cov}(\xi_{i,j},\xi_{k,j}|C_1) = \sigma_{i,k,j}^{(2)} = \frac{1}{w_{i,k,j}^{(2)}} - \frac{1}{w_{i,j}^{(2)}} \times \frac{1}{w_{k,j}^{(2)}}$ if both $i$ and $k$, $i \neq k$, are non-cases in stratum $j$, $\text{cov}(\xi_{i,j},\xi_{k,j}|C_1) = 0$ if $i$ and/or $k$ is a case in stratum $j$, $\text{cov}(\xi_{i,j},\xi_{k,l}|C_1) = 0$ if $i$ and $k$ are in stratum $j$ and stratum $l$, respectively, with $j \neq l$, $\text{var}(\xi_{i,j}|C_1) = \sigma_{i,i,j}^{(2)} \equiv \sigma_{i,j}^{(2)} = \frac{1}{w_{i,j}^{(2)}}\left(1 - \frac{1}{w_{i,j}^{(2)}}\right)$ if $i$ is a non-case in stratum $j$, and $\text{var}(\xi_{i,j}|C_1) = 0$ if $i$ is a case in stratum $j$. In addition, because the third phase of sampling is

Bernoulli, $E(V_{i,j}|C_1,C_2) = \frac{1}{w_{i,j}^{(3)}}$ and $E(V_{i,j}V_{k,j}|C_1,C_2) = \frac{1}{w_{i,j}^{(3)}} \times \frac{1}{w_{k,j}^{(3)}}$ if $i \neq k$, thus

$\text{cov}(V_{i,j}, V_{k,j}|C_1, C_2) = \sigma_{i,k,j}^{(3)} = 0$ if $i \neq k$ and $\text{var}(V_{i,j}|C_1, C_2) = \sigma_{i,i,j}^{(3)} \equiv \sigma_{i,j}^{(3)} = \frac{1}{w_{i,j}^{(3)}}\left(1 - \frac{1}{w_{i,j}^{(3)}}\right)$;

and obviously $E(V_{i,j}V_{k,l}|C_1,C_2) = \frac{1}{w_{i,j}^{(3)}} \times \frac{1}{w_{k,l}^{(3)}}$ and $\text{cov}(V_{i,j}, V_{k,l}|C_1,C_2) = 0$ in stratum $j$ and

stratum $l$, $j \neq l$. For our notation, we will also use $w_{i,k,j} = w_{i,k,j}^{(2)} \times w_{i,j}^{(3)} \times w_{k,j}^{(3)}$, for $i \neq k$, because

$E(\xi_{i,j}\,\xi_{k,j}\,V_{i,j}\,V_{k,j}|C_1) = E\{E(\xi_{i,j}\,\xi_{k,j}\,V_{i,j}\,V_{k,j}|C_1,C_2)|C_1\},$

$= E\{\xi_{i,j}\,\xi_{k,j}\,E(V_{i,j}|C_1,C_2)E(V_{k,j}|C_1,C_2)|C_1\},$

$= \frac{1}{w_{i,j}^{(3)}} \times \frac{1}{w_{k,j}^{(3)}} \times E(\xi_{i,j}\,\xi_{k,j}|C_1) = \frac{1}{w_{i,j}^{(3)}} \times \frac{1}{w_{k,j}^{(3)}} \times \frac{1}{w_{i,k,j}^{(2)}},$

and we will also use $w_{i,k,j}^{(3)} = w_{i,j}^{(3)} \times w_{k,j}^{(3)}$, for $i \neq k$, $w_{i,j} = w_{i,j}^{(2)} \times w_{i,j}^{(3)}$, $\sigma_{i,j} = \frac{1}{w_{i,j}^{(2)}} \times \frac{1}{w_{i,j}^{(3)}}\left(1 - \right.$

$\left.\frac{1}{w_{i,j}^{(2)}} \times \frac{1}{w_{i,j}^{(3)}}\right)$ and $\sigma_{i,k,j} = \frac{1}{w_{i,k,j}} - \frac{1}{w_{i,j}} \times \frac{1}{w_{k,j}} = \frac{1}{w_{i,j}^{(3)}} \times \frac{1}{w_{k,j}^{(3)}} \times \sigma_{i,k,j}^{(2)}$, for $i \neq k$.

We have shown in Web Appendix G.1 that for any $\widetilde{\boldsymbol{\theta}} \in \{\widetilde{\boldsymbol{\gamma}}, \widetilde{\boldsymbol{\beta}}, d\widetilde{\Lambda}_0(t), \widetilde{\Lambda}_0(t), \widetilde{\pi}(\tau_1, \tau_2; x)\}$,

$\boldsymbol{\Delta}_{i,j}(\widetilde{\boldsymbol{\theta}}) = \xi_{i,j}\boldsymbol{IF}_{i,j}^{(2)}(\widetilde{\boldsymbol{\theta}}) + \xi_{i,j}\,V_{i,j}\,\widetilde{w}_{i,j}^{(3)}\boldsymbol{IF}_{i,j}^{(3)}(\widetilde{\boldsymbol{\theta}})$, with $\widetilde{w}_{i,j}^{(3)} = \exp(\widetilde{\boldsymbol{\gamma}}'\boldsymbol{B}_{i,j})$. We have

$E\left\{\sum_{j=1}^J \sum_{i=1}^{n^{(j)}} \boldsymbol{\Delta}_{i,j}(\widetilde{\boldsymbol{\theta}})\,|C_1,C_2\right\} = \sum_{j=1}^J \sum_{i=1}^{n^{(j)}}\left\{\xi_{i,j}\boldsymbol{IF}_{i,j}^{(2)}(\widetilde{\boldsymbol{\theta}}) + E(V_{i,j}|C_1,C_2)\,\xi_{i,j}\,\widetilde{w}_{i,j}^{(3)}\boldsymbol{IF}_{i,j}^{(3)}(\widetilde{\boldsymbol{\theta}})\right\},$

$= \sum_{j=1}^J \sum_{i=1}^{n^{(j)}}\left\{\xi_{i,j}\boldsymbol{IF}_{i,j}^{(2)}(\widetilde{\boldsymbol{\theta}}) + \xi_{i,j}\,\frac{\widetilde{w}_{i,j}^{(3)}}{w_{i,j}^{(3)}}\boldsymbol{IF}_{i,j}^{(3)}(\widetilde{\boldsymbol{\theta}})\right\},$

and we have

$\text{var}\left\{\sum_{j=1}^J \sum_{i=1}^{n^{(j)}} \boldsymbol{\Delta}_{i,j}(\widetilde{\boldsymbol{\theta}})\,|C_1,C_2\right\} =$

$\sum_{j=1}^J \sum_{i=1}^{n^{(j)}} \text{var}(V_{i,j}|C_1,C_2)\left\{\xi_{i,j}\,\widetilde{w}_{i,j}^{(3)}\boldsymbol{IF}_{i,j}^{(3)}(\widetilde{\boldsymbol{\theta}})\right\}\left\{\xi_{i,j}\,\widetilde{w}_{i,j}^{(3)}\boldsymbol{IF}_{i,j}^{(3)}(\widetilde{\boldsymbol{\theta}})\right\}',$

$= \sum_{j=1}^J \sum_{i=1}^{n^{(j)}} \sigma_{i,j}^{(3)}\,\xi_{i,j}\,\widetilde{w}_{i,j}^{(3)}\,\widetilde{w}_{i,j}^{(3)}\boldsymbol{IF}_{i,j}^{(3)}(\widetilde{\boldsymbol{\theta}})\boldsymbol{IF}_{i,j}^{(3)}(\widetilde{\boldsymbol{\theta}})',$

as indeed $\xi_{i,j}IF_{i,j}^{(2)}(\widetilde{\boldsymbol{\theta}})$ is fixed conditional on $C_2$.

Then we have

$$E\left[E\left\{\sum_{j=1}^{J}\sum_{i=1}^{n^{(j)}}\boldsymbol{\Delta}_{i,j}(\widetilde{\boldsymbol{\theta}})\,|C_1,C_2\right\}|C_1\right] = \sum_{j=1}^{J}\sum_{i=1}^{n^{(j)}}E(\xi_{i,j}|C_1)\,IF_{i,j}^{(2)}(\widetilde{\boldsymbol{\theta}}) +$$

$$E(\xi_{i,j}|C_1)\frac{\widetilde{w}_{i,j}^{(3)}}{w_{i,j}^{(3)}}IF_{i,j}^{(3)}(\widetilde{\boldsymbol{\theta}}),$$

$$= \sum_{j=1}^{J}\sum_{i=1}^{n^{(j)}}\frac{1}{w_{i,j}^{(2)}}\left\{IF_{i,j}^{(2)}(\widetilde{\boldsymbol{\theta}}) + \frac{\widetilde{w}_{i,j}^{(3)}}{w_{i,j}^{(3)}}IF_{i,j}^{(3)}(\widetilde{\boldsymbol{\theta}})\right\},$$

and

$$\mathrm{var}\left[E\left\{\sum_{j=1}^{J}\sum_{i=1}^{n^{(j)}}\boldsymbol{\Delta}_{i,j}(\widetilde{\boldsymbol{\theta}})\,|C_1,C_2\right\}|C_1\right] = \mathrm{var}\left\{\sum_{j=1}^{J}\sum_{i=1}^{n^{(j)}}\xi_{i,j}IF_{i,j}^{(2)}(\widetilde{\boldsymbol{\theta}}) + \xi_{i,j}\frac{\widetilde{w}_{i,j}^{(3)}}{w_{i,j}^{(3)}}IF_{i,j}^{(3)}(\widetilde{\boldsymbol{\theta}})\right\},$$

$$= \sum_{j=1}^{J}\mathrm{var}\left[\sum_{i=1}^{n^{(j)}}\xi_{i,j}\left\{IF_{i,j}^{(2)}(\widetilde{\boldsymbol{\theta}}) + \frac{\widetilde{w}_{i,j}^{(3)}}{w_{i,j}^{(3)}}IF_{i,j}^{(3)}(\widetilde{\boldsymbol{\theta}})\right\}\right],$$

$$= \sum_{j=1}^{J}\sum_{i=1}^{n^{(j)}}\sum_{k=1}^{n^{(j)}}\sigma_{i,k,j}^{(2)}\left\{IF_{i,j}^{(2)}(\widetilde{\boldsymbol{\theta}}) + \frac{\widetilde{w}_{i,j}^{(3)}}{w_{i,j}^{(3)}}IF_{i,j}^{(3)}(\widetilde{\boldsymbol{\theta}})\right\}\left\{IF_{k,j}^{(2)}(\widetilde{\boldsymbol{\theta}}) + \frac{\widetilde{w}_{k,j}^{(3)}}{w_{k,j}^{(3)}}IF_{k,j}^{(3)}(\widetilde{\boldsymbol{\theta}})\right\}',$$

and

$$E\left[\mathrm{var}\left\{\sum_{j=1}^{J}\sum_{i=1}^{n^{(j)}}\boldsymbol{\Delta}_{i,j}(\widetilde{\boldsymbol{\theta}})\,|C_1,C_2\right\}|C_1\right] = \sum_{j=1}^{J}\sum_{i=1}^{n^{(j)}}E(\xi_{i,j}|C_1)\,\sigma_{i,j}^{(3)}\,\widetilde{w}_{i,j}^{(3)}\,\widetilde{w}_{i,j}^{(3)}IF_{i,j}^{(3)}(\widetilde{\boldsymbol{\theta}})IF_{i,j}^{(3)}(\widetilde{\boldsymbol{\theta}})',$$

$$= \sum_{j=1}^{J}\sum_{i=1}^{n^{(j)}}\frac{1}{w_{i,j}^{(2)}}\sigma_{i,j}^{(3)}\,\widetilde{w}_{i,j}^{(3)}\,\widetilde{w}_{i,j}^{(3)}IF_{i,j}^{(3)}(\widetilde{\boldsymbol{\theta}})IF_{i,j}^{(3)}(\widetilde{\boldsymbol{\theta}})'.$$

Finally,

$$\mathrm{var}\left(E\left[E\left\{\sum_{j=1}^{J}\sum_{i=1}^{n^{(j)}}\boldsymbol{\Delta}_{i,j}(\widetilde{\boldsymbol{\theta}})\,|C_1,C_2\right\}|C_1\right]\right) = \mathrm{var}\left[\sum_{j=1}^{J}\sum_{i=1}^{n^{(j)}}\frac{1}{w_{i,j}^{(2)}}\left\{IF_{i,j}^{(2)}(\widetilde{\boldsymbol{\theta}}) + \frac{\widetilde{w}_{i,j}^{(3)}}{w_{i,j}^{(3)}}IF_{i,j}^{(3)}(\widetilde{\boldsymbol{\theta}})\right\}\right],$$

which can be estimated by

$$\frac{n}{n-1}\sum_{j=1}^{J}\sum_{i=1}^{n^{(j)}}\left(\frac{1}{w_{i,j}^{(2)}}\right)^2\left\{\xi_{i,j}\,w_{i,j}^{(2)}IF_{i,j}^{(2)}(\widetilde{\boldsymbol{\theta}})IF_{i,j}^{(2)}(\widetilde{\boldsymbol{\theta}})' + 2\,\xi_{i,j}\,V_{i,j}\,\widetilde{w}_{i,j}IF_{i,j}^{(2)}(\widetilde{\boldsymbol{\theta}})IF_{i,j}^{(3)}(\widetilde{\boldsymbol{\theta}})' +\right.$$

$$\left.\xi_{i,j}\,V_{i,j}\,\widetilde{w}_{i,j}IF_{i,j}^{(3)}(\widetilde{\boldsymbol{\theta}})IF_{i,j}^{(3)}(\widetilde{\boldsymbol{\theta}})'\right\},$$

that is by

$$\frac{n}{n-1}\sum_{j=1}^{J}\sum_{i=1}^{n^{(j)}}\frac{1}{w_{i,j}^{(2)}}\Big\{\xi_{i,j}\,IF_{i,j}^{(2)}(\widetilde{\boldsymbol{\theta}})IF_{i,j}^{(2)}(\widetilde{\boldsymbol{\theta}})' + 2\,\xi_{i,j}\,V_{i,j}\,\widetilde{w}_{i,j}^{(3)}IF_{i,j}^{(2)}(\widetilde{\boldsymbol{\theta}})IF_{i,j}^{(3)}(\widetilde{\boldsymbol{\theta}})' +$$

$$\xi_{i,j}\,V_{i,j}\,\widetilde{w}_{i,j}^{(3)}IF_{i,j}^{(3)}(\widetilde{\boldsymbol{\theta}})IF_{i,j}^{(3)}(\widetilde{\boldsymbol{\theta}})'\Big\},$$

or again by

$$\frac{n}{n-1}\sum_{j=1}^{J}\sum_{i=1}^{n^{(j)}}\frac{1}{w_{i,j}^{(2)}}\Big\{\xi_{i,j}\,IF_{i,j}^{(2)}(\widetilde{\boldsymbol{\theta}})IF_{i,j}^{(2)}(\widetilde{\boldsymbol{\theta}})' + 2\,\xi_{i,j}\,V_{i,j}\,\widetilde{w}_{i,j}^{(3)}IF_{i,j}^{(2)}(\widetilde{\boldsymbol{\theta}})IF_{i,j}^{(3)}(\widetilde{\boldsymbol{\theta}})' +$$

$$\frac{1}{\widetilde{w}_{i,j}^{(3)}}\xi_{i,j}\,V_{i,j}\widetilde{w}_{i,j}^{(3)}\widetilde{w}_{i,j}^{(3)}IF_{i,j}^{(3)}(\widetilde{\boldsymbol{\theta}})IF_{i,j}^{(3)}(\widetilde{\boldsymbol{\theta}})'\Big\}.$$

Then

$$\mathrm{E}\left(\mathrm{var}\left[\mathrm{E}\left\{\sum_{j=1}^{J}\sum_{i=1}^{n^{(j)}}\boldsymbol{\Delta}_{i,j}(\widetilde{\boldsymbol{\theta}})\,|C_1,C_2\right\}|C_1\right]\right) = \mathrm{E}\left[\sum_{j=1}^{J}\sum_{i=1}^{n^{(j)}}\sum_{k=1}^{n^{(j)}}\sigma_{i,k,j}^{(2)}\Big\{IF_{i,j}^{(2)}(\widetilde{\boldsymbol{\theta}}) +$$

$$\frac{\widetilde{w}_{i,j}^{(3)}}{w_{i,j}^{(3)}}IF_{i,j}^{(3)}(\widetilde{\boldsymbol{\theta}})\Big\}\Big\{IF_{k,j}^{(2)}(\widetilde{\boldsymbol{\theta}}) + \frac{\widetilde{w}_{k,j}^{(3)}}{w_{k,j}^{(3)}}IF_{k,j}^{(3)}(\widetilde{\boldsymbol{\theta}})\Big\}'\right],$$

that can be estimated by

$$\sum_{j=1}^{J}\sum_{i=1}^{n^{(j)}}\sigma_{i,j}^{(2)}\Big\{\xi_{i,j}\,w_{i,j}^{(2)}IF_{i,j}^{(2)}(\widetilde{\boldsymbol{\theta}})IF_{i,j}^{(2)}(\widetilde{\boldsymbol{\theta}})' + 2\,\xi_{i,j}\,V_{i,j}\,w_{i,j}^{(2)}\,\widetilde{w}_{i,j}^{(3)}\,IF_{i,j}^{(2)}(\widetilde{\boldsymbol{\theta}})IF_{i,j}^{(3)}(\widetilde{\boldsymbol{\theta}})' +$$

$$\xi_{i,j}\,V_{i,j}\,w_{i,j}^{(2)}\,\widetilde{w}_{i,j}^{(3)}\,IF_{i,j}^{(3)}(\widetilde{\boldsymbol{\theta}})IF_{i,j}^{(3)}(\widetilde{\boldsymbol{\theta}})'\Big\} +$$

$$\sum_{j=1}^{J}\sum_{i=1}^{n^{(j)}}\sum_{\substack{k=1,\\k\neq i}}^{n^{(j)}}\sigma_{i,k,j}^{(2)}\Big\{\xi_{i,j}\,\xi_{k,j}\,w_{i,k,j}^{(2)}IF_{i,j}^{(2)}(\widetilde{\boldsymbol{\theta}})IF_{k,j}^{(2)}(\widetilde{\boldsymbol{\theta}})' +$$

$$\xi_{i,j}\,\xi_{k,j}\,V_{k,j}\,w_{i,k,j}^{(2)}\,\widetilde{w}_{k,j}^{(3)}IF_{i,j}^{(2)}(\widetilde{\boldsymbol{\theta}})IF_{k,j}^{(3)}(\widetilde{\boldsymbol{\theta}})' + \xi_{i,j}\,\xi_{k,j}\,V_{i,j}\,w_{i,k,j}^{(2)}\,\widetilde{w}_{i,j}^{(3)}IF_{i,j}^{(3)}(\widetilde{\boldsymbol{\theta}})IF_{k,j}^{(2)}(\widetilde{\boldsymbol{\theta}})' +$$

$$\xi_{i,j}\,\xi_{k,j}\,V_{i,j}\,V_{k,j}\,\widetilde{w}_{i,k,j}IF_{i,j}^{(3)}(\widetilde{\boldsymbol{\theta}})IF_{k,j}^{(3)}(\widetilde{\boldsymbol{\theta}})'\Big\},$$

that is by

$\sum_{j=1}^{J} \sum_{i=1}^{n^{(j)}} \sigma_{i,j}^{(2)} w_{i,j}^{(2)} \left\{ \xi_{i,j} \, IF_{i,j}^{(2)}(\widetilde{\boldsymbol{\theta}}) IF_{i,j}^{(2)}(\widetilde{\boldsymbol{\theta}})' + 2 \, \xi_{i,j} \, V_{i,j} \, \widetilde{w}_{i,j}^{(3)} IF_{i,j}^{(2)}(\widetilde{\boldsymbol{\theta}}) IF_{i,j}^{(3)}(\widetilde{\boldsymbol{\theta}})' + \right.$

$\left. \frac{1}{\widetilde{w}_{i,j}^{(3)}} \xi_{i,j} \, V_{i,j} \, \widetilde{w}_{i,j}^{(3)} \widetilde{w}_{i,j}^{(3)} IF_{i,j}^{(3)}(\widetilde{\boldsymbol{\theta}}) IF_{i,j}^{(3)}(\widetilde{\boldsymbol{\theta}})' \right\} + \sum_{j=1}^{J} \sum_{i=1}^{n^{(j)}} \sum_{\substack{k=1, \\ k \neq i}}^{n^{(j)}} \sigma_{i,k,j}^{(2)} w_{i,k,j}^{(2)} \left\{ \xi_{i,j} \, IF_{i,j}^{(2)}(\widetilde{\boldsymbol{\theta}}) + \right.$

$\left. \xi_{i,j} \, V_{i,j} \, \widetilde{w}_{i,j}^{(3)} IF_{i,j}^{(3)}(\widetilde{\boldsymbol{\theta}}) \right\} \left\{ \xi_{k,j} \, IF_{k,j}^{(2)}(\widetilde{\boldsymbol{\theta}}) + \xi_{k,j} \, V_{k,j} \, \widetilde{w}_{k,j}^{(3)} IF_{i,j}^{(3)}(\widetilde{\boldsymbol{\theta}}) \right\}',$

as indeed $\widetilde{w}_{i,k,j} = w_{i,k,j}^{(2)} \times \widetilde{w}_{i,j}^{(3)} \times \widetilde{w}_{k,j}^{(3)}$, and

$E \left( E \left[ \text{var} \left\{ \sum_{j=1}^{J} \sum_{i=1}^{n^{(j)}} \boldsymbol{\Delta}_{i,j}(\widetilde{\boldsymbol{\theta}}) \mid C_1, C_2 \right\} \mid C_1 \right] \right) = E \left\{ \sum_{j=1}^{J} \sum_{i=1}^{n^{(j)}} \frac{1}{w_{i,j}^{(2)}} \sigma_{i,j}^{(3)} \widetilde{w}_{i,j}^{(3)} \widetilde{w}_{i,j}^{(3)} IF_{i,j}^{(3)}(\widetilde{\boldsymbol{\theta}}) IF_{i,j}^{(3)}(\widetilde{\boldsymbol{\theta}})' \right\},$

that can be estimated by

$\sum_{j=1}^{J} \sum_{i=1}^{n^{(j)}} \widetilde{\sigma}_{i,j}^{(3)} \widetilde{w}_{i,j}^{(3)} \xi_{i,j} \, V_{i,j} \, \widetilde{w}_{i,j}^{(3)} \widetilde{w}_{i,j}^{(3)} IF_{i,j}^{(3)}(\widetilde{\boldsymbol{\theta}}) IF_{i,j}^{(3)}(\widetilde{\boldsymbol{\theta}})'.$

As a result, $\text{var}(\widetilde{\boldsymbol{\theta}})$ can be estimated by

$\frac{n}{n-1} \sum_{j=1}^{J} \sum_{i=1}^{n^{(j)}} \frac{1}{w_{i,j}^{(2)}} \left\{ \xi_{i,j} \, IF_{i,j}^{(2)}(\widetilde{\boldsymbol{\theta}}) IF_{i,j}^{(2)}(\widetilde{\boldsymbol{\theta}})' + 2 \, \xi_{i,j} \, V_{i,j} \, \widetilde{w}_{i,j}^{(3)} IF_{i,j}^{(2)}(\widetilde{\boldsymbol{\theta}}) IF_{i,j}^{(3)}(\widetilde{\boldsymbol{\theta}})' + \right.$

$\left. \frac{1}{\widetilde{w}_{i,j}^{(3)}} \xi_{i,j} \, V_{i,j} \widetilde{w}_{i,j}^{(3)} \widetilde{w}_{i,j}^{(3)} IF_{i,j}^{(3)}(\widetilde{\boldsymbol{\theta}}) IF_{i,j}^{(3)}(\widetilde{\boldsymbol{\theta}})' \right\} + \sum_{j=1}^{J} \sum_{i=1}^{n^{(j)}} \sigma_{i,j}^{(2)} w_{i,j}^{(2)} \left\{ \xi_{i,j} \, IF_{i,j}^{(2)}(\widetilde{\boldsymbol{\theta}}) IF_{i,j}^{(2)}(\widetilde{\boldsymbol{\theta}})' + \right.$

$\left. 2 \, \xi_{i,j} \, V_{i,j} \, \widetilde{w}_{i,j}^{(3)} IF_{i,j}^{(2)}(\widetilde{\boldsymbol{\theta}}) IF_{i,j}^{(3)}(\widetilde{\boldsymbol{\theta}})' + \frac{1}{\widetilde{w}_{i,j}^{(3)}} \xi_{i,j} \, V_{i,j} \, \widetilde{w}_{i,j}^{(3)} \widetilde{w}_{i,j}^{(3)} IF_{i,j}^{(3)}(\widetilde{\boldsymbol{\theta}}) IF_{i,j}^{(3)}(\widetilde{\boldsymbol{\theta}})' \right\} +$

$\sum_{j=1}^{J} \sum_{i=1}^{n^{(j)}} \sum_{\substack{k=1, \\ k \neq i}}^{n^{(j)}} \sigma_{i,k,j}^{(2)} w_{i,k,j}^{(2)} \left\{ \xi_{i,j} \, IF_{i,j}^{(2)}(\widetilde{\boldsymbol{\theta}}) + \xi_{i,j} \, V_{i,j} \, \widetilde{w}_{i,j}^{(3)} IF_{i,j}^{(3)}(\widetilde{\boldsymbol{\theta}}) \right\} \left\{ \xi_{k,j} \, IF_{k,j}^{(2)}(\widetilde{\boldsymbol{\theta}}) + \right.$

$\left. \xi_{k,j} \, V_{k,j} \, \widetilde{w}_{k,j}^{(3)} IF_{i,j}^{(3)}(\widetilde{\boldsymbol{\theta}}) \right\}' + \sum_{j=1}^{J} \sum_{i=1}^{n^{(j)}} \widetilde{\sigma}_{i,j}^{(3)} \widetilde{w}_{i,j}^{(3)} \xi_{i,j} \, V_{i,j} \, \widetilde{w}_{i,j}^{(3)} \, \widetilde{w}_{i,j}^{(3)} IF_{i,j}^{(3)}(\widetilde{\boldsymbol{\theta}}) IF_{i,j}^{(3)}(\widetilde{\boldsymbol{\theta}})'.$

Note that we can rewrite this quantity as

$\frac{n}{n-1} \sum_{j=1}^{J} \sum_{i=1}^{n^{(j)}} \frac{1}{w_{i,j}^{(2)}} \left\{ \xi_{i,j} \, IF_{i,j}^{(2)}(\widetilde{\boldsymbol{\theta}}) IF_{i,j}^{(2)}(\widetilde{\boldsymbol{\theta}})' + 2 \, \xi_{i,j} \, V_{i,j} \, \widetilde{w}_{i,j}^{(3)} \, IF_{i,j}^{(2)}(\widetilde{\boldsymbol{\theta}}) IF_{i,j}^{(3)}(\widetilde{\boldsymbol{\theta}})' + \right.$

$\left. \frac{1}{\widetilde{w}_{i,j}^{(3)}} \xi_{i,j} \, V_{i,j} \widetilde{w}_{i,j}^{(3)} \widetilde{w}_{i,j}^{(3)} IF_{i,j}^{(3)}(\widetilde{\boldsymbol{\theta}}) IF_{i,j}^{(3)}(\widetilde{\boldsymbol{\theta}})' \right\} + \sum_{j=1}^{J} \sum_{i=1}^{n^{(j)}} \sum_{k=1}^{n^{(j)}} \sigma_{i,k,j}^{(2)} w_{i,k,j}^{(2)} \left\{ \xi_{i,j} \, IF_{i,j}^{(2)}(\widetilde{\boldsymbol{\theta}}) + \right.$

$$\xi_{i,j}\, V_{i,j}\, \widetilde{w}_{i,j}^{(3)} IF_{i,j}^{(3)}(\widetilde{\boldsymbol{\theta}})\}\{\xi_{k,j}\, IF_{k,j}^{(2)}(\widetilde{\boldsymbol{\theta}}) + \xi_{k,j}\, V_{k,j}\, \widetilde{w}_{k,j}^{(3)} IF_{i,j}^{(3)}(\widetilde{\boldsymbol{\theta}})\}' +$$

$$\sum_{j=1}^{J} \sum_{i=1}^{n^{(j)}} \frac{1}{w_{i,j}^{(2)}} \widetilde{\sigma}_{i,j}^{(3)} \widetilde{w}_{i,j}^{(3)} \left\{ \xi_{i,j}\, V_{i,j}\, \widetilde{w}_{i,j}^{(3)} \widetilde{w}_{i,j}^{(3)} IF_{i,j}^{(3)}(\widetilde{\boldsymbol{\theta}}) IF_{i,j}^{(3)}(\widetilde{\boldsymbol{\theta}})' \right\},$$

or also as

$$\frac{n}{n-1} \sum_{j=1}^{J} \sum_{i=1}^{n^{(j)}} \frac{1}{w_{i,j}^{(2)}} \left\{ \xi_{i,j}\, IF_{i,j}^{(2)}(\widetilde{\boldsymbol{\theta}}) IF_{i,j}^{(2)}(\widetilde{\boldsymbol{\theta}})' + 2\, \xi_{i,j}\, V_{i,j}\, \widetilde{w}_{i,j}^{(3)}\, IF_{i,j}^{(2)}(\widetilde{\boldsymbol{\theta}}) IF_{i,j}^{(3)}(\widetilde{\boldsymbol{\theta}})' + \right.$$

$$\left. \frac{1}{\widetilde{w}_{i,j}^{(3)}} \xi_{i,j}\, V_{i,j} \widetilde{w}_{i,j}^{(3)} \widetilde{w}_{i,j}^{(3)} IF_{i,j}^{(3)}(\widetilde{\boldsymbol{\theta}}) IF_{i,j}^{(3)}(\widetilde{\boldsymbol{\theta}})' \right\} + \sum_{j=1}^{J} \sum_{i=1}^{n^{(j)}} \sum_{k=1}^{n^{(j)}} \widetilde{\sigma}_{i,k,j}\, \widetilde{w}_{i,k,j} \{\xi_{i,j}\, IF_{i,j}^{(2)}(\widetilde{\boldsymbol{\theta}}) +$$

$$\xi_{i,j}\, V_{i,j}\, \widetilde{w}_{i,j}^{(3)} IF_{i,j}^{(3)}(\widetilde{\boldsymbol{\theta}})\}\{\xi_{k,j}\, IF_{k,j}^{(2)}(\widetilde{\boldsymbol{\theta}}) + \xi_{k,j}\, V_{k,j}\, \widetilde{w}_{k,j}^{(3)} IF_{i,j}^{(3)}(\widetilde{\boldsymbol{\theta}})\}' -$$

$$\sum_{j=1}^{J} \sum_{i=1}^{n^{(j)}} \frac{1}{w_{i,j}^{(2)}} \widetilde{\sigma}_{i,j}^{(3)} \widetilde{w}_{i,j}^{(3)} \left\{ \xi_{i,j}\, IF_{i,j}^{(2)}(\widetilde{\boldsymbol{\theta}}) IF_{i,j}^{(2)}(\widetilde{\boldsymbol{\theta}})' + 2\, \xi_{i,j}\, V_{i,j}\, \widetilde{w}_{i,j}^{(3)}\, IF_{i,j}^{(2)}(\widetilde{\boldsymbol{\theta}}) IF_{i,j}^{(3)}(\widetilde{\boldsymbol{\theta}})' \right\}.$$

## Web Appendix H.    MISSING DATA WHEN THE PHASE-THREE DESIGN SAMPLING PROBABILITIES ARE KNOWN

### H.1    Parameters estimation

Here, we assume that the phase-three design sampling probabilities are known; in that case, the phase-three design weights $w_{i,j}^{(3)}$ and variances $\sigma_{i,j}^{(3)}$ are known too, and do not need to be estimated, $i \in \{1, \ldots, n^{(j)}\}$, $j \in \{1, \ldots, J\}$. From the available data, we estimate the log-relative hazard by solving in $\boldsymbol{\beta}$ the estimating equation

$$\sum_{j=1}^{J} \sum_{i=1}^{n^{(j)}} \int_t V_{i,j}\, w_{i,j}^{(3)} \left\{ \boldsymbol{X}_{i,j} - \frac{S_1(t;\boldsymbol{\beta})}{S_0(t;\boldsymbol{\beta})} \right\} \mathrm{d}N_{i,j}(t) = 0,$$

with $S_0(t;\boldsymbol{\beta}) = \sum_{j=1}^{J} \sum_{k=1}^{n^{(j)}} \xi_{k,j}\, V_{k,j}\, w_{k,j}\, Y_{k,j}(t) \exp(\boldsymbol{\beta}' \boldsymbol{X}_{k,j})$ and $\boldsymbol{S}_1(t;\boldsymbol{\beta}) = \sum_{j=1}^{J} \sum_{k=1}^{n^{(j)}} \xi_{k,j}\, V_{k,j}\, w_{k,j}\, Y_{k,j}(t) \exp(\boldsymbol{\beta}' X_{k,j})\, \boldsymbol{X}_{k,j}$, and we estimate the baseline hazard point mass at time $t$ by $\mathrm{d}\widehat{\Lambda}_0(t; \widehat{\boldsymbol{\beta}}) \equiv \mathrm{d}\widehat{\Lambda}_0(t) = \frac{\sum_{j=1}^{J} \sum_{i=1}^{n^{(j)}} \mathrm{d}N_{i,j}(t)}{S_0(t;\widehat{\boldsymbol{\beta}})}$. We then estimate the cumulative baseline hazard up to time $t$ by $\widehat{\Lambda}_0(t; \widehat{\boldsymbol{\beta}}) \equiv \widehat{\Lambda}_0(t) = \int_0^t \mathrm{d}\widehat{\Lambda}_0(s)$, and we finally estimate the pure risk as

$$\widehat{\pi}(\tau_1, \tau_2; \boldsymbol{x}, \widehat{\boldsymbol{\beta}}, \mathrm{d}\widehat{\Lambda}_0) \equiv \widehat{\pi}(\tau_1, \tau_2; \boldsymbol{x}) = 1 - \exp\left\{-\int_{\tau_1}^{\tau_2} \exp(\widehat{\boldsymbol{\beta}}'\boldsymbol{x}) \mathrm{d}\widehat{\Lambda}_0(t)\right\}.$$

We assume that failure time and case-status are known for all $n$ individuals in the cohort; then, and as in the Main Document, we leave the numerator of the modified non-parametric Breslow estimator unchanged. In addition, observe that the estimating equation given above could be rewritten as $\sum_{j=1}^{J} \sum_{i=1}^{n^{(j)}} \int_t \xi_{i,j} V_{i,j} w_{i,j} \left\{X_{i,j} - \frac{S_1(t;\beta)}{S_0(t;\beta)}\right\} \mathrm{d}N_{i,j}(t) = 0$ and that $\mathrm{d}\widehat{\Lambda}_0(t)$ could be rewritten as $\frac{\sum_{j=1}^{J} \sum_{i=1}^{n^{(j)}} \xi_{i,j} w_{i,j}^{(2)} \mathrm{d}N_{i,j}(t)}{S_0(t;\widehat{\boldsymbol{\beta}})}$, as indeed all the cases are initially included in the stratified case-cohort (i.e., included in the phase-two sample) and have unit phase-two sampling weights, and thus $\xi_{i,j} w_{i,j}^{(2)} = 1$ for any subject $i$ in stratum $j$ such that $\int_t \mathrm{d}N_{i,j}(t) = 1$.

### H.2 Derivation of the influence functions

We let $\boldsymbol{\Delta}_{i,j}(\widehat{\boldsymbol{\theta}})$ denote the influence of subject $i$ in stratum $j$ on $\widehat{\boldsymbol{\theta}}$, $i \in \{1, \ldots, n^{(j)}\}, j \in \{1, \ldots, J\}$, $\widehat{\boldsymbol{\theta}} \in \{\widehat{\boldsymbol{\beta}}, \mathrm{d}\widehat{\Lambda}_0(t), \widehat{\Lambda}_0(t), \widehat{\pi}(\tau_1, \tau_2; \boldsymbol{x})\}$. Because the derivations are similar to that in Web Appendices C.1 and D.2 we do not give details. We have

$$\boldsymbol{\Delta}_{i,j}(\widehat{\boldsymbol{\beta}}) = \xi_{i,j} \boldsymbol{IF}_{i,j}^{(2)}(\widehat{\boldsymbol{\beta}}) + \xi_{i,j} V_{i,j} w_{i,j} \boldsymbol{IF}_{i,j}^{(3)}(\widehat{\boldsymbol{\beta}}),$$

with $\boldsymbol{IF}_{i,j}^{(2)}(\widehat{\boldsymbol{\beta}}) = 0$,

$$\boldsymbol{IF}_{i,j}^{(3)}(\widehat{\boldsymbol{\beta}}) = \left[\sum_{l=1}^{J} \sum_{k=1}^{n^{(l)}} \int_t \xi_{k,l} V_{k,l} w_{k,l} \left\{\frac{S_2(t;\widehat{\boldsymbol{\beta}})}{S_0(t;\widehat{\boldsymbol{\beta}})} - \frac{S_1(t;\widehat{\boldsymbol{\beta}}) S_1(t;\widehat{\boldsymbol{\beta}})'}{S_0(t;\widehat{\boldsymbol{\beta}})^2}\right\} \mathrm{d}N_{k,l}(t)\right]^{-1} \times \left[\int_t \left\{X_{i,j} - \frac{S_1(t;\widehat{\boldsymbol{\beta}})}{S_0(t;\widehat{\boldsymbol{\beta}})}\right\} \left\{\mathrm{d}N_{i,j}(t) - Y_{i,j}(t) \exp(\widehat{\boldsymbol{\beta}}' X_{i,j}) \frac{\sum_{l=1}^{J} \sum_{k=1}^{n^{(l)}} \xi_{k,l} V_{k,l} w_{k,l} \mathrm{d}N_{k,l}(t)}{S_0(t;\widehat{\boldsymbol{\beta}})}\right\}\right],$$

and $S_2(t; \widehat{\boldsymbol{\beta}}) = \sum_{j=1}^{J} \sum_{k=1}^{n^{(j)}} \xi_{k,j} V_{k,j} w_{k,j} Y_{k,j}(t) \exp(\widehat{\boldsymbol{\beta}}' X_{k,j}) X_{k,j} X_{k,j}'$.

Then $\Delta_{i,j}\{d\widehat{\Lambda}_0(t)\} = \xi_{i,j} IF_{i,j}^{(2)}\{d\widehat{\Lambda}_0(t)\} + \xi_{i,j} V_{i,j} w_{i,j} IF_{i,j}^{(3)}\{d\widehat{\Lambda}_0(t)\}$, with $IF_{i,j}^{(2)}\{d\widehat{\Lambda}_0(t)\} = \{S_0(t;\hat{\beta})\}^{-1} dN_{i,j}$ and $IF_{i,j}^{(3)}\{d\widehat{\Lambda}_0(t)\} = -\{S_0(t;\hat{\beta})\}^{-1}\{d\widehat{\Lambda}_0(t) S_1(t;\widehat{\beta})' IF_{i,j}^{(2,3)}(\widehat{\beta}) + d\widehat{\Lambda}_0(t) Y_{i,j}(t) \exp(\widehat{\beta}' X_{i,j})\}$.

Note that $IF_{i,j}^{(2)}\{d\widehat{\Lambda}_0(t)\}$ and $IF_{i,j}^{(3)}\{d\widehat{\Lambda}_0(t)\}$ are linear combinations of the increments $dN_{i,j}(t)$ and $d\widetilde{\Lambda}_0(t)$. Hence $\Delta_{i,j}\left\{\int_{\tau_1}^{\tau_2} d\widehat{\Lambda}_0(t)\right\} = \xi_{i,j} IF_{i,j}^{(2)}\left\{\int_{\tau_1}^{\tau_2} d\widehat{\Lambda}_0(t)\right\} + \xi_{i,j} V_{i,j} w_{i,j} IF_{i,j}^{(3)}\left\{\int_{\tau_1}^{\tau_2} d\widehat{\Lambda}_0(t)\right\}$, with $IF_{i,j}^{(s)}\left\{\int_{\tau_1}^{\tau_2} d\widehat{\Lambda}_0(t)\right\} = \int_{\tau_1}^{\tau_2} IF_{i,j}^{(s)}\{d\widehat{\Lambda}_0(s)\}$, $s \in \{2,3\}$. Finally, $\Delta_{i,j}\{\hat{\pi}(\tau_1,\tau_2;x)\} = \xi_{i,j} IF_{i,j}^{(2)}\{\hat{\pi}(\tau_1,\tau_2;x)\} + \xi_{i,j} V_{i,j} w_{i,j} IF_{i,j}^{(3)}\{\hat{\pi}(\tau_1,\tau_2;x)\}$, with $IF_{i,j}^{(s)}\{\hat{\pi}(\tau_1,\tau_2;x)\} = \left\{\frac{\partial \hat{\pi}(\tau_1,\tau_2;x)}{\partial \beta}\Big|_{\beta=\widehat{\beta}}\right\} IF_{i,j}^{(s)} + \left[\frac{\partial \hat{\pi}(\tau_1,\tau_2;x)}{\partial \{\int_{\tau_1}^{\tau_2} d\Lambda_0(t)\}}\Big|_{d\Lambda_0(t)=d\widehat{\Lambda}_0(t)}\right] \Delta_{i,j}\left\{\int_{\tau_1}^{\tau_2} d\widehat{\Lambda}_0(t)\right\}$, $s \in \{2,3\}$, and with

$\frac{\partial \hat{\pi}(\tau_1,\tau_2;x)}{\partial \beta}\Big|_{\beta=\widehat{\beta}} = \{\sum_{t=\tau_1}^{\tau_2} \hat{\lambda}_0(t) \exp(\widehat{\beta}'x)\}\{1 - \hat{\pi}(\tau_1,\tau_2;x)\} x'$,

$\frac{\partial \widetilde{\pi}(\tau_1,\tau_2;x)}{\partial \Lambda_0(\tau_1)}\Big|_{\Lambda_0(\tau_1)=\widetilde{\Lambda}_0(\tau_1)} = \exp(\widetilde{\beta}'x)\{1 - \widetilde{\pi}(\tau_1,\tau_2;x)\}$,

and $\frac{\partial \hat{\pi}(\tau_1,\tau_2;x)}{\partial \{\int_{\tau_1}^{\tau_2} d\Lambda_0(t)\}}\Big|_{d\Lambda_0(t)=d\widehat{\Lambda}_0(t)} = -\exp(\widehat{\beta}'x)\{1 - \hat{\pi}(\tau_1,\tau_2;x)\}$.

Here, $\Delta_{i,j}(\widehat{\beta})$ is zero if subject $i$ in stratum $j$ has not been sampled in both the second and third phase of sampling, $i \in \{1, \ldots, n^{(j)}\}, j \in \{1, \ldots, J\}$; thus, a case may have a zero influence on $\widehat{\beta}$. On the other hand, $\xi_{i,j} V_{i,j} w_{i,j} IF_{i,j}^{(3)}\{d\widehat{\Lambda}_0(t)\}$, $\xi_{i,j} V_{i,j} w_{i,j} IF_{i,j}^{(3)}\left\{\int_{\tau_1}^{\tau_2} d\widehat{\Lambda}_0(t)\right\}$ and $\xi_{i,j} V_{i,j} w_{i,j} IF_{i,j}^{(3)}\{\hat{\pi}(\tau_1,\tau_2;x)\}$ are also zero if subject $i$ in stratum $j$ has not been sampled in both the second and third phase of sampling, but $\xi_{i,j} IF_{i,j}^{(2)}\{d\widehat{\Lambda}_0(t)\}$, $\xi_{i,j} IF_{i,j}^{(2)}\left\{\int_{\tau_1}^{\tau_2} d\widehat{\Lambda}_0(t)\right\}$ and $\xi_{i,j} IF_{i,j}^{(2)}\{\hat{\pi}(\tau_1,\tau_2;x)\}$ may be non-zero. Indeed, because we choose not to reweight the numerator of the Breslow estimator (i.e., we choose to use the actual failure times of the observed cases), even if case $i$ in stratum $j$ has missing covariate data (i.e., has not been sampled in the third phase of sampling), he/she has a non-zero influence on $d\widehat{\Lambda}_0(t)$, equal to $\{S_0(t;\widehat{\beta})\}^{-1}dN_{i,j}(t)$, at her/his time of failure $t$, and subsequently may have a non-zero influence on $\widehat{\Lambda}_0(t)$ and $\hat{\pi}(\tau_1,\tau_2;x)$.

### H.3 Variance decomposition and estimation from influence functions

For any $\widehat{\theta} \in \{\widehat{\beta}, d\widehat{\Lambda}_0(t), \widehat{\Lambda}_0(t), \hat{\pi}(\tau_1,\tau_2;x)\}$, the variance based on the influences is $\text{var}\{\sum_{j=1}^{J}\sum_{i=1}^{n^{(j)}}\Delta_{i,j}(\widehat{\theta})\}$. Using the law of total covariance and the law of total expectation, it can be decomposed as $\text{var}\left(\text{E}\left[\text{E}\{\sum_{j=1}^{J}\sum_{i=1}^{n^{(j)}}\Delta_{i,j}(\widehat{\theta})|C_1,C_2\}|C_1\right]\right) +$

$\text{E}\left(\text{var}\left[\text{E}\{\sum_{j=1}^{J}\sum_{i=1}^{n^{(j)}}\Delta_{i,j}(\widehat{\theta})|C_1,C_2\}|C_1\right]\right) + \text{E}\left(\text{E}\left[\text{var}\{\sum_{j=1}^{J}\sum_{i=1}^{n^{(j)}}\Delta_{i,j}(\widehat{\theta})|C_1,C_2\}|C_1\right]\right)$,

where $C_1$ denote the information from the whole cohort, and $C_2$ denote the information from the phase-two sample.

We have shown in Web Appendix H.2 that $\Delta_{i,j}(\widehat{\theta}) = \xi_{i,j}\mathbf{IF}^{(2)}_{i,j}(\widehat{\theta}) + \xi_{i,j}V_{i,j}w_{i,j}\mathbf{IF}^{(3)}_{i,j}(\widehat{\theta})$, with $\mathbf{IF}^{(2)}_{i,j}(\widehat{\beta}) = 0$. Using similar arguments as in Web Appendix G.2, we have

$\text{E}\{\sum_{j=1}^{J}\sum_{i=1}^{n^{(j)}}\Delta_{i,j}(\widehat{\theta})|C_1,C_2\} = \sum_{j=1}^{J}\sum_{i=1}^{n^{(j)}}\{\xi_{i,j}\mathbf{IF}^{(2)}_{i,j}(\widehat{\theta}) +$

$\text{E}(V_{i,j}w^{(3)}_{i,j}|C_1,C_2)\xi_{i,j}w^{(2)}_{i,j}\mathbf{IF}^{(3)}_{i,j}(\widehat{\theta})\}$,

$= \sum_{j=1}^{J}\sum_{i=1}^{n^{(j)}}\xi_{i,j}\{\mathbf{IF}^{(2)}_{i,j}(\widehat{\theta}) + w^{(2)}_{i,j}\mathbf{IF}^{(3)}_{i,j}(\widehat{\theta})\}$,

and we have

$\text{var}\{\sum_{j=1}^{J}\sum_{i=1}^{n^{(j)}}\Delta_{i,j}(\widehat{\theta})|C_1,C_2\} =$

$\sum_{j=1}^{J}\sum_{i=1}^{n^{(j)}}\text{var}(V_{i,j}|C_1,C_2)\{\xi_{i,j}w_{i,j}\mathbf{IF}^{(3)}_{i,j}(\widehat{\theta})\}\{\xi_{i,j}w_{i,j}\mathbf{IF}^{(3)}_{i,j}(\widehat{\theta})\}'$,

$= \sum_{j=1}^{J}\sum_{i=1}^{n^{(j)}}\sigma^{(3)}_{i,j}\xi_{i,j}w_{i,j}w_{i,j}\mathbf{IF}^{(3)}_{i,j}(\widehat{\theta})\mathbf{IF}^{(3)}_{i,j}(\widehat{\theta})'$,

as indeed $\xi_{i,j} IF_{i,j}^{(2)}(\widehat{\boldsymbol{\theta}})$ is fixed conditional on $C_2$.

Then we have

$$\mathrm{E}\left[\mathrm{E}\left\{\sum_{j=1}^{J}\sum_{i=1}^{n^{(j)}}\Delta_{i,j}(\widehat{\boldsymbol{\theta}})\,|C_1,C_2\right\}|C_1\right] = \sum_{j=1}^{J}\sum_{i=1}^{n^{(j)}}\mathrm{E}(\xi_{i,j}\,|C_1)\left\{IF_{i,j}^{(2)}(\widehat{\boldsymbol{\theta}}) + w_{i,j}^{(2)} IF_{i,j}^{(3)}(\widehat{\boldsymbol{\theta}})\right\},$$

$$= \sum_{j=1}^{J}\sum_{i=1}^{n^{(j)}}\left\{\frac{1}{w_{i,j}^{(2)}} IF_{i,j}^{(2)}(\widehat{\boldsymbol{\theta}}) + IF_{i,j}^{(3)}(\widehat{\boldsymbol{\theta}})\right\},$$

and

$$\mathrm{var}\left[\mathrm{E}\left\{\sum_{j=1}^{J}\sum_{i=1}^{n^{(j)}}\Delta_{i,j}(\widehat{\boldsymbol{\theta}})\,|C_1,C_2\right\}|C_1\right] = \sum_{j=1}^{J}\mathrm{var}\left[\sum_{i=1}^{n^{(j)}}\xi_{i,j}\left\{IF_{i,j}^{(2)}(\widehat{\boldsymbol{\theta}}) + w_{i,j}^{(2)} IF_{i,j}^{(3)}(\widehat{\boldsymbol{\theta}})\right\}\right],$$

$$= \sum_{j=1}^{J}\sum_{i=1}^{n^{(j)}}\sum_{k=1}^{n^{(j)}}\sigma_{i,k,j}^{(2)}\left\{IF_{i,j}^{(2)}(\widehat{\boldsymbol{\theta}}) + w_{i,j}^{(2)} IF_{i,j}^{(3)}(\widehat{\boldsymbol{\theta}})\right\}\left\{IF_{k,j}^{(2)}(\widehat{\boldsymbol{\theta}}) + w_{k,j}^{(2)} IF_{k,j}^{(3)}(\widehat{\boldsymbol{\theta}})\right\}',$$

$$= \sum_{j=1}^{J}\sum_{i=1}^{n^{(j)}}\sum_{k=1}^{n^{(j)}}\sigma_{i,k,j}^{(2)}\,w_{i,j}^{(2)} w_{k,j}^{(2)} IF_{i,j}^{(3)}(\widehat{\boldsymbol{\theta}}) IF_{k,j}^{(3)}(\widehat{\boldsymbol{\theta}})',$$

as indeed any pair of individuals $i$ and $k$ in stratum $j$, $j \in \{1,\dots,J\}$, such that $\sigma_{i,k,j}^{(2)} \neq 0$ is necessarily such that $IF_{i,j}^{(2)}(\widehat{\boldsymbol{\theta}}) = IF_{k,j}^{(3)}(\widehat{\boldsymbol{\theta}}) = 0$, and

$$\mathrm{E}\left[\mathrm{var}\left\{\sum_{j=1}^{J}\sum_{i=1}^{n^{(j)}}\Delta_{i,j}(\widehat{\boldsymbol{\theta}})\,|C_1,C_2\right\}|C_1\right] = \sum_{j=1}^{J}\sum_{i=1}^{n^{(j)}}\mathrm{E}(\xi_{i,j}|C_1)\,\sigma_{i,j}^{(3)}\,w_{i,j}\,w_{i,j} IF_{i,j}^{(3)}(\widehat{\boldsymbol{\theta}}) IF_{i,j}^{(3)}(\widehat{\boldsymbol{\theta}})',$$

$$= \sum_{j=1}^{J}\sum_{i=1}^{n^{(j)}}\sigma_{i,j}^{(3)}\,w_{i,j}^{(3)} w_{i,j}\,IF_{i,j}^{(3)}(\widehat{\boldsymbol{\theta}}) IF_{i,j}^{(3)}(\widehat{\boldsymbol{\theta}})'.$$

Finally,

$$\mathrm{var}\left(\mathrm{E}\left[\mathrm{E}\left\{\sum_{j=1}^{J}\sum_{i=1}^{n^{(j)}}\Delta_{i,j}(\widehat{\boldsymbol{\theta}})\,|C_1,C_2\right\}|C_1\right]\right) = \mathrm{var}\left[\sum_{j=1}^{J}\sum_{i=1}^{n^{(j)}}\frac{1}{w_{i,j}^{(2)}} IF_{i,j}^{(2)}(\widehat{\boldsymbol{\theta}}) + IF_{i,j}^{(3)}(\widehat{\boldsymbol{\theta}})\right],$$

and can be estimated by

$$\frac{n}{n-1}\sum_{j=1}^{J}\sum_{i=1}^{n^{(j)}}\left\{\xi_{i,j}\,\frac{1}{w_{i,j}^{(2)}} IF_{i,j}^{(2)}(\widehat{\boldsymbol{\theta}}) IF_{i,j}^{(2)}(\widehat{\boldsymbol{\theta}})' + 2\,\xi_{i,j}\,V_{i,j}\,\frac{w_{i,j}}{w_{i,j}^{(2)}}\,IF_{i,j}^{(2)}(\widehat{\boldsymbol{\theta}}) IF_{i,j}^{(3)}(\widehat{\boldsymbol{\theta}})' + \right.$$

$$\left.\xi_{i,j}\,V_{i,j}\,w_{i,j}\,IF_{i,j}^{(3)}(\widehat{\boldsymbol{\theta}}) IF_{i,j}^{(3)}(\widehat{\boldsymbol{\theta}})'\right\},$$

that is by

$$\frac{n}{n-1}\sum_{j=1}^{J}\sum_{i=1}^{n^{(j)}}\left\{\frac{1}{w_{i,j}^{(2)}}\xi_{i,j}\,IF_{i,j}^{(2)}(\widehat{\boldsymbol{\theta}})IF_{i,j}^{(2)}(\widehat{\boldsymbol{\theta}})' + 2\frac{1}{w_{i,j}^{(2)}}\xi_{i,j}\,V_{i,j}\,w_{i,j}\,IF_{i,j}^{(2)}(\widehat{\boldsymbol{\theta}})IF_{i,j}^{(3)}(\widehat{\boldsymbol{\theta}})' + \right.$$

$$\left.\frac{1}{w_{i,j}}\xi_{i,j}\,V_{i,j}\,w_{i,j}\,w_{i,j}IF_{i,j}^{(3)}(\widehat{\boldsymbol{\theta}})IF_{i,j}^{(3)}(\widehat{\boldsymbol{\theta}})'\right\},$$

and

$$\mathrm{E}\left(\mathrm{var}\left[\mathrm{E}\left\{\sum_{j=1}^{J}\sum_{i=1}^{n^{(j)}}\boldsymbol{\Delta}_{i,j}(\widehat{\boldsymbol{\theta}})\,|C_1,C_2\right\}|C_1\right]\right) =$$

$$\mathrm{E}\left[\sum_{j=1}^{J}\sum_{i=1}^{n^{(j)}}\sum_{k=1}^{n^{(j)}}\sigma_{i,k,j}^{(2)}\,w_{i,j}^{(2)}w_{k,j}^{(2)}IF_{i,j}^{(3)}(\widehat{\boldsymbol{\theta}})IF_{k,j}^{(3)}(\widehat{\boldsymbol{\theta}})'\right],$$

that can be estimated by

$$\sum_{j=1}^{J}\sum_{i=1}^{n^{(j)}}\sum_{k=1}^{n^{(j)}}\sigma_{i,k,j}^{(2)}\,w_{i,k,j}\,\xi_{i,j}\,\xi_{k,j}\,V_{i,j}\,V_{k,j}\,w_{i,j}^{(2)}w_{k,j}^{(2)}IF_{i,j}^{(3)}(\widehat{\boldsymbol{\theta}})IF_{k,j}^{(3)}(\widehat{\boldsymbol{\theta}})',$$

that is by

$$\sum_{j=1}^{J}\sum_{i=1}^{n^{(j)}}\sigma_{i,j}^{(2)}\frac{w_{i,j}^{(2)}}{w_{i,j}^{(3)}}\xi_{i,j}\,V_{i,j}\,w_{i,j}w_{i,j}IF_{i,j}^{(3)}(\widehat{\boldsymbol{\theta}})IF_{i,j}^{(3)}(\widehat{\boldsymbol{\theta}})' +$$

$$\sum_{j=1}^{J}\sum_{i=1}^{n^{(j)}}\sum_{\substack{k=1,\\k\neq i}}^{n^{(j)}}\sigma_{i,k,j}^{(2)}\,w_{i,k,j}^{(2)}\,\xi_{i,j}\,V_{i,j}\,\xi_{k,j}\,V_{k,j}\,w_{i,j}w_{k,j}IF_{i,j}^{(3)}(\widehat{\boldsymbol{\theta}})\,IF_{i,j}^{(3)}(\widehat{\boldsymbol{\theta}})',$$

and

$$\mathrm{E}\left(\mathrm{E}\left[\mathrm{var}\left\{\sum_{j=1}^{J}\sum_{i=1}^{n^{(j)}}\boldsymbol{\Delta}_{i,j}(\widehat{\boldsymbol{\theta}})\,|C_1,C_2\right\}|C_1\right]\right) = \mathrm{E}\left\{\sum_{j=1}^{J}\sum_{i=1}^{n^{(j)}}\sigma_{i,j}^{(3)}\,w_{i,j}^{(3)}\,w_{i,j}\,IF_{i,j}^{(3)}(\widehat{\boldsymbol{\theta}})IF_{i,j}^{(3)}(\widehat{\boldsymbol{\theta}})'\right\},$$

that can be estimated by

$$\sum_{j=1}^{J}\sum_{i=1}^{n^{(j)}}\xi_{i,j}\,V_{i,j}\,w_{i,j}\,\sigma_{i,j}^{(3)}\,w_{i,j}^{(3)}\,w_{i,j}IF_{i,j}^{(3)}(\widehat{\boldsymbol{\theta}})IF_{i,j}^{(3)}(\widehat{\boldsymbol{\theta}})',$$

that is by

$$\sum_{j=1}^{J}\sum_{i=1}^{n^{(j)}}\sigma_{i,j}^{(3)}w_{i,j}^{(3)}\,\xi_{i,j}\,V_{i,j}\,w_{i,j}\,w_{i,j}IF_{i,j}^{(3)}(\widehat{\boldsymbol{\theta}})\,IF_{i,j}^{(3)}(\widehat{\boldsymbol{\theta}})'.$$

As a result, $\mathrm{var}(\widehat{\boldsymbol{\theta}})$ can be estimated by

$$\frac{n}{n-1}\sum_{j=1}^{J}\sum_{i=1}^{n^{(j)}}\left\{\frac{1}{w_{i,j}^{(2)}}\xi_{i,j}\,IF_{i,j}^{(2)}(\widehat{\boldsymbol{\theta}})IF_{i,j}^{(2)}(\widehat{\boldsymbol{\theta}})' + 2\frac{1}{w_{i,j}^{(2)}}\xi_{i,j}\,V_{i,j}\,w_{i,j}\,IF_{i,j}^{(2)}(\widehat{\boldsymbol{\theta}})IF_{i,j}^{(3)}(\widehat{\boldsymbol{\theta}})' + \right.$$

$$\left.\frac{1}{w_{i,j}}\xi_{i,j}\,V_{i,j}\,w_{i,j}\,w_{i,j}IF_{i,j}^{(3)}(\widehat{\boldsymbol{\theta}})IF_{i,j}^{(3)}(\widehat{\boldsymbol{\theta}})'\right\} +$$

$$\sum_{j=1}^{J}\sum_{i=1}^{n^{(j)}}\sigma_{i,j}^{(2)}\frac{w_{i,j}^{(2)}}{w_{i,j}^{(3)}}\xi_{i,j}\,V_{i,j}\,w_{i,j}w_{i,j}IF_{i,j}^{(3)}(\widehat{\boldsymbol{\theta}})IF_{i,j}^{(3)}(\widehat{\boldsymbol{\theta}})' +$$

$$\sum_{j=1}^{J}\sum_{i=1}^{n^{(j)}}\sum_{\substack{k=1\\k\neq i}}^{n^{(j)}}\sigma_{i,k,j}^{(2)}\,w_{i,k,j}^{(2)}\,\xi_{i,j}\,V_{i,j}\,\xi_{k,j}\,V_{k,j}\,w_{i,j}w_{k,j}IF_{i,j}^{(3)}(\widehat{\boldsymbol{\theta}})\,IF_{i,j}^{(3)}(\widehat{\boldsymbol{\theta}})' +$$

$$\sum_{j=1}^{J}\sum_{i=1}^{n^{(j)}}\sigma_{i,j}^{(3)}w_{i,j}^{(3)}\,\xi_{i,j}\,V_{i,j}\,w_{i,j}\,w_{i,j}IF_{i,j}^{(3)}(\widehat{\boldsymbol{\theta}})\,IF_{i,j}^{(3)}(\widehat{\boldsymbol{\theta}})',$$

that we can rewrite

$$\frac{n}{n-1}\sum_{j=1}^{J}\sum_{i=1}^{n^{(j)}}\left\{\frac{1}{w_{i,j}^{(2)}}\xi_{i,j}\,IF_{i,j}^{(2)}(\widehat{\boldsymbol{\theta}})IF_{i,j}^{(2)}(\widehat{\boldsymbol{\theta}})' + 2\frac{1}{w_{i,j}^{(2)}}\xi_{i,j}\,V_{i,j}\,w_{i,j}\,IF_{i,j}^{(2)}(\widehat{\boldsymbol{\theta}})IF_{i,j}^{(3)}(\widehat{\boldsymbol{\theta}})' + \right.$$

$$\left.\frac{1}{w_{i,j}}\xi_{i,j}\,V_{i,j}\,w_{i,j}\,w_{i,j}IF_{i,j}^{(3)}(\widehat{\boldsymbol{\theta}})IF_{i,j}^{(3)}(\widehat{\boldsymbol{\theta}})'\right\} + \qquad (4)$$

$$\sum_{j=1}^{J}\sum_{i=1}^{n^{(j)}}\sum_{k=1}^{n^{(j)}}\sigma_{i,k,j}\,w_{i,k,j}\,\xi_{i,j}\,V_{i,j}\,\xi_{k,j}\,V_{k,j}\,w_{i,j}w_{k,j}\,IF_{i,j}^{(3)}(\widehat{\boldsymbol{\theta}})\,IF_{i,j}^{(3)}(\widehat{\boldsymbol{\theta}})',$$

as indeed $\sigma_{i,j}^{(2)}\frac{w_{i,j}^{(2)}}{w_{i,j}^{(3)}} + \sigma_{i,j}^{(3)}w_{i,j}^{(3)} = 1 - \frac{1}{w_{i,j}} = \sigma_{i,j}\,w_{i,j}$, and $\sigma_{i,k,j}^{(2)}\,w_{i,k,j}^{(2)} = \sigma_{i,k,j}^{(2)} \times \frac{1}{w_{i,j}^{(3)}} \times \frac{1}{w_{k,j}^{(3)}} \times$

$w_{i,k,j}^{(2)} \times w_{i,j}^{(3)} \times w_{k,j}^{(3)} = \sigma_{i,k,j}w_{i,k,j}$, $i,k \in \{1,\dots,n^{(j)}\}, i \neq k, j \in \{1,\dots,J\}$.

Note, when we assume that the $w_i^{(3)}$ are known, and because the third phase of sampling is Bernoulli, we could collapse the phase-two and phase-three into a single sampling phase. The variance estimate given above in Equation (4) and written with only two components, corresponds to the form we would obtain directly when the second and third phases of sampling are collapsed into a single phase of sampling.

# Web Appendix I.   SIMULATIONS WITH MISSING DATA

## I.1 Simulation design

We simulated cohorts with $n \in \{5 \times 10^3, 10^4\}$ as in Section 7 in the Main Document. We sampled independently across the $J = 4$ strata, defined by $W = 0 \times I(X_1 \geq 0, X_2 = 0) + 1 \times I(X_1 < 0, X_2 < 2) + 2 \times I(X_1 \geq 0, X_2 > 0) + 3 \times I(X_1 < 0, X_2 = 2)$, fixed numbers of individuals, $m^{(j)} = \left\lfloor \frac{\lambda_0 \times 10 \times E\{\exp(\beta_1 X_1 + \beta_2 X_2 + \beta_3 X_3) | W = j\}}{1 - \lambda_0 \times 10 \times E\{\exp(\beta_1 X_1 + \beta_2 X_2 + \beta_3 X_3) | W = j\}} \times E(n^{(j)}) \times K + \frac{1}{2} \right\rfloor$, with $K \in \{2, 4\}$, $j \in \{0,1,2,3\}$, and where $\lfloor \ \rfloor$ is the floor function. These individuals, in addition to the remaining cases, constituted the phase-two sample. In parallel, we sampled individuals from the $J^{(3)} = 2$ strata defined by case status, with phase-three sampling probabilities $\boldsymbol{\pi}^{(3)} \in \{(0.9, 0.8), (0.98, 0.9)\}$. In other words, we assumed that cases always had a higher probability of missing covariate information, for example due to stored blood samples being previously used. Finally, we regarded the individuals who were in both samples, i.e., individuals with complete covariate data, as the phase-three sample.

Phase-two sampling design weights were computed as in Section 3.3 in the Main Document, and phase-three sampling design weights were given by the inverse of the phase-three sampling probabilities. The overall sampling designs weights were then obtained from their product. However, in practice, the phase-three sampling design weights, $\frac{1}{\pi^{(3)}}$, are usually unknown. We thus estimated the phase-three sampling design weights as in Section 7.2 in the Main Document; the overall estimated sampling designs weights were then obtained from the product with the known phase-two sampling design weights.

For each scenario, we simulated 5,000 cohorts. We estimated the log-relative hazard $\boldsymbol{\beta} = (\beta_1, \beta_2, \beta_3)'$ and pure risks $\pi(\tau_1, \tau_2; \boldsymbol{x})$ in time interval $(\tau_1, \tau_2] = (0,8]$ and for covariate profiles $x \in \{(-1, 1, -0.6)', (1, -1, 0.6)', (1, 1, 0.6)'\}$, using: the stratified case-cohort with estimated design weights (SCC.Est); the stratified case-cohort with true design weights (SCC.True); the unstratified case-cohort with estimated design weights (USCC.Est); and the unstratified case-cohort with true design weights (USCC.True). For each simulated realization, we then estimated their variances. For SCC.Est, we used: the variance estimate with superpopulation, phase-two and phase-three variance components ($\hat{V}$) from Equation (22) in Section 5.5 the Main Document or in Web Appendix G.2; and the robust variance estimate ($\hat{V}_{\text{Robust}}$) computed as in Section 3.3 or Section 4.3 in the Main Document but with the influences given in Section 5.4 in the Main document or in Web Appendix G.1. For SCC.True, we used: the variance estimate with superpopulation, phase-two and phase-three variance components ($\hat{V}$) from Equation (4) in Web Appendix H.3; and $\hat{V}_{\text{Robust}}$ computed as in Section 3.3 or Section 4.3 in the Main Document but with the influences given in Web Appendix H.2. For USCC.Est and USCC.True, we used similar variance estimates, with $J = 1$. An alternative was also to treat the estimated design weights as if they were the true known ones, and to estimate the variance as in Web Appendix H.3 (SCC.Naive and USCC.Naive). As a point of reference, we also estimated these parameters using the data from the whole cohort (Cohort).

### I.2 Simulation results

**WEB TABLE 29** to **WEB TABLE 40** display the coverages of 95% CIs, and **WEB TABLE 41** to **WEB TABLE 52** display the mean of estimated variances and empirical variances, for $\beta_1$, $\beta_2$, $\beta_3$ and $\log\{\pi(\tau_1, \tau_2; x)\}$, $(\tau_1, \tau_2] = (0,8]$ and $x \in \{(-1, 1, -0.6)', (1, -1, 0.6)', (1, 1, 0.6)'\}$, respectively. For brevity, we left out results for $n = 10{,}000$ and $p_Y = 0.1$. Again, the robust variance overestimated the variance and yielded supra-nominal confidence interval coverage for log-relative hazards and pure risks with stratified designs in most of the scenarios, and for pure risk with unstratified designs in most of the scenarios. On the other hand, variance estimation properly accounting for the sampling features yielded proper coverage in most of the scenarios. Using the estimated design weights led to similar performance as using the true design weights. In particular, treating the estimated weights as if they were the true known ones, and using the simpler variance formula given in Web Appendix H.3, usually led to similar performance as using the more complex influence-based variance given in Equation (22) in the Main Document. Therefore, accounting for the variability from estimation of the phase-three weights may not be necessary. Using estimated weights in SCC.Est and SCC.Naive (respectively USCC.Est and USCC.Naive) yielded negligible, if any, gain of efficiency compared to SCC.True (respectively USCC.True) in the present simulations.

| Cohort | SCC.True | | SCC.Est | | SCC.Naive | | $n$ | $K$ | $p_Y$ | $\pi^{(3)}$ | $\beta_1$ |
|---|---|---|---|---|---|---|---|---|---|---|---|
| | $\hat{V}_{Robust}$ | $\hat{V}$ | $\hat{V}_{Robust}$ | $\hat{V}$ | $\hat{V}_{Robust}$ | $\hat{V}$ | | | | | |
| 0.9504 | 0.969* | 0.9574* | 0.969* | 0.9574* | 0.969* | 0.9574* | 5000 | 2 | 0.02 | (0.9,0.8) | -0.2 |
| 0.9504 | 0.9694* | 0.9554 | 0.9694* | 0.9554 | 0.9694* | 0.9554 | 5000 | 2 | 0.02 | (0.98,0.9) | -0.2 |
| 0.943* | 0.9534 | 0.9474 | 0.9536 | 0.9474 | 0.9536 | 0.9474 | 5000 | 4 | 0.02 | (0.9,0.8) | -0.2 |
| 0.943* | 0.9568* | 0.948 | 0.957* | 0.948 | 0.957* | 0.948 | 5000 | 4 | 0.02 | (0.98,0.9) | -0.2 |
| 0.9478 | 0.9652* | 0.953 | 0.9652* | 0.953 | 0.9652* | 0.953 | 10000 | 2 | 0.02 | (0.9,0.8) | -0.2 |
| 0.9478 | 0.9664* | 0.9534 | 0.9662* | 0.9534 | 0.9662* | 0.9534 | 10000 | 2 | 0.02 | (0.98,0.9) | -0.2 |
| 0.9482 | 0.9542 | 0.9462 | 0.9542 | 0.9462 | 0.9542 | 0.9462 | 10000 | 4 | 0.02 | (0.9,0.8) | -0.2 |
| 0.9482 | 0.9572* | 0.9474 | 0.9572* | 0.9474 | 0.9572* | 0.9474 | 10000 | 4 | 0.02 | (0.98,0.9) | -0.2 |
| 0.9486 | 0.9638* | 0.9548 | 0.9638* | 0.9546 | 0.9638* | 0.9546 | 5000 | 2 | 0.05 | (0.9,0.8) | -0.2 |
| 0.9486 | 0.96* | 0.9472 | 0.96* | 0.947 | 0.96* | 0.947 | 5000 | 2 | 0.05 | (0.98,0.9) | -0.2 |
| 0.9462 | 0.9506 | 0.944 | 0.9506 | 0.9442 | 0.9506 | 0.9442 | 5000 | 4 | 0.05 | (0.9,0.8) | -0.2 |
| 0.9462 | 0.955 | 0.9474 | 0.9548 | 0.9476 | 0.9548 | 0.9476 | 5000 | 4 | 0.05 | (0.98,0.9) | -0.2 |
| 0.9486 | 0.9654* | 0.9536 | 0.9654* | 0.9536 | 0.9654* | 0.9536 | 10000 | 2 | 0.05 | (0.9,0.8) | -0.2 |
| 0.9486 | 0.966* | 0.9498 | 0.966* | 0.95 | 0.966* | 0.95 | 10000 | 2 | 0.05 | (0.98,0.9) | -0.2 |
| 0.948 | 0.9554 | 0.9496 | 0.9554 | 0.9496 | 0.9554 | 0.9496 | 10000 | 4 | 0.05 | (0.9,0.8) | -0.2 |
| 0.948 | 0.9594* | 0.9514 | 0.9596* | 0.9514 | 0.9596* | 0.9514 | 10000 | 4 | 0.05 | (0.98,0.9) | -0.2 |
| 0.949 | 0.9574* | 0.947 | 0.9574* | 0.9474 | 0.9574* | 0.9474 | 5000 | 2 | 0.1 | (0.9,0.8) | -0.2 |
| 0.949 | 0.9592* | 0.9484 | 0.9594* | 0.9484 | 0.9594* | 0.9484 | 5000 | 2 | 0.1 | (0.98,0.9) | -0.2 |
| 0.9494 | 0.952 | 0.9486 | 0.9518 | 0.9484 | 0.9518 | 0.9484 | 5000 | 4 | 0.1 | (0.9,0.8) | -0.2 |
| 0.9494 | 0.954 | 0.9494 | 0.9536 | 0.9494 | 0.9536 | 0.9494 | 5000 | 4 | 0.1 | (0.98,0.9) | -0.2 |

**WEB TABLE 29-** Coverage of 95% CIs for log-relative hazard parameter $\beta_1$ from stratified sampling using different methods of analysis and variance estimation, for various probabilities of missing covariate data, in 5,000 simulated cohorts. * indicates coverage outside the expected interval [0.9440; 0.9560].

| Cohort | USCC.True | | USCC.Est | | USCC.Naive | | $n$ | $K$ | $p_Y$ | $\pi^{(3)}$ | $\beta_1$ |
|---|---|---|---|---|---|---|---|---|---|---|---|
| | $\hat{V}_{\text{Robust}}$ | $\hat{V}$ | $\hat{V}_{\text{Robust}}$ | $\hat{V}$ | $\hat{V}_{\text{Robust}}$ | $\hat{V}$ | | | | | |
| 0.9504 | 0.9576* | 0.9576* | 0.9576* | 0.9578* | 0.9576* | 0.9578* | 5000 | 2 | 0.02 | (0.9,0.8) | -0.2 |
| 0.9504 | 0.9564* | 0.9566* | 0.956 | 0.9566* | 0.956 | 0.9566* | 5000 | 2 | 0.02 | (0.98,0.9) | -0.2 |
| 0.943* | 0.9514 | 0.9514 | 0.9514 | 0.9514 | 0.9514 | 0.9514 | 5000 | 4 | 0.02 | (0.9,0.8) | -0.2 |
| 0.943* | 0.9486 | 0.9486 | 0.9486 | 0.9486 | 0.9486 | 0.9486 | 5000 | 4 | 0.02 | (0.98,0.9) | -0.2 |
| 0.9478 | 0.9542 | 0.9542 | 0.9542 | 0.9542 | 0.9542 | 0.9542 | 10000 | 2 | 0.02 | (0.9,0.8) | -0.2 |
| 0.9478 | 0.9512 | 0.9512 | 0.9512 | 0.9512 | 0.9512 | 0.9512 | 10000 | 2 | 0.02 | (0.98,0.9) | -0.2 |
| 0.9482 | 0.9458 | 0.9458 | 0.9458 | 0.946 | 0.9458 | 0.946 | 10000 | 4 | 0.02 | (0.9,0.8) | -0.2 |
| 0.9482 | 0.948 | 0.948 | 0.948 | 0.948 | 0.948 | 0.948 | 10000 | 4 | 0.02 | (0.98,0.9) | -0.2 |
| 0.9486 | 0.9504 | 0.9504 | 0.9506 | 0.9508 | 0.9506 | 0.9508 | 5000 | 2 | 0.05 | (0.9,0.8) | -0.2 |
| 0.9486 | 0.9506 | 0.9506 | 0.9506 | 0.9506 | 0.9506 | 0.9506 | 5000 | 2 | 0.05 | (0.98,0.9) | -0.2 |
| 0.9462 | 0.9492 | 0.9492 | 0.9492 | 0.9492 | 0.9492 | 0.9492 | 5000 | 4 | 0.05 | (0.9,0.8) | -0.2 |
| 0.9462 | 0.9472 | 0.9472 | 0.9472 | 0.9472 | 0.9472 | 0.9472 | 5000 | 4 | 0.05 | (0.98,0.9) | -0.2 |
| 0.9486 | 0.9474 | 0.9476 | 0.9476 | 0.9476 | 0.9476 | 0.9476 | 10000 | 2 | 0.05 | (0.9,0.8) | -0.2 |
| 0.9486 | 0.9502 | 0.9504 | 0.9502 | 0.9504 | 0.9502 | 0.9504 | 10000 | 2 | 0.05 | (0.98,0.9) | -0.2 |
| 0.948 | 0.9514 | 0.9514 | 0.9512 | 0.9512 | 0.9512 | 0.9512 | 10000 | 4 | 0.05 | (0.9,0.8) | -0.2 |
| 0.948 | 0.95 | 0.95 | 0.95 | 0.95 | 0.95 | 0.95 | 10000 | 4 | 0.05 | (0.98,0.9) | -0.2 |
| 0.949 | 0.9506 | 0.9506 | 0.9508 | 0.9508 | 0.9508 | 0.9508 | 5000 | 2 | 0.1 | (0.9,0.8) | -0.2 |
| 0.949 | 0.95 | 0.95 | 0.9498 | 0.9498 | 0.9498 | 0.9498 | 5000 | 2 | 0.1 | (0.98,0.9) | -0.2 |
| 0.9494 | 0.9482 | 0.9482 | 0.9484 | 0.9484 | 0.9484 | 0.9484 | 5000 | 4 | 0.1 | (0.9,0.8) | -0.2 |
| 0.9494 | 0.9496 | 0.9496 | 0.9496 | 0.9496 | 0.9496 | 0.9496 | 5000 | 4 | 0.1 | (0.98,0.9) | -0.2 |

**WEB TABLE 30-** Coverage of 95% CIs for log-relative hazard parameter $\beta_1$ from unstratified sampling using different methods of analysis and variance estimation, for various probabilities of missing covariate data, in 5,000 simulated cohorts. * indicates coverage outside the expected interval [0.9440; 0.9560].

| Cohort | SCC.True | | SCC.Est | | SCC.Naive | | $n$ | $K$ | $p_Y$ | $\pi^{(3)}$ | $\beta_2$ |
| --- | --- | --- | --- | --- | --- | --- | --- | --- | --- | --- | --- |
| | $\hat{V}_{Robust}$ | $\hat{V}$ | $\hat{V}_{Robust}$ | $\hat{V}$ | $\hat{V}_{Robust}$ | $\hat{V}$ | | | | | |
| 0.9464 | 0.9632* | 0.9486 | 0.9634* | 0.9488 | 0.9634* | 0.9488 | 5000 | 2 | 0.02 | (0.9,0.8) | 0.25 |
| 0.9464 | 0.9662* | 0.947 | 0.9662* | 0.9468 | 0.9662* | 0.947 | 5000 | 2 | 0.02 | (0.98,0.9) | 0.25 |
| 0.9508 | 0.9614* | 0.9524 | 0.9614* | 0.9524 | 0.9614* | 0.9524 | 5000 | 4 | 0.02 | (0.9,0.8) | 0.25 |
| 0.9508 | 0.9614* | 0.95 | 0.9614* | 0.95 | 0.9614* | 0.95 | 5000 | 4 | 0.02 | (0.98,0.9) | 0.25 |
| 0.9474 | 0.9676* | 0.9502 | 0.9676* | 0.9502 | 0.9676* | 0.9502 | 10000 | 2 | 0.02 | (0.9,0.8) | 0.25 |
| 0.9474 | 0.9682* | 0.9512 | 0.9682* | 0.9512 | 0.9682* | 0.9512 | 10000 | 2 | 0.02 | (0.98,0.9) | 0.25 |
| 0.9488 | 0.9612* | 0.949 | 0.961* | 0.949 | 0.961* | 0.949 | 10000 | 4 | 0.02 | (0.9,0.8) | 0.25 |
| 0.9488 | 0.9634* | 0.951 | 0.9636* | 0.9512 | 0.9636* | 0.9512 | 10000 | 4 | 0.02 | (0.98,0.9) | 0.25 |
| 0.9508 | 0.9658* | 0.9542 | 0.9658* | 0.9546 | 0.9658* | 0.9546 | 5000 | 2 | 0.05 | (0.9,0.8) | 0.25 |
| 0.9508 | 0.9714* | 0.9574* | 0.9712* | 0.9574* | 0.9712* | 0.9574* | 5000 | 2 | 0.05 | (0.98,0.9) | 0.25 |
| 0.9438* | 0.9546 | 0.9466 | 0.9546 | 0.9464 | 0.9546 | 0.9464 | 5000 | 4 | 0.05 | (0.9,0.8) | 0.25 |
| 0.9438* | 0.9556 | 0.9474 | 0.9556 | 0.9478 | 0.9556 | 0.9478 | 5000 | 4 | 0.05 | (0.98,0.9) | 0.25 |
| 0.95 | 0.9594* | 0.9464 | 0.9594* | 0.9464 | 0.9594* | 0.9464 | 10000 | 2 | 0.05 | (0.9,0.8) | 0.25 |
| 0.95 | 0.9652* | 0.9506 | 0.965* | 0.9506 | 0.965* | 0.9506 | 10000 | 2 | 0.05 | (0.98,0.9) | 0.25 |
| 0.9458 | 0.959* | 0.9506 | 0.9588* | 0.9504 | 0.9588* | 0.9504 | 10000 | 4 | 0.05 | (0.9,0.8) | 0.25 |
| 0.9458 | 0.9584* | 0.9498 | 0.9582* | 0.9498 | 0.9582* | 0.9498 | 10000 | 4 | 0.05 | (0.98,0.9) | 0.25 |
| 0.9492 | 0.963* | 0.9492 | 0.9628* | 0.9488 | 0.9628* | 0.9488 | 5000 | 2 | 0.1 | (0.9,0.8) | 0.25 |
| 0.9492 | 0.9664* | 0.9514 | 0.9664* | 0.9514 | 0.9664* | 0.9514 | 5000 | 2 | 0.1 | (0.98,0.9) | 0.25 |
| 0.954 | 0.954 | 0.9494 | 0.9542 | 0.9496 | 0.9542 | 0.9496 | 5000 | 4 | 0.1 | (0.9,0.8) | 0.25 |
| 0.954 | 0.9552 | 0.9496 | 0.9554 | 0.9494 | 0.9556 | 0.9494 | 5000 | 4 | 0.1 | (0.98,0.9) | 0.25 |

**WEB TABLE 31-** Coverage of 95% CIs for log-relative hazard parameter $\beta_2$ from stratified sampling using different methods of analysis and variance estimation, for various probabilities of missing covariate data, in 5,000 simulated cohorts. * indicates coverage outside the expected interval [0.9440; 0.9560].

| Cohort | USCC.True | | USCC.Est | | USCC.Naive | | $n$ | $K$ | $p_Y$ | $\pi^{(3)}$ | $\beta_2$ |
|---|---|---|---|---|---|---|---|---|---|---|---|
| | $\hat{V}_{Robust}$ | $\hat{V}$ | $\hat{V}_{Robust}$ | $\hat{V}$ | $\hat{V}_{Robust}$ | $\hat{V}$ | | | | | |
| 0.9464 | 0.9484 | 0.9486 | 0.9484 | 0.9486 | 0.9484 | 0.9486 | 5000 | 2 | 0.02 | (0.9,0.8) | 0.25 |
| 0.9464 | 0.95 | 0.9506 | 0.95 | 0.9506 | 0.95 | 0.9506 | 5000 | 2 | 0.02 | (0.98,0.9) | 0.25 |
| 0.9508 | 0.9518 | 0.9518 | 0.9518 | 0.9518 | 0.9518 | 0.9518 | 5000 | 4 | 0.02 | (0.9,0.8) | 0.25 |
| 0.9508 | 0.9534 | 0.9534 | 0.9534 | 0.9534 | 0.9534 | 0.9534 | 5000 | 4 | 0.02 | (0.98,0.9) | 0.25 |
| 0.9474 | 0.9512 | 0.9512 | 0.9512 | 0.9512 | 0.9512 | 0.9512 | 10000 | 2 | 0.02 | (0.9,0.8) | 0.25 |
| 0.9474 | 0.9486 | 0.9486 | 0.9486 | 0.9488 | 0.9486 | 0.9488 | 10000 | 2 | 0.02 | (0.98,0.9) | 0.25 |
| 0.9488 | 0.951 | 0.951 | 0.9508 | 0.951 | 0.9508 | 0.951 | 10000 | 4 | 0.02 | (0.9,0.8) | 0.25 |
| 0.9488 | 0.9532 | 0.9532 | 0.953 | 0.9532 | 0.953 | 0.9532 | 10000 | 4 | 0.02 | (0.98,0.9) | 0.25 |
| 0.9508 | 0.9514 | 0.9516 | 0.9518 | 0.952 | 0.9518 | 0.952 | 5000 | 2 | 0.05 | (0.9,0.8) | 0.25 |
| 0.9508 | 0.9538 | 0.954 | 0.9538 | 0.9538 | 0.9538 | 0.9538 | 5000 | 2 | 0.05 | (0.98,0.9) | 0.25 |
| 0.9438* | 0.9472 | 0.9472 | 0.947 | 0.947 | 0.947 | 0.947 | 5000 | 4 | 0.05 | (0.9,0.8) | 0.25 |
| 0.9438* | 0.945 | 0.9452 | 0.9454 | 0.9454 | 0.9454 | 0.9454 | 5000 | 4 | 0.05 | (0.98,0.9) | 0.25 |
| 0.95 | 0.95 | 0.95 | 0.95 | 0.95 | 0.95 | 0.95 | 10000 | 2 | 0.05 | (0.9,0.8) | 0.25 |
| 0.95 | 0.9482 | 0.9482 | 0.9482 | 0.9482 | 0.9482 | 0.9482 | 10000 | 2 | 0.05 | (0.98,0.9) | 0.25 |
| 0.9458 | 0.954 | 0.954 | 0.9538 | 0.9538 | 0.9538 | 0.9538 | 10000 | 4 | 0.05 | (0.9,0.8) | 0.25 |
| 0.9458 | 0.9522 | 0.9522 | 0.952 | 0.952 | 0.952 | 0.952 | 10000 | 4 | 0.05 | (0.98,0.9) | 0.25 |
| 0.9492 | 0.9488 | 0.949 | 0.9488 | 0.9488 | 0.9488 | 0.9488 | 5000 | 2 | 0.1 | (0.9,0.8) | 0.25 |
| 0.9492 | 0.9504 | 0.9504 | 0.9504 | 0.9504 | 0.9504 | 0.9504 | 5000 | 2 | 0.1 | (0.98,0.9) | 0.25 |
| 0.954 | 0.9494 | 0.9494 | 0.9492 | 0.9492 | 0.9492 | 0.9492 | 5000 | 4 | 0.1 | (0.9,0.8) | 0.25 |
| 0.954 | 0.95 | 0.95 | 0.9498 | 0.9498 | 0.9498 | 0.9498 | 5000 | 4 | 0.1 | (0.98,0.9) | 0.25 |

**WEB TABLE 32-** Coverage of 95% CIs for log-relative hazard parameter $\beta_2$ from unstratified sampling using different methods of analysis and variance estimation, for various probabilities of missing covariate data, in 5,000 simulated cohorts. * indicates coverage outside the expected interval [0.9440; 0.9560].

| Cohort | SCC.True | | SCC.Est | | SCC.Naive | | $n$ | $K$ | $p_Y$ | $\pi^{(3)}$ | $\beta_3$ |
|---|---|---|---|---|---|---|---|---|---|---|---|
| | $\hat{V}_{Robust}$ | $\hat{V}$ | $\hat{V}_{Robust}$ | $\hat{V}$ | $\hat{V}_{Robust}$ | $\hat{V}$ | | | | | |
| 0.9456 | 0.9524 | 0.9526 | 0.9524 | 0.9526 | 0.9524 | 0.9526 | 5000 | 2 | 0.02 | (0.9,0.8) | -0.3 |
| 0.9456 | 0.9538 | 0.9548 | 0.9534 | 0.9546 | 0.9534 | 0.9546 | 5000 | 2 | 0.02 | (0.98,0.9) | -0.3 |
| 0.9412* | 0.9492 | 0.9494 | 0.9492 | 0.9496 | 0.9492 | 0.9496 | 5000 | 4 | 0.02 | (0.9,0.8) | -0.3 |
| 0.9412* | 0.9462 | 0.9462 | 0.9462 | 0.9462 | 0.9462 | 0.9462 | 5000 | 4 | 0.02 | (0.98,0.9) | -0.3 |
| 0.9478 | 0.9484 | 0.949 | 0.9486 | 0.9492 | 0.9486 | 0.9492 | 10000 | 2 | 0.02 | (0.9,0.8) | -0.3 |
| 0.9478 | 0.9522 | 0.9524 | 0.9522 | 0.9522 | 0.9522 | 0.9522 | 10000 | 2 | 0.02 | (0.98,0.9) | -0.3 |
| 0.9448 | 0.9514 | 0.9514 | 0.951 | 0.951 | 0.951 | 0.951 | 10000 | 4 | 0.02 | (0.9,0.8) | -0.3 |
| 0.9448 | 0.949 | 0.949 | 0.949 | 0.949 | 0.949 | 0.949 | 10000 | 4 | 0.02 | (0.98,0.9) | -0.3 |
| 0.945 | 0.949 | 0.949 | 0.9488 | 0.949 | 0.9488 | 0.949 | 5000 | 2 | 0.05 | (0.9,0.8) | -0.3 |
| 0.945 | 0.944 | 0.9442 | 0.944 | 0.9442 | 0.944 | 0.9442 | 5000 | 2 | 0.05 | (0.98,0.9) | -0.3 |
| 0.952 | 0.9556 | 0.9556 | 0.956 | 0.956 | 0.956 | 0.956 | 5000 | 4 | 0.05 | (0.9,0.8) | -0.3 |
| 0.952 | 0.9558 | 0.9558 | 0.9558 | 0.9558 | 0.9558 | 0.9558 | 5000 | 4 | 0.05 | (0.98,0.9) | -0.3 |
| 0.9462 | 0.952 | 0.9518 | 0.9524 | 0.9524 | 0.9524 | 0.9524 | 10000 | 2 | 0.05 | (0.9,0.8) | -0.3 |
| 0.9462 | 0.9488 | 0.949 | 0.9488 | 0.9488 | 0.9488 | 0.9488 | 10000 | 2 | 0.05 | (0.98,0.9) | -0.3 |
| 0.9532 | 0.9514 | 0.9514 | 0.9514 | 0.9514 | 0.9514 | 0.9514 | 10000 | 4 | 0.05 | (0.9,0.8) | -0.3 |
| 0.9532 | 0.949 | 0.949 | 0.9486 | 0.9486 | 0.9486 | 0.9486 | 10000 | 4 | 0.05 | (0.98,0.9) | -0.3 |
| 0.9512 | 0.9494 | 0.9496 | 0.9494 | 0.9496 | 0.9496 | 0.9496 | 5000 | 2 | 0.1 | (0.9,0.8) | -0.3 |
| 0.9512 | 0.9476 | 0.9476 | 0.9476 | 0.9476 | 0.9476 | 0.9476 | 5000 | 2 | 0.1 | (0.98,0.9) | -0.3 |
| 0.9518 | 0.9506 | 0.9506 | 0.9502 | 0.9502 | 0.9502 | 0.9502 | 5000 | 4 | 0.1 | (0.9,0.8) | -0.3 |
| 0.9518 | 0.9512 | 0.9512 | 0.9514 | 0.9514 | 0.9514 | 0.9514 | 5000 | 4 | 0.1 | (0.98,0.9) | -0.3 |

**WEB TABLE 33-** Coverage of 95% CIs for log-relative hazard parameter $\beta_3$ from stratified sampling using different methods of analysis and variance estimation, for various probabilities of missing covariate data, in 5,000 simulated cohorts. * indicates coverage outside the expected interval [0.9440; 0.9560].

| Cohort | USCC.True | | USCC.Est | | USCC.Naive | | $n$ | $K$ | $p_Y$ | $\pi^{(3)}$ | $\beta_3$ |
|---|---|---|---|---|---|---|---|---|---|---|---|
| | $\hat{V}_{Robust}$ | $\hat{V}$ | $\hat{V}_{Robust}$ | $\hat{V}$ | $\hat{V}_{Robust}$ | $\hat{V}$ | | | | | |
| 0.9456 | 0.951 | 0.9512 | 0.951 | 0.9512 | 0.951 | 0.9512 | 5000 | 2 | 0.02 | (0.9,0.8) | -0.3 |
| 0.9456 | 0.9522 | 0.9522 | 0.9522 | 0.9522 | 0.9522 | 0.9522 | 5000 | 2 | 0.02 | (0.98,0.9) | -0.3 |
| 0.9412 | 0.9456 | 0.9458 | 0.9456 | 0.9456 | 0.9456 | 0.9456 | 5000 | 4 | 0.02 | (0.9,0.8) | -0.3 |
| 0.9412 | 0.9468 | 0.947 | 0.9468 | 0.947 | 0.9468 | 0.947 | 5000 | 4 | 0.02 | (0.98,0.9) | -0.3 |
| 0.9478 | 0.9516 | 0.9516 | 0.9514 | 0.9518 | 0.9514 | 0.9518 | 10000 | 2 | 0.02 | (0.9,0.8) | -0.3 |
| 0.9478 | 0.9496 | 0.9496 | 0.9496 | 0.9496 | 0.9496 | 0.9496 | 10000 | 2 | 0.02 | (0.98,0.9) | -0.3 |
| 0.9448 | 0.9488 | 0.9488 | 0.9488 | 0.9488 | 0.9488 | 0.9488 | 10000 | 4 | 0.02 | (0.9,0.8) | -0.3 |
| 0.9448 | 0.947 | 0.947 | 0.947 | 0.947 | 0.947 | 0.947 | 10000 | 4 | 0.02 | (0.98,0.9) | -0.3 |
| 0.945 | 0.947 | 0.947 | 0.9468 | 0.9468 | 0.9468 | 0.9468 | 5000 | 2 | 0.05 | (0.9,0.8) | -0.3 |
| 0.945 | 0.9508 | 0.9508 | 0.9508 | 0.9508 | 0.9508 | 0.9508 | 5000 | 2 | 0.05 | (0.98,0.9) | -0.3 |
| 0.952 | 0.949 | 0.949 | 0.9488 | 0.9488 | 0.9488 | 0.9488 | 5000 | 4 | 0.05 | (0.9,0.8) | -0.3 |
| 0.952 | 0.954 | 0.954 | 0.954 | 0.954 | 0.954 | 0.954 | 5000 | 4 | 0.05 | (0.98,0.9) | -0.3 |
| 0.9462 | 0.9482 | 0.9482 | 0.9482 | 0.9482 | 0.9482 | 0.9482 | 10000 | 2 | 0.05 | (0.9,0.8) | -0.3 |
| 0.9462 | 0.9474 | 0.9476 | 0.9472 | 0.9472 | 0.9472 | 0.9472 | 10000 | 2 | 0.05 | (0.98,0.9) | -0.3 |
| 0.9532 | 0.9546 | 0.9546 | 0.9544 | 0.9544 | 0.9544 | 0.9544 | 10000 | 4 | 0.05 | (0.9,0.8) | -0.3 |
| 0.9532 | 0.9574* | 0.9574* | 0.9574* | 0.9574* | 0.9574* | 0.9574* | 10000 | 4 | 0.05 | (0.98,0.9) | -0.3 |
| 0.9512 | 0.9558 | 0.9558 | 0.956 | 0.956 | 0.956 | 0.956 | 5000 | 2 | 0.1 | (0.9,0.8) | -0.3 |
| 0.9512 | 0.9554 | 0.9554 | 0.9556 | 0.9556 | 0.9556 | 0.9556 | 5000 | 2 | 0.1 | (0.98,0.9) | -0.3 |
| 0.9518 | 0.95 | 0.95 | 0.9496 | 0.9496 | 0.9496 | 0.9496 | 5000 | 4 | 0.1 | (0.9,0.8) | -0.3 |
| 0.9518 | 0.9498 | 0.9498 | 0.9496 | 0.9496 | 0.9496 | 0.9496 | 5000 | 4 | 0.1 | (0.98,0.9) | -0.3 |

**WEB TABLE 34-** Coverage of 95% CIs for log-relative hazard parameter $\beta_3$ from unstratified sampling using different methods of analysis and variance estimation, for various probabilities of missing covariate data, in 5,000 simulated cohorts. * indicates coverage outside the expected interval [0.9440; 0.9560].

| Cohort | SCC.True | | SCC.Est | | SCC.Naive | | $n$ | $K$ | $p_Y$ | $\pi^{(3)}$ | $\log\{\pi(\tau_1,\tau_2;x)\}$ |
| --- | --- | --- | --- | --- | --- | --- | --- | --- | --- | --- | --- |
| | $\hat{V}_{\text{Robust}}$ | $\hat{V}$ | $\hat{V}_{\text{Robust}}$ | $\hat{V}$ | $\hat{V}_{\text{Robust}}$ | $\hat{V}$ | | | | | |
| 0.949 | 0.9724* | 0.9548 | 0.9716* | 0.9554 | 0.973* | 0.9566 | 5000 | 2 | 0.02 | (0.9,0.8) | -3.948 |
| 0.949 | 0.9714* | 0.9524 | 0.9706* | 0.9522 | 0.9708* | 0.9528 | 5000 | 2 | 0.02 | (0.98,0.9) | -3.948 |
| 0.9494 | 0.9594* | 0.9486 | 0.96* | 0.9498 | 0.9608* | 0.9512 | 5000 | 4 | 0.02 | (0.9,0.8) | -3.948 |
| 0.9494 | 0.9598* | 0.9492 | 0.9596* | 0.949 | 0.9596* | 0.949 | 5000 | 4 | 0.02 | (0.98,0.9) | -3.948 |
| 0.949 | 0.9688* | 0.9508 | 0.9692* | 0.9508 | 0.9698* | 0.953 | 10000 | 2 | 0.02 | (0.9,0.8) | -3.948 |
| 0.949 | 0.969* | 0.9498 | 0.9684* | 0.9498 | 0.9688* | 0.9504 | 10000 | 2 | 0.02 | (0.98,0.9) | -3.948 |
| 0.9476 | 0.9572* | 0.9446 | 0.9564* | 0.9444 | 0.9582* | 0.946 | 10000 | 4 | 0.02 | (0.9,0.8) | -3.948 |
| 0.9476 | 0.9606* | 0.9474 | 0.962* | 0.9476 | 0.9622* | 0.9482 | 10000 | 4 | 0.02 | (0.98,0.9) | -3.948 |
| 0.9532 | 0.9688* | 0.9528 | 0.9696* | 0.9552 | 0.9716* | 0.9564* | 5000 | 2 | 0.05 | (0.9,0.8) | -3.046 |
| 0.9532 | 0.9712* | 0.955 | 0.9708* | 0.9556 | 0.9712* | 0.9558 | 5000 | 2 | 0.05 | (0.98,0.9) | -3.046 |
| 0.9438* | 0.9586* | 0.95 | 0.9578* | 0.9498 | 0.959* | 0.9508 | 5000 | 4 | 0.05 | (0.9,0.8) | -3.046 |
| 0.9438* | 0.9566* | 0.9472 | 0.9554 | 0.947 | 0.9558 | 0.947 | 5000 | 4 | 0.05 | (0.98,0.9) | -3.046 |
| 0.9496 | 0.9702* | 0.9544 | 0.9712* | 0.954 | 0.9722* | 0.955 | 10000 | 2 | 0.05 | (0.9,0.8) | -3.046 |
| 0.9496 | 0.9708* | 0.954 | 0.9704* | 0.955 | 0.9706* | 0.955 | 10000 | 2 | 0.05 | (0.98,0.9) | -3.046 |
| 0.9474 | 0.9578* | 0.9486 | 0.958* | 0.9488 | 0.9592* | 0.95 | 10000 | 4 | 0.05 | (0.9,0.8) | -3.046 |
| 0.9474 | 0.9562* | 0.9464 | 0.9566* | 0.947 | 0.957* | 0.948 | 10000 | 4 | 0.05 | (0.98,0.9) | -3.046 |
| 0.9482 | 0.9622* | 0.9484 | 0.9634* | 0.949 | 0.9642* | 0.9524 | 5000 | 2 | 0.1 | (0.9,0.8) | -2.377 |
| 0.9482 | 0.963* | 0.9494 | 0.964* | 0.9484 | 0.9642* | 0.9492 | 5000 | 2 | 0.1 | (0.98,0.9) | -2.377 |
| 0.9488 | 0.9522 | 0.9478 | 0.9522 | 0.9472 | 0.9536 | 0.9484 | 5000 | 4 | 0.1 | (0.9,0.8) | -2.377 |
| 0.9488 | 0.9526 | 0.9492 | 0.9542 | 0.9512 | 0.9542 | 0.9514 | 5000 | 4 | 0.1 | (0.98,0.9) | -2.377 |

**WEB TABLE 35-** Coverage of 95% CIs for pure risk parameter $\log\{\pi(\tau_1,\tau_2;x)\}$ with $x=(-1,1,-0.6)'$, from stratified sampling using different methods of analysis and variance estimation, for various probabilities of missing covariate data, in 5,000 simulated cohorts. * indicates coverage outside the expected interval [0.9440; 0.9560].

| Cohort | USCC.True | | USCC.Est | | USCC.Naive | | $n$ | $K$ | $p_Y$ | $\pi^{(3)}$ | $\log\{\pi(\tau_1,\tau_2;x)\}$ |
|---|---|---|---|---|---|---|---|---|---|---|---|
| | $\hat{V}_{\text{Robust}}$ | $\hat{V}$ | $\hat{V}_{\text{Robust}}$ | $\hat{V}$ | $\hat{V}_{\text{Robust}}$ | $\hat{V}$ | | | | | |
| 0.949 | 0.9638* | 0.952 | 0.966* | 0.9532 | 0.9668* | 0.9546 | 5000 | 2 | 0.02 | (0.9,0.8) | -3.948 |
| 0.949 | 0.9658* | 0.951 | 0.9652* | 0.953 | 0.9656* | 0.9536 | 5000 | 2 | 0.02 | (0.98,0.9) | -3.948 |
| 0.9494 | 0.959* | 0.951 | 0.9584* | 0.9522 | 0.9596* | 0.9534 | 5000 | 4 | 0.02 | (0.9,0.8) | -3.948 |
| 0.9494 | 0.959* | 0.9516 | 0.9592* | 0.952 | 0.9592* | 0.9522 | 5000 | 4 | 0.02 | (0.98,0.9) | -3.948 |
| 0.949 | 0.9624* | 0.9518 | 0.962* | 0.951 | 0.9634* | 0.9532 | 10000 | 2 | 0.02 | (0.9,0.8) | -3.948 |
| 0.949 | 0.9628* | 0.9506 | 0.9628* | 0.9524 | 0.9632* | 0.9528 | 10000 | 2 | 0.02 | (0.98,0.9) | -3.948 |
| 0.9476 | 0.9556 | 0.9486 | 0.9548 | 0.9492 | 0.9556 | 0.9504 | 10000 | 4 | 0.02 | (0.9,0.8) | -3.948 |
| 0.9476 | 0.956 | 0.9474 | 0.9554 | 0.9476 | 0.9556 | 0.9478 | 10000 | 4 | 0.02 | (0.98,0.9) | -3.948 |
| 0.9532 | 0.9668* | 0.9576* | 0.9672* | 0.9568* | 0.9688* | 0.9586* | 5000 | 2 | 0.05 | (0.9,0.8) | -3.046 |
| 0.9532 | 0.9688* | 0.9566* | 0.9682* | 0.957* | 0.9682* | 0.9574* | 5000 | 2 | 0.05 | (0.98,0.9) | -3.046 |
| 0.9438* | 0.947 | 0.9384* | 0.946 | 0.9394* | 0.9466 | 0.9404* | 5000 | 4 | 0.05 | (0.9,0.8) | -3.046 |
| 0.9438* | 0.947 | 0.9402* | 0.9462 | 0.9392* | 0.9462 | 0.9394* | 5000 | 4 | 0.05 | (0.98,0.9) | -3.046 |
| 0.9496 | 0.9634* | 0.9526 | 0.9628* | 0.9506 | 0.9638* | 0.9522 | 10000 | 2 | 0.05 | (0.9,0.8) | -3.046 |
| 0.9496 | 0.9636* | 0.9522 | 0.9644* | 0.9524 | 0.9644* | 0.9526 | 10000 | 2 | 0.05 | (0.98,0.9) | -3.046 |
| 0.9474 | 0.9568* | 0.9508 | 0.958* | 0.9504 | 0.9584* | 0.9514 | 10000 | 4 | 0.05 | (0.9,0.8) | -3.046 |
| 0.9474 | 0.956 | 0.952 | 0.9558 | 0.9516 | 0.9564* | 0.952 | 10000 | 4 | 0.05 | (0.98,0.9) | -3.046 |
| 0.9482 | 0.9572* | 0.9454 | 0.9584* | 0.9494 | 0.9598* | 0.9506 | 5000 | 2 | 0.1 | (0.9,0.8) | -2.377 |
| 0.9482 | 0.9582* | 0.9446 | 0.9588* | 0.9448 | 0.9588* | 0.9452 | 5000 | 2 | 0.1 | (0.98,0.9) | -2.377 |
| 0.9488 | 0.9484 | 0.9426 | 0.9498 | 0.9446 | 0.9498 | 0.9452 | 5000 | 4 | 0.1 | (0.9,0.8) | -2.377 |
| 0.9488 | 0.9528 | 0.9492 | 0.9532 | 0.9492 | 0.9532 | 0.9494 | 5000 | 4 | 0.1 | (0.98,0.9) | -2.377 |

**WEB TABLE 36-** Coverage of 95% CIs for pure risk parameter $\log\{\pi(\tau_1,\tau_2;x)\}$ with $x = (-1, 1, -0.6)'$, from unstratified sampling using different methods of analysis and variance estimation, for various probabilities of missing covariate data, in 5,000 simulated cohorts. * indicates coverage outside the expected interval [0.9440; 0.9560].

| Cohort | SCC.True | | SCC.Est | | SCC.Naive | | $n$ | $K$ | $p_Y$ | $\pi^{(3)}$ | $\log\{\pi(\tau_1,\tau_2;x)\}$ |
|---|---|---|---|---|---|---|---|---|---|---|---|
| | $\hat{V}_{\text{Robust}}$ | $\hat{V}$ | $\hat{V}_{\text{Robust}}$ | $\hat{V}$ | $\hat{V}_{\text{Robust}}$ | $\hat{V}$ | | | | | |
| 0.9484 | 0.967* | 0.9488 | 0.9664* | 0.9488 | 0.9664* | 0.949 | 5000 | 2 | 0.02 | (0.9,0.8) | -5.201 |
| 0.9484 | 0.971* | 0.9498 | 0.9716* | 0.9514 | 0.9716* | 0.9514 | 5000 | 2 | 0.02 | (0.98,0.9) | -5.201 |
| 0.9448 | 0.9606* | 0.9522 | 0.96* | 0.9518 | 0.96* | 0.9522 | 5000 | 4 | 0.02 | (0.9,0.8) | -5.201 |
| 0.9448 | 0.9608* | 0.95 | 0.9606* | 0.9506 | 0.9606* | 0.9506 | 5000 | 4 | 0.02 | (0.98,0.9) | -5.201 |
| 0.95 | 0.9662* | 0.9472 | 0.9664* | 0.9474 | 0.9668* | 0.9482 | 10000 | 2 | 0.02 | (0.9,0.8) | -5.201 |
| 0.95 | 0.9698* | 0.9492 | 0.9694* | 0.9492 | 0.9694* | 0.9494 | 10000 | 2 | 0.02 | (0.98,0.9) | -5.201 |
| 0.9494 | 0.9608* | 0.9502 | 0.9618* | 0.9508 | 0.9618* | 0.951 | 10000 | 4 | 0.02 | (0.9,0.8) | -5.201 |
| 0.9494 | 0.965* | 0.9538 | 0.965* | 0.9542 | 0.965* | 0.9542 | 10000 | 4 | 0.02 | (0.98,0.9) | -5.201 |
| 0.949 | 0.9694* | 0.9498 | 0.969* | 0.9496 | 0.969* | 0.9502 | 5000 | 2 | 0.05 | (0.9,0.8) | -4.289 |
| 0.949 | 0.9692* | 0.9502 | 0.969* | 0.9502 | 0.969* | 0.9504 | 5000 | 2 | 0.05 | (0.98,0.9) | -4.289 |
| 0.9432* | 0.9614* | 0.9516 | 0.9626* | 0.9514 | 0.9626* | 0.9514 | 5000 | 4 | 0.05 | (0.9,0.8) | -4.289 |
| 0.9432* | 0.957* | 0.9478 | 0.957* | 0.9476 | 0.957* | 0.9476 | 5000 | 4 | 0.05 | (0.98,0.9) | -4.289 |
| 0.944 | 0.9656* | 0.949 | 0.9662* | 0.9474 | 0.9662* | 0.9478 | 10000 | 2 | 0.05 | (0.9,0.8) | -4.289 |
| 0.944 | 0.97* | 0.9492 | 0.9708* | 0.9508 | 0.971* | 0.9508 | 10000 | 2 | 0.05 | (0.98,0.9) | -4.289 |
| 0.9486 | 0.9556 | 0.9498 | 0.9564* | 0.95 | 0.9568* | 0.9502 | 10000 | 4 | 0.05 | (0.9,0.8) | -4.289 |
| 0.9486 | 0.9588* | 0.9486 | 0.9596* | 0.949 | 0.9596* | 0.949 | 10000 | 4 | 0.05 | (0.98,0.9) | -4.289 |
| 0.9526 | 0.9652* | 0.9474 | 0.9646* | 0.9494 | 0.9652* | 0.9498 | 5000 | 2 | 0.1 | (0.9,0.8) | -3.602 |
| 0.9526 | 0.9664* | 0.9494 | 0.9666* | 0.95 | 0.9668* | 0.95 | 5000 | 2 | 0.1 | (0.98,0.9) | -3.602 |
| 0.9462 | 0.9562* | 0.9498 | 0.9572* | 0.9484 | 0.9574* | 0.9486 | 5000 | 4 | 0.1 | (0.9,0.8) | -3.602 |
| 0.9462 | 0.9538 | 0.9474 | 0.9538 | 0.948 | 0.9538 | 0.948 | 5000 | 4 | 0.1 | (0.98,0.9) | -3.602 |

**WEB TABLE 37-** Coverage of 95% CIs for pure risk parameter $\log\{\pi(\tau_1,\tau_2;x)\}$ with $x=(1,-1,0.6)'$, from stratified sampling using different methods of analysis and variance estimation, for various probabilities of missing covariate data, in 5,000 simulated cohorts. * indicates coverage outside the expected interval [0.9440; 0.9560].

| Cohort | USCC.True | | USCC.Est | | USCC.Naive | | $n$ | $K$ | $p_Y$ | $\pi^{(3)}$ | $\log\{\pi(\tau_1,\tau_2;x)\}$ |
|---|---|---|---|---|---|---|---|---|---|---|---|
| | $\hat{V}_{\text{Robust}}$ | $\hat{V}$ | $\hat{V}_{\text{Robust}}$ | $\hat{V}$ | $\hat{V}_{\text{Robust}}$ | $\hat{V}$ | | | | | |
| 0.9484 | 0.9578* | 0.9552 | 0.9584* | 0.9552 | 0.9584* | 0.9556 | 5000 | 2 | 0.02 | (0.9,0.8) | -5.201 |
| 0.9484 | 0.9554 | 0.9528 | 0.9556 | 0.9536 | 0.9556 | 0.9536 | 5000 | 2 | 0.02 | (0.98,0.9) | -5.201 |
| 0.9448 | 0.9478 | 0.9462 | 0.946 | 0.9452 | 0.9466 | 0.9454 | 5000 | 4 | 0.02 | (0.9,0.8) | -5.201 |
| 0.9448 | 0.948 | 0.9466 | 0.9482 | 0.9466 | 0.9482 | 0.9466 | 5000 | 4 | 0.02 | (0.98,0.9) | -5.201 |
| 0.95 | 0.956 | 0.951 | 0.956 | 0.952 | 0.9562* | 0.9522 | 10000 | 2 | 0.02 | (0.9,0.8) | -5.201 |
| 0.95 | 0.9564* | 0.953 | 0.9552 | 0.953 | 0.9554 | 0.953 | 10000 | 2 | 0.02 | (0.98,0.9) | -5.201 |
| 0.9494 | 0.9534 | 0.9518 | 0.9526 | 0.951 | 0.9532 | 0.9512 | 10000 | 4 | 0.02 | (0.9,0.8) | -5.201 |
| 0.9494 | 0.953 | 0.9516 | 0.953 | 0.9516 | 0.953 | 0.9516 | 10000 | 4 | 0.02 | (0.98,0.9) | -5.201 |
| 0.949 | 0.9524 | 0.9508 | 0.9518 | 0.9508 | 0.9518 | 0.951 | 5000 | 2 | 0.05 | (0.9,0.8) | -4.289 |
| 0.949 | 0.9544 | 0.952 | 0.9546 | 0.9518 | 0.9546 | 0.9518 | 5000 | 2 | 0.05 | (0.98,0.9) | -4.289 |
| 0.9432* | 0.9476 | 0.9462 | 0.9486 | 0.9474 | 0.9488 | 0.9476 | 5000 | 4 | 0.05 | (0.9,0.8) | -4.289 |
| 0.9432* | 0.9512 | 0.9498 | 0.9502 | 0.9488 | 0.9502 | 0.9488 | 5000 | 4 | 0.05 | (0.98,0.9) | -4.289 |
| 0.944 | 0.952 | 0.9504 | 0.9526 | 0.9504 | 0.9528 | 0.9508 | 10000 | 2 | 0.05 | (0.9,0.8) | -4.289 |
| 0.944 | 0.9534 | 0.95 | 0.954 | 0.951 | 0.954 | 0.951 | 10000 | 2 | 0.05 | (0.98,0.9) | -4.289 |
| 0.9486 | 0.9508 | 0.9496 | 0.9522 | 0.951 | 0.9524 | 0.9514 | 10000 | 4 | 0.05 | (0.9,0.8) | -4.289 |
| 0.9486 | 0.952 | 0.9504 | 0.9522 | 0.9504 | 0.9522 | 0.9504 | 10000 | 4 | 0.05 | (0.98,0.9) | -4.289 |
| 0.9526 | 0.9546 | 0.952 | 0.9546 | 0.9522 | 0.9548 | 0.9526 | 5000 | 2 | 0.1 | (0.9,0.8) | -3.602 |
| 0.9526 | 0.9586* | 0.9566* | 0.9596* | 0.9562* | 0.9596* | 0.9564* | 5000 | 2 | 0.1 | (0.98,0.9) | -3.602 |
| 0.9462 | 0.9504 | 0.9492 | 0.9514 | 0.9508 | 0.9518 | 0.9512 | 5000 | 4 | 0.1 | (0.9,0.8) | -3.602 |
| 0.9462 | 0.9512 | 0.9504 | 0.9514 | 0.95 | 0.9514 | 0.95 | 5000 | 4 | 0.1 | (0.98,0.9) | -3.602 |

**WEB TABLE 38-** Coverage of 95% CIs for pure risk parameter $\log\{\pi(\tau_1,\tau_2;x)\}$ with $x = (1,-1,0.6)'$, from unstratified sampling using different methods of analysis and variance estimation, for various probabilities of missing covariate data, in 5,000 simulated cohorts. * indicates coverage outside the expected interval [0.9440; 0.9560]

| Cohort | SCC.True | | SCC.Est | | SCC.Naive | | $n$ | $K$ | $p_Y$ | $\pi^{(3)}$ | $\log\{\pi(\tau_1,\tau_2;x)\}$ |
| --- | --- | --- | --- | --- | --- | --- | --- | --- | --- | --- | --- |
| | $\hat{V}_{\text{Robust}}$ | $\hat{V}$ | $\hat{V}_{\text{Robust}}$ | $\hat{V}$ | $\hat{V}_{\text{Robust}}$ | $\hat{V}$ | | | | | |
| 0.944 | 0.962 | 0.952 | 0.964 | 0.952 | 0.9644 | 0.9522 | 5000 | 2 | 0.02 | (0.9,0.8) | -4.702 |
| 0.944 | 0.9652 | 0.949 | 0.9652 | 0.9484 | 0.9658 | 0.949 | 5000 | 2 | 0.02 | (0.98,0.9) | -4.702 |
| 0.9426 | 0.955 | 0.9498 | 0.9546 | 0.9496 | 0.9548 | 0.9504 | 5000 | 4 | 0.02 | (0.9,0.8) | -4.702 |
| 0.9426 | 0.9542 | 0.9476 | 0.954 | 0.9468 | 0.954 | 0.947 | 5000 | 4 | 0.02 | (0.98,0.9) | -4.702 |
| 0.9458 | 0.9598 | 0.9498 | 0.9614 | 0.9486 | 0.9618 | 0.949 | 10000 | 2 | 0.02 | (0.9,0.8) | -4.702 |
| 0.9458 | 0.9678 | 0.9524 | 0.9666 | 0.951 | 0.9666 | 0.9514 | 10000 | 2 | 0.02 | (0.98,0.9) | -4.702 |
| 0.9514 | 0.9602 | 0.9528 | 0.96 | 0.953 | 0.9604 | 0.9534 | 10000 | 4 | 0.02 | (0.9,0.8) | -4.702 |
| 0.9514 | 0.9594 | 0.952 | 0.9602 | 0.9522 | 0.9602 | 0.9526 | 10000 | 4 | 0.02 | (0.98,0.9) | -4.702 |
| 0.9468 | 0.96 | 0.9456 | 0.9598 | 0.9462 | 0.9604 | 0.9476 | 5000 | 2 | 0.05 | (0.9,0.8) | -3.793 |
| 0.9468 | 0.9558 | 0.9452 | 0.9564 | 0.944 | 0.9564 | 0.944 | 5000 | 2 | 0.05 | (0.98,0.9) | -3.793 |
| 0.9526 | 0.9554 | 0.9512 | 0.9572 | 0.9522 | 0.9576 | 0.9528 | 5000 | 4 | 0.05 | (0.9,0.8) | -3.793 |
| 0.9526 | 0.9594 | 0.9536 | 0.9602 | 0.9534 | 0.9602 | 0.9534 | 5000 | 4 | 0.05 | (0.98,0.9) | -3.793 |
| 0.9464 | 0.9658 | 0.9546 | 0.9654 | 0.9538 | 0.9662 | 0.9546 | 10000 | 2 | 0.05 | (0.9,0.8) | -3.793 |
| 0.9464 | 0.9654 | 0.952 | 0.9648 | 0.9522 | 0.9648 | 0.9526 | 10000 | 2 | 0.05 | (0.98,0.9) | -3.793 |
| 0.9532 | 0.9602 | 0.9562 | 0.9592 | 0.9548 | 0.9594 | 0.955 | 10000 | 4 | 0.05 | (0.9,0.8) | -3.793 |
| 0.9532 | 0.9616 | 0.956 | 0.9616 | 0.954 | 0.9618 | 0.9542 | 10000 | 4 | 0.05 | (0.98,0.9) | -3.793 |
| 0.9532 | 0.9616 | 0.9542 | 0.9614 | 0.9532 | 0.9632 | 0.9542 | 5000 | 2 | 0.1 | (0.9,0.8) | -3.111 |
| 0.9532 | 0.9638 | 0.951 | 0.9638 | 0.9518 | 0.964 | 0.9524 | 5000 | 2 | 0.1 | (0.98,0.9) | -3.111 |
| 0.9532 | 0.9556 | 0.951 | 0.956 | 0.952 | 0.9564 | 0.9522 | 5000 | 4 | 0.1 | (0.9,0.8) | -3.111 |
| 0.9532 | 0.9556 | 0.9512 | 0.9546 | 0.9518 | 0.9546 | 0.9518 | 5000 | 4 | 0.1 | (0.98,0.9) | -3.111 |

**WEB TABLE 39-** Coverage of 95% CIs for pure risk parameter $\log\{\pi(\tau_1,\tau_2;x)\}$ with $x = (1,1,0.6)'$, from stratified sampling using different methods of analysis and variance estimation, for various probabilities of missing covariate data, in 5,000 simulated cohorts. * indicates coverage outside the expected interval [0.9440; 0.9560].

| Cohort | USCC.True | | USCC.Est | | USCC.Naive | | $n$ | $K$ | $p_Y$ | $\pi^{(3)}$ | $\log\{\pi(\tau_1,\tau_2;\boldsymbol{x})\}$ |
|---|---|---|---|---|---|---|---|---|---|---|---|
| | $\hat{V}_{\text{Robust}}$ | $\hat{V}$ | $\hat{V}_{\text{Robust}}$ | $\hat{V}$ | $\hat{V}_{\text{Robust}}$ | $\hat{V}$ | | | | | |
| 0.944 | 0.9604* | 0.9524 | 0.9578* | 0.9534 | 0.9584* | 0.9538 | 5000 | 2 | 0.02 | (0.9,0.8) | -4.702 |
| 0.944 | 0.959* | 0.954 | 0.9588* | 0.9536 | 0.9588* | 0.9536 | 5000 | 2 | 0.02 | (0.98,0.9) | -4.702 |
| 0.9426* | 0.95 | 0.947 | 0.9488 | 0.9466 | 0.9492 | 0.9468 | 5000 | 4 | 0.02 | (0.9,0.8) | -4.702 |
| 0.9426* | 0.9518 | 0.9474 | 0.9512 | 0.947 | 0.9512 | 0.947 | 5000 | 4 | 0.02 | (0.98,0.9) | -4.702 |
| 0.9458 | 0.9608* | 0.9554 | 0.9608* | 0.957* | 0.9616* | 0.958* | 10000 | 2 | 0.02 | (0.9,0.8) | -4.702 |
| 0.9458 | 0.9608* | 0.9548 | 0.9614* | 0.9544 | 0.9614* | 0.9546 | 10000 | 2 | 0.02 | (0.98,0.9) | -4.702 |
| 0.9514 | 0.956 | 0.9516 | 0.9542 | 0.9504 | 0.9542 | 0.951 | 10000 | 4 | 0.02 | (0.9,0.8) | -4.702 |
| 0.9514 | 0.9572* | 0.9528 | 0.9558 | 0.952 | 0.956 | 0.952 | 10000 | 4 | 0.02 | (0.98,0.9) | -4.702 |
| 0.9468 | 0.9526 | 0.948 | 0.9556 | 0.9494 | 0.9562* | 0.95 | 5000 | 2 | 0.05 | (0.9,0.8) | -3.793 |
| 0.9468 | 0.9536 | 0.9478 | 0.953 | 0.948 | 0.953 | 0.948 | 5000 | 2 | 0.05 | (0.98,0.9) | -3.793 |
| 0.9526 | 0.9538 | 0.9512 | 0.9554 | 0.9532 | 0.9554 | 0.954 | 5000 | 4 | 0.05 | (0.9,0.8) | -3.793 |
| 0.9526 | 0.9604* | 0.9586* | 0.9604* | 0.958 | 0.9604* | 0.958 | 5000 | 4 | 0.05 | (0.98,0.9) | -3.793 |
| 0.9464 | 0.9586* | 0.9522 | 0.9584* | 0.9532 | 0.9594* | 0.9536 | 10000 | 2 | 0.05 | (0.9,0.8) | -3.793 |
| 0.9464 | 0.9596* | 0.9538 | 0.9602* | 0.9542 | 0.9602* | 0.9542 | 10000 | 2 | 0.05 | (0.98,0.9) | -3.793 |
| 0.9532 | 0.9564* | 0.9546 | 0.958* | 0.9566* | 0.9586* | 0.9566* | 10000 | 4 | 0.05 | (0.9,0.8) | -3.793 |
| 0.9532 | 0.9564* | 0.9536 | 0.9572* | 0.953 | 0.9574* | 0.953 | 10000 | 4 | 0.05 | (0.98,0.9) | -3.793 |
| 0.9532 | 0.9566* | 0.9522 | 0.9578* | 0.9536 | 0.9586* | 0.9544 | 5000 | 2 | 0.1 | (0.9,0.8) | -3.111 |
| 0.9532 | 0.9562* | 0.9518 | 0.9558 | 0.9518 | 0.956 | 0.9518 | 5000 | 2 | 0.1 | (0.98,0.9) | -3.111 |
| 0.9532 | 0.9474 | 0.9454 | 0.9484 | 0.947 | 0.9486 | 0.9476 | 5000 | 4 | 0.1 | (0.9,0.8) | -3.111 |
| 0.9532 | 0.954 | 0.952 | 0.9532 | 0.9522 | 0.9532 | 0.9522 | 5000 | 4 | 0.1 | (0.98,0.9) | -3.111 |

**WEB TABLE 40-** Coverage of 95% CIs for pure risk parameter $\log\{\pi(\tau_1,\tau_2;\boldsymbol{x})\}$ with $\boldsymbol{x}=(1,1,0.6)'$, from unstratified sampling using different methods of analysis and variance estimation, for various probabilities of missing covariate data, in 5,000 simulated cohorts. * indicates coverage outside the expected interval [0.9440; 0.9560].

| Cohort | | SCC.True | | | SCC.Est | | | SCC.Naive | | | $n$ | $K$ | $p_Y$ | $\pi^{(3)}$ | $\beta_1$ |
|---|---|---|---|---|---|---|---|---|---|---|---|---|---|---|---|
| Empir var | $\hat{V}_{Robust}$ | Empir var | $\hat{V}_{Robust}$ | $\hat{V}$ | Empir var | $\hat{V}_{Robust}$ | $\hat{V}$ | Empir var | $\hat{V}_{Robust}$ | $\hat{V}$ | | | | | |
| 0.0135 | 0.014 | 0.0224 | 0.0261 | 0.0232 | 0.0224 | 0.0261 | 0.0232 | 0.0224 | 0.0261 | 0.0232 | 5000 | 2 | 0.02 | (0.9,0.8) | -0.2 |
| 0.0135 | 0.014 | 0.0195 | 0.0232 | 0.0202 | 0.0195 | 0.0232 | 0.0202 | 0.0195 | 0.0232 | 0.0202 | 5000 | 2 | 0.02 | (0.98,0.9) | -0.2 |
| 0.0141 | 0.0139 | 0.0207 | 0.0214 | 0.02 | 0.0207 | 0.0214 | 0.02 | 0.0207 | 0.0214 | 0.02 | 5000 | 4 | 0.02 | (0.9,0.8) | -0.2 |
| 0.0141 | 0.0139 | 0.0181 | 0.019 | 0.0176 | 0.0181 | 0.019 | 0.0176 | 0.0181 | 0.019 | 0.0176 | 5000 | 4 | 0.02 | (0.98,0.9) | -0.2 |
| 0.007 | 0.007 | 0.0111 | 0.0124 | 0.0109 | 0.0111 | 0.0124 | 0.0109 | 0.0111 | 0.0124 | 0.0109 | 10000 | 2 | 0.02 | (0.9,0.8) | -0.2 |
| 0.007 | 0.007 | 0.0096 | 0.0111 | 0.0096 | 0.0096 | 0.0111 | 0.0096 | 0.0096 | 0.0111 | 0.0096 | 10000 | 2 | 0.02 | (0.98,0.9) | -0.2 |
| 0.0071 | 0.0069 | 0.01 | 0.0104 | 0.0097 | 0.01 | 0.0104 | 0.0097 | 0.01 | 0.0104 | 0.0097 | 10000 | 4 | 0.02 | (0.9,0.8) | -0.2 |
| 0.0071 | 0.0069 | 0.0086 | 0.0093 | 0.0086 | 0.0086 | 0.0093 | 0.0086 | 0.0086 | 0.0093 | 0.0086 | 10000 | 4 | 0.02 | (0.98,0.9) | -0.2 |
| 0.0056 | 0.0056 | 0.0084 | 0.0095 | 0.0085 | 0.0084 | 0.0095 | 0.0085 | 0.0084 | 0.0095 | 0.0085 | 5000 | 2 | 0.05 | (0.9,0.8) | -0.2 |
| 0.0056 | 0.0056 | 0.0075 | 0.0085 | 0.0075 | 0.0075 | 0.0085 | 0.0075 | 0.0075 | 0.0085 | 0.0075 | 5000 | 2 | 0.05 | (0.98,0.9) | -0.2 |
| 0.0058 | 0.0056 | 0.0078 | 0.0081 | 0.0077 | 0.0078 | 0.0081 | 0.0077 | 0.0078 | 0.0081 | 0.0077 | 5000 | 4 | 0.05 | (0.9,0.8) | -0.2 |
| 0.0058 | 0.0056 | 0.0069 | 0.0072 | 0.0068 | 0.0069 | 0.0072 | 0.0068 | 0.0069 | 0.0072 | 0.0068 | 5000 | 4 | 0.05 | (0.98,0.9) | -0.2 |
| 0.0028 | 0.0028 | 0.0041 | 0.0047 | 0.0042 | 0.0041 | 0.0047 | 0.0042 | 0.0041 | 0.0047 | 0.0042 | 10000 | 2 | 0.05 | (0.9,0.8) | -0.2 |
| 0.0028 | 0.0028 | 0.0036 | 0.0042 | 0.0037 | 0.0036 | 0.0042 | 0.0037 | 0.0036 | 0.0042 | 0.0037 | 10000 | 2 | 0.05 | (0.98,0.9) | -0.2 |
| 0.0028 | 0.0028 | 0.0038 | 0.004 | 0.0038 | 0.0038 | 0.004 | 0.0038 | 0.0038 | 0.004 | 0.0038 | 10000 | 4 | 0.05 | (0.9,0.8) | -0.2 |
| 0.0028 | 0.0028 | 0.0034 | 0.0036 | 0.0034 | 0.0034 | 0.0036 | 0.0034 | 0.0034 | 0.0036 | 0.0034 | 10000 | 4 | 0.05 | (0.98,0.9) | -0.2 |
| 0.0029 | 0.0029 | 0.0042 | 0.0045 | 0.0041 | 0.0042 | 0.0045 | 0.0041 | 0.0042 | 0.0045 | 0.0041 | 5000 | 2 | 0.1 | (0.9,0.8) | -0.2 |
| 0.0029 | 0.0029 | 0.0037 | 0.0041 | 0.0037 | 0.0037 | 0.0041 | 0.0037 | 0.0037 | 0.0041 | 0.0037 | 5000 | 2 | 0.1 | (0.98,0.9) | -0.2 |
| 0.0029 | 0.0029 | 0.0038 | 0.0039 | 0.0038 | 0.0038 | 0.0039 | 0.0038 | 0.0038 | 0.0039 | 0.0038 | 5000 | 4 | 0.1 | (0.9,0.8) | -0.2 |
| 0.0029 | 0.0029 | 0.0033 | 0.0035 | 0.0033 | 0.0033 | 0.0035 | 0.0033 | 0.0033 | 0.0035 | 0.0033 | 5000 | 4 | 0.1 | (0.98,0.9) | -0.2 |

**WEB TABLE 41-** Empirical variance and mean of estimated variances of log-relative hazard parameter $\beta_1$ from stratified sampling using different methods of analysis and variance estimation, for various probabilities of missing covariate data, in 5,000 simulated cohorts.

| Cohort | | USCC.True | | | USCC.Est | | | USCC.Naive | | | $n$ | $K$ | $p_Y$ | $\pi^{(3)}$ | $\beta_1$ |
|---|---|---|---|---|---|---|---|---|---|---|---|---|---|---|---|
| Empir var | $\hat{V}_{\text{Robust}}$ | Empir var | $\hat{V}_{\text{Robust}}$ | $\hat{V}$ | Empir var | $\hat{V}_{\text{Robust}}$ | $\hat{V}$ | Empir var | $\hat{V}_{\text{Robust}}$ | $\hat{V}$ | | | | | |
| 0.0135 | 0.014 | 0.0264 | 0.0273 | 0.0273 | 0.0264 | 0.0273 | 0.0273 | 0.0264 | 0.0273 | 0.0273 | 5000 | 2 | 0.02 | (0.9,0.8) | -0.2 |
| 0.0135 | 0.014 | 0.0235 | 0.0242 | 0.0243 | 0.0235 | 0.0242 | 0.0243 | 0.0235 | 0.0242 | 0.0243 | 5000 | 2 | 0.02 | (0.98,0.9) | -0.2 |
| 0.0141 | 0.0139 | 0.0223 | 0.0221 | 0.0221 | 0.0223 | 0.0221 | 0.0221 | 0.0223 | 0.0221 | 0.0221 | 5000 | 4 | 0.02 | (0.9,0.8) | -0.2 |
| 0.0141 | 0.0139 | 0.0197 | 0.0196 | 0.0196 | 0.0197 | 0.0196 | 0.0196 | 0.0197 | 0.0196 | 0.0196 | 5000 | 4 | 0.02 | (0.98,0.9) | -0.2 |
| 0.007 | 0.007 | 0.0129 | 0.0128 | 0.0128 | 0.0129 | 0.0128 | 0.0128 | 0.0129 | 0.0128 | 0.0128 | 10000 | 2 | 0.02 | (0.9,0.8) | -0.2 |
| 0.007 | 0.007 | 0.0115 | 0.0115 | 0.0115 | 0.0115 | 0.0115 | 0.0115 | 0.0115 | 0.0115 | 0.0115 | 10000 | 2 | 0.02 | (0.98,0.9) | -0.2 |
| 0.0071 | 0.0069 | 0.0111 | 0.0106 | 0.0106 | 0.0111 | 0.0106 | 0.0106 | 0.0111 | 0.0106 | 0.0106 | 10000 | 4 | 0.02 | (0.9,0.8) | -0.2 |
| 0.0071 | 0.0069 | 0.0099 | 0.0095 | 0.0095 | 0.0099 | 0.0095 | 0.0095 | 0.0099 | 0.0095 | 0.0095 | 10000 | 4 | 0.02 | (0.98,0.9) | -0.2 |
| 0.0056 | 0.0056 | 0.0097 | 0.0098 | 0.0098 | 0.0097 | 0.0098 | 0.0098 | 0.0097 | 0.0098 | 0.0098 | 5000 | 2 | 0.05 | (0.9,0.8) | -0.2 |
| 0.0056 | 0.0056 | 0.0087 | 0.0088 | 0.0088 | 0.0087 | 0.0088 | 0.0088 | 0.0087 | 0.0088 | 0.0088 | 5000 | 2 | 0.05 | (0.98,0.9) | -0.2 |
| 0.0058 | 0.0056 | 0.0084 | 0.0083 | 0.0083 | 0.0084 | 0.0083 | 0.0083 | 0.0084 | 0.0083 | 0.0083 | 5000 | 4 | 0.05 | (0.9,0.8) | -0.2 |
| 0.0058 | 0.0056 | 0.0075 | 0.0074 | 0.0074 | 0.0075 | 0.0074 | 0.0074 | 0.0075 | 0.0074 | 0.0074 | 5000 | 4 | 0.05 | (0.98,0.9) | -0.2 |
| 0.0028 | 0.0028 | 0.0049 | 0.0049 | 0.0049 | 0.0049 | 0.0049 | 0.0049 | 0.0049 | 0.0049 | 0.0049 | 10000 | 2 | 0.05 | (0.9,0.8) | -0.2 |
| 0.0028 | 0.0028 | 0.0044 | 0.0044 | 0.0044 | 0.0044 | 0.0044 | 0.0044 | 0.0044 | 0.0044 | 0.0044 | 10000 | 2 | 0.05 | (0.98,0.9) | -0.2 |
| 0.0028 | 0.0028 | 0.004 | 0.0041 | 0.0041 | 0.004 | 0.0041 | 0.0041 | 0.004 | 0.0041 | 0.0041 | 10000 | 4 | 0.05 | (0.9,0.8) | -0.2 |
| 0.0028 | 0.0028 | 0.0036 | 0.0037 | 0.0037 | 0.0036 | 0.0037 | 0.0037 | 0.0036 | 0.0037 | 0.0037 | 10000 | 4 | 0.05 | (0.98,0.9) | -0.2 |
| 0.0029 | 0.0029 | 0.0047 | 0.0047 | 0.0047 | 0.0047 | 0.0047 | 0.0047 | 0.0047 | 0.0047 | 0.0047 | 5000 | 2 | 0.1 | (0.9,0.8) | -0.2 |
| 0.0029 | 0.0029 | 0.0041 | 0.0042 | 0.0042 | 0.0041 | 0.0042 | 0.0042 | 0.0041 | 0.0042 | 0.0042 | 5000 | 2 | 0.1 | (0.98,0.9) | -0.2 |
| 0.0029 | 0.0029 | 0.004 | 0.004 | 0.004 | 0.004 | 0.004 | 0.004 | 0.004 | 0.004 | 0.004 | 5000 | 4 | 0.1 | (0.9,0.8) | -0.2 |
| 0.0029 | 0.0029 | 0.0036 | 0.0035 | 0.0035 | 0.0036 | 0.0035 | 0.0035 | 0.0036 | 0.0035 | 0.0035 | 5000 | 4 | 0.1 | (0.98,0.9) | -0.2 |

**WEB TABLE 42-** Empirical variance and mean of estimated variances of log-relative hazard parameter $\beta_1$ from unstratified sampling using different methods of analysis and variance estimation, for various probabilities of missing covariate data, in 5,000 simulated cohorts.

| Cohort | | SCC.True | | | SCC.Est | | | SCC.Naive | | | $n$ | $K$ | $p_Y$ | $\pi^{(3)}$ | $\beta_2$ |
|---|---|---|---|---|---|---|---|---|---|---|---|---|---|---|---|
| Empir var | $\hat{V}_{Robust}$ | Empir var | $\hat{V}_{Robust}$ | $\hat{V}$ | Empir var | $\hat{V}_{Robust}$ | $\hat{V}$ | Empir var | $\hat{V}_{Robust}$ | $\hat{V}$ | | | | | |
| 0.021 | 0.0197 | 0.0321 | 0.0353 | 0.0303 | 0.0321 | 0.0353 | 0.0303 | 0.0321 | 0.0353 | 0.0303 | 5000 | 2 | 0.02 | (0.9,0.8) | 0.25 |
| 0.021 | 0.0197 | 0.0281 | 0.0314 | 0.0264 | 0.0281 | 0.0314 | 0.0264 | 0.0281 | 0.0314 | 0.0264 | 5000 | 2 | 0.02 | (0.98,0.9) | 0.25 |
| 0.02 | 0.0198 | 0.0274 | 0.0298 | 0.0275 | 0.0274 | 0.0298 | 0.0275 | 0.0274 | 0.0298 | 0.0275 | 5000 | 4 | 0.02 | (0.9,0.8) | 0.25 |
| 0.02 | 0.0198 | 0.0243 | 0.0265 | 0.0241 | 0.0243 | 0.0265 | 0.0241 | 0.0243 | 0.0265 | 0.0241 | 5000 | 4 | 0.02 | (0.98,0.9) | 0.25 |
| 0.0097 | 0.0098 | 0.0145 | 0.017 | 0.0145 | 0.0145 | 0.017 | 0.0145 | 0.0145 | 0.017 | 0.0145 | 10000 | 2 | 0.02 | (0.9,0.8) | 0.25 |
| 0.0097 | 0.0098 | 0.0126 | 0.0151 | 0.0127 | 0.0126 | 0.0151 | 0.0127 | 0.0126 | 0.0151 | 0.0127 | 10000 | 2 | 0.02 | (0.98,0.9) | 0.25 |
| 0.0099 | 0.0097 | 0.0135 | 0.0144 | 0.0133 | 0.0135 | 0.0144 | 0.0133 | 0.0135 | 0.0144 | 0.0133 | 10000 | 4 | 0.02 | (0.9,0.8) | 0.25 |
| 0.0099 | 0.0097 | 0.0119 | 0.0129 | 0.0117 | 0.0119 | 0.0129 | 0.0117 | 0.0119 | 0.0129 | 0.0117 | 10000 | 4 | 0.02 | (0.98,0.9) | 0.25 |
| 0.0079 | 0.0079 | 0.0116 | 0.0132 | 0.0114 | 0.0116 | 0.0132 | 0.0114 | 0.0116 | 0.0132 | 0.0114 | 5000 | 2 | 0.05 | (0.9,0.8) | 0.25 |
| 0.0079 | 0.0079 | 0.01 | 0.0118 | 0.01 | 0.01 | 0.0118 | 0.01 | 0.01 | 0.0118 | 0.01 | 5000 | 2 | 0.05 | (0.98,0.9) | 0.25 |
| 0.0083 | 0.0079 | 0.0109 | 0.0113 | 0.0106 | 0.0109 | 0.0113 | 0.0106 | 0.0109 | 0.0113 | 0.0106 | 5000 | 4 | 0.05 | (0.9,0.8) | 0.25 |
| 0.0083 | 0.0079 | 0.0096 | 0.0101 | 0.0093 | 0.0096 | 0.0101 | 0.0093 | 0.0096 | 0.0101 | 0.0093 | 5000 | 4 | 0.05 | (0.98,0.9) | 0.25 |
| 0.004 | 0.0039 | 0.0058 | 0.0065 | 0.0056 | 0.0058 | 0.0065 | 0.0056 | 0.0058 | 0.0065 | 0.0056 | 10000 | 2 | 0.05 | (0.9,0.8) | 0.25 |
| 0.004 | 0.0039 | 0.0051 | 0.0058 | 0.005 | 0.0051 | 0.0058 | 0.005 | 0.0051 | 0.0058 | 0.005 | 10000 | 2 | 0.05 | (0.98,0.9) | 0.25 |
| 0.004 | 0.0039 | 0.0053 | 0.0056 | 0.0052 | 0.0053 | 0.0056 | 0.0052 | 0.0053 | 0.0056 | 0.0052 | 10000 | 4 | 0.05 | (0.9,0.8) | 0.25 |
| 0.004 | 0.0039 | 0.0046 | 0.005 | 0.0046 | 0.0046 | 0.005 | 0.0046 | 0.0046 | 0.005 | 0.0046 | 10000 | 4 | 0.05 | (0.98,0.9) | 0.25 |
| 0.004 | 0.004 | 0.0056 | 0.0063 | 0.0056 | 0.0056 | 0.0063 | 0.0056 | 0.0056 | 0.0063 | 0.0056 | 5000 | 2 | 0.1 | (0.9,0.8) | 0.25 |
| 0.004 | 0.004 | 0.0049 | 0.0056 | 0.005 | 0.0049 | 0.0056 | 0.005 | 0.0049 | 0.0056 | 0.005 | 5000 | 2 | 0.1 | (0.98,0.9) | 0.25 |
| 0.004 | 0.004 | 0.0052 | 0.0054 | 0.0052 | 0.0052 | 0.0054 | 0.0052 | 0.0052 | 0.0054 | 0.0052 | 5000 | 4 | 0.1 | (0.9,0.8) | 0.25 |
| 0.004 | 0.004 | 0.0046 | 0.0049 | 0.0046 | 0.0046 | 0.0049 | 0.0046 | 0.0046 | 0.0049 | 0.0046 | 5000 | 4 | 0.1 | (0.98,0.9) | 0.25 |

**WEB TABLE 43-** Empirical variance and mean of estimated variances of log-relative hazard parameter $\beta_2$ from stratified sampling using different methods of analysis and variance estimation, for various probabilities of missing covariate data, in 5,000 simulated cohorts.

| Cohort | | USCC.True | | | USCC.Est | | | USCC.Naive | | | $n$ | $K$ | $p_Y$ | $\pi^{(3)}$ | $\beta_2$ |
|---|---|---|---|---|---|---|---|---|---|---|---|---|---|---|---|
| Empir var | $\hat{V}_{Robust}$ | Empir var | $\hat{V}_{Robust}$ | $\hat{V}$ | Empir var | $\hat{V}_{Robust}$ | $\hat{V}$ | Empir var | $\hat{V}_{Robust}$ | $\hat{V}$ | | | | | |
| 0.021 | 0.0197 | 0.0368 | 0.0357 | 0.0357 | 0.0368 | 0.0357 | 0.0357 | 0.0368 | 0.0357 | 0.0357 | 5000 | 2 | 0.02 | (0.9,0.8) | 0.25 |
| 0.021 | 0.0197 | 0.0332 | 0.0318 | 0.0318 | 0.0332 | 0.0318 | 0.0318 | 0.0332 | 0.0318 | 0.0318 | 5000 | 2 | 0.02 | (0.98,0.9) | 0.25 |
| 0.02 | 0.0198 | 0.0301 | 0.03 | 0.03 | 0.0301 | 0.03 | 0.03 | 0.0301 | 0.03 | 0.03 | 5000 | 4 | 0.02 | (0.9,0.8) | 0.25 |
| 0.02 | 0.0198 | 0.0268 | 0.0267 | 0.0267 | 0.0268 | 0.0267 | 0.0267 | 0.0268 | 0.0267 | 0.0267 | 5000 | 4 | 0.02 | (0.98,0.9) | 0.25 |
| 0.0097 | 0.0098 | 0.017 | 0.0171 | 0.0171 | 0.017 | 0.0171 | 0.0171 | 0.017 | 0.0171 | 0.0171 | 10000 | 2 | 0.02 | (0.9,0.8) | 0.25 |
| 0.0097 | 0.0098 | 0.0152 | 0.0153 | 0.0153 | 0.0152 | 0.0153 | 0.0153 | 0.0152 | 0.0153 | 0.0153 | 10000 | 2 | 0.02 | (0.98,0.9) | 0.25 |
| 0.0099 | 0.0097 | 0.0144 | 0.0145 | 0.0145 | 0.0144 | 0.0145 | 0.0145 | 0.0144 | 0.0145 | 0.0145 | 10000 | 4 | 0.02 | (0.9,0.8) | 0.25 |
| 0.0099 | 0.0097 | 0.0129 | 0.0129 | 0.0129 | 0.0129 | 0.0129 | 0.0129 | 0.0129 | 0.0129 | 0.0129 | 10000 | 4 | 0.02 | (0.98,0.9) | 0.25 |
| 0.0079 | 0.0079 | 0.0131 | 0.0132 | 0.0133 | 0.0131 | 0.0132 | 0.0133 | 0.0131 | 0.0132 | 0.0133 | 5000 | 2 | 0.05 | (0.9,0.8) | 0.25 |
| 0.0079 | 0.0079 | 0.0119 | 0.0118 | 0.0119 | 0.0119 | 0.0118 | 0.0119 | 0.0119 | 0.0118 | 0.0119 | 5000 | 2 | 0.05 | (0.98,0.9) | 0.25 |
| 0.0083 | 0.0079 | 0.0117 | 0.0113 | 0.0113 | 0.0117 | 0.0113 | 0.0113 | 0.0117 | 0.0113 | 0.0113 | 5000 | 4 | 0.05 | (0.9,0.8) | 0.25 |
| 0.0083 | 0.0079 | 0.0104 | 0.0101 | 0.0101 | 0.0104 | 0.0101 | 0.0101 | 0.0104 | 0.0101 | 0.0101 | 5000 | 4 | 0.05 | (0.98,0.9) | 0.25 |
| 0.004 | 0.0039 | 0.0066 | 0.0065 | 0.0065 | 0.0066 | 0.0065 | 0.0065 | 0.0066 | 0.0065 | 0.0065 | 10000 | 2 | 0.05 | (0.9,0.8) | 0.25 |
| 0.004 | 0.0039 | 0.0059 | 0.0058 | 0.0058 | 0.0059 | 0.0058 | 0.0058 | 0.0059 | 0.0058 | 0.0058 | 10000 | 2 | 0.05 | (0.98,0.9) | 0.25 |
| 0.004 | 0.0039 | 0.0055 | 0.0056 | 0.0056 | 0.0055 | 0.0056 | 0.0056 | 0.0055 | 0.0056 | 0.0056 | 10000 | 4 | 0.05 | (0.9,0.8) | 0.25 |
| 0.004 | 0.0039 | 0.005 | 0.005 | 0.005 | 0.005 | 0.005 | 0.005 | 0.005 | 0.005 | 0.005 | 10000 | 4 | 0.05 | (0.98,0.9) | 0.25 |
| 0.004 | 0.004 | 0.0063 | 0.0063 | 0.0063 | 0.0063 | 0.0063 | 0.0063 | 0.0063 | 0.0063 | 0.0063 | 5000 | 2 | 0.1 | (0.9,0.8) | 0.25 |
| 0.004 | 0.004 | 0.0055 | 0.0057 | 0.0057 | 0.0055 | 0.0057 | 0.0057 | 0.0055 | 0.0057 | 0.0057 | 5000 | 2 | 0.1 | (0.98,0.9) | 0.25 |
| 0.004 | 0.004 | 0.0054 | 0.0055 | 0.0055 | 0.0054 | 0.0055 | 0.0055 | 0.0054 | 0.0055 | 0.0055 | 5000 | 4 | 0.1 | (0.9,0.8) | 0.25 |
| 0.004 | 0.004 | 0.0049 | 0.0049 | 0.0049 | 0.0049 | 0.0049 | 0.0049 | 0.0049 | 0.0049 | 0.0049 | 5000 | 4 | 0.1 | (0.98,0.9) | 0.25 |

**WEB TABLE 44-** Empirical variance and mean of estimated variances of log-relative hazard parameter $\beta_2$ from unstratified sampling using different methods of analysis and variance estimation, for various probabilities of missing covariate data, in 5,000 simulated cohorts.

| Cohort | | SCC.True | | | SCC.Est | | | SCC.Naive | | | $n$ | $K$ | $p_Y$ | $\pi^{(3)}$ | $\beta_3$ |
|---|---|---|---|---|---|---|---|---|---|---|---|---|---|---|---|
| Empir var | $\hat{V}_{\text{Robust}}$ | Empir var | $\hat{V}_{\text{Robust}}$ | $\hat{V}$ | Empir var | $\hat{V}_{\text{Robust}}$ | $\hat{V}$ | Empir var | $\hat{V}_{\text{Robust}}$ | $\hat{V}$ | | | | | |
| 0.0137 | 0.0136 | 0.027 | 0.0262 | 0.0263 | 0.027 | 0.0262 | 0.0263 | 0.027 | 0.0262 | 0.0263 | 5000 | 2 | 0.02 | (0.9,0.8) | -0.3 |
| 0.0137 | 0.0136 | 0.0236 | 0.0233 | 0.0234 | 0.0236 | 0.0233 | 0.0234 | 0.0236 | 0.0233 | 0.0234 | 5000 | 2 | 0.02 | (0.98,0.9) | -0.3 |
| 0.0141 | 0.0136 | 0.0218 | 0.0214 | 0.0214 | 0.0218 | 0.0214 | 0.0214 | 0.0218 | 0.0214 | 0.0214 | 5000 | 4 | 0.02 | (0.9,0.8) | -0.3 |
| 0.0141 | 0.0136 | 0.0194 | 0.019 | 0.019 | 0.0194 | 0.019 | 0.019 | 0.0194 | 0.019 | 0.019 | 5000 | 4 | 0.02 | (0.98,0.9) | -0.3 |
| 0.0069 | 0.0068 | 0.0126 | 0.0124 | 0.0125 | 0.0126 | 0.0124 | 0.0125 | 0.0126 | 0.0124 | 0.0125 | 10000 | 2 | 0.02 | (0.9,0.8) | -0.3 |
| 0.0069 | 0.0068 | 0.011 | 0.0111 | 0.0111 | 0.011 | 0.0111 | 0.0111 | 0.011 | 0.0111 | 0.0111 | 10000 | 2 | 0.02 | (0.98,0.9) | -0.3 |
| 0.0068 | 0.0068 | 0.0104 | 0.0103 | 0.0103 | 0.0104 | 0.0103 | 0.0103 | 0.0104 | 0.0103 | 0.0103 | 10000 | 4 | 0.02 | (0.9,0.8) | -0.3 |
| 0.0068 | 0.0068 | 0.0094 | 0.0092 | 0.0092 | 0.0094 | 0.0092 | 0.0092 | 0.0094 | 0.0092 | 0.0092 | 10000 | 4 | 0.02 | (0.98,0.9) | -0.3 |
| 0.0057 | 0.0055 | 0.01 | 0.0096 | 0.0096 | 0.01 | 0.0096 | 0.0096 | 0.01 | 0.0096 | 0.0096 | 5000 | 2 | 0.05 | (0.9,0.8) | -0.3 |
| 0.0057 | 0.0055 | 0.009 | 0.0086 | 0.0086 | 0.009 | 0.0086 | 0.0086 | 0.009 | 0.0086 | 0.0086 | 5000 | 2 | 0.05 | (0.98,0.9) | -0.3 |
| 0.0054 | 0.0056 | 0.0078 | 0.0081 | 0.0081 | 0.0078 | 0.0081 | 0.0081 | 0.0078 | 0.0081 | 0.0081 | 5000 | 4 | 0.05 | (0.9,0.8) | -0.3 |
| 0.0054 | 0.0056 | 0.007 | 0.0072 | 0.0072 | 0.007 | 0.0072 | 0.0072 | 0.007 | 0.0072 | 0.0072 | 5000 | 4 | 0.05 | (0.98,0.9) | -0.3 |
| 0.0028 | 0.0028 | 0.0046 | 0.0047 | 0.0047 | 0.0046 | 0.0047 | 0.0047 | 0.0046 | 0.0047 | 0.0047 | 10000 | 2 | 0.05 | (0.9,0.8) | -0.3 |
| 0.0028 | 0.0042 | 0.0042 | 0.0042 | 0.0042 | 0.0042 | 0.0042 | 0.0042 | 0.0042 | 0.0042 | 0.0042 | 10000 | 2 | 0.05 | (0.98,0.9) | -0.3 |
| 0.0028 | 0.0039 | 0.0039 | 0.004 | 0.004 | 0.0039 | 0.004 | 0.004 | 0.0039 | 0.004 | 0.004 | 10000 | 4 | 0.05 | (0.9,0.8) | -0.3 |
| 0.0028 | 0.0035 | 0.0035 | 0.0036 | 0.0036 | 0.0035 | 0.0036 | 0.0036 | 0.0035 | 0.0036 | 0.0036 | 10000 | 4 | 0.05 | (0.98,0.9) | -0.3 |
| 0.0028 | 0.0044 | 0.0044 | 0.0045 | 0.0045 | 0.0044 | 0.0045 | 0.0045 | 0.0044 | 0.0045 | 0.0045 | 5000 | 2 | 0.1 | (0.9,0.8) | -0.3 |
| 0.0028 | 0.004 | 0.004 | 0.004 | 0.004 | 0.004 | 0.004 | 0.004 | 0.004 | 0.004 | 0.004 | 5000 | 2 | 0.1 | (0.98,0.9) | -0.3 |
| 0.0028 | 0.0038 | 0.0038 | 0.0038 | 0.0038 | 0.0038 | 0.0038 | 0.0038 | 0.0038 | 0.0038 | 0.0038 | 5000 | 4 | 0.1 | (0.9,0.8) | -0.3 |
| 0.0028 | 0.0034 | 0.0034 | 0.0034 | 0.0034 | 0.0034 | 0.0034 | 0.0034 | 0.0034 | 0.0034 | 0.0034 | 5000 | 4 | 0.1 | (0.98,0.9) | -0.3 |

**WEB TABLE 45-** Empirical variance and mean of estimated variances of log-relative hazard parameter $\beta_3$ from stratified sampling using different methods of analysis and variance estimation, for various probabilities of missing covariate data, in 5,000 simulated cohorts.

| Cohort | | USCC.True | | | USCC.Est | | | USCC.Naive | | | $n$ | $K$ | $p_Y$ | $\pi^{(3)}$ | $\beta_3$ |
|---|---|---|---|---|---|---|---|---|---|---|---|---|---|---|---|
| Empir var | $\hat{V}_{\text{Robust}}$ | Empir var | $\hat{V}_{\text{Robust}}$ | $\hat{V}$ | Empir var | $\hat{V}_{\text{Robust}}$ | $\hat{V}$ | Empir var | $\hat{V}_{\text{Robust}}$ | $\hat{V}$ | | | | | |
| 0.0137 | 0.0136 | 0.028 | 0.0274 | 0.0274 | 0.028 | 0.0274 | 0.0274 | 0.028 | 0.0274 | 0.0274 | 5000 | 2 | 0.02 | (0.9,0.8) | -0.3 |
| 0.0137 | 0.0136 | 0.0249 | 0.0244 | 0.0244 | 0.0249 | 0.0244 | 0.0244 | 0.0249 | 0.0244 | 0.0244 | 5000 | 2 | 0.02 | (0.98,0.9) | -0.3 |
| 0.0141 | 0.0136 | 0.0232 | 0.0219 | 0.0219 | 0.0232 | 0.0219 | 0.0219 | 0.0232 | 0.0219 | 0.0219 | 5000 | 4 | 0.02 | (0.9,0.8) | -0.3 |
| 0.0141 | 0.0136 | 0.0205 | 0.0195 | 0.0195 | 0.0205 | 0.0195 | 0.0195 | 0.0205 | 0.0195 | 0.0195 | 5000 | 4 | 0.02 | (0.98,0.9) | -0.3 |
| 0.0069 | 0.0068 | 0.0131 | 0.0129 | 0.0129 | 0.0131 | 0.0129 | 0.0129 | 0.0131 | 0.0129 | 0.0129 | 10000 | 2 | 0.02 | (0.9,0.8) | -0.3 |
| 0.0069 | 0.0068 | 0.0119 | 0.0115 | 0.0115 | 0.0119 | 0.0115 | 0.0115 | 0.0119 | 0.0115 | 0.0115 | 10000 | 2 | 0.02 | (0.98,0.9) | -0.3 |
| 0.0068 | 0.0068 | 0.0108 | 0.0105 | 0.0105 | 0.0108 | 0.0105 | 0.0105 | 0.0108 | 0.0105 | 0.0105 | 10000 | 4 | 0.02 | (0.9,0.8) | -0.3 |
| 0.0068 | 0.0068 | 0.0096 | 0.0094 | 0.0094 | 0.0096 | 0.0094 | 0.0094 | 0.0096 | 0.0094 | 0.0094 | 10000 | 4 | 0.02 | (0.98,0.9) | -0.3 |
| 0.0057 | 0.0055 | 0.0101 | 0.0099 | 0.0099 | 0.0101 | 0.0099 | 0.0099 | 0.0101 | 0.0099 | 0.0099 | 5000 | 2 | 0.05 | (0.9,0.8) | -0.3 |
| 0.0057 | 0.0055 | 0.0089 | 0.0088 | 0.0089 | 0.0089 | 0.0088 | 0.0089 | 0.0089 | 0.0088 | 0.0089 | 5000 | 2 | 0.05 | (0.98,0.9) | -0.3 |
| 0.0054 | 0.0056 | 0.008 | 0.0082 | 0.0082 | 0.008 | 0.0082 | 0.0082 | 0.008 | 0.0082 | 0.0082 | 5000 | 4 | 0.05 | (0.9,0.8) | -0.3 |
| 0.0054 | 0.0056 | 0.0071 | 0.0073 | 0.0074 | 0.0071 | 0.0073 | 0.0074 | 0.0071 | 0.0073 | 0.0074 | 5000 | 4 | 0.05 | (0.98,0.9) | -0.3 |
| 0.0028 | 0.0028 | 0.0049 | 0.0049 | 0.0049 | 0.0049 | 0.0049 | 0.0049 | 0.0049 | 0.0049 | 0.0049 | 10000 | 2 | 0.05 | (0.9,0.8) | -0.3 |
| 0.0028 | 0.0042 | 0.0045 | 0.0044 | 0.0044 | 0.0045 | 0.0044 | 0.0044 | 0.0045 | 0.0044 | 0.0044 | 10000 | 2 | 0.05 | (0.98,0.9) | -0.3 |
| 0.0028 | 0.0039 | 0.004 | 0.0041 | 0.0041 | 0.004 | 0.0041 | 0.0041 | 0.004 | 0.0041 | 0.0041 | 10000 | 4 | 0.05 | (0.9,0.8) | -0.3 |
| 0.0028 | 0.0035 | 0.0035 | 0.0036 | 0.0036 | 0.0035 | 0.0036 | 0.0036 | 0.0035 | 0.0036 | 0.0036 | 10000 | 4 | 0.05 | (0.98,0.9) | -0.3 |
| 0.0028 | 0.0044 | 0.0044 | 0.0046 | 0.0046 | 0.0044 | 0.0046 | 0.0046 | 0.0044 | 0.0046 | 0.0046 | 5000 | 2 | 0.1 | (0.9,0.8) | -0.3 |
| 0.0028 | 0.004 | 0.0039 | 0.0042 | 0.0042 | 0.0039 | 0.0042 | 0.0042 | 0.0039 | 0.0042 | 0.0042 | 5000 | 2 | 0.1 | (0.98,0.9) | -0.3 |
| 0.0028 | 0.0038 | 0.0039 | 0.0039 | 0.0039 | 0.0039 | 0.0039 | 0.0039 | 0.0039 | 0.0039 | 0.0039 | 5000 | 4 | 0.1 | (0.9,0.8) | -0.3 |
| 0.0028 | 0.0034 | 0.0034 | 0.0035 | 0.0035 | 0.0034 | 0.0035 | 0.0035 | 0.0034 | 0.0035 | 0.0035 | 5000 | 4 | 0.1 | (0.98,0.9) | -0.3 |

**WEB TABLE 46-** Empirical variance and mean of estimated variances of log-relative hazard parameter $\beta_3$ from unstratified sampling using different methods of analysis and variance estimation, for various probabilities of missing covariate data, in 5,000 simulated cohorts.

| Cohort | | SCC.True | | | SCC.Est | | | SCC.Naive | | | $n$ | $K$ | $p_Y$ | $\pi^{(3)}$ | $\log\{\pi(\tau_1,\tau_2;x)\}$ |
|---|---|---|---|---|---|---|---|---|---|---|---|---|---|---|---|
| Empir var | $\widehat{V}_{\text{Robust}}$ | Empir var | $\widehat{V}_{\text{Robust}}$ | $\widehat{V}$ | Empir var | $\widehat{V}_{\text{Robust}}$ | $\widehat{V}$ | Empir var | $\widehat{V}_{\text{Robust}}$ | $\widehat{V}$ | | | | | |
| 0.0256 | 0.025 | 0.0337 | 0.0398 | 0.0334 | 0.0332 | 0.0393 | 0.0329 | 0.0332 | 0.0398 | 0.0334 | 5000 | 2 | 0.02 | (0.9,0.8) | -3.948 |
| 0.0256 | 0.025 | 0.0312 | 0.0371 | 0.0307 | 0.0311 | 0.037 | 0.0306 | 0.0311 | 0.0371 | 0.0307 | 5000 | 2 | 0.02 | (0.98,0.9) | -3.948 |
| 0.0255 | 0.0248 | 0.031 | 0.0332 | 0.0301 | 0.0307 | 0.0329 | 0.0299 | 0.0307 | 0.0332 | 0.0301 | 5000 | 4 | 0.02 | (0.9,0.8) | -3.948 |
| 0.0255 | 0.0248 | 0.0289 | 0.0311 | 0.0281 | 0.0288 | 0.0311 | 0.028 | 0.0288 | 0.0311 | 0.0281 | 5000 | 4 | 0.02 | (0.98,0.9) | -3.948 |
| 0.0123 | 0.0123 | 0.0162 | 0.0191 | 0.0159 | 0.0159 | 0.0188 | 0.0156 | 0.0159 | 0.0191 | 0.0159 | 10000 | 2 | 0.02 | (0.9,0.8) | -3.948 |
| 0.0123 | 0.0123 | 0.015 | 0.0179 | 0.0147 | 0.0149 | 0.0179 | 0.0147 | 0.0149 | 0.0179 | 0.0147 | 10000 | 2 | 0.02 | (0.98,0.9) | -3.948 |
| 0.0127 | 0.0122 | 0.0154 | 0.0161 | 0.0146 | 0.0152 | 0.016 | 0.0145 | 0.0152 | 0.0161 | 0.0146 | 10000 | 4 | 0.02 | (0.9,0.8) | -3.948 |
| 0.0127 | 0.0122 | 0.0142 | 0.0152 | 0.0137 | 0.0141 | 0.0152 | 0.0137 | 0.0141 | 0.0152 | 0.0137 | 10000 | 4 | 0.02 | (0.98,0.9) | -3.948 |
| 0.0095 | 0.0096 | 0.012 | 0.0144 | 0.0122 | 0.0118 | 0.0142 | 0.012 | 0.0118 | 0.0144 | 0.0122 | 5000 | 2 | 0.05 | (0.9,0.8) | -3.046 |
| 0.0095 | 0.0096 | 0.0111 | 0.0135 | 0.0114 | 0.0111 | 0.0134 | 0.0113 | 0.0111 | 0.0135 | 0.0114 | 5000 | 2 | 0.05 | (0.98,0.9) | -3.046 |
| 0.0103 | 0.0097 | 0.0117 | 0.0122 | 0.0113 | 0.0116 | 0.0121 | 0.0112 | 0.0116 | 0.0122 | 0.0113 | 5000 | 4 | 0.05 | (0.9,0.8) | -3.046 |
| 0.0103 | 0.0097 | 0.0111 | 0.0115 | 0.0106 | 0.0111 | 0.0115 | 0.0106 | 0.0111 | 0.0115 | 0.0106 | 5000 | 4 | 0.05 | (0.98,0.9) | -3.046 |
| 0.0047 | 0.0048 | 0.0059 | 0.0071 | 0.006 | 0.0058 | 0.007 | 0.0059 | 0.0058 | 0.0071 | 0.006 | 10000 | 2 | 0.05 | (0.9,0.8) | -3.046 |
| 0.0047 | 0.0048 | 0.0055 | 0.0067 | 0.0056 | 0.0055 | 0.0067 | 0.0056 | 0.0055 | 0.0067 | 0.0056 | 10000 | 2 | 0.05 | (0.98,0.9) | -3.046 |
| 0.0049 | 0.0048 | 0.0056 | 0.006 | 0.0056 | 0.0056 | 0.006 | 0.0055 | 0.0056 | 0.006 | 0.0056 | 10000 | 4 | 0.05 | (0.9,0.8) | -3.046 |
| 0.0049 | 0.0048 | 0.0053 | 0.0057 | 0.0053 | 0.0053 | 0.0057 | 0.0052 | 0.0053 | 0.0057 | 0.0053 | 10000 | 4 | 0.05 | (0.98,0.9) | -3.046 |
| 0.0048 | 0.0047 | 0.0059 | 0.0065 | 0.0058 | 0.0058 | 0.0065 | 0.0057 | 0.0058 | 0.0065 | 0.0058 | 5000 | 2 | 0.1 | (0.9,0.8) | -2.377 |
| 0.0048 | 0.0047 | 0.0055 | 0.0062 | 0.0054 | 0.0055 | 0.0061 | 0.0054 | 0.0055 | 0.0062 | 0.0054 | 5000 | 2 | 0.1 | (0.98,0.9) | -2.377 |
| 0.0048 | 0.0047 | 0.0055 | 0.0056 | 0.0054 | 0.0055 | 0.0056 | 0.0053 | 0.0055 | 0.0056 | 0.0054 | 5000 | 4 | 0.1 | (0.9,0.8) | -2.377 |
| 0.0048 | 0.0047 | 0.0051 | 0.0053 | 0.0051 | 0.0051 | 0.0053 | 0.0051 | 0.0051 | 0.0053 | 0.0051 | 5000 | 4 | 0.1 | (0.98,0.9) | -2.377 |

**WEB TABLE 47-** Empirical variance and mean of estimated variances of pure risk parameter $\log\{\pi(\tau_1,\tau_2;x)\}$ with $x=(-1,1,-0.6)'$, from stratified sampling using different methods of analysis and variance estimation, for various probabilities of missing covariate data, in 5,000 simulated cohorts.

| Cohort | | USCC.True | | | USCC.Est | | | USCC.Naive | | | $n$ | $K$ | $p_Y$ | $\pi^{(3)}$ | $\log\{\pi(\tau_1,\tau_2;x)\}$ |
|---|---|---|---|---|---|---|---|---|---|---|---|---|---|---|---|
| Empir var | $\hat{V}_{Robust}$ | Empir var | $\hat{V}_{Robust}$ | $\hat{V}$ | Empir var | $\hat{V}_{Robust}$ | $\hat{V}$ | Empir var | $\hat{V}_{Robust}$ | $\hat{V}$ | | | | | |
| 0.0256 | 0.025 | 0.038 | 0.042 | 0.0375 | 0.0374 | 0.0415 | 0.037 | 0.0374 | 0.042 | 0.0375 | 5000 | 2 | 0.02 | (0.9,0.8) | -3.948 |
| 0.0256 | 0.025 | 0.0352 | 0.0391 | 0.0347 | 0.0351 | 0.039 | 0.0346 | 0.0351 | 0.0391 | 0.0347 | 5000 | 2 | 0.02 | (0.98,0.9) | -3.948 |
| 0.0255 | 0.0248 | 0.0325 | 0.0343 | 0.0322 | 0.0321 | 0.0341 | 0.0319 | 0.0321 | 0.0343 | 0.0322 | 5000 | 4 | 0.02 | (0.9,0.8) | -3.948 |
| 0.0255 | 0.0248 | 0.0304 | 0.0322 | 0.03 | 0.0303 | 0.0321 | 0.03 | 0.0303 | 0.0322 | 0.03 | 5000 | 4 | 0.02 | (0.98,0.9) | -3.948 |
| 0.0123 | 0.0123 | 0.018 | 0.0201 | 0.0179 | 0.0178 | 0.0199 | 0.0176 | 0.0178 | 0.0201 | 0.0179 | 10000 | 2 | 0.02 | (0.9,0.8) | -3.948 |
| 0.0123 | 0.0123 | 0.0167 | 0.0188 | 0.0166 | 0.0167 | 0.0188 | 0.0165 | 0.0167 | 0.0188 | 0.0166 | 10000 | 2 | 0.02 | (0.98,0.9) | -3.948 |
| 0.0127 | 0.0122 | 0.0162 | 0.0166 | 0.0155 | 0.0161 | 0.0165 | 0.0154 | 0.0161 | 0.0166 | 0.0155 | 10000 | 4 | 0.02 | (0.9,0.8) | -3.948 |
| 0.0127 | 0.0122 | 0.0153 | 0.0156 | 0.0146 | 0.0153 | 0.0156 | 0.0145 | 0.0153 | 0.0156 | 0.0146 | 10000 | 4 | 0.02 | (0.98,0.9) | -3.948 |
| 0.0095 | 0.0096 | 0.0132 | 0.0151 | 0.0136 | 0.0131 | 0.0149 | 0.0134 | 0.0131 | 0.0151 | 0.0136 | 5000 | 2 | 0.05 | (0.9,0.8) | -3.046 |
| 0.0095 | 0.0096 | 0.0125 | 0.0142 | 0.0126 | 0.0125 | 0.0141 | 0.0126 | 0.0125 | 0.0142 | 0.0126 | 5000 | 2 | 0.05 | (0.98,0.9) | -3.046 |
| 0.0103 | 0.0097 | 0.0127 | 0.0126 | 0.0119 | 0.0126 | 0.0125 | 0.0118 | 0.0126 | 0.0126 | 0.0119 | 5000 | 4 | 0.05 | (0.9,0.8) | -3.046 |
| 0.0103 | 0.0097 | 0.0119 | 0.0119 | 0.0112 | 0.0119 | 0.0119 | 0.0112 | 0.0119 | 0.0119 | 0.0112 | 5000 | 4 | 0.05 | (0.98,0.9) | -3.046 |
| 0.0047 | 0.0048 | 0.0066 | 0.0075 | 0.0067 | 0.0065 | 0.0074 | 0.0066 | 0.0065 | 0.0075 | 0.0067 | 10000 | 2 | 0.05 | (0.9,0.8) | -3.046 |
| 0.0047 | 0.0048 | 0.0061 | 0.007 | 0.0062 | 0.0061 | 0.007 | 0.0062 | 0.0061 | 0.007 | 0.0062 | 10000 | 2 | 0.05 | (0.98,0.9) | -3.046 |
| 0.0049 | 0.0048 | 0.0058 | 0.0062 | 0.0059 | 0.0058 | 0.0062 | 0.0058 | 0.0058 | 0.0062 | 0.0059 | 10000 | 4 | 0.05 | (0.9,0.8) | -3.046 |
| 0.0049 | 0.0048 | 0.0056 | 0.0059 | 0.0055 | 0.0056 | 0.0059 | 0.0055 | 0.0056 | 0.0059 | 0.0055 | 10000 | 4 | 0.05 | (0.98,0.9) | -3.046 |
| 0.0048 | 0.0047 | 0.0066 | 0.0069 | 0.0063 | 0.0065 | 0.0068 | 0.0062 | 0.0065 | 0.0069 | 0.0063 | 5000 | 2 | 0.1 | (0.9,0.8) | -2.377 |
| 0.0048 | 0.0047 | 0.0061 | 0.0065 | 0.0059 | 0.0061 | 0.0065 | 0.0059 | 0.0061 | 0.0065 | 0.0059 | 5000 | 2 | 0.1 | (0.98,0.9) | -2.377 |
| 0.0048 | 0.0047 | 0.0057 | 0.0058 | 0.0056 | 0.0057 | 0.0057 | 0.0055 | 0.0057 | 0.0058 | 0.0056 | 5000 | 4 | 0.1 | (0.9,0.8) | -2.377 |
| 0.0048 | 0.0047 | 0.0054 | 0.0055 | 0.0053 | 0.0054 | 0.0055 | 0.0052 | 0.0054 | 0.0055 | 0.0053 | 5000 | 4 | 0.1 | (0.98,0.9) | -2.377 |

**WEB TABLE 48-** Empirical variance and mean of estimated variances of pure risk parameter $\log\{\pi(\tau_1,\tau_2;x)\}$ with $x=(-1,1,-0.6)'$, from unstratified sampling using different methods of analysis and variance estimation, for various probabilities of missing covariate data, in 5,000 simulated cohorts.

|  | Cohort |  | SCC.True |  |  | SCC.Est |  |  | SCC.Naive |  |  | $n$ | $K$ | $p_Y$ | $\pi^{(3)}$ | $\log\{\pi(\tau_1,\tau_2;x)\}$ |
|---|---|---|---|---|---|---|---|---|---|---|---|---|---|---|---|---|
|  | Empir var | $\hat{V}_{Robust}$ | Empir var | $\hat{V}_{Robust}$ | $\hat{V}$ | Empir var | $\hat{V}_{Robust}$ | $\hat{V}$ | Empir var | $\hat{V}_{Robust}$ | $\hat{V}$ |  |  |  |  |  |
| | 0.1304 | 0.126 | 0.1913 | 0.216 | 0.1829 | 0.1904 | 0.2154 | 0.1823 | 0.1904 | 0.216 | 0.1829 | 5000 | 2 | 0.02 | (0.9,0.8) | -5.201 |
| | 0.1304 | 0.126 | 0.1644 | 0.1933 | 0.1602 | 0.1642 | 0.1932 | 0.16 | 0.1642 | 0.1933 | 0.1602 | 5000 | 2 | 0.02 | (0.98,0.9) | -5.201 |
| | 0.1311 | 0.126 | 0.1699 | 0.1831 | 0.1672 | 0.1697 | 0.1828 | 0.1669 | 0.1697 | 0.183 | 0.1672 | 5000 | 4 | 0.02 | (0.9,0.8) | -5.201 |
| | 0.1311 | 0.126 | 0.1528 | 0.1642 | 0.1484 | 0.1527 | 0.1642 | 0.1483 | 0.1527 | 0.1642 | 0.1484 | 5000 | 4 | 0.02 | (0.98,0.9) | -5.201 |
| | 0.0623 | 0.062 | 0.0891 | 0.1031 | 0.0868 | 0.0888 | 0.1028 | 0.0865 | 0.0888 | 0.1031 | 0.0868 | 10000 | 2 | 0.02 | (0.9,0.8) | -5.201 |
| | 0.0623 | 0.062 | 0.0779 | 0.0928 | 0.0764 | 0.0778 | 0.0928 | 0.0764 | 0.0778 | 0.0928 | 0.0764 | 10000 | 2 | 0.02 | (0.98,0.9) | -5.201 |
| | 0.0617 | 0.0618 | 0.08 | 0.0886 | 0.0807 | 0.0797 | 0.0884 | 0.0805 | 0.0797 | 0.0886 | 0.0807 | 10000 | 4 | 0.02 | (0.9,0.8) | -5.201 |
| | 0.0617 | 0.0618 | 0.0707 | 0.0797 | 0.0718 | 0.0707 | 0.0797 | 0.0718 | 0.0707 | 0.0797 | 0.0718 | 10000 | 4 | 0.02 | (0.98,0.9) | -5.201 |
| | 0.0504 | 0.0496 | 0.0691 | 0.0796 | 0.0679 | 0.0689 | 0.0794 | 0.0677 | 0.0689 | 0.0796 | 0.0679 | 5000 | 2 | 0.05 | (0.9,0.8) | -4.289 |
| | 0.0504 | 0.0496 | 0.0621 | 0.0719 | 0.0601 | 0.0621 | 0.0718 | 0.0601 | 0.0621 | 0.0719 | 0.0601 | 5000 | 2 | 0.05 | (0.98,0.9) | -4.289 |
| | 0.0519 | 0.0498 | 0.0661 | 0.0691 | 0.0638 | 0.0659 | 0.069 | 0.0637 | 0.0659 | 0.0691 | 0.0638 | 5000 | 4 | 0.05 | (0.9,0.8) | -4.289 |
| | 0.0519 | 0.0498 | 0.0589 | 0.0623 | 0.057 | 0.0588 | 0.0623 | 0.057 | 0.0588 | 0.0623 | 0.057 | 5000 | 4 | 0.05 | (0.98,0.9) | -4.289 |
| | 0.0246 | 0.0247 | 0.0335 | 0.0393 | 0.0335 | 0.0334 | 0.0392 | 0.0333 | 0.0334 | 0.0393 | 0.0335 | 10000 | 2 | 0.05 | (0.9,0.8) | -4.289 |
| | 0.0246 | 0.0247 | 0.0294 | 0.0355 | 0.0297 | 0.0294 | 0.0355 | 0.0297 | 0.0294 | 0.0355 | 0.0297 | 10000 | 2 | 0.05 | (0.98,0.9) | -4.289 |
| | 0.0248 | 0.0247 | 0.0318 | 0.0341 | 0.0315 | 0.0317 | 0.0341 | 0.0315 | 0.0317 | 0.0341 | 0.0315 | 10000 | 4 | 0.05 | (0.9,0.8) | -4.289 |
| | 0.0248 | 0.0247 | 0.0281 | 0.0308 | 0.0282 | 0.028 | 0.0308 | 0.0282 | 0.028 | 0.0308 | 0.0282 | 10000 | 4 | 0.05 | (0.98,0.9) | -4.289 |
| | 0.0246 | 0.0248 | 0.0328 | 0.0376 | 0.0329 | 0.0325 | 0.0375 | 0.0328 | 0.0325 | 0.0376 | 0.0329 | 5000 | 2 | 0.1 | (0.9,0.8) | -3.602 |
| | 0.0246 | 0.0248 | 0.029 | 0.034 | 0.0293 | 0.0289 | 0.034 | 0.0293 | 0.0289 | 0.034 | 0.0293 | 5000 | 2 | 0.1 | (0.98,0.9) | -3.602 |
| | 0.0253 | 0.0248 | 0.0313 | 0.0328 | 0.031 | 0.0312 | 0.0328 | 0.031 | 0.0312 | 0.0328 | 0.031 | 5000 | 4 | 0.1 | (0.9,0.8) | -3.602 |
| | 0.0253 | 0.0248 | 0.0283 | 0.0296 | 0.0278 | 0.0283 | 0.0296 | 0.0278 | 0.0283 | 0.0296 | 0.0278 | 5000 | 4 | 0.1 | (0.98,0.9) | -3.602 |

**WEB TABLE 49-** Empirical variance and mean of estimated variances of pure risk parameter $\log\{\pi(\tau_1,\tau_2;x)\}$ with $x=(1,-1,0.6)'$, from stratified sampling using different methods of analysis and variance estimation, for various probabilities of missing covariate data, in 5,000 simulated cohorts.

| Cohort | | USCC.True | | | USCC.Est | | | USCC.Naive | | | $n$ | $K$ | $p_Y$ | $\pi^{(3)}$ | $\log\{\pi(\tau_1,\tau_2;\boldsymbol{x})\}$ |
|---|---|---|---|---|---|---|---|---|---|---|---|---|---|---|---|
| Empir var | $\hat{V}_{\text{Robust}}$ | Empir var | $\hat{V}_{\text{Robust}}$ | $\hat{V}$ | Empir var | $\hat{V}_{\text{Robust}}$ | $\hat{V}$ | Empir var | $\hat{V}_{\text{Robust}}$ | $\hat{V}$ | | | | | |
| 0.1304 | 0.126 | 0.2136 | 0.2179 | 0.2135 | 0.2133 | 0.2174 | 0.2129 | 0.2133 | 0.2179 | 0.2135 | 5000 | 2 | 0.02 | (0.9,0.8) | -5.201 |
| 0.1304 | 0.126 | 0.1925 | 0.1951 | 0.1906 | 0.1924 | 0.195 | 0.1905 | 0.1924 | 0.1951 | 0.1906 | 5000 | 2 | 0.02 | (0.98,0.9) | -5.201 |
| 0.1311 | 0.126 | 0.1909 | 0.1855 | 0.1833 | 0.1904 | 0.1852 | 0.183 | 0.1904 | 0.1854 | 0.1833 | 5000 | 4 | 0.02 | (0.9,0.8) | -5.201 |
| 0.1311 | 0.126 | 0.1718 | 0.166 | 0.1639 | 0.1716 | 0.166 | 0.1638 | 0.1716 | 0.166 | 0.1638 | 5000 | 4 | 0.02 | (0.98,0.9) | -5.201 |
| 0.0623 | 0.062 | 0.0988 | 0.1027 | 0.1004 | 0.0986 | 0.1025 | 0.1002 | 0.0986 | 0.1027 | 0.1004 | 10000 | 2 | 0.02 | (0.9,0.8) | -5.201 |
| 0.0623 | 0.062 | 0.0893 | 0.0925 | 0.0902 | 0.0893 | 0.0924 | 0.0902 | 0.0893 | 0.0925 | 0.0902 | 10000 | 2 | 0.02 | (0.98,0.9) | -5.201 |
| 0.0617 | 0.0618 | 0.0874 | 0.0886 | 0.0875 | 0.0873 | 0.0884 | 0.0873 | 0.0873 | 0.0886 | 0.0875 | 10000 | 4 | 0.02 | (0.9,0.8) | -5.201 |
| 0.0617 | 0.0618 | 0.0784 | 0.0796 | 0.0784 | 0.0784 | 0.0795 | 0.0784 | 0.0784 | 0.0796 | 0.0784 | 10000 | 4 | 0.02 | (0.98,0.9) | -5.201 |
| 0.0504 | 0.0496 | 0.0773 | 0.0788 | 0.0772 | 0.0769 | 0.0786 | 0.077 | 0.0769 | 0.0788 | 0.0772 | 5000 | 2 | 0.05 | (0.9,0.8) | -4.289 |
| 0.0504 | 0.0496 | 0.0698 | 0.0712 | 0.0695 | 0.0697 | 0.0711 | 0.0695 | 0.0697 | 0.0712 | 0.0695 | 5000 | 2 | 0.05 | (0.98,0.9) | -4.289 |
| 0.0519 | 0.0498 | 0.0703 | 0.0689 | 0.0681 | 0.0702 | 0.0687 | 0.068 | 0.0702 | 0.0688 | 0.0681 | 5000 | 4 | 0.05 | (0.9,0.8) | -4.289 |
| 0.0519 | 0.0498 | 0.0624 | 0.062 | 0.0613 | 0.0623 | 0.062 | 0.0613 | 0.0623 | 0.062 | 0.0613 | 5000 | 4 | 0.05 | (0.98,0.9) | -4.289 |
| 0.0246 | 0.0247 | 0.0379 | 0.039 | 0.0381 | 0.0377 | 0.0389 | 0.038 | 0.0377 | 0.039 | 0.0381 | 10000 | 2 | 0.05 | (0.9,0.8) | -4.289 |
| 0.0246 | 0.0247 | 0.0341 | 0.0352 | 0.0344 | 0.034 | 0.0352 | 0.0344 | 0.034 | 0.0352 | 0.0344 | 10000 | 2 | 0.05 | (0.98,0.9) | -4.289 |
| 0.0248 | 0.0247 | 0.0333 | 0.034 | 0.0336 | 0.0333 | 0.0339 | 0.0335 | 0.0333 | 0.034 | 0.0336 | 10000 | 4 | 0.05 | (0.9,0.8) | -4.289 |
| 0.0248 | 0.0247 | 0.0302 | 0.0307 | 0.0303 | 0.0302 | 0.0306 | 0.0303 | 0.0302 | 0.0307 | 0.0303 | 10000 | 4 | 0.05 | (0.98,0.9) | -4.289 |
| 0.0246 | 0.0248 | 0.0365 | 0.0371 | 0.0365 | 0.0364 | 0.037 | 0.0363 | 0.0364 | 0.0371 | 0.0365 | 5000 | 2 | 0.1 | (0.9,0.8) | -3.602 |
| 0.0246 | 0.0248 | 0.0324 | 0.0336 | 0.0329 | 0.0324 | 0.0335 | 0.0329 | 0.0324 | 0.0336 | 0.0329 | 5000 | 2 | 0.1 | (0.98,0.9) | -3.602 |
| 0.0253 | 0.0248 | 0.0326 | 0.0326 | 0.0323 | 0.0325 | 0.0325 | 0.0323 | 0.0325 | 0.0326 | 0.0323 | 5000 | 4 | 0.1 | (0.9,0.8) | -3.602 |
| 0.0253 | 0.0248 | 0.0296 | 0.0294 | 0.0292 | 0.0296 | 0.0294 | 0.0292 | 0.0296 | 0.0294 | 0.0292 | 5000 | 4 | 0.1 | (0.98,0.9) | -3.602 |

**WEB TABLE 50-** Empirical variance and mean of estimated variances of pure risk parameter $\log\{\pi(\tau_1,\tau_2;\boldsymbol{x})\}$ with $\boldsymbol{x}=(1,-1,0.6)'$, from unstratified sampling using different methods of analysis and variance estimation, for various probabilities of missing covariate data, in 5,000 simulated cohorts.

| Cohort | | SCC.True | | | SCC.Est | | | SCC.Naive | | | $n$ | $K$ | $p_Y$ | $\pi^{(3)}$ | $\log\{\pi(\tau_1,\tau_2;\boldsymbol{x})\}$ |
|---|---|---|---|---|---|---|---|---|---|---|---|---|---|---|---|
| Empir var | $\hat{V}_{Robust}$ | Empir var | $\hat{V}_{Robust}$ | $\hat{V}$ | Empir var | $\hat{V}_{Robust}$ | $\hat{V}$ | Empir var | $\hat{V}_{Robust}$ | $\hat{V}$ | | | | | |
| 0.0569 | 0.0559 | 0.0876 | 0.0938 | 0.0847 | 0.0867 | 0.0932 | 0.0841 | 0.0867 | 0.0938 | 0.0847 | 5000 | 2 | 0.02 | (0.9,0.8) | -4.702 |
| 0.0569 | 0.0559 | 0.0771 | 0.0849 | 0.0758 | 0.0768 | 0.0848 | 0.0757 | 0.0768 | 0.0849 | 0.0758 | 5000 | 2 | 0.02 | (0.98,0.9) | -4.702 |
| 0.0577 | 0.0559 | 0.0762 | 0.0793 | 0.075 | 0.0763 | 0.0791 | 0.0747 | 0.0763 | 0.0793 | 0.075 | 5000 | 4 | 0.02 | (0.9,0.8) | -4.702 |
| 0.0577 | 0.0559 | 0.07 | 0.0721 | 0.0678 | 0.0699 | 0.0721 | 0.0677 | 0.0699 | 0.0721 | 0.0678 | 5000 | 4 | 0.02 | (0.98,0.9) | -4.702 |
| 0.0274 | 0.0277 | 0.0401 | 0.0444 | 0.0398 | 0.0399 | 0.0441 | 0.0396 | 0.0399 | 0.0444 | 0.0398 | 10000 | 2 | 0.02 | (0.9,0.8) | -4.702 |
| 0.0274 | 0.0277 | 0.0357 | 0.0404 | 0.0358 | 0.0355 | 0.0403 | 0.0358 | 0.0355 | 0.0404 | 0.0358 | 10000 | 2 | 0.02 | (0.98,0.9) | -4.702 |
| 0.0267 | 0.0276 | 0.0354 | 0.0382 | 0.036 | 0.0353 | 0.0381 | 0.0359 | 0.0353 | 0.0382 | 0.036 | 10000 | 4 | 0.02 | (0.9,0.8) | -4.702 |
| 0.0267 | 0.0276 | 0.0319 | 0.0349 | 0.0327 | 0.0319 | 0.0348 | 0.0326 | 0.0319 | 0.0349 | 0.0327 | 10000 | 4 | 0.02 | (0.98,0.9) | -4.702 |
| 0.0228 | 0.0221 | 0.0317 | 0.034 | 0.0308 | 0.0314 | 0.0338 | 0.0306 | 0.0314 | 0.034 | 0.0308 | 5000 | 2 | 0.05 | (0.9,0.8) | -3.793 |
| 0.0228 | 0.0221 | 0.0291 | 0.0311 | 0.0278 | 0.0291 | 0.031 | 0.0278 | 0.0291 | 0.0311 | 0.0278 | 5000 | 2 | 0.05 | (0.98,0.9) | -3.793 |
| 0.0221 | 0.0222 | 0.0285 | 0.0296 | 0.0282 | 0.0283 | 0.0295 | 0.0281 | 0.0283 | 0.0296 | 0.0282 | 5000 | 4 | 0.05 | (0.9,0.8) | -3.793 |
| 0.0221 | 0.0222 | 0.0257 | 0.0271 | 0.0256 | 0.0256 | 0.0271 | 0.0256 | 0.0256 | 0.0271 | 0.0256 | 5000 | 4 | 0.05 | (0.98,0.9) | -3.793 |
| 0.011 | 0.011 | 0.0147 | 0.0167 | 0.0151 | 0.0147 | 0.0166 | 0.015 | 0.0147 | 0.0167 | 0.0151 | 10000 | 2 | 0.05 | (0.9,0.8) | -3.793 |
| 0.011 | 0.011 | 0.0133 | 0.0153 | 0.0137 | 0.0133 | 0.0153 | 0.0136 | 0.0133 | 0.0153 | 0.0137 | 10000 | 2 | 0.05 | (0.98,0.9) | -3.793 |
| 0.0108 | 0.011 | 0.0136 | 0.0146 | 0.0139 | 0.0136 | 0.0146 | 0.0138 | 0.0136 | 0.0146 | 0.0139 | 10000 | 4 | 0.05 | (0.9,0.8) | -3.793 |
| 0.0108 | 0.011 | 0.0123 | 0.0134 | 0.0126 | 0.0123 | 0.0134 | 0.0126 | 0.0123 | 0.0134 | 0.0126 | 10000 | 4 | 0.05 | (0.98,0.9) | -3.793 |
| 0.0108 | 0.011 | 0.0142 | 0.0158 | 0.0145 | 0.0141 | 0.0157 | 0.0144 | 0.0141 | 0.0158 | 0.0145 | 5000 | 2 | 0.1 | (0.9,0.8) | -3.111 |
| 0.0108 | 0.011 | 0.013 | 0.0145 | 0.0132 | 0.013 | 0.0145 | 0.0132 | 0.013 | 0.0145 | 0.0132 | 5000 | 2 | 0.1 | (0.98,0.9) | -3.111 |
| 0.0108 | 0.011 | 0.0133 | 0.0139 | 0.0134 | 0.0133 | 0.0139 | 0.0134 | 0.0133 | 0.0139 | 0.0134 | 5000 | 4 | 0.1 | (0.9,0.8) | -3.111 |
| 0.0108 | 0.011 | 0.0122 | 0.0127 | 0.0122 | 0.0121 | 0.0127 | 0.0122 | 0.0121 | 0.0127 | 0.0122 | 5000 | 4 | 0.1 | (0.98,0.9) | -3.111 |

**WEB TABLE 51-** Empirical variance and mean of estimated variances of pure risk parameter $\log\{\pi(\tau_1,\tau_2;\boldsymbol{x})\}$ with $\boldsymbol{x}=(1,1,0.6)'$, from stratified sampling using different methods of analysis and variance estimation, for various probabilities of missing covariate data, in 5,000 simulated cohorts.

|  | Cohort | | USCC.True | | | USCC.Est | | | USCC.Naive | | | $n$ | $K$ | $p_Y$ | $\pi^{(3)}$ | $\log\{\pi(\tau_1,\tau_2;x)\}$ |
|---|---|---|---|---|---|---|---|---|---|---|---|---|---|---|---|---|
| | Empir var | $\hat{V}_{Robust}$ | Empir var | $\hat{V}_{Robust}$ | $\hat{V}$ | Empir var | $\hat{V}_{Robust}$ | $\hat{V}$ | Empir var | $\hat{V}_{Robust}$ | $\hat{V}$ | | | | | |
| | 0.0569 | 0.0559 | 0.0922 | 0.0962 | 0.0917 | 0.0916 | 0.0957 | 0.0911 | 0.0916 | 0.0962 | 0.0917 | 5000 | 2 | 0.02 | (0.9,0.8) | -4.702 |
| | 0.0569 | 0.0559 | 0.083 | 0.087 | 0.0825 | 0.0828 | 0.0869 | 0.0824 | 0.0828 | 0.087 | 0.0825 | 5000 | 2 | 0.02 | (0.98,0.9) | -4.702 |
| | 0.0577 | 0.0559 | 0.0822 | 0.0808 | 0.0786 | 0.082 | 0.0805 | 0.0783 | 0.082 | 0.0808 | 0.0786 | 5000 | 4 | 0.02 | (0.9,0.8) | -4.702 |
| | 0.0577 | 0.0559 | 0.074 | 0.0732 | 0.071 | 0.0738 | 0.0731 | 0.0709 | 0.0738 | 0.0732 | 0.071 | 5000 | 4 | 0.02 | (0.98,0.9) | -4.702 |
| | 0.0274 | 0.0277 | 0.0418 | 0.0449 | 0.0426 | 0.0415 | 0.0446 | 0.0423 | 0.0415 | 0.0449 | 0.0426 | 10000 | 2 | 0.02 | (0.9,0.8) | -4.702 |
| | 0.0274 | 0.0277 | 0.038 | 0.041 | 0.0387 | 0.038 | 0.0409 | 0.0386 | 0.038 | 0.041 | 0.0387 | 10000 | 2 | 0.02 | (0.98,0.9) | -4.702 |
| | 0.0267 | 0.0276 | 0.0371 | 0.0385 | 0.0374 | 0.037 | 0.0384 | 0.0373 | 0.037 | 0.0385 | 0.0374 | 10000 | 4 | 0.02 | (0.9,0.8) | -4.702 |
| | 0.0267 | 0.0276 | 0.0334 | 0.0351 | 0.034 | 0.0333 | 0.0351 | 0.034 | 0.0333 | 0.0351 | 0.034 | 10000 | 4 | 0.02 | (0.98,0.9) | -4.702 |
| | 0.0228 | 0.0221 | 0.0332 | 0.0343 | 0.0327 | 0.033 | 0.0341 | 0.0325 | 0.033 | 0.0343 | 0.0327 | 5000 | 2 | 0.05 | (0.9,0.8) | -3.793 |
| | 0.0228 | 0.0221 | 0.0301 | 0.0313 | 0.0297 | 0.0301 | 0.0313 | 0.0297 | 0.0301 | 0.0313 | 0.0297 | 5000 | 2 | 0.05 | (0.98,0.9) | -3.793 |
| | 0.0221 | 0.0222 | 0.0285 | 0.0299 | 0.0292 | 0.0284 | 0.0298 | 0.0291 | 0.0284 | 0.0299 | 0.0292 | 5000 | 4 | 0.05 | (0.9,0.8) | -3.793 |
| | 0.0221 | 0.0222 | 0.0261 | 0.0273 | 0.0266 | 0.0261 | 0.0272 | 0.0265 | 0.0261 | 0.0273 | 0.0266 | 5000 | 4 | 0.05 | (0.98,0.9) | -3.793 |
| | 0.011 | 0.011 | 0.016 | 0.0169 | 0.016 | 0.0159 | 0.0168 | 0.0159 | 0.0159 | 0.0169 | 0.016 | 10000 | 2 | 0.05 | (0.9,0.8) | -3.793 |
| | 0.011 | 0.011 | 0.0145 | 0.0154 | 0.0146 | 0.0145 | 0.0154 | 0.0146 | 0.0145 | 0.0154 | 0.0146 | 10000 | 2 | 0.05 | (0.98,0.9) | -3.793 |
| | 0.0108 | 0.011 | 0.0139 | 0.0147 | 0.0144 | 0.0138 | 0.0147 | 0.0143 | 0.0138 | 0.0147 | 0.0144 | 10000 | 4 | 0.05 | (0.9,0.8) | -3.793 |
| | 0.0108 | 0.011 | 0.0127 | 0.0135 | 0.0131 | 0.0127 | 0.0134 | 0.0131 | 0.0127 | 0.0135 | 0.0131 | 10000 | 4 | 0.05 | (0.98,0.9) | -3.793 |
| | 0.0108 | 0.011 | 0.0148 | 0.016 | 0.0153 | 0.0147 | 0.0159 | 0.0152 | 0.0147 | 0.016 | 0.0153 | 5000 | 2 | 0.1 | (0.9,0.8) | -3.111 |
| | 0.0108 | 0.011 | 0.0135 | 0.0146 | 0.014 | 0.0135 | 0.0146 | 0.0139 | 0.0135 | 0.0146 | 0.014 | 5000 | 2 | 0.1 | (0.98,0.9) | -3.111 |
| | 0.0108 | 0.011 | 0.0137 | 0.014 | 0.0137 | 0.0137 | 0.0139 | 0.0137 | 0.0137 | 0.014 | 0.0137 | 5000 | 4 | 0.1 | (0.9,0.8) | -3.111 |
| | 0.0108 | 0.011 | 0.0123 | 0.0128 | 0.0126 | 0.0123 | 0.0128 | 0.0125 | 0.0123 | 0.0128 | 0.0126 | 5000 | 4 | 0.1 | (0.98,0.9) | -3.111 |

**WEB TABLE 52-** Empirical variance and mean of estimated variances of pure risk parameter $\log\{\pi(\tau_1,\tau_2;x)\}$ with $x = (1, 1, 0.6)'$, from unstratified sampling using different methods of analysis and variance estimation, for various probabilities of missing covariate data, in 5,000 simulated cohorts.

# Web Appendix J.   STEP BY STEP PSEUDO CODE TO OBTAIN $\hat{V}$ FOR SCC AND SCC.Calib IN TABLE 3 IN THE MAIN DOCUMENT

```
Load the case cohort data set
Load the cohort data set

-----------------------------------------------------------------------------
A. Estimation using the stratified case cohort with design weights ----------

1. Run the coxph model to estimate the log-relative hazard from the case-cohort
data with design weights

2. Estimate the baseline hazard non parametrically from the cohort data with
design weights and log-relative hazard estimate obtained in step A.2

3. Estimate the pure risk from the relative hazard and baseline hazard estimates
obtained in steps A.1 and A.2

4. Compute the influences of the case-cohort individuals on the parameter
estimates

5. Estimate the variances of the parameter estimates from the influences
obtained in step A.4

-----------------------------------------------------------------------------
B. Estimation using the stratified case cohort with calibrated weights -------

1. Build a regression model to predict the phase-two covariates

2. Impute the phase-two covariates on the whole cohort using the model built
in step B.1
```

3. Run the coxph model to estimate the log-relative hazard from the imputed cohort data obtained in step B.2

4. Compute the influences of the cohort individuals on the log-relative hazard estimate obtained in step B.3 to construct the auxiliary variables proposed by Breslow et al. (2009)

5. Calibrate the design weights against the influences obtained in step B.4

6. Run the coxph model to estimate the log-relative hazard from the case-cohort data with the calibrated weights obtained in step 5

7. Compute the total follow-up time on the pure-risk time interval on the whole cohort

8. Compute the relative hazard on the imputed cohort data, using the log-relative hazard estimate obtained in step 6.

9. Multiply the estimated relative hazard obtained in step B.8 to the total follow-up time obtained in step 7, to get construct an additional auxiliary variable, as proposed by Shin et al. (2020)

10. Calibrate the design weights against the influences obtained in steps B.4 and B.9

11. Run the coxph model to estimate the log-relative hazard from the case-cohort data with calibrated weights obtained in step B.10

12. Estimate the baseline hazard non parametrically from the cohort data with calibrated weights obtained in step B.10 and log-relative hazard estimate obtained in step B.11

12. Estimate the pure risk from the relative hazard and baseline hazard estimates obtained in steps B.11 and B.12

12. Compute the influences of the case-cohort individuals on the parameter estimates, taking calibration into account

13. Estimate the variances of the parameter estimates from the influences obtained in step B.12

## DATA AVAILABILITY STATEMENT

R code and functions used for the simulations in Web Appendices E and I are available in the Supporting Information of this article and on GitHub at https://github.com/Etievant/CaseCohort. We are not authorized to release the clinical data used in Web Appendix F.